\documentclass{aa}  
\usepackage{graphicx}
\usepackage{graphics}
\usepackage[varg]{txfonts}
\usepackage[colorlinks=true,allcolors=blue]{hyperref} 
\usepackage{enumerate}
\usepackage{amsmath}
\usepackage{comment}
\usepackage[para]{threeparttable}
\begin{document}

   \title{The physical properties of local (U)LIRGs: a comparison with nearby early- and late-type galaxies}

   \subtitle{}

   \author{
            E.-D. Paspaliaris\inst{1,2}\fnmsep\thanks{\email{epaspal@noa.gr}}\and
            E.M. Xilouris\inst{1}\and
            A. Nersesian\inst{1,3}\and
            V.A. Masoura\inst{1,2}\and
            M. Plionis\inst{1,2}\and
            I. Georgantopoulos\inst{1} \and \\
            S. Bianchi\inst{4} \and
            S. Katsioli\inst{1,5} \and
            G. Mountrichas\inst{6}
            }

   \institute{National Observatory of Athens, Institute for Astronomy, Astrophysics, Space Applications and Remote Sensing, Ioannou Metaxa and Vasileos Pavlou GR-15236, Athens, Greece \and
   Department of Astrophysics, Astronomy \& Mechanics, School of Physics, Aristotle University of Thessaloniki \and
   Sterrenkundig Observatorium, Universiteit Gent, Krijgslaan 281 S9, 9000 Gent, Belgium \and
   INAF - Osservatorio Astrofisico di Arcetri, Largo E. Fermi 5, 50125, Florence, Italy \and
   Department of Astrophysics, Astronomy \& Mechanics, Faculty of Physics, University of Athens, Panepistimiopolis, GR15784 Zografos, Athens, Greece \and
   Instituto de Fisica de Cantabria (CSIC-Universidad de Cantabria), Avenida de los Castros, 39005 Santander, Spain
   }

   \date{Received / Accepted}

% \abstract{}{}{}{}{} 
% 5 {} token are mandatory
 
  \abstract
   {}
   {In order to pinpoint the place of the (ultra-) luminous infrared galaxies [(U)LIRGs] in the local universe we examine the properties of a sample of 67 such nearby systems and compare them with those of 268 early- and 542 late-type, well studied, galaxies from the DustPedia database.}
   {We make use of multi-wavelength photometric data (from the ultra-violet to the sub-millimetre), culled from the literature, and the CIGALE Spectral Energy Distribution fitting code, to extract the physical parameters of each system.
   The median Spectral Energy Distributions as well as the values of the derived parameters were compared to those of the local early- and late-type galaxies. In addition to that, (U)LIRGs were divided into seven classes, according to the merging stage of each system, and variations in the derived parameters were investigated.}
   {(U)LIRGs occupy the `high-end' on the dust mass, stellar mass, and star-formation rate plane in the local Universe with median values of 5.2$\times10^7~\text{M}_{\odot}$, 6.3$\times10^{10}~\text{M}_{\odot}$ and 52$~\text{M}_{\odot}$yr$^{-1}$, respectively.
   The median value of the dust temperature in (U)LIRGs is 32 K, which is higher compared to both the early-type (28 K) and the late-type (22 K) galaxies.
   The dust emission in PDR regions in (U)LIRGs is 11.7\% of the total dust luminosity, significantly higher than early-type (1.6\%) and the late-type (5.2\%) galaxies. Small differences, in the derived parameters, are seen for the seven merging classes of our sample of (U)LIRGs with the most evident one being on the star-formation rate, where in systems in late merging stages (`M3' and `M4') the median \textit{SFR} reaches up to 99 M$_{\odot}$ yr$^{-1}$ compared to 26 M$_{\odot}$ yr$^{-1}$ for the isolated ones. In contrast to the local early- and late-type galaxies where the old stars are the dominant source of the stellar emission, the young stars in (U)LIRGs contribute with 64\% of their luminosity to the total stellar luminosity. 
   The fraction of the stellar luminosity absorbed by the dust is extremely high in (U)LIRGs (78\%) compared to 7\% and 25\% in early- and late-type galaxies, respectively. The fraction of the stellar luminosity used to heat up the dust grains is very high in (U)LIRGs, for both stellar components (92\% and 56\%, for the young and the old stellar populations, respectively) while 74\% of the dust emission comes from the young stars.}
   {}


   \maketitle
%
%-------------------------------------------------------------------

%%%%%%%%%%%%%%%%%%%%%%%%%%%%%%%%%%%% INTRODUCTION
\section{Introduction}

   A multi-wavelength approach is needed for a comprehensive study of galaxies. Each region of the electromagnetic spectrum provides unique information about the different building blocks of galaxies (stars, dust, and gas) and their physical properties. These building blocks constituting the baryonic matter in galaxies are not in isolation, but, instead, are constantly interacting with each other, modifying the stellar content and the inter-stellar medium (ISM). This evolutionary process is imprinted on the galaxy's spectral energy distribution (SED). Studying the SEDs of galaxies is, thus, a key procedure towards understanding galaxy formation and evolution.
   
   One property, often related to the degree of the current star-formation is the infrared (IR) luminosity \mbox{($L_\text{IR}$ = $L_{8-1000~\mu\text{m}}$)}, which, in most of the cases, is dominated by the emission of dust grains heated by the interstellar radiation field (ISRF). Early-type galaxies (elliptical and lenticular galaxies) exhibit very low to moderate IR-luminosities (\mbox{$L_\text{IR}$ < 10$^{9}$~L$_{\odot}$}), while the star-forming spiral galaxies are brighter in the infrared \citep[\mbox{10$^{9}$ < $L_\text{IR}$/L$_{\odot}$ < 10$^{11}$};][]{2006asup.book..285L}. Imaging studies of (Ultra) Luminous Infrared Galaxies [(U)LIRGs; \mbox{$L_\text{IR}$ > 10$^{11}$~L$_{\odot}$}] in the local Universe, indicate a more violent kind of formation of these systems with the majority of them exhibiting signs of galaxy interactions \citep{1996ARA&A..34..749S, 1997A&A...326..537D}. \citet{1987AJ.....94..831A} concluded that more than $\sim70\%$ of such systems show signs of interactions originating from merging events between their parent galaxies. Later studies of IR-luminous galaxies \citep[][]{1990A&A...231L..19M, 1991AJ....101..434H, 1996MNRAS.279..477C} indicate even larger fractions ($\sim90\%$) of the sources in their samples being mergers, while, other studies concluded in fractions of such sources lower than 70$\%$ \citep{1989MNRAS.240..329L, 1991MNRAS.252..593Z, 1994MNRAS.267..253L}. In all cases, however, it becomes clear that galaxy merging is a key process that is responsible for the high IR-luminosity and drives the high level of star-formation activity observed in these systems, making them a unique class of objects in the local Universe.
   
   %%%%%%%%%%%%%%%%%%%%%%%%%%%%%%%%%%%%% FIGURE 1
   \begin{figure}[t!]
   \centering
   \includegraphics[width=0.5\textwidth]{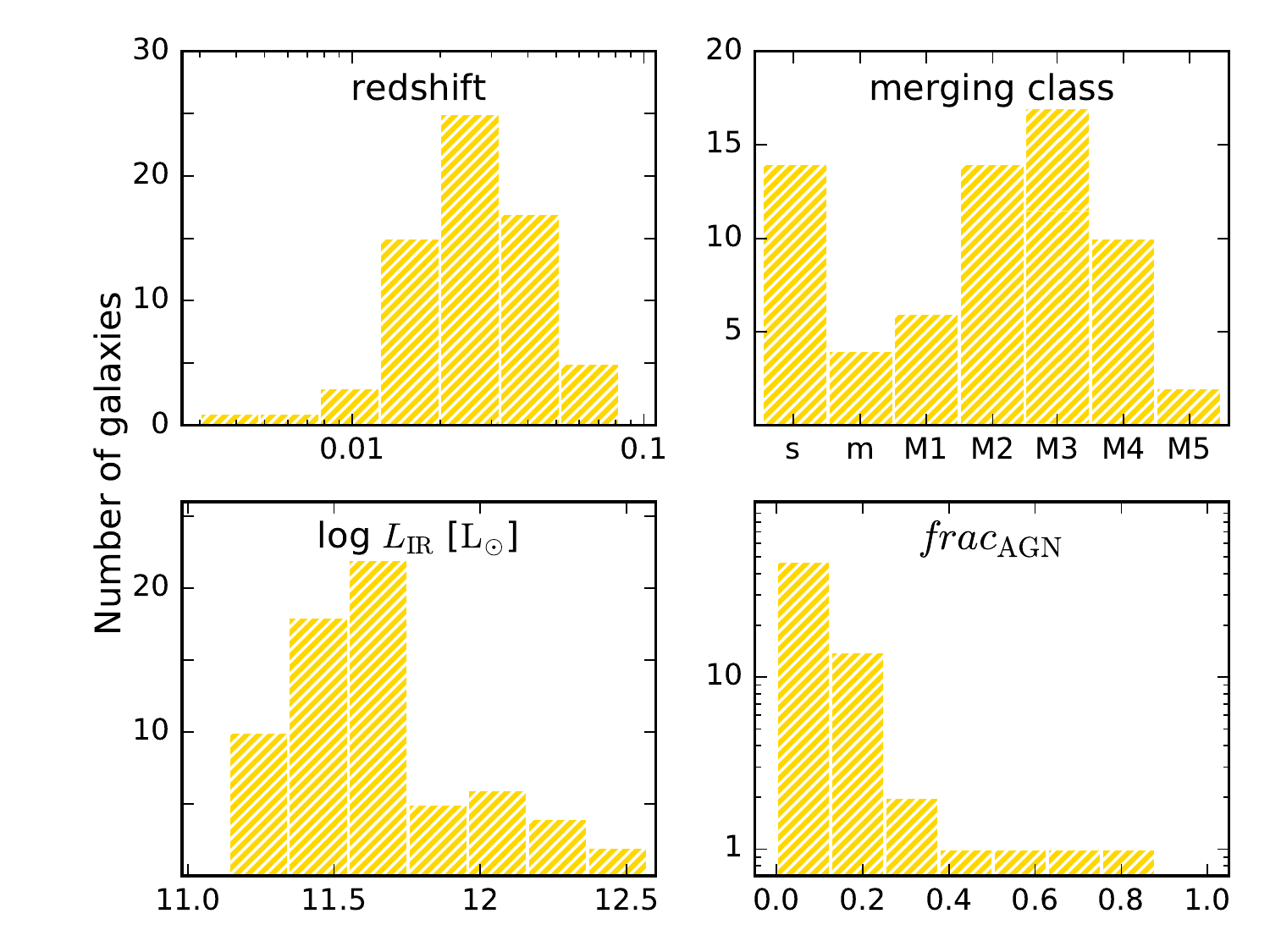}
   \caption{Distributions of the basic properties of our galaxy sample. The redshift distribution is            plotted in the top-left panel and the number of sources per merging class in the top-right panel. In the bottom-left panel the infrared luminosity distribution, based on IRAS observations \citep{2003AJ....126.1607S} is shown, while, the bolometric AGN fractions calculated in \citet{2017ApJ...846...32D}, are plotted in the bottom-right panel.
   \label{distributions_literature}}
   \end{figure}
%%%%%%%%%%%%%%%%%%%%%%%%%%%%%%%%%%%%%%%% TABLE 1
   \begin{table*}[h!]
   \caption{Properties of the (U)LIRGs in our sample.}
   \label{tab:properties1}    
   \tiny
   \begin{center}
   \vspace{-0.5cm}\begin{tabular}{l|l|c|c|c|c|c|c}
\hline 
   \multicolumn{1}{l|}{IRAS name} &
   \multicolumn{1}{l|}{alternative name} &
   \multicolumn{1}{c|}{redshift} &
   \multicolumn{1}{c|}{merging class} &
   \multicolumn{1}{c|}{log ($L_\text{IR}$) [L$_{\odot}$]~$^{(a)}$} &
   \multicolumn{1}{c|}{$frac_\text{AGN}$~$^{(b)}$} &
   \multicolumn{1}{c|}{bands; $\lambda$$<$3 {$\mu$}m} &
   \multicolumn{1}{c}{bands; $\lambda$$>$3 {$\mu$}m} \\
\hline
\hline
   F00085-1223 & NGC 0034 & 0.0196 & M4 & 11.49 & 0.04 $\pm$ 0.02 & 9 & 17\\
   F00163-1039 & MCG -02-01-051/2 & 0.0272 & M2 & 11.48 & 0.07 $\pm$ 0.05 & 8 & 17\\
   F00402-2349 & NGC 0232 & 0.0222 & M2 & 11.44 & 0.09 $\pm$ 0.03 & 6 & 17\\
   F01053-1746 & IC 1623A/B & 0.0201 & M3 & 11.71 & 0.12 $\pm$ 0.04 & 9 & 15\\
   F01076-1707 & MCG -03-04-014 & 0.0335 & s & 11.65 & 0.07 $\pm$ 0.04 & 9 & 17\\
   F01173+1405 & CGCG 436-030 & 0.0312 & M2 & 11.69 & 0.09 $\pm$ 0.02 & 9 & 18\\
   F01364-1042 & IRAS F01364-1042 & 0.0483 & M3 & 11.85 & 0.05 $\pm$ 0.03 & 9 & 11\\
   F01417+1651 & III Zw 035 & 0.0279 & M3 & 11.64 & 0.00 $\pm$ 0.00 & 9 & 17\\
   F01484+2220 & NGC 0695 & 0.0325 & m & 11.68 & 0.09 $\pm$ 0.05 & 9 & 18\\
   F02281-0309 & NGC 0958 & 0.0191 & s & 11.20 & 0.20 $\pm$ 0.07 & 9 & 18\\
   F02435+1253 & UGC 02238 & 0.0219 & M4 & 11.33 & 0.10 $\pm$ 0.06 & 8 & 18\\
   F02512+1446 & UGC 02369 & 0.0319 & M2 & 11.67 & 0.09 $\pm$ 0.05 & 7 & 18\\
   F03359+1523 & IRAS F03359+1523 & 0.0354 & M3 & 11.55 & ... & 7 & 17\\
   F04097+0525 & UGC 02982 & 0.0177 & s & 11.20 & 0.11 $\pm$ 0.08 & 9 & 18\\
   F04191-1855 & ESO 550-IG025 & 0.0321 & M2 & 11.51 & ... & 9 & 16\\
   F04315-0840 & NGC 1614 & 0.0159 & m & 11.65 & 0.12 $\pm$ 0.08 & 8 & 19\\
   F05189-2524 & IRAS F05189-2524 & 0.0426 & M5 & 12.16 & 0.60 $\pm$ 0.07 & 10 & 15\\
   F06107+7822 & NGC 2146 & 0.0030 & M5 & 11.07 & 0.01 $\pm$ 0.00 & 5 & 11\\
   F08354+2555 & NGC 2623 & 0.0185 & M4 & 11.60 & 0.10 $\pm$ 0.03 & 10 & 18\\
   F08572+3915 & IRAS F08572+3915 & 0.0584 & M3 & 12.16 & 0.47 $\pm$ 0.09 & 10 & 18\\
   F09126+4432 & UGC 04881 & 0.0395 & M3 & 11.74 & 0.05 $\pm$ 0.03 & 8 & 17\\
   F09320+6134 & UGC 05101 & 0.0394 & M4 & 12.01 & 0.25 $\pm$ 0.05 & 10 & 16\\
   F09333+4841 & MCG +08-18-012/3 & 0.0259 & M1 & 11.34 & 0.20 $\pm$ 0.00 & 8 & 18\\
   F09437+0317 & IC 0563/4 & 0.0200 & M1 & 11.23 & 0.09 $\pm$ 0.06 & 9 & 18\\
   F10015-0614 & NGC 3110 & 0.0169 & M1 & 11.37 & 0.10 $\pm$ 0.06 & 8 & 18\\
   F10173+0828 & IRAS F10173+0828 & 0.0491 & M4 & 11.86 & 0.04 $\pm$ 0.02 & 9 & 13\\
   F10257-4339 & NGC 3256 & 0.0094 & M3 & 11.56 & ... & 4 & 10\\
   F10565+2448 & IRAS F10565+2448 & 0.0431 & M2 & 12.08 & 0.04 $\pm$ 0.01 & 8 & 18\\
   F11011+4107 & MCG +07-23-019 & 0.0345 & M2 & 11.62 & 0.03 $\pm$ 0.02 & 9 & 16\\
   F11186-0242 & CGCG 011-076 & 0.0249 & m & 11.43 & 0.18 $\pm$ 0.09 & 9 & 19\\
   F11231+1456 & IC 2810A/B & 0.0340 & M1 & 11.64 & 0.05 $\pm$ 0.01 & 9 & 18\\
   F11257+5850 & NGC 3690/IC 694 & 0.0103 & M3 & 11.93 & 0.04 $\pm$ 0.02 & 9 & 14\\
   F12112+0305 & IRAS F12112+0305 & 0.0733 & M3 & 12.36 & 0.06 $\pm$ 0.03 & 7 & 13\\
   F12224-0624 & IRAS F12224-0624 & 0.0264 & s & 11.36 & 0.20 $\pm$ 0.10 & 7 & 13\\
   F12540+5708 & UGC 08058 & 0.0422 & M4 & 12.57 & 0.275$^{\star}$ & 8 & 12\\
   F12590+2934 & NGC 4922 & 0.0236 & M2 & 11.38 & 0.17 $\pm$ 0.05 & 8 & 18\\
   F13001-2339 & ESO 507-G070 & 0.0217 & M3 & 11.56 & 0.03 $\pm$ 0.01 & 9 & 17\\
   F13126+2453 & IC 0860 & 0.0112 & s & 11.14 & 0.06 $\pm$ 0.03 & 7 & 16\\
   F13136+6223 & VV 250 & 0.0311 & M2 & 11.81 & 0.05 $\pm$ 0.03 & 7 & 14\\
   F13182+3424 & UGC 08387 & 0.0233 & M4 & 11.73 & 0.03 $\pm$ 0.01 & 9 & 18\\
   F13188+0036 & NGC 5104 & 0.0186 & s & 11.27 & 0.10 $\pm$ 0.04 & 8 & 18\\
   F13197-1627 & MCG -03-34-064 & 0.0165 & s & 11.28 & 0.88 $\pm$ 0.04 & 8 & 15\\
   F13229-2934 & NGC 5135 & 0.0137 & s & 11.30 & 0.24 $\pm$ 0.06 & 9 & 14\\
   F13362+4831 & NGC 5256 & 0.0278 & M3 & 11.56 & 0.23 $\pm$ 0.07 & 9 & 18\\
   F13373+0105 & NGC 5257/8 & 0.0226 & M2 & 11.62 & 0.11 $\pm$ 0.05 & 9 & 17\\
   F13428+5608 & UGC 08696 & 0.0378 & M4 & 12.21 & 0.31 $\pm$ 0.08 & 10 & 16\\
   F14179+4927 & CGCG 247-020 & 0.0257 & s & 11.39 & 0.06 $\pm$ 0.03 & 8 & 19\\
   F14348-1447 & IRAS F14348-1447 & 0.0827 & M3 & 12.39 & 0.09 $\pm$ 0.05 & 10 & 14\\
   F14547+2449 & VV 340 & 0.0337 & M1 & 11.74 & 0.16 $\pm$ 0.05 & 9 & 18\\
   F15107+0724 & CGCG 049-057 & 0.0130 & s & 11.35 & 0.04 $\pm$ 0.02 & 8 & 12\\
   F15163+4255 & VV 705 & 0.0398 & M3 & 11.92 & ... & 9 & 19\\
   F15250+3608 & IRAS F15250+3608 & 0.0552 & M4 & 12.08 & 0.18 $\pm$ 0.16 & 8 & 18\\
   F15327+2340 & UGC 09913 & 0.0181 & M4 & 12.28 & 0.09 $\pm$ 0.05 & 10 & 14\\
   F16104+5235 & NGC 6090 & 0.0298 & M3 & 11.58 & 0.05 $\pm$ 0.03 & 9 & 17\\
   F16284+0411 & CGCG 052-037 & 0.0245 & s & 11.45 & 0.07 $\pm$ 0.04 & 9 & 18\\
   F16577+5900 & NGC 6285/6 & 0.0184 & M2 & 11.37 & 0.08 $\pm$ 0.04 & 8 & 17\\
   F17132+5313 & IRAS F17132+5313 & 0.0509 & M3 & 11.96 & 0.02 $\pm$ 0.0 & 9 & 17\\
   F22287-1917 & ESO 602-G025 & 0.0250 & s & 11.34 & 0.14 $\pm$ 0.04 & 9 & 17\\
   F22491-1808 & IRAS F22491-1808 & 0.0778 & M3 & 12.20 & 0.02 $\pm$ 0.02 & 10 & 15\\
   F23007+0836 & NGC 7469/IC 5283 & 0.0163 & M2 & 11.65 & 0.24 $\pm$ 0.06 & 10 & 17\\
   F23024+1916 & CGCG 453-062 & 0.0251 & s & 11.38 & 0.08 $\pm$ 0.03 & 8 & 18\\
   F23133-4251 & NGC 7552 & 0.0054 & s & 11.03 & 0.08 $\pm$ 0.05 & 5 & 7\\
   F23135+2517 & IC 5298 & 0.0274 & m & 11.60 & 0.33 $\pm$ 0.05 & 9 & 18\\
   F23157-0441 & NGC 7592 & 0.0244 & M3 & 11.40 & 0.20 $\pm$ 0.06 & 9 & 18\\
   F23254+0830 & NGC 7674 & 0.0289 & M2 & 11.56 & 0.74 $\pm$ 0.07 & 10 & 17\\
   F23488+1949 & NGC 7770/1 & 0.0143 & M2 & 11.40 & 0.16 $\pm$ 0.04 & 10 & 17\\
   F23488+2018 & MRK 0331 & 0.0185 & M1 & 11.50 & 0.03 $\pm$ 0.01 & 9 & 15\\
\hline
   \end{tabular}
   \begin{tablenotes}
   \begin{flushleft}
   %\\
   $^{(a)}$ L$_{IR}$ \citep[from][]{2003AJ....126.1607S}.
   $^{(b)}$ Average bolometric AGN fractions by \citet{2017ApJ...846...32D}, or $^{\star}$ in the range 5-1000 ${\mu}m$, by \citet{2006MNRAS.366..767F} %$^{(c)}$ Number of fitted observations, provided by \citet{2012ApJS..203....9U}, \citet{2017ApJS..229...25C} and the DustPedia database.\\
   \end{flushleft}
   \end{tablenotes}
   \label{table1}
   \end{center}
   \end{table*}
   The mid-infrared (MIR) part of the SED is a very complex regime where different components of the galaxy contribute to the emission. The SED in these wavelengths can be a mixture of light originating from old stellar populations, emission of dust grains (mainly heated by newly formed stars), as well as radiation emitted by an active galactic nucleus (AGN). In (U)LIRGs, the MIR spectra can probe the conditions in their star-forming regions, to which they owe their high luminosities. Although galaxy interactions can trigger and enhance nuclear activity in local (U)LIRGs, star-formation processes dominate the MIR emission in the majority of these systems \citep{1998ApJ...498..579G, 2011ApJ...730...28P, 2013ApJS..206....1S}. The UV and optical radiation emitted by the newly formed stars, in dusty molecular clouds, is absorbed and then re-emitted by dust in infrared wavelengths. Dust emission is seen either through emission features (e.g. Polycyclic Aromatic Hydrocarbons; PAHs) or continuum emission originating from the photo-dissociation regions (PDRs), as well as dust diffusely distributed throughout the galaxy.

   One way to decompose the SED of a galaxy and extract useful information on the different components contributing their emission is by using SED-fitting codes. Such codes usually make use of templates for the emission of different components combined in such a way so that energy is conserved. Assuming two different stellar populations (old and young stars) and a certain variation of the star-formation activity through cosmic time (i.e. the star-formation history; SFH), several characteristic properties of the galaxies can be determined by fitting their observed SEDs. Such properties are stellar and dust masses, luminosities, and the star-formation rate (\textit{SFR}). In addition to the stellar and dust components, an AGN component can be considered when needed. The decomposition of the galactic SEDs has lead to versatile studies of the properties of galaxies \citep{2011A&A...525A.150G, 2014A&A...562A..15M, 2016A&A...589A..11P, 2017A&A...597A..51V}, and more specifically, studies about the attenuation processes \citep{2013A&A...554A..14B, 2018ApJ...859...11S} or the contribution of the different stellar populations in the heating of the dust and the dependence on the morphological type \citep{2018A&A...620A.112B, 2019A&A...624A..80N}.

   In this paper we model the SEDs of 67 IR-luminous local galaxies, in order to derive their physical properties and to investigate how these properties compare with other types of galaxies in the local Universe. We explore how several physical properties, such as the \textit{SFR}, the stellar mass and the dust mass, dust temperature, and PDR luminosity vary with galaxy type. Dividing (U)LIRGs in different classes, according to the interaction of their parent galaxies, we investigate the evolution of all the aforementioned parameters with merging stage. Finally, we examine the relative contributions of the different stellar populations to the bolometric luminosity of galaxies and their role in the dust heating. By comparing our results with a recent study on passive and star-forming galaxies \citep{2019A&A...624A..80N}, we determine the role of each stellar population in the dust heating and for the different types of galaxies in the local universe.

   This paper is structured as follows. The sample analysed in this work is presented in Sect.~\ref{sec:sample} followed by a description of the SED fitting method in Sect.~\ref{sec:fitting_method}. The results of our analysis as well as a comparison of the properties of the (U)LIRGs in our sample with those of `normal' galaxies in the local universe are explored in Sec.~\ref{sec:local_universe}. The evolution of the physical properties of (U)LIRGs with merging stage is investigated in Sect.~\ref{sec:merg-evol}, while the different stellar populations and their role in the heating of the dust grains is discussed in Sec.~\ref{sec:dust_heating}. Our findings are summarised in Sec.~\ref{sec:sum_conc}. Finally, a mock analysis, performed by CIGALE, is presented in Appendix~\ref{ap:mock}, the best-fit SED models are shown in Appendix~\ref{appendix_figs}, a comparison with other studies is performed in Appendix~\ref{appC:comparison}, while the cumulative distributions of various physical parameters are presented in Appendix~\ref{cumulative}.

%--------------------------------------------------------------------
\section{The sample}\label{sec:sample}

   Our main sample consists of 67 local (U)LIRGs (see Table~\ref{table1}) drawn from the Great Observatories All-sky LIRG Survey \citep[GOALS;][]{2009PASP..121..559A}. All sources have sufficient wavelength coverage with an average of 8 observations in the UV/optical/NIR ($<3 \mu$m) part of the spectrum and 16 in the MIR/FIR/submm ($>3 \mu$m) part of the spectrum (see Table~\ref{table1}, but also the corresponding SEDs in Fig.~\ref{all_SEDs}). This allows us to conduct a full SED fitting study and derive valuable information on the intrinsic properties of the sources. 

   All galaxies have been observed by the \textit{Herschel Space Observatory} \citep[HSO;][]{2010A&A...518L...1P} with the corresponding photometry published in \citet{2017ApJS..229...25C}. In this study, total system fluxes (if systems of galaxies consist of more than one galaxy) and component fluxes (where possible) were computed for all six  \textit{Herschel} bands [PACS (70, 100, and 160 $\mu$m) and SPIRE (250, 350, and 500 $\mu$m)]. The photometric apertures were carefully chosen, first by visual inspection, and, subsequently, by plotting the curve of growth and checking that all of the flux was included in.   

   Apart from the Herschel photometry, the rest of the multiwavelength data were compiled from two literature resources, the photometry presented in \citet{2012ApJS..203....9U} (for 64 galaxies) and the photometry presented in \citet{2018A&A...609A..37C} (for 3 galaxies; F06107+7822, F10257-4339, and F23133-4251, in the DustPedia\footnote{http://dustpedia.astro.noa.gr/} database). All the details for the data assembly and treatment are discussed in the relevant papers. In summary, the Space-based observations include Ultraviolet data from GALEX \citep[][FUV; $\lambda_\text{eff}=0.15~\mu$m, and NUV; $\lambda_\text{eff}=0.23~\mu$m]{2007ApJS..173..682M}, MIR data from \textit{Spitzer}$/IRAC$ \citep[][3.6, 4.6, 5.4, and 8 $\mu$m]{2004ApJS..154....1W} and FIR data from \textit{Spitzer}$/MIPS$ \citep[][24, 70, and 160 $\mu$m]{2004ApJS..154....1W}. $IRAS$ data at 12, 25, 60, and 100 $\mu$m as published in the Revised Bright Galaxy Sample \citep[RBGS;][]{2003AJ....126.1607S} have also been incorporated into our analysis. Concerning the ground-based data, the main source in optical wavebands are observations taken with the University of Hawaii (UH) 2.2 m Telescope on Mauna Kea  with the remaining optical data compiled from the literature and the NASA/IPAC Extragalactic Database \citep[NED; see][and references therein]{2012ApJS..203....9U}. NIR J, H, K$_\text{s}$ were extracted from the Two Micron All Sky Survey \citep[2MASS;][]{2006AJ....131.1163S}. Submillimeter data at 850 $\mu$m and 450 $\mu$m, obtained using the Submillimeter Common User Bolometer Array (SCUBA) at the James Clerk Maxwell Telescope were taken from \citet{2000MNRAS.315..115D} and \citet{2001MNRAS.327..697D}. In both of these studies, a careful aperture-matched photometry was performed to the datasets, securing a consistent photometry among the different photometric bands. In the case of \citet{2012ApJS..203....9U} masked photometry has been extracted from the images taken at effective wavelengths 0.15 $\mu$m $< \lambda_{eff} <$ 8 $\mu$m and at MIPS 24 $\mu$m band with the masks defined based on isophotes in the median- and boxcar-smoothed $I$-band images at the surface brightness limit of 24.5 mag arcsec$^{-2}$ so that the global flux from tidal debris as well as individual components within these merger systems are encapsulated. For the remaining wavelengths, masks would not improve the precision of the total flux, due to lack in resolution, so circular or elliptical apertures were used. In \citet{2018A&A...609A..37C} the photometry and uncertainty estimation, in all bands, is done within a master elliptical aperture which is found after combining the elliptical apertures fitted in each band, separately, in such a way so that the different beam sizes are taken into account. 

   Out of these 67 systems, the vast majority (56) are LIRGs (L$_\text{IR} > 10^{11}~$L$_{\odot}$), whereas 11 are ULIRGs (L$_\text{IR} > 10^{12}~$L$_{\odot}$). Their redshifts range from  0.003 to 0.083 with a mean value of 0.029, while $\sim80\%$ of the galaxies are in the redshift range between 0.01 and 0.04. Their IR luminosities \citep[based on $IRAS$ photometry, by][]{2003AJ....126.1607S}, as well as their redshift distribution are presented in Fig.~\ref{distributions_literature} (see also Table~\ref{table1}).

   Since merging is a key process of the regulation of the properties of these systems, we also wanted to analyse the properties of our sample considering their different merging stages. For the classification of their stage we have used the categorization adopted in \citet{2016ApJ...825..128L}. In that paper, optical imaging data from the GOALS $HST$ sample \citep[PID: 10592, PI: Evans;][]{2013ApJ...768..102K} as well as optical $I$-band images obtained with the UH 2.2 m telescope on Mauna Kea by \citet{2004PhDT........18I} were used to visually classify the galaxies. Several criteria were imposed to conduct the classification, all discussed in detail in  \citet{2016ApJ...825..128L}. For the sake of completeness we, briefly, repeat here the broad characteristics of the seven different merging classes along with their abbreviation:
   \vspace{-0.7 cm}
   \[
    \begin{array}{lp{0.8\linewidth}}
        \bf{s} & \textit{Single galaxies:} galaxies that show no current sign of an interaction or merger event. \\
        \bf{m} & \textit{Minor mergers:} interacting pairs with estimated mass ratios > 4:1. \\
        \bf{M1} & \textit{Major merger-stage 1:} pairs that appear to be on their initial approach and have no prominent tidal features; with low relative velocity and galaxy separation, $\Delta$ V < 250~km~$s^{-1}$,~n$_{sep}$ < 75~kpc. \\
        \bf{M2} & \textit{Major merger-stage 2:} interacting galaxy pairs with obvious tidal bridges and tails \citep{1972ApJ...178..623T} or other disturbances consistent with having already undergone a first close passage. \\
        \bf{M3} & \textit{Major  merger-stage 3:} merging galaxies with multiple nuclei. These systems have distinct nuclei in disturbed, overlapping disks, along with visible tidal tails. \\
        \bf{M4} & \textit{Major merger-stage 4:} galaxies with apparent single nuclei  and  obvious tidal tails. The galaxy nuclei have n$_{sep}$ $\le$ 2~kpc. \\
        \bf{M5} & \textit{Major merger-stage 5:} galaxies which appear to be evolved merger remnants. These galaxies have diffuse envelopes which may exhibit shells or other fine structures \citep{1992AJ....104.1039S} and a single, possibly off-center nucleus. These merger remnants no longer have bright tidal tails.
    \end{array}
   \]
   Examples of galaxies in the different classes are illustrated in Fig.~1 of \citet{2016ApJ...825..128L}.
   
   %%%%%%%%%%%%%%%%%%%%%%%%%%%%%%%%%%%%% TABLE 2
   \begin{table*}[t!]
   \caption{The grid of input parameters used in \textsc{CIGALE} for the best-fit model computation. The combination of these parameters results to 14,515,200 models per redshift bin. With 9 redshift bins used, 130,636,800 models are computed in total. The parameters in the brackets indicate the exact naming of each module as it is used in \textsc{CIGALE}.}
   \begin{center}
   \scalebox{0.9}{
   \begin{threeparttable}
   \begin{tabular}{lc}
\hline
\hline
   \textbf{\textit{Parameters}} & \textbf{\textit{Values}}\\
\hline
\hline
   \textbf{star-formation history} & \textbf{delayed} + \textbf{burst} + \textbf{quenching} $^\mathrm{(i)}$\\
   & [[sfhdelayedbq]] \\
\hline
   $\tau_\mathrm{main}$, e-folding time [Myr] & 1000, 3000, 5000, 10000\\
   $t_\mathrm{gal}$, galaxy age [Myr] & 4500, 7000, 9500, 12000\\
   $t_\mathrm{bq}$, quenching or bursting age [Myr] & 10, 20, 30, 100\\
   $r_\mathrm{SFR}$, ratio of the \textit{SFR} after/before quenching or bursting age & 1.0, 3.16, 10.0, 20.0, 100.0, 1000.0, 1200.0\\
\hline
\hline
   \textbf{stellar population model} & \textbf{Bruzual \& Charlot}$^\mathrm{(ii)}$\\
   &[[bc03]] \\
\hline
   initial mass function & Salpeter$^\mathrm{(iii)}$ \\
   metallicity & 0.02\\
\hline
\hline
   \textbf{dust attenuation} & \textbf{modified Calzetti}$^\mathrm{(iv)}$\\
   &[[dustatt$\_$calzleit]]\\
\hline
    \textit{E(B-V)}, colour excess of the young stellar population & 0.20, 0.29, 0.44, 0.66, 1.0  \\
   \textit{E(B-V)}$_\mathrm{old}$/\textit{E(B-V)}$_\mathrm{young}$, reduction factor for \textit{E(B-V)} & 0.25, 0.50, 0.75\\
   $\delta$, power law index of the attenuation curve & -0.5, -0.25, 0.0\\
\hline
\hline
   \textbf{dust grain model} & \textbf{\textsc{THEMIS}}$^\mathrm{(v)}$\\ &[[themis]] \\
\hline
   $q_\mathrm{hac}$, fraction of small hydrocarbon solids & 0.02, 0.17\\
   $U_\mathrm{min}$, minimum radiation field & 5, 10, 25, 50, 80\\
   $\alpha$, power law index of the radiation field & 2.0\\
   $\gamma$, fraction illuminated from $U_\mathrm{min}$ to $U_\mathrm{max}$ & 0.01\\
\hline
\hline
   \textbf{active nucleus model} & \textbf{Skirtor}$^\mathrm{(vi)}$\\ & [[skirtor2016]] \\
\hline
   $\tau_{9.7}$, optical depth at 9.7$~\mu$m & 3.0, 7.0 \\
   pl, torus density radial parameter & 1.0\\
   q, torus density angular parameter & 1.0\\
   oa, angle between the equatorial plan and edge of the torus [deg] & 40\\
   R, ratio of outer to inner radius & 20\\
   Mcl, fraction of total dust mass inside clumps [\%] & 97\\
   i, inclination (viewing angle) [deg] & 30 (type 1), 70 (type 2) \\
   AGN fraction & 0.0, 0.1, 0.2, 0.3, 0.4, 0.5, 0.6, 0.7, 0.85\\
   extinction law of polar dust & SMC\\
   E(B-V) of polar dust & 0, 0.8\\
   polar dust temperature [K] & 100\\
   polar dust emissivity index & 1.6\\
\hline
\hline
   \end{tabular}
   \begin{tablenotes}
   References: (i) \citet{2016A&A...585A..43C}; (ii) \citet{2003MNRAS.344.1000B}; (iii) \citet{1955ApJ...121..161S}; (iv) \citet{2000ApJ...533..682C}; (v) \citet{2017A&A...602A..46J}; (vi) \citet{2012MNRAS.420.2756S, 2016MNRAS.458.2288S}.
   \end{tablenotes}
   \end{threeparttable}}
   \label{tab:grid}
   \end{center}
   \end{table*}
%%%%%%%%%%%%%%%%%%%%%%%%%%%%%%%%%%%%%%%%%%%%%%%%%%%%%%%%%%%%
   
   Two of the galaxies in our sample (namely, F10173+0828 and F01076-1707) were classified as ambiguous (amb) in \citet{2016ApJ...825..128L}. F10173+0828 appears to have a single nucleus with faint tidal tails which would have resulted in an `M4' classification, however, a second smaller disturbed galaxy (SDSS CGB24551.1) can be a possible companion but with no reported redshift. F01076-1707 was previously classified as a non-interacting galaxy by \citet{2013ApJS..206....1S}. The galaxy consists of a diffuse disk and a possible disconnected diffuse tail ($\sim$37 kpc to the West), though, it is unclear if this diffuse structure is the remnant of a tidal tail. In order to ease the analysis we have chosen to keep their most obvious classification in place (i.e. `M4' for F10173+0828 and `s' for F01076-1707) although we caution the reader that these two galaxies may be missclassified. The three galaxies drawn from the DustPedia database have been classified by us, by visually inspecting their optical images. F06107+7822 shows signs of evolved merger remnant with a single nucleus and was thus classified as `M5'. F10257-4339 shows a system of merging galaxies with two distinct nuclei in disturbed overlapping disks and with obvious tidal tails, thus classified as `M3'. Finally, F23133-4251 shows no current signs of interaction or merger events and thus classified as `s'.
   
   The classification of the systems in our sample is presented in Fig.~\ref{distributions_literature} (see also Table~\ref{table1}). In total, our sample consists of 14 `s', 4 `m', 6 `M1', 14 `M2', 17 `M3', 10 `M4', and 2 `M5' systems. We caution the reader that a few merging classes are underrepresented (e.g. `m' and `M5') so their statistical significance in the following analysis may be questionable. In the subsequent SED analysis, systems with multiple components are treated as one system \citep[a similar approach already presented in][]{2012ApJS..203....9U}. Even though this concept is correct (if multiple systems are bound dynamically they should be treated as one) it possesses some limitations on the validity of the SED fitting, especially for multiple systems with the parent galaxies in large distances from each other (e.g., `M1' systems) where the galaxies (with different properties) are forced to be treated as one entity. In such systems we assume that the SED modelling performed, produced an average description of each multiple system.
   
   Since nearby (U)LIRGs constitute a separate class of galactic systems in the local Universe (which, by definition, show high IR luminosities and enhanced \textit{SFR}) we need to have a reference sample of `normal' galaxies [in the sense of being less active in forming stars; either passive (ETGs) or low level star-forming galaxies (LTGs)] to compare with. The most complete, to date, sample of local galaxies is that of DustPedia \citep[][]{2017PASP..129d4102D, 2018A&A...609A..37C}. This sample includes 875 galaxies (with distances of less than 40 Mpc) all having \textit{Herschel} detections, with \textit{D}$_{25}$ > 1$\arcmin$ (\textit{D}$_{25}$ being the major axis isophote at which optical surface brightness falls beneath 25 mag arcsec$^{-2}$). Additionally, the DustPedia sample includes, mostly, isolated galaxies that have a detected stellar component; WISE observations at 3.4 $\mu$m are the deepest all-sky data sensitive to the stellar component of galaxies, and hence provide the most consistent way of implementing this stellar detection requirement. For full sample details see \citet{2017PASP..129d4102D}, and \citet{2018A&A...609A..37C}. Furthermore, it contains galaxies of various morphologies parametrized by the Hubble Stage \citep[T;][]{2014A&A...570A..13M} ranging from T=-5 (pure elliptical galaxies) to T=10 (irregular galaxies). Throughout our analysis we consider as ETGs galaxies with T<0.5 and as LTGs galaxies with T$\geq$0.5. Out of these 875 galaxies, 814 were successfully modelled using the \textsc{CIGALE} (Code Investigating GALaxy Emission) SED fitting tool \citep{2019A&A...622A.103B} with only three of them (NGC2146, NGC3256, NGC7552) overlapping with our current sample, since they are LIRGs. The products of this modelling are available in the DustPedia database with a full description given in \citet{2019A&A...624A..80N}.

\section{SED modelling}\label{sec:fitting_method}

%%%%%%%%%%%%%%%%%%%%%%%%%%%%%%%%%%%%%%%% TABLE 3
\begin{table*}[h!]
\caption{CIGALE-derived physical properties of the (U)LIRGs in our sample} %This table is published in its entirety in the electronic version of A\&A.}
\label{tab:properties2}    
\tiny
\centering
\begin{tabular}{l|c|c|c|c|c|c}
\hline
  \multicolumn{1}{c|}{id} &
  \multicolumn{1}{c|}{$\log~M_\text{star}$ [M$_{\odot}$]} &
  \multicolumn{1}{c|}{$\log~M_\text{dust}$ [M$_{\odot}$]} &
  \multicolumn{1}{c|}{$T_\text{dust}$ [K]} &
  \multicolumn{1}{c|}{\textit{SFR} [M$_{\odot}$~yr$^{-1}$]} &
  \multicolumn{1}{c|}{$frac_\mathrm{AGN}$} &
  \multicolumn{1}{c}{reduced $\chi^2$} \\
\hline
\hline
  F00085-1223 & 10.66 $\pm$ 0.12 & 7.18 $\pm$ 0.03 & 38.96 $\pm$ 0.34 & 52.01 $\pm$ 8.37 & 0.10 $\pm$ 0.00 & 2.08\\
  F00163-1039 & 10.70 $\pm$ 0.13 & 7.62 $\pm$ 0.02 & 31.91 $\pm$ 0.12 & 51.65 $\pm$ 8.25 & 0.19 $\pm$ 0.03 & 2.45\\
  F00402-2349 & 11.04 $\pm$ 0.14 & 7.71 $\pm$ 0.02 & 31.91 $\pm$ 0.07 & 50.18 $\pm$ 11.89 & 0.10 $\pm$ 0.01 & 1.66\\
  F01053-1746 & 10.54 $\pm$ 0.14 & 7.83 $\pm$ 0.02 & 31.91 $\pm$ 0.11 & 73.38 $\pm$ 8.21 & 0.20 $\pm$ 0.01 & 1.93\\
  F01076-1707 & 11.02 $\pm$ 0.13 & 7.82 $\pm$ 0.02 & 31.91 $\pm$ 0.01 & 65.80 $\pm$ 10.64 & 0.10 $\pm$ 0.00 & 3.13\\
  F01173+1405 & 10.57 $\pm$ 0.12 & 7.39 $\pm$ 0.02 & 38.97 $\pm$ 0.29 & 101.46 $\pm$ 13.56 & 0.10 $\pm$ 0.01 & 1.13\\
  F01364-1042 & 10.46 $\pm$ 0.07 & 7.70 $\pm$ 0.05 & 35.61 $\pm$ 1.03 & 120.59 $\pm$ 17.46 & 0.00 $\pm$ 0.02 & 6.22\\
  F01417+1651 & 10.30 $\pm$ 0.14 & 7.65 $\pm$ 0.02 & 31.93 $\pm$ 0.31 & 51.23 $\pm$ 9.93 & 0.00 $\pm$ 0.00 & 6.61\\
  F01484+2220 & 11.21 $\pm$ 0.09 & 8.24 $\pm$ 0.02 & 27.25 $\pm$ 0.34 & 76.02 $\pm$ 12.14 & 0.10 $\pm$ 0.01 & 3.11\\
  F02281-0309 & 11.39 $\pm$ 0.08 & 8.13 $\pm$ 0.02 & 24.16 $\pm$ 0.01 & 19.99 $\pm$ 5.11 & 0.10 $\pm$ 0.01 & 4.62\\
  F02435+1253 & 10.82 $\pm$ 0.14 & 7.90 $\pm$ 0.02 & 27.24 $\pm$ 0.01 & 38.21 $\pm$ 3.91 & 0.10 $\pm$ 0.00 & 4.67\\
  F02512+1446 & 11.28 $\pm$ 0.10 & 7.79 $\pm$ 0.02 & 31.91 $\pm$ 0.06 & 53.46 $\pm$ 10.01 & 0.10 $\pm$ 0.01 & 1.67\\
  F03359+1523 & 10.22 $\pm$ 0.03 & 7.46 $\pm$ 0.02 & 35.97 $\pm$ 0.01 & 76.94 $\pm$ 3.85 & 0.00 $\pm$ 0.00 & 2.01\\
  F04097+0525 & 10.74 $\pm$ 0.08 & 8.05 $\pm$ 0.02 & 24.20 $\pm$ 0.36 & 22.80 $\pm$ 2.97 & 0.00 $\pm$ 0.00 & 4.79\\
  F04191-1855 & 11.12 $\pm$ 0.04 & 7.90 $\pm$ 0.14 & 30.32 $\pm$ 2.09 & 50.52 $\pm$ 12.12 & 0.04 $\pm$ 0.05 & 2.94\\
  F04315-0840 & 10.52 $\pm$ 0.12 & 7.32 $\pm$ 0.02 & 39.00 $\pm$ 0.08 & 64.85 $\pm$ 3.24 & 0.10 $\pm$ 0.00 & 5.34\\
  F05189-2524 & 11.01 $\pm$ 0.12 & 7.63 $\pm$ 0.02 & 39.00 $\pm$ 0.03 & 118.41 $\pm$ 6.98 & 0.40 $\pm$ 0.02 & 2.76\\
  F06107+7822 & 10.78 $\pm$ 0.08 & 7.36 $\pm$ 0.02 & 31.93 $\pm$ 0.38 & 24.09 $\pm$ 2.26 & 0.10 $\pm$ 0.01 & 2.57\\
  F08354+2555 & 10.53 $\pm$ 0.05 & 7.58 $\pm$ 0.02 & 31.91 $\pm$ 0.01 & 51.97 $\pm$ 2.65 & 0.00 $\pm$ 0.00 & 5.65\\
  F08572+3915 & 10.00 $\pm$ 0.07 & 7.01 $\pm$ 0.07 & 38.68 $\pm$ 0.88 & 43.05 $\pm$ 2.77 & 0.85 $\pm$ 0.01 & 6.82\\
  F09126+4432 & 11.08 $\pm$ 0.09 & 7.91 $\pm$ 0.04 & 31.89 $\pm$ 0.26 & 104.39 $\pm$ 11.61 & 0.00 $\pm$ 0.00 & 4.77\\
  F09320+6134 & 10.78 $\pm$ 0.15 & 8.17 $\pm$ 0.10 & 31.77 $\pm$ 0.68 & 187.97 $\pm$ 9.40 & 0.10 $\pm$ 0.00 & 3.98\\
  F09333+4841 & 10.50 $\pm$ 0.13 & 7.50 $\pm$ 0.02 & 31.94 $\pm$ 0.41 & 42.74 $\pm$ 4.22 & 0.10 $\pm$ 0.01 & 1.79\\
  F09437+0317 & 11.03 $\pm$ 0.05 & 8.07 $\pm$ 0.02 & 24.16 $\pm$ 0.03 & 19.80 $\pm$ 3.67 & 0.10 $\pm$ 0.00 & 3.80\\
  F10015-0614 & 10.90 $\pm$ 0.11 & 7.88 $\pm$ 0.02 & 27.24 $\pm$ 0.01 & 34.53 $\pm$ 6.51 & 0.10 $\pm$ 0.00 & 2.87\\
  F10173+0828 & 10.28 $\pm$ 0.04 & 7.67 $\pm$ 0.10 & 34.31 $\pm$ 1.95 & 87.20 $\pm$ 14.64 & 0.03 $\pm$ 0.05 & 7.87\\
  F10257-4339 & 10.92 $\pm$ 0.16 & 7.69 $\pm$ 0.02 & 36.00 $\pm$ 0.36 & 111.47 $\pm$ 19.99 & 0.00 $\pm$ 0.00 & 2.68\\
  F10565+2448 & 10.82 $\pm$ 0.08 & 7.95 $\pm$ 0.02 & 35.97 $\pm$ 0.01 & 243.71 $\pm$ 12.19 & 0.00 $\pm$ 0.00 & 1.46\\
  F11011+4107 & 10.61 $\pm$ 0.08 & 7.85 $\pm$ 0.02 & 31.91 $\pm$ 0.02 & 99.86 $\pm$ 4.99 & 0.00 $\pm$ 0.01 & 1.62\\
  F11186-0242 & 10.80 $\pm$ 0.06 & 7.85 $\pm$ 0.02 & 27.24 $\pm$ 0.01 & 20.03 $\pm$ 1.21 & 0.20 $\pm$ 0.02 & 2.17\\
  F11231+1456 & 11.12 $\pm$ 0.02 & 7.89 $\pm$ 0.09 & 31.78 $\pm$ 0.64 & 94.60 $\pm$ 4.98 & 0.00 $\pm$ 0.00 & 1.73\\
  F11257+5850 & 10.45 $\pm$ 0.09 & 7.47 $\pm$ 0.02 & 39.01 $\pm$ 0.02 & 92.45 $\pm$ 9.10 & 0.10 $\pm$ 0.00 & 3.05\\
  F12112+0305 & 10.73 $\pm$ 0.02 & 8.28 $\pm$ 0.02 & 31.91 $\pm$ 0.01 & 257.73 $\pm$ 12.89 & 0.00 $\pm$ 0.00 & 3.68\\
  F12224-0624 &  9.63 $\pm$ 0.35 & 7.70 $\pm$ 0.02 & 27.29 $\pm$ 0.40 & 20.05 $\pm$ 3.92 & 0.00 $\pm$ 0.00 & 6.98\\
  F12540+5708 & 11.36 $\pm$ 0.34 & 8.08 $\pm$ 0.03 & 39.00 $\pm$ 0.08 & 389.78 $\pm$ 64.04 & 0.25 $\pm$ 0.05 & 2.92\\
  F12590+2934 & 11.36 $\pm$ 0.04 & 7.30 $\pm$ 0.02 & 31.91 $\pm$ 0.02 & 20.87 $\pm$ 2.41 & 0.30 $\pm$ 0.02 & 4.47\\
  F13001-2339 & 11.07 $\pm$ 0.05 & 7.57 $\pm$ 0.02 & 31.91 $\pm$ 0.01 & 45.14 $\pm$ 3.24 & 0.00 $\pm$ 0.00 & 2.15\\
  F13126+2453 &  9.91 $\pm$ 0.13 & 6.98 $\pm$ 0.02 & 31.91 $\pm$ 0.01 & 10.47 $\pm$ 1.81 & 0.00 $\pm$ 0.00 & 9.46\\
  F13136+6223 & 10.67 $\pm$ 0.16 & 7.43 $\pm$ 0.03 & 38.97 $\pm$ 0.31 & 107.50 $\pm$ 16.18 & 0.10 $\pm$ 0.01 & 2.61\\
  F13182+3424 & 10.50 $\pm$ 0.06 & 7.85 $\pm$ 0.02 & 31.91 $\pm$ 0.06 & 96.71 $\pm$ 4.84 & 0.00 $\pm$ 0.00 & 1.97\\
  F13188+0036 & 10.98 $\pm$ 0.12 & 7.70 $\pm$ 0.02 & 27.24 $\pm$ 0.02 & 19.44 $\pm$ 1.60 & 0.10 $\pm$ 0.01 & 2.26\\
  F13197-1627 & 11.01 $\pm$ 0.03 & 7.27 $\pm$ 0.04 & 24.19 $\pm$ 0.35 & 2.73 $\pm$ 0.99 & 0.85 $\pm$ 0.03 & 2.98\\
  F13229-2934 & 11.09 $\pm$ 0.07 & 7.86 $\pm$ 0.02 & 27.24 $\pm$ 0.01 & 29.01 $\pm$ 5.01 & 0.10 $\pm$ 0.01 & 1.45\\
  F13362+4831 & 11.05 $\pm$ 0.10 & 7.66 $\pm$ 0.04 & 32.09 $\pm$ 0.95 & 50.56 $\pm$ 10.06 & 0.10 $\pm$ 0.00 & 1.75\\
  F13373+0105 & 11.31 $\pm$ 0.10 & 8.15 $\pm$ 0.02 & 27.24 $\pm$ 0.01 & 56.03 $\pm$ 9.96 & 0.10 $\pm$ 0.00 & 2.27\\
  F13428+5608 & 11.03 $\pm$ 0.08 & 7.96 $\pm$ 0.04 & 36.18 $\pm$ 0.86 & 241.61 $\pm$ 17.51 & 0.00 $\pm$ 0.00 & 4.77\\
  F14179+4927 & 10.39 $\pm$ 0.02 & 7.23 $\pm$ 0.02 & 35.98 $\pm$ 0.22 & 38.06 $\pm$ 1.90 & 0.00 $\pm$ 0.00 & 2.06\\
  F14348-1447 & 10.93 $\pm$ 0.02 & 8.19 $\pm$ 0.03 & 35.95 $\pm$ 0.25 & 410.82 $\pm$ 20.54 & 0.00 $\pm$ 0.00 & 2.73\\
  F14547+2449 & 11.51 $\pm$ 0.07 & 8.31 $\pm$ 0.02 & 27.24 $\pm$ 0.01 & 75.01 $\pm$ 9.80 & 0.00 $\pm$ 0.00 & 4.98\\
  F15107+0724 & 10.37 $\pm$ 0.07 & 7.73 $\pm$ 0.02 & 27.24 $\pm$ 0.03 & 26.44 $\pm$ 2.30 & 0.00 $\pm$ 0.00 & 9.66\\
  F15163+4255 & 10.69 $\pm$ 0.18 & 7.71 $\pm$ 0.02 & 36.01 $\pm$ 0.39 & 115.36 $\pm$ 21.30 & 0.10 $\pm$ 0.00 & 2.21\\
  F15250+3608 & 10.53 $\pm$ 0.03 & 7.60 $\pm$ 0.03 & 38.78 $\pm$ 0.73 & 163.96 $\pm$ 10.16 & 0.11 $\pm$ 0.03 & 4.72\\
  F15327+2340 & 10.84 $\pm$ 0.03 & 8.58 $\pm$ 0.02 & 24.16 $\pm$ 0.01 & 100.62 $\pm$ 5.03 & 0.00 $\pm$ 0.00 & 14.69\\
  F16104+5235 & 10.79 $\pm$ 0.12 & 7.66 $\pm$ 0.08 & 32.72 $\pm$ 1.78 & 53.97 $\pm$ 8.35 & 0.10 $\pm$ 0.00 & 2.82\\
  F16284+0411 & 10.84 $\pm$ 0.02 & 7.60 $\pm$ 0.02 & 31.91 $\pm$ 0.01 & 49.98 $\pm$ 2.50 & 0.10 $\pm$ 0.00 & 1.74\\
  F16577+5900 & 10.83 $\pm$ 0.09 & 8.20 $\pm$ 0.05 & 24.42 $\pm$ 0.96 & 27.54 $\pm$ 6.83 & 0.00 $\pm$ 0.00 & 5.21\\
  F17132+5313 & 10.76 $\pm$ 0.13 & 8.02 $\pm$ 0.02 & 31.94 $\pm$ 0.43 & 100.44 $\pm$ 14.29 & 0.10 $\pm$ 0.00 & 3.94\\
  F22287-1917 & 10.99 $\pm$ 0.11 & 7.88 $\pm$ 0.02 & 27.24 $\pm$ 0.18 & 26.21 $\pm$ 5.36 & 0.19 $\pm$ 0.04 & 2.41\\
  F22491-1808 & 10.63 $\pm$ 0.02 & 7.86 $\pm$ 0.04 & 36.24 $\pm$ 0.95 & 205.01 $\pm$ 10.25 & 0.00 $\pm$ 0.00 & 5.90\\
  F23007+0836 & 11.25 $\pm$ 0.03 & 7.73 $\pm$ 0.02 & 31.91 $\pm$ 0.01 & 48.15 $\pm$ 3.90 & 0.30 $\pm$ 0.00 & 2.97\\
  F23024+1916 & 10.77 $\pm$ 0.08 & 7.63 $\pm$ 0.07 & 31.83 $\pm$ 0.50 & 39.53 $\pm$ 6.74 & 0.00 $\pm$ 0.00 & 2.67\\
  F23133-4251 & 10.77 $\pm$ 0.13 & 7.43 $\pm$ 0.10 & 31.59 $\pm$ 1.01 & 30.04 $\pm$ 13.21 & 0.08 $\pm$ 0.10 & 0.68\\
  F23135+2517 & 11.00 $\pm$ 0.08 & 7.48 $\pm$ 0.05 & 35.88 $\pm$ 0.51 & 67.58 $\pm$ 11.71 & 0.05 $\pm$ 0.05 & 3.23\\
  F23157-0441 & 10.73 $\pm$ 0.10 & 7.54 $\pm$ 0.02 & 31.91 $\pm$ 0.01 & 38.12 $\pm$ 5.75 & 0.17 $\pm$ 0.04 & 1.62\\
  F23254+0830 & 11.41 $\pm$ 0.02 & 8.06 $\pm$ 0.02 & 24.17 $\pm$ 0.09 & 22.52 $\pm$ 3.18 & 0.60 $\pm$ 0.01 & 2.35\\
  F23488+1949 & 11.29 $\pm$ 0.04 & 8.02 $\pm$ 0.02 & 27.24 $\pm$ 0.01 & 30.22 $\pm$ 3.82 & 0.10 $\pm$ 0.00 & 1.66\\
  F23488+2018 & 10.65 $\pm$ 0.03 & 7.47 $\pm$ 0.02 & 35.96 $\pm$ 0.14 & 64.60 $\pm$ 3.23 & 0.00 $\pm$ 0.00 & 2.76\\
\hline\end{tabular}
\end{table*}

   In order to exploit the information hidden in the SEDs of the galaxies we make use of \textsc{CIGALE} \citep[see][and references therein]{2019A&A...622A.103B}. \textsc{CIGALE} models the SED of each galaxy, by selecting a suitable set of parameters, through a Bayesian approach, that best represents the observed SED. A basic principle of the code is the conservation of the energy between the amount absorbed by the dust and that re-emitted by the dust grains in infrared and submm wavelengths \citep{2014ASPC..485..347R}. In the current study we use the version 2020.0 of CIGALE. Beyond many improvements and optimisations related to the diminution of the computational time and the estimation of the physical properties, the most important new feature is related to the AGN model used. In addition to the active nucleus model formulated by \citet{2006MNRAS.366..767F}, the SKIRTOR module \citep{2012MNRAS.420.2756S,2016MNRAS.458.2288S} has been added. Following recent theoretical and observational studies \citep{2009ApJ...707.1550N, 2012ApJ...754...45I, 2012MNRAS.420.2756S, 2019ApJ...883..110T}, SKIRTOR assumes a two-phase dusty clumpy torus, where most of the dust has high density and is clumpy, while the rest is smoothly distributed. SKIRTOR is based on the 3D radiative transfer code SKIRT \citep{2011ApJS..196...22B, 2015A&C.....9...20C}. In addition, this version of CIGALE includes not only torus obscuration, but also obscuration by dust settled along polar directions, allowing for a more precise treatment of both type-1 and type-2 AGN cases \citep{2012MNRAS.427.3103B, 2012ApJ...759....6E, 2017MNRAS.472.3854S, 2019MNRAS.484.3334S, 2018ApJ...866...92L}.
   
   The stellar radiation field is built based on the \citet{2003MNRAS.344.1000B} population synthesis model and a \citet{1955ApJ...121..161S} initial mass function (IMF). After the fitting procedure, the stellar emission can be decomposed into two distinct populations, an old and a young, depending on the SFH of each galaxy. To account for the starlight attenuation by dust, a modified \citet{2000ApJ...533..682C} starburst attenuation law is applied \citep{2009A&A...507.1793N}, to the intrinsic spectra of the different stellar populations. For the dust emission properties we adopted the \textsc{THEMIS} model \citep{2017A&A...602A..46J}. In addition, we use a flexible SFH that gets an analytical expression of a delayed SFH allowing for an instantaneous burst (or quenching) of the star-formation activity (see, \citealt{2015A&A...576A..10C} and \citealt{2019A&A...624A..80N}). Since all of the galaxies in our sample are actively forming stars we only use the bursting mode of the delayed SFH (over the last 10 to 100 Myr) with six cases of instantaneous bursts (at various levels) and one case with no burst. The nuclear activity of local (U)LIRGs has already been investigated by others \citep[e.g.,][]{2010MNRAS.405.2505N, 2017MNRAS.468.1273R, 2019ApJ...871..166U}. In a series of two papers [C-GOALS I \citep{2011A&A...529A.106I} and C-GOALS II \citep{ 2018A&A...620A.140T}], a total of 107 (U)LIRGs of the GOALS sample, were observed with the $Chandra$ X-ray Observatory revealing AGN signatures for $31\pm5\%$ and $38\pm7\%$ of the two sub-samples respectively. A careful treatment of the AGN component is, therefore, necessary for the purposes of the current study. To account for a realistic description of the AGN emission we make use of the SKIRTOR module, that also includes polar-dust extinction. The parameter space of this module follows the parametrization presented in \citet{2020MNRAS.491..740Y}, except for the values of polar dust colour excess, E(B-V), where only two values were used (0, 0.8). This was indicated by \citet{2020arXiv201109220M}, who find that CIGALE is not very sensitive in this parameter.
   
   Concerning the input parameters used in CIGALE, except from those defining the AGN module, we adopt a similar parameter space as the one used in \citet{2019A&A...624A..80N}, modified accordingly so that it is able to successfully model the SEDs of the systems in the current sample. By performing several test-runs with CIGALE, we were able to minimize the number of values that control the SFH and the dust modules, in favour of the SKIRTOR module, taking into account the computational demands imposed by the total number of parameters. These test-runs indicated values of the parameters that were not used by CIGALE and thus were excluded from the parameter space, resulting in the final input parameter grid listed in Table \ref{tab:grid}. The attenuation law considered here \citep{2000ApJ...533..682C}  was also tested against that of the widely adopted \citet{CF200} attenuation law resulting in small differences, and within the uncertainties, on the shape of the SEDs and on the output parameters.
   
   In order to assess whether or not physical properties can actually be estimated in a  reliable  way we have performed a mock analysis by using the relevant feature built in CIGALE. The results of the mock analysis are presented and discussed in Appendix~\ref{ap:mock}. The modelled SEDs of all the systems in our sample are presented in Fig.~\ref{all_SEDs}. To further evaluate the CIGALE-derived properties, we perform a comparison with other studies in Appendix~\ref{appC:comparison}.

\section{The physical properties and SEDs of local (U)LIRGs} \label{sec:local_universe}   
   
   The values and associated uncertainties of the physical properties ($M_\text{star}$, $M_\text{dust}$, $T_\text{dust}$, \textit{SFR}, $frac_\text{AGN}$), as derived by \textsc{CIGALE}, for each source in our sample, along with the reduced $\chi^2$ of the fit, are listed in Table~\ref{tab:properties2}. (U)LIRGs are systems of enhanced IR luminosity and, as such, are expected to show up as the dustiest and most actively star-forming galaxies in the local Universe (as already shown in previous studies, see, e.g. \citealt{2010A&A...523A..78D} and \citealt{2012ApJS..203....9U}).  The mean dust mass of the galaxies in our sample is  \mbox{$\langle M_\text{dust}\rangle = 7.2\times10^7~\text{M}_{\odot}$ ranging from \mbox{$M_\text{dust} = 9.5\times10^6~\text{M}_{\odot}$}} (for F13126+2453) to \mbox{$M_\text{dust} = 3.8\times10^8~\text{M}_{\odot}$} (for F15327+2340). Concerning the current \textit{SFR} we compute a mean value of $\langle \textit{SFR}\rangle = 81~\text{M}_{\odot}$yr$^{-1}$ ranging from 2.7~M$_{\odot}$yr$^{-1}$ (for F13197-1627) to 410.8~M$_{\odot}$yr$^{-1}$ for \mbox{(F14348-1447)}. As a comparison, the dustier and more actively star-forming `normal' galaxies in the local Universe (those of morphological classes Sb-Sc) show an average dust mass of $2\times10^7~\text{M}_{\odot}$ and an average \textit{SFR} of $2.4~\text{M}_{\odot}$yr$^{-1}$ \citep{ 2019A&A...624A..80N}. (U)LIRGs are also amongst the most massive galaxies in the local Universe with a mean value of their stellar mass of \mbox{$\langle M_\text{star}\rangle = 8.6\times10^{10}~\text{M}_{\odot}$} ranging from \mbox{$M_\text{star} = 4.3\times10^9~\text{M}_{\odot}$} (for F12224-0624) to \mbox{$M_\text{star} = 3.2\times10^{11}~\text{M}_{\odot}$} (for F14547+2449). Local elliptical galaxies, being the most massive amongst the `normal' galaxies, show a mean stellar mass of $8.3\times10^{10}~\text{M}_{\odot}$ \citep{2019A&A...624A..80N}.
   
%%%%%%%%%%%%%%%%%%%%%%%%%%%%%%%%%%%%% FIGURE 2
   \begin{figure*}[t!]
   \centering
   \includegraphics[width=\textwidth]{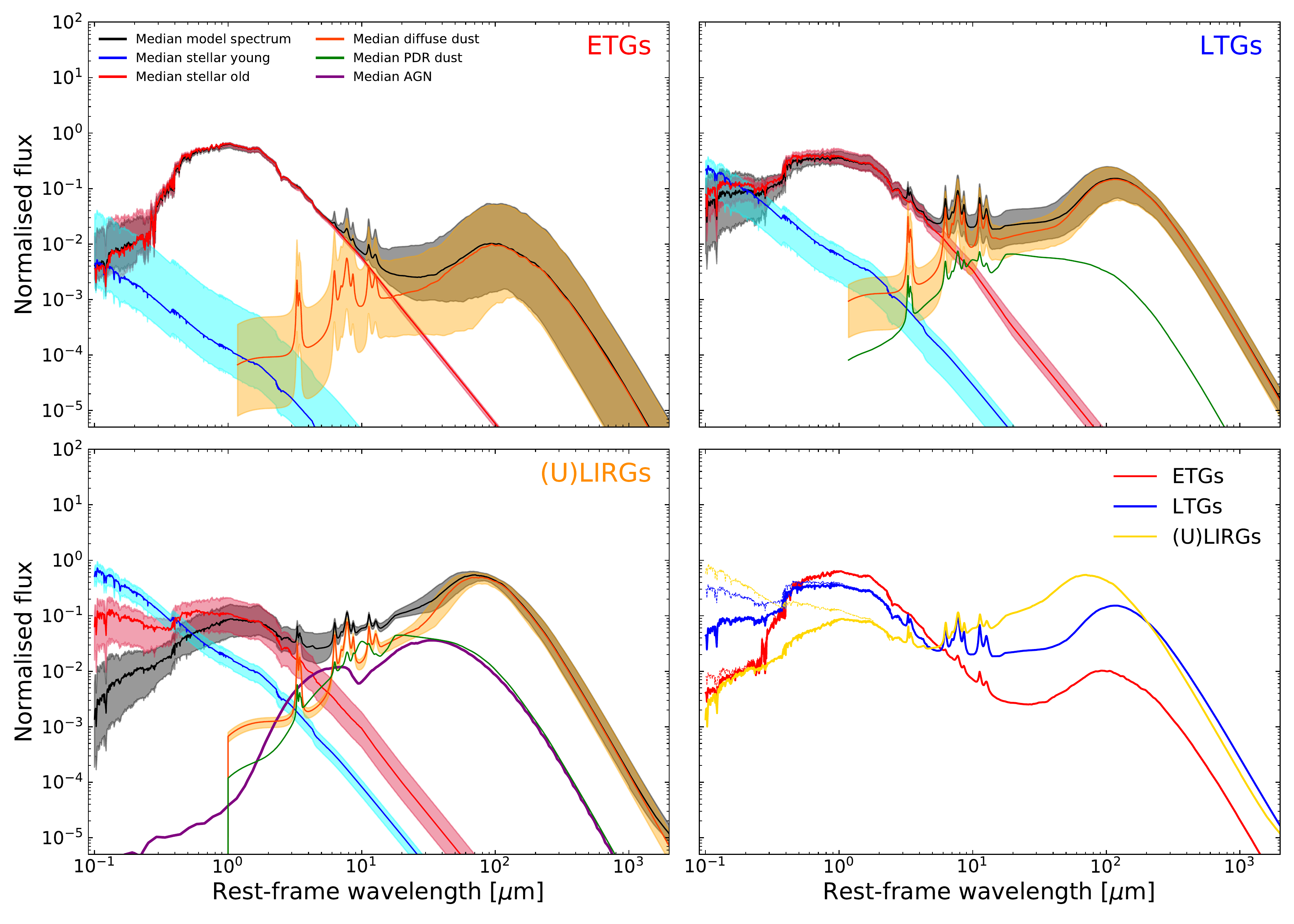}
   \caption{Template SEDs of the DustPedia ETGs (top-left panel), LTGs (top-right panel) and the local (U)LIRGs (bottom-left panel), as modelled by \textsc{CIGALE}. The black curves indicate the median SEDs, while the blue and the red curves denote the median unattenuated SEDs of the young and old stellar populations, respectively. The median distribution of the diffuse dust is shown by the orange curve and the emission from the PDRs is shown by the green curve. The purple curve, in the bottom-left panel, indicates the median AGN distribution. The shaded areas cover the range between the 16$^{th}$ and 84$^{th}$ percentiles to the median values.
   For clarity reasons, since the PDR and the AGN components may vary substantially, we choose not to present the percentile widths of the curves.
   The median SEDs of the three different types of local galaxies are plotted in the bottom-right panel. Red curve represents the median, observed, SED of ETGs, blue curve represents the one of LTGs, while the yellow curve corresponds to the median SED of (U)LIRGs. The thin, dotted, curves show the unattenuated total stellar emission for the different classes. The reader is also referred to the corresponding median SEDs in \citet{2018A&A...620A.112B} and \citet{2019A&A...624A..80N}.
   \label{stacks_types}}
   \end{figure*}

\subsection{The SEDs of (U)LIRGs}\label{SED}

   The SEDs of galaxies have been proven to be valuable assets for characterising their content in stars and dust, as well as their current star-formation activity. Even by visual inspection of the SEDs one can spot the differences between different populations and this is what we do next by comparing, in a statistically significant and systematic way, the SEDs of local ETGs, LTGs, and (U)LIRGs.

   In our analysis we have modelled the SEDs of the 67 local (U)LIRGs in our sample, using \textsc{CIGALE}, and compare them with those of 268 Early-Type Galaxies (ETGs) and 542 Late-Type Galaxies (LTGs), already modelled in \cite{2019A&A...624A..80N}. This comparison is shown in Fig.~\ref{stacks_types} with the median SEDs of ETGs, LTGs, and (U)LIRGs on the top-left, top-right, and bottom-left panels, respectively. In the bottom-right panel we present the median SEDs of ETGs, LTGs, and (U)LIRGs (red, blue, and yellow curves respectively) so that a direct comparison can be made. For each of the three classes, both the attenuated and the unattenuated SEDs are also shown (thick solid lines and thin dotted lines, respectively, in the UV-optical part of the SED). The most obvious change among them is the differences in the IR and UV wavelength regimes of the SEDs, and more specifically, at the peak of the dust emission (around 100~$\mu$m) and the peak of the stellar emission (around 1~$\mu$m).
   
   What is clearly visible is that the dust emission shows a large variation with the peak being higher by 1.17 dex for LTGs (compared to ETGs), and higher by 0.56 dex for (U)LIRGs (compared to LTGs). This already dictates the differences in dust content among the three populations with ETGs being the most devoid and (U)LIRGs the abundant ones. The slight shift in the peak (in the wavelength axis) seen in the dust SED inclines towards a variation in dust temperature with LTGs being the cooler ones and (U)LIRGs the warmest ones. The actual wavelength where the peak of the FIR emission happens is at 96, 116, and 70 $\mu$m for ETGs, LTGs, and (U)LIRGs, respectively. The dust emission, as treated by CIGALE, consists of two components, one accounting for the diffuse dust and one accounting for the PDRs, mostly associated with the star-forming regions \citep{1999RvMP...71..173H}. From Fig.~\ref{stacks_types} (green curves) we see that the significance of the dust in the PDR regions increases from ETGs (having negligible, practically non existent emission) to LTGs (with a significant, but still, less dominant contribution compared to the diffuse dust) to (U)LIRGs (where the PDR emission is comparable to the diffuse dust emission  in the 5-30~$\mu$m wavelength range, with a clear effect on the shape of the SED in this region). The shape of the SED in this wavelength range can also be affected by the presence of an AGN (purple curve), though it is only a few galaxies ($\sim$12\%) in our sample with strong or moderate AGN activity as revealed by CIGALE, with \mbox{$frac_\text{AGN} >$ 0.2.} Such examples are F13197-1627 and F23254+0830 where a large part of their SEDs (even at FIR wavelengths) is dominated by the AGN component (see their SEDs in Fig.~\ref{all_SEDs}). It is worth noting though that in the MIR wavelength range, where both the emission from the PDRs and the AGN contribute, a degeneracy between the two components may be present, especially in cases of strong AGNs.
   
   The, observed, stellar SED, on the other hand, shows the opposite trend, with the peak (measured at 1 $\mu$m) being the highest for ETGs, slightly lower (by 0.23 dex) for LTGs and much lower (by 1.95 dex) for (U)LIRGs (compared to ETGs). This happens, of course, when the dust effects are considered, and so stellar light is re-processed under different levels of attenuation, depending on the dust content in different galaxy populations. This picture changes substantially when the unattenuated stellar light is considered (see the thin dotted lines in the SEDs in the bottom-right panel in Fig.~\ref{stacks_types}). The most obvious change is seen in the FUV (0.15 $\mu$m) where the (U)LIRGs show the highest emission, followed by the LTGs (lower by 0.37 dex) and with the ETGs showing about two orders of magnitude less emission (1.95 dex) compared to (U)LIRGs. The effects of dust attenuation are better seen in the UV wavelengths, with the ETGs showing only a very small change between attenuated and unattenuated curves (a difference of 0.28 dex at 0.15 $\mu$m; a combination of, both, small amounts of dust and small fraction of young stars). These effects are more prominent for LTGs where a decent amount of dust is present (a change by 0.78 dex), while it is severe for (U)LIRGs (a change of 2.44 dex between attenuated and unattenuated curves) where large amounts of dust dim the stellar light. What is also interesting is that, in (U)LIRGs, the attenuation by the dust starts to become significant shortwards of $\sim$2$~\mu\text{m}$ and so both stellar populations (old and young) are significantly affected (better seen in the bottom-left panel of Fig.~\ref{stacks_types}). For LTGs the dust effects on the stellar populations are seen shortwards of $\sim$1$~\mu\text{m}$, but significantly affects only the young stars (see the top-right panel of Fig.~\ref{stacks_types}), while for ETGs, the effect is negligible, making its presence obvious in wavelengths shortwards of $\sim$0.2$~\mu\text{m}$ with a small change in the emission of the young stars (top-left panel of Fig.~\ref{stacks_types}). 

\subsection{Local (U)LIRGs: a comparison with early- and late-type galaxies}\label{place}   

\begin{figure}[t!]
   \centering
   \includegraphics[width=0.5\textwidth]{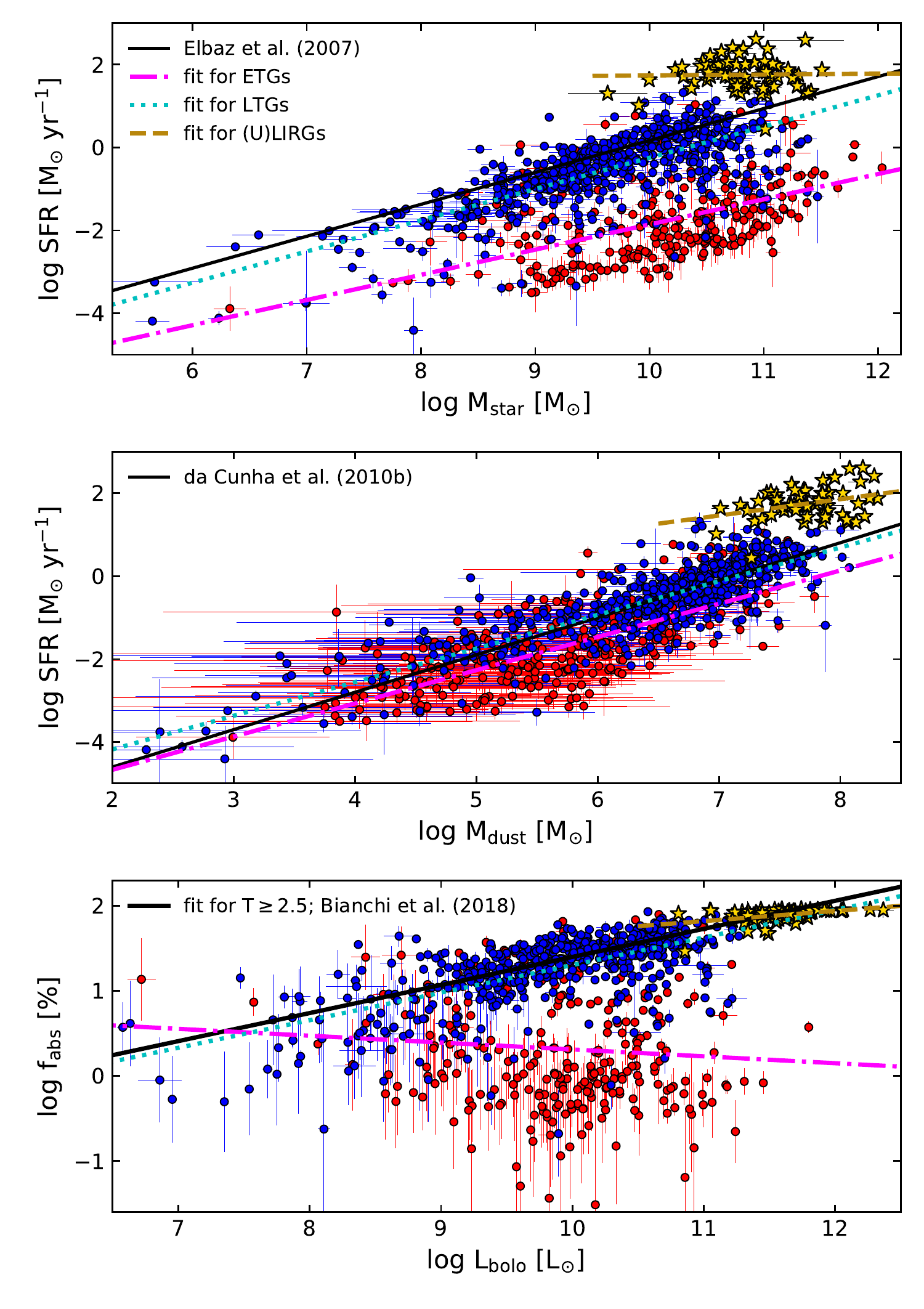}
   \caption{The correlations of \textit{SFR} with stellar mass and dust mass (upper and middle panels respectively), and between $f_\text{abs}$ and $L_\text{bolo}$ (bottom panel; see the text for the definition of the parameters). The DustPedia ETGs and LTGs are shown with red and blue circles, respectively, while yellow stars are the local (U)LIRGs in our sample. All values are plotted along with their corresponding uncertainties. In all plots, magenta dash-dotted, cyan dotted, and yellow dashed lines are the linear fits to the ETGs, LTGs, and (U)LIRGs respectively. The black solid lines correspond to the best fits found in \citet{2007A&A...468...33E} to the 0.015$\leq$z$\leq$0.1 SDSS star-forming galaxies (top panel), to the best fit found in \citet{2010MNRAS.403.1894D} for a sample of low redshift galaxies (middle panel), and to the best fit found in \citet{2018A&A...620A.112B} for the DustPedia galaxies with T $\geq$ 2.5 (bottom panel).}
   \label{fig:main-seq}
   \end{figure}

   The existence of a very tight correlation between the \textit{SFR} and the stellar mass for star-forming galaxies (often referred to as the `main-sequence') is well established, not only for local galaxies \citep[see, e.g.][and references therein]{2007A&A...468...33E, 2012ApJ...754L..29W, 2019A&A...626A..63D} but also for galaxies at high redshifts \citep[see, e.g.][and references therein]{2004MNRAS.351.1151B,2007A&A...468...33E,2011ApJ...742...96W,2012ApJ...760....6M,2015ApJS..219....8C,2018A&A...615A.146P}. Local galaxies of various morphologies (ranging from pure ellipticals to irregulars) occupy different loci in the \textit{SFR}/$M_\text{star}$ plane with the most prominent ones being the two distinct sequences, one for the star-forming galaxies (with high \textit{SFR}s) and one for the more relaxed systems (with low \textit{SFR}s), with a third population occupying the space in between (see, e.g., \cite{2019A&A...626A..63D}, and references therein). This is presented in Fig.~\ref{fig:main-seq} (top panel) with the blue circles being the LTGs (galaxies with T$\geq$0.5) and the red circles being the ETGs (galaxies with T<0.5). A linear regression to the LTGs (cyan dotted line) gives 
   \begin{equation*}
      log(\textit{SFR}[M_{\odot} yr^{-1}]) = 0.75~log(\textit{M}_{star}[M_{\odot}]) - 7.78,
   \end{equation*}
   with a Spearman's correlation coefficient \citep{10.2307/1412159} of $\rho=0.69$, indicating a moderate positive correlation, while the same relation for ETGs (magenta dash-dotted line) is
   \begin{equation*}
      log(\textit{SFR}[M_{\odot} yr^{-1}]) = 0.61~log(\textit{M}_{star}[M_{\odot}]) - 7.93,
   \end{equation*}
   \noindent
   with $\rho=0.44$, indicating a moderate positive correlation, as well. For comparison we overplot the relation for SDSS star-forming galaxies at 0.015$\leq$z$\leq$0.1 (black solid line) found in \citet{2007A&A...468...33E}. Overall we find a good agreement between our fit to the star-forming galaxies and that of \citet{2007A&A...468...33E}, with the slopes differing slightly (0.77 in \citet{2007A&A...468...33E} compared to 0.75 in the current study) and with an offset of 0.4 dex at $M_\text{star}$= 5$\times10^9$ M$_{\odot}$, which is to be expected, though, since the galaxies in \citet{2007A&A...468...33E} extend to larger redshifts. The evolution with redshift is well studied by several works \citep[e. g.][]{Whitaker2014,Johnston2015,Schreiber2015}.
   
   Completing the picture in the local Universe, we include the nearest (U)LIRGs (yellow stars). What is immediately evident is that these systems are amongst the most massive (with stellar masses above \mbox{$\sim10^{10}$ M$_{\odot}$}) and the most actively star-forming ones, with a clear threshold in \textit{SFR} above \mbox{$\sim10$ M$_{\odot}~yr^{-1}$}, covering the high end of the main-sequence. A linear regression to this population gives
   \begin{equation*}
      log(\textit{SFR}[M_{\odot} yr^{-1}]) = 0.02~log(\textit{M}_{star}[M_{\odot}]) + 1.52
   \end{equation*}
   \noindent
   (yellow dashed line). For this population of local galaxies we see an almost flat relation between $M_{\rm star}$ and \textit{SFR} (slope of 0.02) indicating no evident correlation. This is also supported by the value of the Spearman's correlation coefficient being very close to zero ($\rho=-0.1$) indicating a very weak correlation. Such a behaviour is to be expected since the parameter space for this population of galaxies is somehow limited with their values bounded by their high-end $M_{\rm star}$ and \textit{SFR} extremes. Similar results are also presented in previous studies analysing samples of local luminous galaxies, e.g., \citet{2010A&A...523A..78D}, and \citet{2014ApJ...797...54K}
   
%%%%%%%%%%%%%%%%%%%%%%%%%%%%%%%%%%%%% FIGURE 3
   
   A similar behaviour is seen, even more clearly, by comparing the \textit{SFR} with the dust mass (middle panel in Fig.~\ref{fig:main-seq}). Here, the different symbols are as in the top panel. It is obvious that (U)LIRGs are a separate population of local galaxies with both the \textit{SFR} (as mentioned above) and the dust mass occupying the high ends of these quantities (with $M_\text{dust}$ being above \mbox{$\sim 10^7$ M$_{\odot}$}). A linear regression to the DustPedia LTGs in the $M_\text{dust}$ \textit{SFR} plane gives
   \begin{equation*}
      log(\textit{SFR}[M_{\odot} yr^{-1}]) = 0.81~ log(\textit{M}_\text{dust}[M_{\odot}]) - 5.8,
   \end{equation*}
   \noindent
   (cyan dotted line) with a Spearman's correlation coefficient of $\rho=0.8$, indicating a strong correlation, while this relation for the ETGs becomes 
   \begin{equation*}
      log(\textit{SFR}[M_{\odot} yr^{-1}]) = 0.8~ log(\textit{M}_\text{dust}[M_{\odot}]) - 6.27,
   \end{equation*}
   (magenta dash-dotted line) with $\rho=0.67$ indicating a moderate correlation.
   \noindent
   Our findings are in fair agreement with the relation found in \citet{2010MNRAS.403.1894D} studying a sample of star-forming galaxies from the SDSS at z$\leq$0.22 (see the black solid line in the middle panel of Fig.~\ref{fig:main-seq}) with the slope of the linear regression in their sample (0.9) being slightly higher than the one we find (0.81 for the LTGs).
   
   As expected, local (U)LIRGs occupy a different regime of the \textit{SFR}-$M_\mathrm{dust}$ space, which is also indicated by the following best-fit relation,
   \begin{equation*}
      log(\textit{SFR}[M_{\odot} yr^{-1}]) = 0.39~ log(\textit{M}_\text{dust}[M_{\odot}]) - 1.29
   \end{equation*}
   \noindent
   (yellow dashed line) with the slope (0.39) indicating a much flatter distribution with respect to the LTGs and ETGs (slopes of 0.81 and 0.8 respectively). This is also supported by the Spearman's correlation coefficient being very low ($\rho=0.22$) indicating a weak monotonic increase of \textit{SFR} with $M_{\rm dust}$.
   
   Although local (U)LIRGs exhibit increased \textit{SFR} with respect to the normal star-forming galaxies (by more than an order of a magnitude; see the transition from blue circles to yellow stars in the top and middle panels of Fig.~\ref{fig:main-seq}) this is not the case for the ratio of dust luminosity to the bolometric luminosity ($f_\text{abs}=L_\text{dust}/L_\text{bolo}$). This quantity \citep[discussed in detail in][]{2018A&A...620A.112B} indicates the significance of the dust in galaxies and the effectiveness of the dust grains in absorbing the stellar radiation (a combination of the total amount of dust, the geometry and the strength of the ISRF), which, in turn, is re-emitted at FIR/submm wavelengths. This is shown in the bottom panel of Fig.~\ref{fig:main-seq} with $f_\text{abs}$ plotted against the bolometric luminosity ($L_\text{bolo}$). Here, the different symbols are as in the top panel. A linear regression to the data gives
   \begin{equation*}
      log(\textit{f}_\text{abs}[\%]) = 0.32~log(\textit{L}_\text{bolo}[L_{\odot}]) - 1.94,
   \end{equation*}
   for LTGs (cyan dotted line) and,
   \begin{equation*}
      log(\textit{f}_\text{abs}[\%]) = -0.08~log(\textit{L}_\text{bolo}[L_{\odot}]) + 1.12,
   \end{equation*}
   for ETGs (magenta dash-dotted line) with Spearman's correlation coefficients of 0.62 and -0.07, respectively, indicating a moderate correlation of $f_\text{abs}$ increasing with $L_\text{bolo}$ for LTGs, but a very weak decrease for ETGs. The black solid line is the best-fit linear relation for LTGs with T$\geq$2.5, as originally presented by \citet{2018A&A...620A.112B}, almost identical to what we find (including galaxies with 0.5$\leq$T$<$2.5). Even though the local (U)LIRGs occupy the high end of the bolometric luminosity, their $f_\text{abs}$ values show a smooth transition from the respective values of the star-forming galaxies (see the differences among the different styles of data-points in the bottom panel of Fig.~\ref{fig:main-seq}), reaching a plateau for large values of $L_\text{bolo}$. This plateau is expected since the $f_\text{abs}$ cannot exceed 100\% and is also indicated by the slope of the best-fit to the data (yellow dashed line),
   \begin{equation*}
      log(\textit{f}_\text{abs}[\%]) = 0.12~log(\textit{L}_\text{bolo}[L_{\odot}]) + 0.51.
   \end{equation*}
   but also from the Spearman's coefficient of the data (0.46) indicating a moderate correlation of $f_\text{abs}$ increasing with $L_\text{bolo}$. A more detailed comparison of the values of $f_\text{abs}$ between the different populations of local galaxies will follow (Sect.~\ref{sec:dust_heating}).
   
\begin{figure}[t!]
   \centering
   \includegraphics[width=0.5\textwidth]{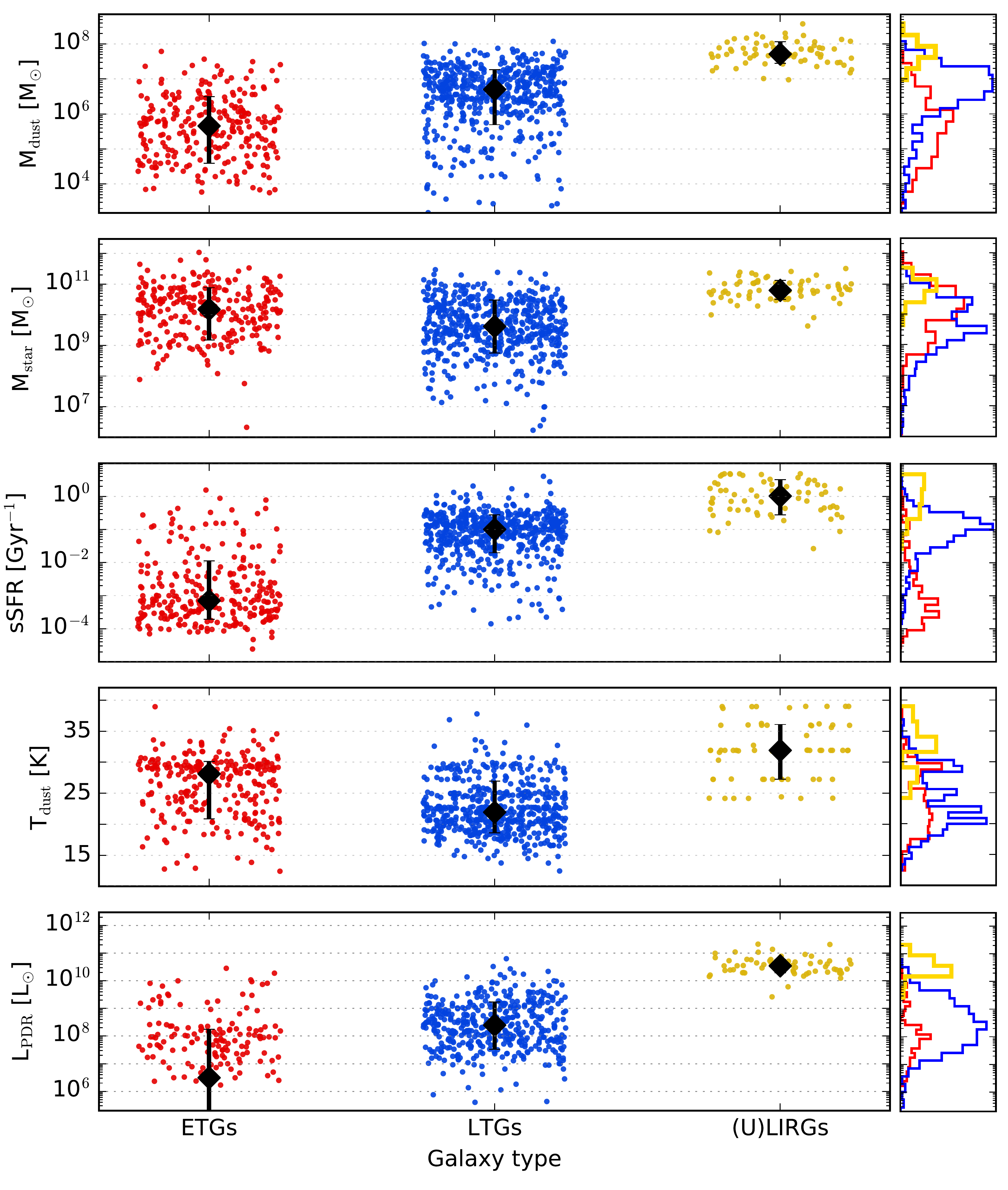}
   \caption{Physical properties, as derived by \textsc{CIGALE}, for the different galaxy types. From top to the bottom $M_\text{dust}$, $M_\text{star}$, \textit{sSFR}, $T_\text{dust}$ and $L_\text{PDR}$ are plotted as a function of galaxy type. Each red and blue dot corresponds to an individual ETG and LTG galaxy, while yellow dots correspond to (U)LIRGs respectively. Black diamonds stand for the median values per galaxy type, while the associated 16$^{th}$ and 84$^{th}$ percentile ranges are indicated with error-bars. Side plots of the distributions of each galaxy type, for all the physical properties are also presented, following the same colouring as for the dots. In the bottom panel ($L_\text{PDR}$) the median value of ETGs is dragged down to lower values because CIGALE predicts no PDR contribution for some sources (see the text for more details).}
   \label{fig:type-props}
   \end{figure}

   In Fig.~\ref{fig:type-props} we have plotted several physical properties of the three samples of galaxies (ETGs, LTGs, and (U)LIRGs) so that we can visualise and quantify the differences. In all cases, red, blue, and yellow symbols correspond to ETGs, LTGs, and (U)LIRGs respectively. The right-hand side plots in each panel show the histograms of the relevant quantity for the three galaxy populations. In addition, in  Appendix~\ref{cumulative}, we have plotted the relevant cumulative distributions of the parameters for each galaxy population (Fig.~\ref{fig:cum_types}) and report the p-values (Tab.~\ref{tab:KStest}) of the Kolmogorov-Smirnov (KS hereafter) tests \cite{smirnov1948}.
   
%%%%%%%%%%%%%%%%%%%%%%%%%%%%%%%%%%%%% FIGURE 4
   
   In the top panel we show the variation in dust mass. As already stated earlier in Sect.~\ref{SED}, the shape of the SEDs already indicate an increase in the dust mass, with ETGs being the most devoid of dust, followed by the LTGs and with a sharp rise of the dust mass for (U)LIRGs. The median values of the three populations are $4.6\times10^5$ M$_{\odot}$, $5\times10^6$ M$_{\odot}$, and 5.2$\times10^7$ M$_{\odot}$ for ETGs, LTGs, and (U)LIRGs respectively. As can also be seen from the histograms on the side plot, but also from the cumulative distributions in Fig.~\ref{fig:cum_types}, $M_\text{dust}$ shows a very different distribution among the three populations occupying different ranges of the parameter.
   
   The second panel, from the top, shows the stellar mass variation among the three populations. It turns out that, although there is a large overlap in masses between ETGs and LTGs, there is a systematic trend with LTGs being the less massive ones and with ETGs and (U)LIRGs  the most massive ones. Median values of the stellar mass are $1.5\times10^{10}$ M$_{\odot}$, $4.2\times10^9$ M$_{\odot}$ and 6.3$\times10^{10}$ M$_{\odot}$ for ETGs, LTGs, and (U)LIRGs respectively. The historgams on the side plot, but also from the cumulative distributions in Fig.~\ref{fig:cum_types}, show that $M_\text{star}$ shows different distribution among the three populations (especially between the (U)LIRGs and the other two). In interacting galaxies, like most of the (U)LIRGs in our sample, high values of stellar mass are expected. This happens not only because of the aggregation of the stellar mass from the individual galaxies that undergo merging, but also due to the intense star-formation of these systems lasting hundreds to thousands of Myrs. This leads to the formation of stars of tens to hundreds of solar masses per year, as also predicted by numerical and hydrodynamical simulations \citep[][]{2008A&A...492...31D, 2009ApJ...696..348W, 2013MNRAS.430.1901H}. Comparing the stellar and dust masses, we find that the dust-to-stellar mass ratio is the lowest in ETGs (a median value of $2.9\times10^{-5}$), while it is very similar in LTGs and (U)LIRGs (median values of $8.9\times10^{-4}$ and 7.5$\times$10$^{-4}$, respectively). This further supports the argument that the enhanced star-formation activity in local (U)LIRGs is mainly driven by external processes (interactions) rather than their intrinsic properties.
   
   The specific star-formation rate (\textit{sSFR}) is defined as the current SFR over the stellar mass of the galaxy (\textit{SFR}/$M_\text{star}$). A useful interpretation of this quantity is to think of $M_\text{star}$ as the cumulative result of the star-formation that occurred in the past. Under this assumption, \textit{sSFR} is then a measure of how intensively the galaxy forms stars now compared to how it used to form stars in the past. In our analysis we have calculated the \textit{sSFR} for the three populations and presented in the third, from the top, panel. We clearly see that the three kinds of systems show a very different behaviour with ETGs having very low values of \textit{sSFR} (a median value of $7\times10^{-4}$ Gyr$^{-1}$), followed by the LTGs (with a median value of 0.1 Gyr$^{-1}$), with (U)LIRGs having very high values (a median value of 1 Gyr$^{-1}$). Three distinct distributions are also seen in the side histograms but also from the quite deviant cumulative distributions in Fig.~\ref{fig:cum_types} This behaviour is mainly driven by the variation in \textit{SFR} among the three populations (on average 0.3, 1.2, 81 M$_{\odot}$ yr$^{-1}$ for ETGs, LTGs, and (U)LIRGs, respectively) but also from the variation in stellar mass discussed above. This already shows how actively local (U)LIRGs are currently forming stars compared to earlier stages in their lives compared to ETGs which show a \textit{sSFR} of more than three orders of magnitude less.
   
   As already stated earlier, Sect.~\ref{SED}, the shape of the dust emission SED indicates a notable dust temperature variation among the three different galaxy populations. Following a more quantitative approach we approximate the dust temperature by:
   \begin{equation} \label{eq:aniano}
      T^\text{CIGALE}_\text{dust} = T_0 U^{(1/(4+\beta))}_\text{min},
   \end{equation}
   \noindent \citep{2012ApJ...756..138A, 2019A&A...624A..80N} with $U_\text{min}$ being the minimum level of the interstellar radiation field that is able to heat the dust (in our analysis calculated by CIGALE for each galaxy), $T_0$ the measured, in the solar neighbourhood, dust temperature (18.3 K), and $\beta$ the dust emissivity index [with a value of 1.79 accounting for the \textsc{THEMIS} dust grain model \citep{2017A&A...602A..46J}, assumed here]. The dust temperature for each galaxy, for the three populations, is plotted in the fourth, from the top, panel in Fig.~\ref{fig:type-props}, with the median values, for each galaxy population, indicated by black squares. These median values are 28, 22, and 32 K for ETGs, LTGs, and (U)LIRGs respectively, confirming the earlier findings (Sect.~\ref{SED}) that ETGs and (U)LIRGs can heat the dust into higher temperatures, compared to LTGs where dust is cooler. It is interesting to notice, though, that the distributions of the dust temperature for ETGs and LTGs are quite similar. This can be confirmed by the histograms on the side plot in Fig.~\ref{fig:type-props} but also by the cumulative distribution in Fig.~\ref{fig:cum_types}. Although the dust in ETGs is, on average, hotter than in LTGs (median values of 28 and 22 K respectively) their range and distributions are very similar, something that may indicate a similar mechanism of dust heating. The distribution of the dust temperature in (U)LIRGs, though, is very different from the other two. Although ETGs and (U)LIRGs may heat up the dust grains into similar temperature levels (at least comparing their median values), the source of heating is quite different, with the ETGs using, mostly, their intense NIR radiation field of the old stars while (U)LIRGs use the intense UV radiation field of the young stars to heat up the dust. In LTGs the radiation field is milder heating the diffuse dust into cooler temperatures.

   As already indicated by the different SEDs (Fig.~\ref{stacks_types}) the PDR part of the dust emission becomes an important contributor to the energy output in (U)LIRGs as compared to LTGs and ETGs. This is what we show in the bottom panel in Fig.~\ref{fig:type-props} with the PDR luminosity for each galaxy, as predicted by CIGALE, plotted for the three different galaxy populations. As in the panels above, the black squares indicate the median values in each galaxy population. We have to notice here, though, that in the case of ETGs, CIGALE predicts no PDR contribution for 129 sources (out of 268), dragging the median value to lower values. Our analysis suggests a median PDR luminosity of $3.1\times10^6$ L$_{\odot}$, $1.1\times10^8$ L$_{\odot}$, and 3.5$\times10^{10}$ L$_{\odot}$ for ETGs, LTGs, and (U)LIRGs respectively, which, as expected, shows the lack of the PDR emission in ETGs and the strength of this component in (U)LIRGs. The historgams on the side plot, but also from the cumulative distributions in Fig.~\ref{fig:cum_types}, show that $L_{\rm PDR}$ shows different distribution among the three populations (especially between the (U)LIRGs and the other two). The importance of this component is made more clear when comparing the PDR luminosities ($L_{\rm PDR}$) with the total dust luminosities ($L_{\rm PDR}+$ $L_{\rm diffuse}$). Our analysis suggests that the median values of the ratios of the PDR-to-total dust luminosity increases from 1.6\% for ETGs, to 5.2\% for LTGs, to 11.7\% for (U)LIRGs.
   
   As also presented in Appendix~\ref{cumulative} almost all the KS p-values are $\leq$ 0.15 indicating that there are no two populations that are drawn from the same parent sample, for all the parameters studied. This means that these three galaxy populations in the local universe are totally unrelated in terms of their fundamental physical parameters. The only exception is with the dust temperature between the ETGs and the LTGs which show a high p-value (0.82) indicating large probability that this parameter shows common characteristics between these two galaxy populations. This comes in contrast to the behaviour of local (U)LIRGs which show very low p-values ($<0.15$) when comparing their dust temperature with those of the other two populations, probably indicating that the mechanism of merging events (which is evident in most of these systems) may result in a different efficiency of the way that the dust grains are heated.
   
%%%%%%%%%%%%%%%%%%%%%%%%%%%%%%%%%%%%% FIGURE 5
   \begin{figure*}[t!]
   \includegraphics[width=1\textwidth]{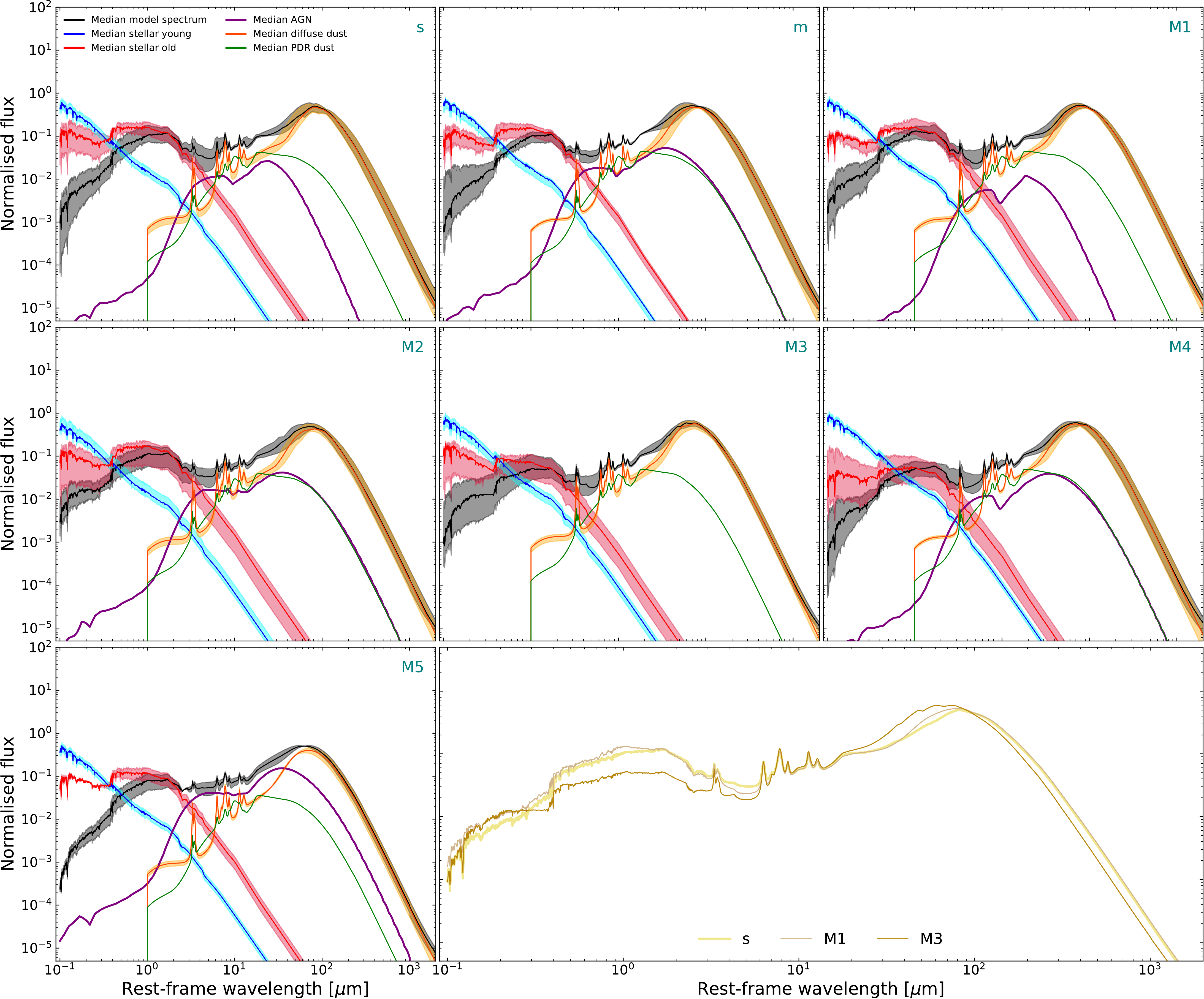}
   \caption{Median SEDs of the local (U)LIRGs, separated in classes based on their merging stage. Classes~`s', `m' and `M1' are presented in the three top panels (from left to right), classes~`M2', `M3, and `M4' in the three middle panels, while in the bottom-left panel the median SED of the `M5' class is plotted. The curves and the shaded areas are coloured in the same manner as in \mbox{Fig.~\ref{stacks_types}}. The bottom-right panel shows the comparison among the median SEDs of `s', `M1' and `M3' classes, allowing for the differences to be spotted.}
   \label{stacks_mergers}
   \end{figure*}
   
\section{Evolution of the physical properties of local (U)LIRGs with merging stage}\label{sec:merg-evol}

   Several simulation studies model the interactions between galaxies and predict the way that \textit{SFR} evolves with time (the SFH).  In \citet{2005MNRAS.361..776S} the authors perform numerical simulations to model the feedback from stars and black holes in galaxies consisting of gas and stellar disks, central bulges and surrounded by dark matter halos, using a Salpeter type initial mass function (IMF). The derived SFH (see their Fig.~14) predicts that at the pre-merging stage (0.71~Gyr) the \textit{SFR} is $\sim10~\text{M}_\odot$~yr$^{-1}$ but during coalescence the \textit{SFR} may raise up to $50~\text{M}_\odot$~yr$^{-1}$ followed by a decrease. \citet{2006MNRAS.373.1013C} investigate the influence of several feedback parameters and models on the \textit{SFR}, for Sbc-like gas-rich galaxies. They find a pre-merging \textit{SFR} of $\sim 30~\text{M}_\odot$~yr$^{-1}$, peaking at $\sim 75~\text{M}_\odot$~yr$^{-1}$ during the coalescence and then decreases, at the post-merging era, to \mbox{$\sim 10~\text{M}_\odot$~yr$^{-1}$}. \citet{2008A&A...492...31D} use data-sets of numerical simulations aiming at studying starburst episodes by galaxy mergers. They use co-planar Sa and Sb galaxies in direct and retrograde encounters, and with varying parameters, like gas fraction, galaxy relative velocity and minimum separation. For different cases the relative \textit{SFR} during the coalescence is at least 2 times higher than during the pre- and the post-merging periods (see their Fig.~4). In a more recent study, \citet{2013MNRAS.430.1901H}, a realistic hydrodynamic simulation investigates the star-formation in galaxy mergers with Milky Way-like galaxies at z=0. The authors find \textit{SFR} $\sim 3~\text{M}_\odot$~yr$^{-1}$ before the main merging event peaking at $\sim 50~\text{M}_\odot$~yr$^{-1}$ and finally, dropping to $\sim 5~\text{M}_\odot$~yr$^{-1}$ after coalescence. With an infrared luminosity of 6.2$\times10^{10}$ L$_{\odot}$ the Antennae Galaxies (Arp~244) is the nearest IR-bright and perhaps the youngest prototypical galaxy-galaxy merger system \citep{2001ApJ...548..172G}. Even if, strictly speaking, Arp~244 is not a LIRG \citep[unless a Virgocentric flow distance of 29.5 Mpc is considered, leading to $L_\text{IR} = 1.3\times10^{11}$ L$_{\odot}$;][]{2001ApJ...548..172G} it is an interesting system that its merging sequence has been studied and modelled in detail \citep[using a hydrodynamical simulation;][]{2015MNRAS.446.2038R}. In this simulation two main bursts of star-formation are predicted, the first starburst occuring at $\sim$ 20 Myr after the first pericentric passage while the second episode of star-formation taking place at $\sim$ 170 Myr after the first pericentric passage. During the first starburst the SFR can reach up to $\sim$ 80 M$_{\odot}$ yr$^{-1}$ which gets higher, up to $\sim$ 110 M$_{\odot}$ yr$^{-1}$, during the second starburst event.
   
   The simulations mentioned above may be derived for specific parameter spaces and specific systems, yet, they are indicative of what to expect during a merging event. In our analysis it is not possible to trace the history of the recent star-formation for each galaxy system but we do see a `snapshot', along this timeline, for each system, depending on how far the interaction has evolved. The rate that stars are formed during merging events depends on the exact content of the interacting galaxies in gas, stars and dust \citep{1978ApJ...219...46L, 1996giis.book.....K, 2006asup.book..115S}, as well as the morphological types \citep{1968ApJ...151..547T}, and the relative inclinations and velocities \citep{2008A&A...492...31D} of the merging galaxies. The grouping that we have adopted for the seven merging stages (see Sect.~\ref{sec:sample} and Table~\ref{table1}) allows us to have a broad representation of different stages along the merging sequence (from single galaxies, to galaxies in long-separated distances and galaxies in coalescence). We take advantage of this grouping and investigate possible differences in the SEDs and physical properties of the systems in our sample. We caution, however, the reader again, as we have already done in Sect.~\ref{sec:sample}, that some classes (e.g. `m' and 'M5') are underrepresented and the robustness on the derived quantities may be questionable.
   
\begin{figure}[t!]
   \centering
   \includegraphics[width=0.5\textwidth]{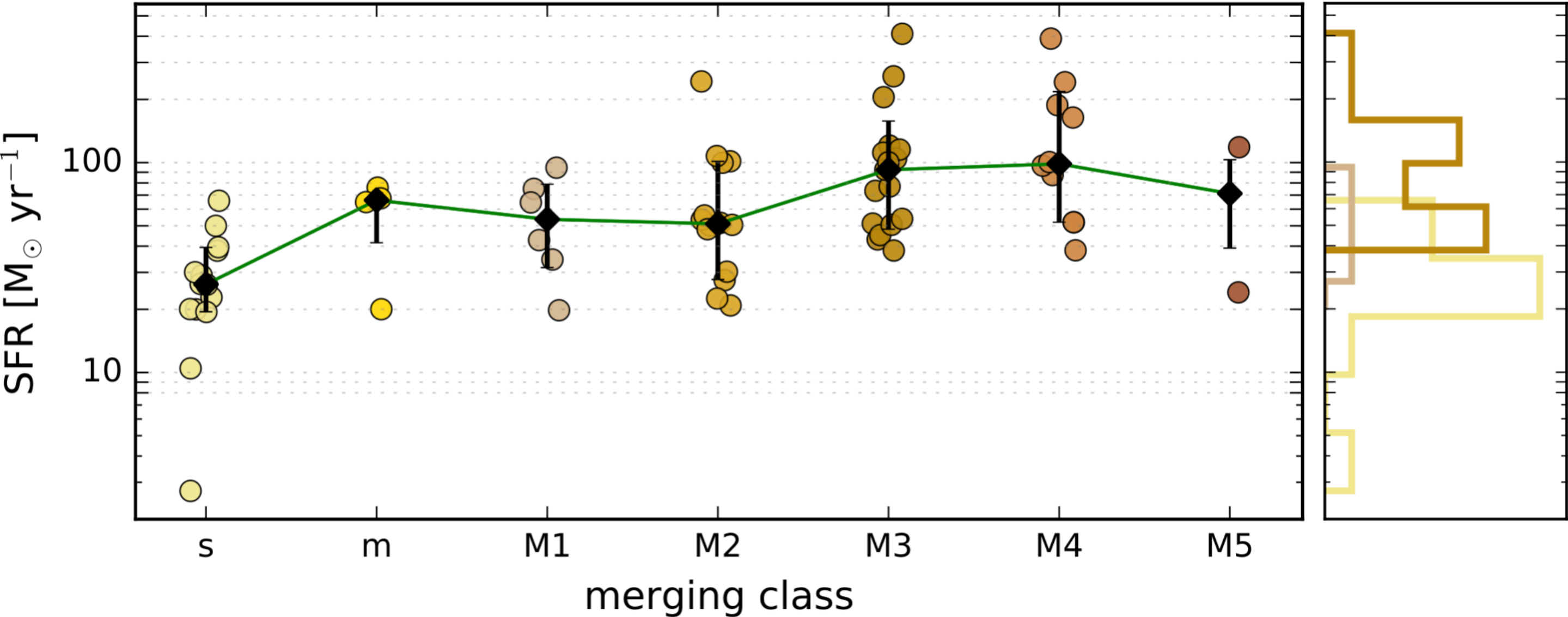}
   \caption{\textit{SFR} of (U)LIRGs in different merging stages, as derived from \textsc{CIGALE}. Each circle corresponds to an individual source. Black diamonds stand for the median values per merging class, while error bars indicate the range between the 16$^{th}$ and 84$^{th}$ percentiles from the median. Side plots of the distributions of class `s', `M1' and `M3' systems, with the corresponding colour, are also presented.
   \label{sfr_per_class}}
   \end{figure}

\begin{figure}[t!]
   \centering
   \vspace{-0.12cm}\includegraphics[width=0.5\textwidth]{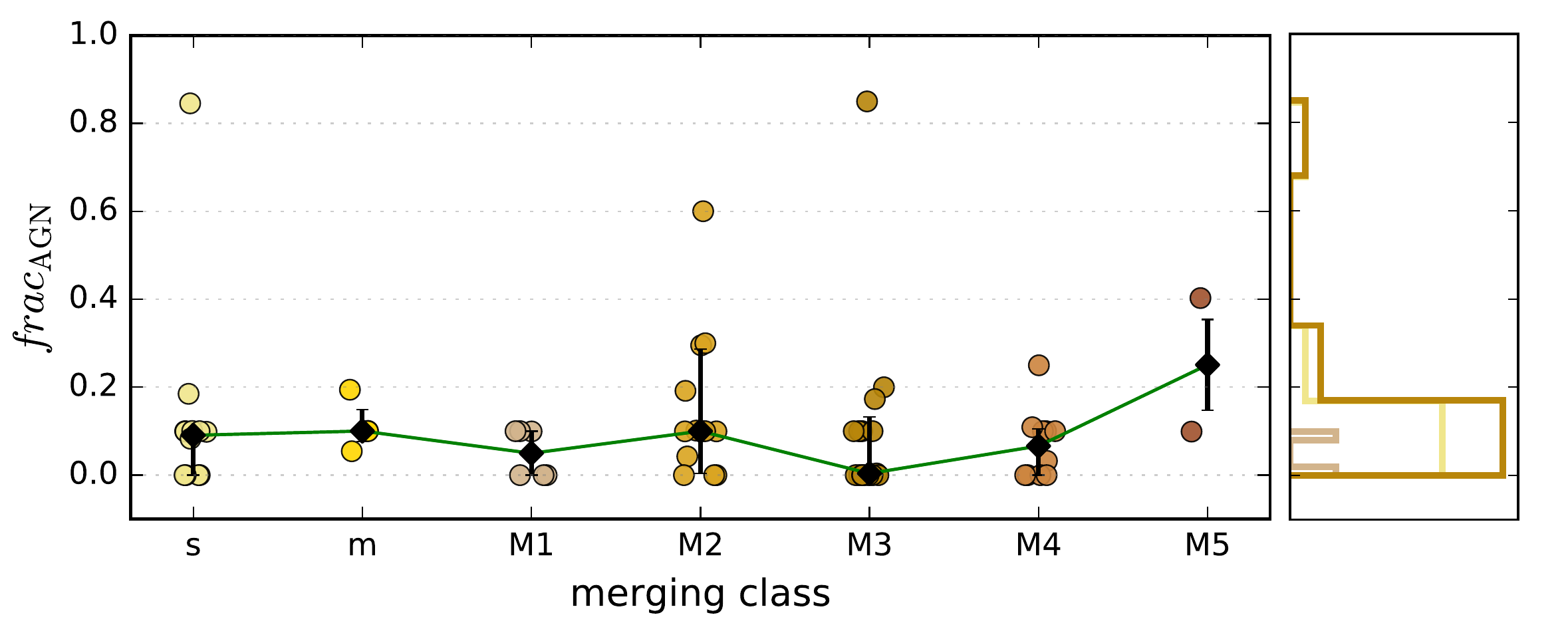}
   \caption{AGN fraction of (U)LIRGs in different merging stages, as derived from \textsc{CIGALE}, along with distributions for the class `s', `M1' and `M3' systems. The style of markers is the same as in Fig.~\ref{sfr_per_class}
   \label{fagn_per_class}}
   \end{figure}

\begin{figure}[t!]
   \centering
   \includegraphics[width=0.5\textwidth]{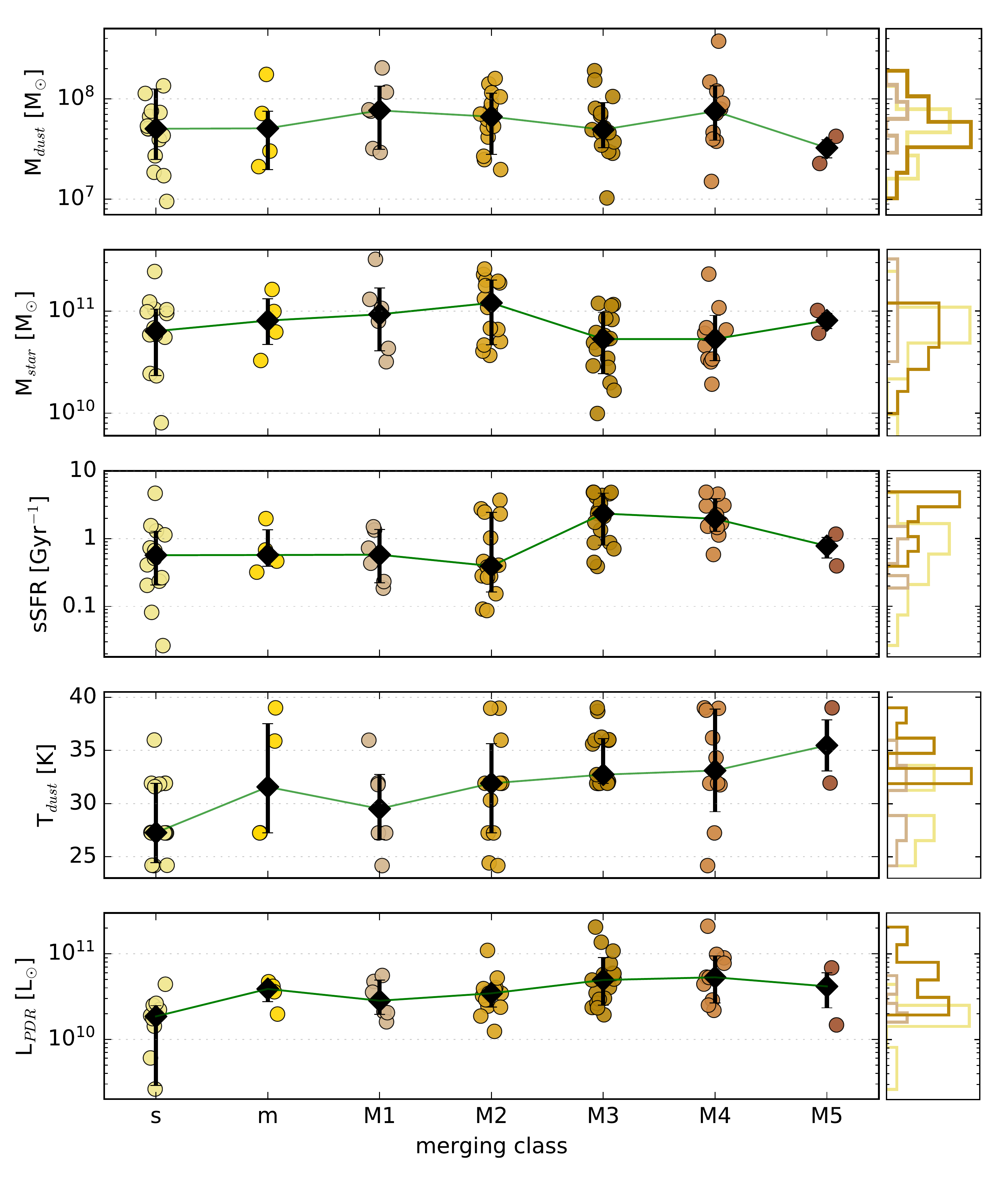}
   \caption{Physical properties, as derived by \textsc{CIGALE}, for the four different merging classes of (U)LIRGs. From top to the bottom $M_\text{dust}$, $M_\text{star}$, \textit{sSFR}, $T_\text{dust}$ and $L_\text{PDR}$ are plotted as a function of merging class. Each circle (following the colour coding of Fig.~\ref{sfr_per_class}) corresponds to an individual system. Black diamonds stand for the median values per galaxy type, while the associated 16$^{th}$ and 84$^{th}$ percentile ranges indicated with error-bars. Distributions of class `s', `M1' and `M3' systems, with the corresponding colour, are also presented as side plots.
   \label{fig:clas-props}}
   \end{figure}

   The different SEDs of the seven merging classes are presented in Fig.~\ref{stacks_mergers} with classes `s', `m', and `M1' on the top-row panels, classes `M2', `M3', and `M4' on the middle row-panels, and class `M5' on the bottom-left panel. For each sub-class the different components are also indicated in the same manner as in Fig.~\ref{stacks_types}. The bottom-right panel, on the other hand, compares the total SEDs of the different merging classes. In order to avoid confusion in the plot we choose to plot three of the merging classes only, namely, `s', `M1' and `M3' which give a broader description of totally isolated galaxies, galaxies in the first stages of the merging, and galaxies in advanced merging, respectively. The different merging classes are shown with different colours as indicated in the inset of this panel. It is true that the differences are very small but sufficient to explain the variations seen in the physical parameters (see Figs.~\ref{sfr_per_class}, \ref{fagn_per_class}, and \ref{fig:clas-props} and the discussion related to these figures). Comparing the SEDs in the bottom-right panel, one can spot two evident differences, a variation in the FUV to the MIR ($\sim 10~\mu$m) wavelength range, and a small shift of the peak of the FIR emission and the Rayleigh-Jeans tail of the dust emission. These differences already indicate differences in stellar masses, dust masses, dust temperatures and \textit{SFR} which are discussed, in detail, below where the variation in the different parameters with merging stage is investigated.
   
   The change in \textit{SFR} with the merging class is visualised in Fig.~\ref{sfr_per_class}. Here, the yellow-coloured circles are the values for individual sources in each sub-class with the median value indicated as black squares. The error-bars bracket the 16$^{th}$ and 84$^{th}$ percentiles from the median. The green line connects the median values and indicates the general trend. The side-plots show the histograms of the \textit{SFR} for the three merging classes, `s', `M1', and `M3'. We see that, although there is a large scatter in each sub-class among the different sources, there is a clear trend with the maximum median \textit{SFR} occurring at sources of class `M4' (99 M$_{\odot}$ yr$^{-1}$), followed by class `M3' (93 M$_{\odot}$ yr$^{-1}$), with the lowest \textit{SFR} at class `s' (26 M$_{\odot}$ yr$^{-1}$) and class `M2' (51 M$_{\odot}$ yr$^{-1}$) with the rest of the classes obtaining intermediate median values (66 M$_{\odot}$ yr$^{-1}$, 54 M$_{\odot}$ yr$^{-1}$, and 71 M$_{\odot}$ yr$^{-1}$, for classes `m', `M1', and `M5' respectively). This general behaviour is to be expected since galaxies in `s' and `M2' classes are more relaxed systems [either separated (class `M2') or totally isolated (class `s')] with their \textit{SFR} being driven mainly by internal processes or by past minor merging events which are not as powerful mechanisms as the tidal disruptions that take place during major merger interactions. It should be noted, however, that if the nuclear \textit{SFR} is considered, as opposed to the global \textit{SFR} treated here, a more obvious, increasing, trend with merging stage is evident \citep{2019ApJ...871..166U}. The change in \textit{SFR} is also evident in the individual SEDs of the different sub-classes. Looking carefully at the median young-stellar SEDs in each plot of Fig.~\ref{stacks_mergers} (the blue curves) we see that there is an obvious enhancement of the young stellar population for class `M3' and `M4' systems (2$^{nd}$ and 3$^{rd}$ middle panel, from the left), followed by class `M1' systems (top-right panel), compared to class `s' and `M5' systems (upper-left and bottom-left panels) which show the lowest content in young stars. This can be appreciated either by comparing the maximum values of the blue curves, or by comparing the relative contribution of the young and old stellar components (comparison of the blue and red curves, more evident in the MIR wavelength range). This picture is in accordance with previous studies. Already from the IRAS era, observations suggested that interactions in merging systems enhance the rate that stars are formed \citep{1987ARA&A..25..187S, 1987AJ.....93.1011K}. In \citet{1996ARA&A..34..749S} the authors find a clear maximum in the infrared luminosity produced by LIRGs in the stage where their nuclei merge. Similarly, \citet{2011AJ....141..100H} report that LIRGs in late merging stages posses total IR luminosities larger by a factor of two than pre- or non-merging LIRGs. The median \textit{SFR} per merging class, extracted from our SED modelling, follows a similar trend as the simulations suggest. Despite the large scatter in the \textit{SFR} all the simulations seem to suggest an enhancement in \textit{SFR} close to coalescence (our merging classes `M3' and `M4') with lower \textit{SFRs} in the other stages of the interaction where the parent galaxies are either isolated `s' or apart from each other (`M1' and `M2') or, even, systems that have been evolved to isolated galaxies (class `M5'). This trend is also evident when comparing the p-values of the KS tests for all combinations of the merging types (see Table~\ref{tab:KStest}). If we do not consider types of `M5', we see that `s' types show very different distributions from all the rest giving low p-values in \textit{SFR}, while for the rest of the combinations give low p-values with `M3' and `M4' types indicating very different distributions. The combination of `M3' and `M4' though give a p-value of 0.84 suggesting very similar distributions (with a probability that they are coming from the same parent population of 84\%).
   
   %%%%%%%%%%%%%%%%%%%%%%%%%%%%%%%%%%%%% FIGURE 6
   
   The strength of the AGN in the systems of our sample, as derived by CIGALE, is presented, by merging class, in Fig.~\ref{fagn_per_class}. We see in this plot that, although the values of the AGN fraction are, in general low ($\leq$ $0.2$) there is a larger scatter of the values in `M2' and `M3' types with two of the three strongest AGNs ($frac_\text{AGN}$$>$$0.6$) appearing at these groups. We are aware that the AGN fraction range is limited and probably a larger sample is needed. This is also shown with the KS tests performed where all p-values values in $frac_\text{AGN}$ being relatively high (see Table~\ref{tab:KStest}), indicating that this parameter shows high probability that the distributions come from the same parent distribution in all combinations of merging stages. Nevertheless, our results agree with those presented in \citet{2011ApJ...730...28P}, where they report no strong trends of the $frac_\mathrm{AGN}$ with merging stage, but they observe an increase in the number of AGN dominated sources in the latest stages. In the current work, we do not find any trend of $frac_\mathrm{AGN}$ with merging stage and systems with strong AGN ($frac_\mathrm{AGN}$$>$$0.2$) lie in stages later than `M2' (with F13197-1627 being the only exception, since it is a `s' system). It is also worth noting that `M3' sources exhibit lower median emission by the AGN component. This is obvious in Fig.~\ref{stacks_mergers}, where the median AGN template is absent in this merging class, while there is an indication (given the small number of sources) of a mild dip in the median value of $frac_\mathrm{AGN}$ of the same class of objects in Fig.~\ref{fagn_per_class}. \citet{2005MNRAS.361..776S}, found that during the main merging event starburst and AGN co-exist in interacting systems. They also claim that tidal forces are able not only to trigger a nuclear starburst but also to fuel rapid growth of the black holes. Thus, although the interacting system is both starburst and AGN, it is likely that the AGN activity would be obscured by gas and dust that surrounds the nucleus. At later stages, when outflows remove the dense gas layers during the final stages of the coalescence, the remnant could be visible as an AGN. Our results, and more specifically the absence of the AGN component in class `M3', may indicate a similar scenario.
   
   Apart from the current \textit{SFR} and the AGN fraction it is interesting to investigate how other physical properties vary for galaxies of different merging stages. We do that in Fig.~\ref{fig:clas-props} where the change in $M_\text{dust}$, $M_\text{star}$, \textit{sSFR}, $T_\text{dust}$, and $L_\text{PDR}$ with merging class is plotted (top to bottom). On the side-plots the histograms of the parameters for three merging classes, `s', `M1', and `M3', are also presented. The symbols are as in Fig.~\ref{sfr_per_class}. Looking at the measurements of the individual systems we see that there is a large scatter. The median values, though, suggest a few trends with merging stage which worth investigating further. 
   
   The median values of $M_\text{dust}$ suggest that this parameter remains practically unchanged with merging stage, with median values of \mbox{5.0$\times10^7$ M$_{\odot}$},  \mbox{5.1$\times10^7$ M$_{\odot}$}, \mbox{7.7$\times10^7$ M$_{\odot}$}, \mbox{6.6$\times10^7$ M$_{\odot}$}, \mbox{4.9$\times10^7$ M$_{\odot}$}, \mbox{7.5$\times10^7$ M$_{\odot}$} and \mbox{3.3$\times10^7$ M$_{\odot}$} for merging classes `s', `m' and `M1' to `M5' respectively. From the side, histogram plots, but also from the cumulative distributions in Appendix~\ref{cumulative} one can see that all the distributions of $M_\text{dust}$ are very similar and well within the scatter of each individual group. This can also be confirmed by the respective p-values of the KS tests with all of them being large, for all combinations of merging stages (see Table~\ref{tab:KStest}), indicating large probability that all the distributions of the dust mass may originate from the same parent distribution.

%%%%%%%%%%%%%%%%%%%%%%%%%%%%%%%%%%%%% FIGURE 7
   
   $M_\text{star}$ shows a mild change with objects with merging stages `M3' and `M4' being less massive. The median values are \mbox{6.4$\times10^{10}$ M$_{\odot}$}, \mbox{8.1$\times10^{10}$ M$_{\odot}$}, \mbox{9.3$\times10^{10}$ M$_{\odot}$}, \mbox{1.2$\times10^{11}$ M$_{\odot}$}, \mbox{5.3$\times10^{10}$ M$_{\odot}$}, \mbox{5.3$\times10^{10}$ M$_{\odot}$} and \mbox{8.1$\times10^{10}$ M$_{\odot}$} for merging classes `s', `m' and `M1' to `M5'. Comparison of the median SEDs in Fig.~\ref{stacks_mergers} suggest that class `M3' and `M4' systems show a deficit in the old stellar population (evident in the NIR wavelength range) compared to the rest of the merging classes (this difference is more clearly seen in the bottom-left panel of Fig.~\ref{stacks_mergers} by comparing the SED of `M3' with the other two SEDs). This is better seen (and will be discussed later) in Fig.~\ref{lum_over_bolo} (left panel) where the histograms of the relative contribution of the old and young stellar populations are presented. The deficit of the old stars in these systems (responsible for the bulk of the stellar mass) results in the slightly lower stellar mass observed. This kind of trend is also confirmed by comparing the p-values of the KS tests for $M_\text{star}$ in Table~\ref{tab:KStest}. We see that all combinations of merging stages exhibit relatively large p-values providing high probabilities that the distributions originate from the same parent distribution with the exception being the p-vales of `M2' with `M3' and `M4' which show low values indicating very different distributions.
   
   The clearest, and most significant, change (compared to the rest of the parameters examined here) is seen in the \textit{sSFR}. The median values of this parameter are \mbox{0.57 Gyr$^{-1}$},  \mbox{0.57 Gyr$^{-1}$}, \mbox{0.58 Gyr$^{-1}$}, \mbox{0.39 Gyr$^{-1}$}, \mbox{2.35 Gyr$^{-1}$}, \mbox{1.96 Gyr$^{-1}$} and \mbox{0.78 Gyr$^{-1}$} for merging classes `s', `m' and `M1' to `M5' respectively. The combination of the enhanced \textit{SFR} (Fig.~\ref{sfr_per_class}) and deficit in stellar mass (Fig.~\ref{fig:clas-props}) of classes `M3' and `M4' systems make these merging classes differentiate from the rest. Considering that \textit{sSFR} is a measure of the current over the past \textit{SFR} suggests that these classes of objects (undergoing, or, have gone through, a major merging event) show the most active current star-formation activity. This effect is clearly seen in the derived p-values of the KS tests with only the combinations including classes `M3' and `M4' showing low values indicating that their distributions differ substantially from the rest. The p-value, on the other hand, of the combination of these two merging classes is large (0.87) indicating high confidence that these two distributions are similar. This is also supported by the dust-to-stellar mass ratio that is comparable in all the merging classes 6.7$\times10^{-4}$, 8.6$\times10^{-4}$, 8.1$\times10^{-4}$, 5.9$\times10^{-4}$, 1.1$\times10^{-3}$, 1.2$\times10^{-3}$, 4.0$\times10^{-4}$, for class `s', `m' and `M1' to `M5' objects respectively, indicating that the variance in the star-formation activity is closer related to the merging stage rather than the stellar and dust content of the galaxies.
   
%%%%%%%%%%%%%%%%%%%%%%%%%%%%%%%%%%%%% FIGURE 8
   
   A mild change in the dust temperature is also obvious with the more relaxed, isolated, galaxies showing colder dust temperatures. The median dust temperatures are 27.3 K, 31.6 K, 29.5 K, 31.9 K, 32.7 K, 33.1 K and 35.5 K, for each merging class, from `s' to `m' and `M1' to `M5', respectively. The colder dust temperature that is seen in `s' systems can also be explained from the SEDs. As can be seen from the lower-right panel in Fig.~\ref{stacks_mergers}, following the merging stage evolution from `s' to `M1' and `M3', a shift of the dust peak is obvious towards shorter wavelengths which translates to hotter temperatures. The p-values of the KS tests for the dust temperature are generally high, indicating large probability that the distributions may originate from the same parent population (Table~\ref{tab:KStest}) with only a few exceptions, mainly involving `M3' and `M4' merging classes.
   
   Finally, the PDR luminosity varies slightly for different merging stages with values of \mbox{1.9$\times10^{10}$ L$_{\odot}$}, \mbox{3.9$\times10^{10}$ L$_{\odot}$}, \mbox{2.8$\times10^{10}$ L$_{\odot}$}, \mbox{3.5$\times10^{10}$ L$_{\odot}$}, \mbox{4.9$\times10^{10}$ L$_{\odot}$}, \mbox{5.3$\times10^{10}$ L$_{\odot}$} and \mbox{4.2$\times10^{10}$ L$_{\odot}$}, for each merging class from `s' to `M5'. A slight enhancement of $L_{\rm PDR}$ is seen in merging classes `M3' and `M4' compared to the rest. This difference is more notable when comparing the distributions of `M2' class systems with `M3' and `M4' with the p-values in the KS tests giving very small values indicating different populations.
   
%%%%%%%%%%%%%%%%%%%%%%%%%%%%%%%%%%%%% FIGURE 9
   \begin{figure}[t!]
   \centering
   \includegraphics[width=0.45\textwidth]{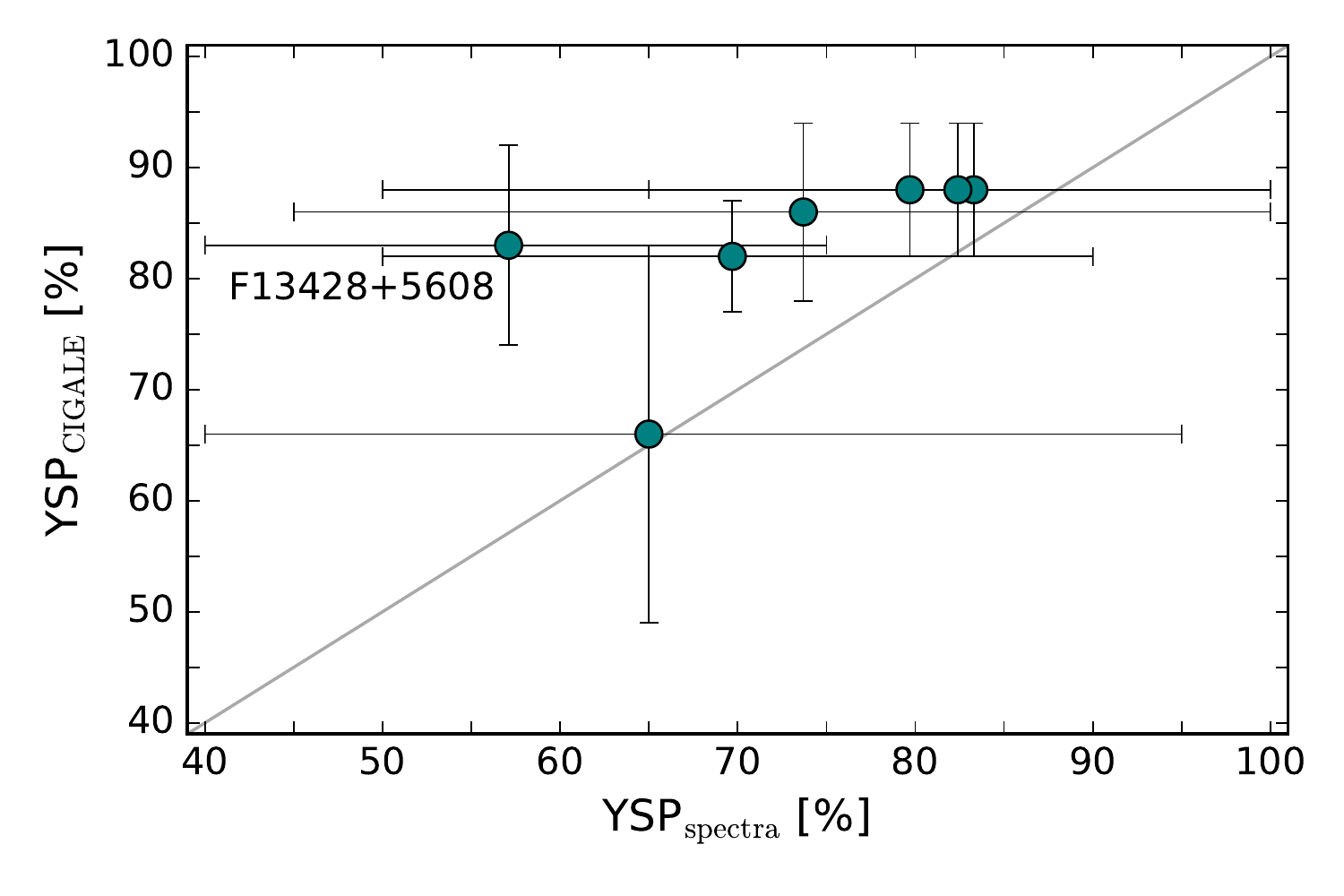}
   \caption{Comparison of the fraction of the young stellar population in seven systems in common between the sample studied here (YSP$_{\rm{CIGALE}}$) and the sample of \citet{2009MNRAS.400.1139R}. In \citet{2009MNRAS.400.1139R} the young stellar populations (YSP$_{\rm{spectra}}$) have been derived through spectral synthesis population modelling of optical long-slit spectra (see the text for more details). 
   \label{YSP}}
   \end{figure}

\section{Old and young stellar populations in (U)LIRGs and their role in dust heating}\label{sec:dust_heating}

%%%%%%%%%%%%%%%%%%%%%%%%%%%%%%%%%%%%% FIGURE 10
   \begin{figure*}[t!]
   \centering
   \includegraphics[width=\textwidth]{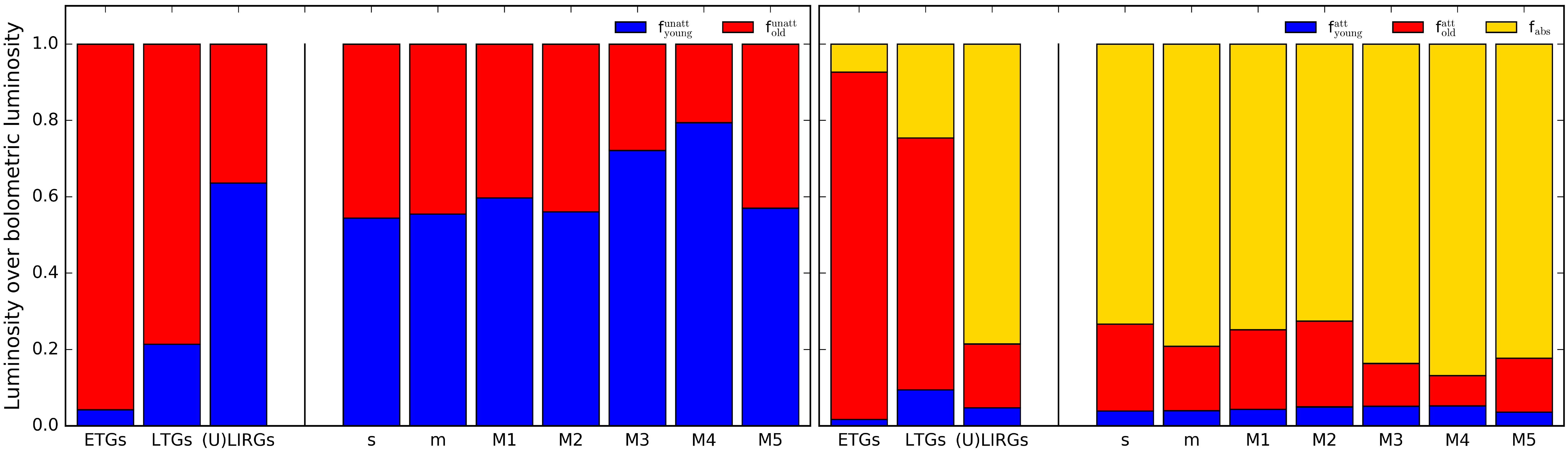}
   \caption{Left panel: The contribution of the old (red) and young (blue) stellar populations to the unattenuated luminosity, over the bolometric luminosity, per galaxy type and (U)LIRGs merging class. Right panel: The contribution of the old and young stellar populations to the attenuated luminosity, over the bolometric luminosity, per galaxy type and (U)LIRGs merging class (same colours as in left panel), together with the ratio of the dust luminosity to the bolometric luminosity (yellow). 
   \label{lum_over_bolo}}
   \end{figure*}
   
   In \cite{2019A&A...624A..80N} the authors explored the different stellar populations in local galaxies and their role in dust heating. The SEDs of 814 galaxies of various morphological types [parametrized with their Hubble Stage (T)], ranging from pure ellipticals (T=-5) to irregular galaxies (T=10) were modelled with CIGALE, in the same way as we do with the current sample. One of the main findings of that study is that the luminosity of ETGs is dominated by the emission of the old stars with only a small contribution (maximum of $\sim$ 10\% at T=0) from young stars. For later types (T=0 to 5) there is a gradual rise in the contribution of the young stars, with respect to the bolometric luminosity, reaching about 25\%, while it stays roughly constant for morphological types of T$>$5. In addition to that, the role of the two different stellar populations (old and young) to the dust heating was investigated for the various morphological types with Sb (T=3) being the most efficient galaxies in the dust heating. In these galaxies, the young stars donate up to $\sim 77\%$ of their luminosity to the dust heating while this fraction is $\sim 24\%$ for the old stars. In what follows, we extend their analysis to local (U)LIRGs, using exactly the same methodology, and compare with the `normal' local galaxies.
   
   Although the use of SED modelling can provide us with useful information on the stellar populations in galaxies it is not as robust as the use of optical spectra where the imprints of the stellar populations can be recognised in the form of various emission lines. In \citet{2009MNRAS.400.1139R} the authors use long-slit spectroscopy of 36 (U)LIRGs (with z $<$ 0.15) to extract the relative contribution of the old and the young stellar populations. In that study, the spectra were fitted with the stellar population synthesis (SSP) models of \citep{2003MNRAS.344.1000B} in three combinations including young stellar populations, old stellar populations and a power-law, whenever appropriate, accounting for a possible AGN component. Their combination, including a young stellar population ($t_{\rm{YSP}} \le 2$ Gyr) and an old stellar population of age 12.5 Gyr, is the one that better mimics the parametrization used of our approach. This allows us for a comparison between the two methods (SED modelling and optical spectroscopy) on the derivation of the stellar populations. The fractions of the young stellar populations of the seven sources in common between the two samples (F08572+3915, F12112+0305, F12540+5708, F13428+5608, F14348-1447, F15327+2340, F22491-1808) are shown in Fig.~\ref{YSP}, where the ones derived from the spectra (YSP$_{\rm{spectra}}$) and those derived from CIGALE (YSP$_{\rm{CIGALE}}$) are compared. Since YSP$_{\rm{spectra}}$ is extracted from several positions in each system, the mean value is considered, while the minimum and maximum values of YSP$_{\rm{spectra}}$ define the uncertainty in this value (the error-bars in the plot). From this plot we see that, although the scatter is large and also the uncertainty in each source is large, there is a clear trend of the sources lining up (within the errors) the one-to-one line (with the exception of F13428+5608, which, considering only the central 5kpc aperture, and neglecting the outer apertures including its long tidal tail, results in a YSP$_\text{spectra}$ of 75\%, much closer to the value of 83\% derived by CIGALE). This indicates that the values derived from the two methods are comparable, given the very different approaches. Furthermore, the median values of the two samples also show to compare well. For the 36 (U)LIRGs in the sample of \citet{2009MNRAS.400.1139R} the median value for the young stellar population is \mbox{74 $\pm$ 13 $\%$}, while, for the 67 sources in our sample this fraction is 64 $\pm$ 18 $\%$.
   
   The different stellar populations (old and young) in our systems are presented, and compared with the local `normal' galaxies, in Fig.~\ref{lum_over_bolo}. In the left panel of Fig.~\ref{lum_over_bolo} the histograms of the unattenuated luminosities of both stellar components to the bolometric luminosity of each galaxy ($f^\text{unatt}_\text{old} = L^\text{unatt}_\text{old}/L_\text{bolo}$, and $f^\text{unatt}_\text{young} = L^\text{unatt}_\text{young}/L_\text{bolo}$, where $L_\text{bolo} = L^\text{unatt}_\text{old} + L^\text{unatt}_\text{young}$) are plotted. In this plot the red and blue histograms indicate the mean values of $f^\text{unatt}_\text{old}$ and $f^\text{unatt}_\text{young}$ respectively. In the leftmost sub-panel the relative contribution of the young and old stellar populations in ETGs, LTGs, and (U)LIRGs are compared, while these  values for each of the seven merging subclasses of the (U)LIRGs sample are indicated in the rightmost sub-panel. The exact numbers are presented in Table~\ref{tab:old_young_dust}. The most striking feature from this plot is the large increase in $f^\text{unatt}_\text{young}$ for (U)LIRGs compared to local ETGs and LTGs. As already stated in \cite{2019A&A...624A..80N} the old stars are the prominent luminosity source in ETGs and LTGs (with mean values of $f^\text{unatt}_\text{old}$ of 96\% and 79\% respectively) while, this picture now reverses in the case of (U)LIRGs with 64\% of their bolometric luminosity originating from young stars. This increase (by a factor $\sim$3) in the luminosity of the young stellar component is, of course, the result of the intense star-formation that takes place in these systems, mostly, due to merging events. The fraction of the young stars in all, seven, merging sub-classes stays around the mean value of 64\% but it is class `M3' and `M4' systems that show the highest fraction (72\% and 79\% respectively) of young stars following the trend in \textit{SFR} (Fig.~\ref{sfr_per_class}). All the related values are given in Table~\ref{tab:old_young_dust}.
   
   The right panel in Fig.~\ref{lum_over_bolo} shows the effects of dust in the stellar populations discussed previously. Here, the ratio of the dust-attenuated luminosity of the old stellar population ($f^\text{att}_\text{old} = L^\text{att}_\text{old}/L_\text{bolo}$) and of the young stellar population ($f^\text{att}_\text{young} = L^\text{att}_\text{young}/L_\text{bolo}$) to the bolometric luminosity is plotted with red and blue colours respectively, while the fraction of the dust-absorbed luminosity ($f_\text{abs}$; see also Sect.~\ref{place}) is indicated with yellow colour. The leftmost sub-panel shows the comparison of these quantities for ETGs, LTGs, and (U)LIRGs, while the rightmost sub-panel shows the comparison among the different merging sub-classes. What is evident from this plot is the large effect that dust has on the energy budget of (U)LIRGs compared to ETGs and LTGs. $f_\text{abs}$ changes from very low (7\%) for ETGs, to moderate (25\%) for LTGs, to very high (78\%) for (U)LIRGs. This can be explained, mainly, by the higher dust mass that is detected in (U)LIRGs (upper panel in Fig.~\ref{fig:type-props}) and, especially, the dust associated with the PDR regions (bottom panel in Fig.~\ref{fig:type-props}). The dramatic effect of the dust on the stellar populations is clearly seen by comparing the leftmost sub-panels of each panel in Fig.~\ref{lum_over_bolo}. For (U)LIRGs we see that the fraction of the young stars is absorbed so heavily that it goes from 64\% in the unattenuated case, to 5\% in the case where absorption by dust is considered. It is also worth mentioning that the attenuated fraction of the luminosity of the young stars is higher for LTGs (9\%) compared to 2\% and 5\% in ETGs and (U)LIRGs. Classes `M3' and `M4' systems show the highest mean $f_\text{abs}$ values (84\% and 87\%, respectively), with classes `s' and `M2' having the lowest mean $f_\text{abs}$ values (73\%). All the related values are given in Table~\ref{tab:old_young_dust}.

%%%%%%%%%%%%%%%%%%%%%%%%%%%%%%%%%%%%% FIGURE 11
   \begin{figure}[t!]
   \centering
   \includegraphics[width=0.5\textwidth]{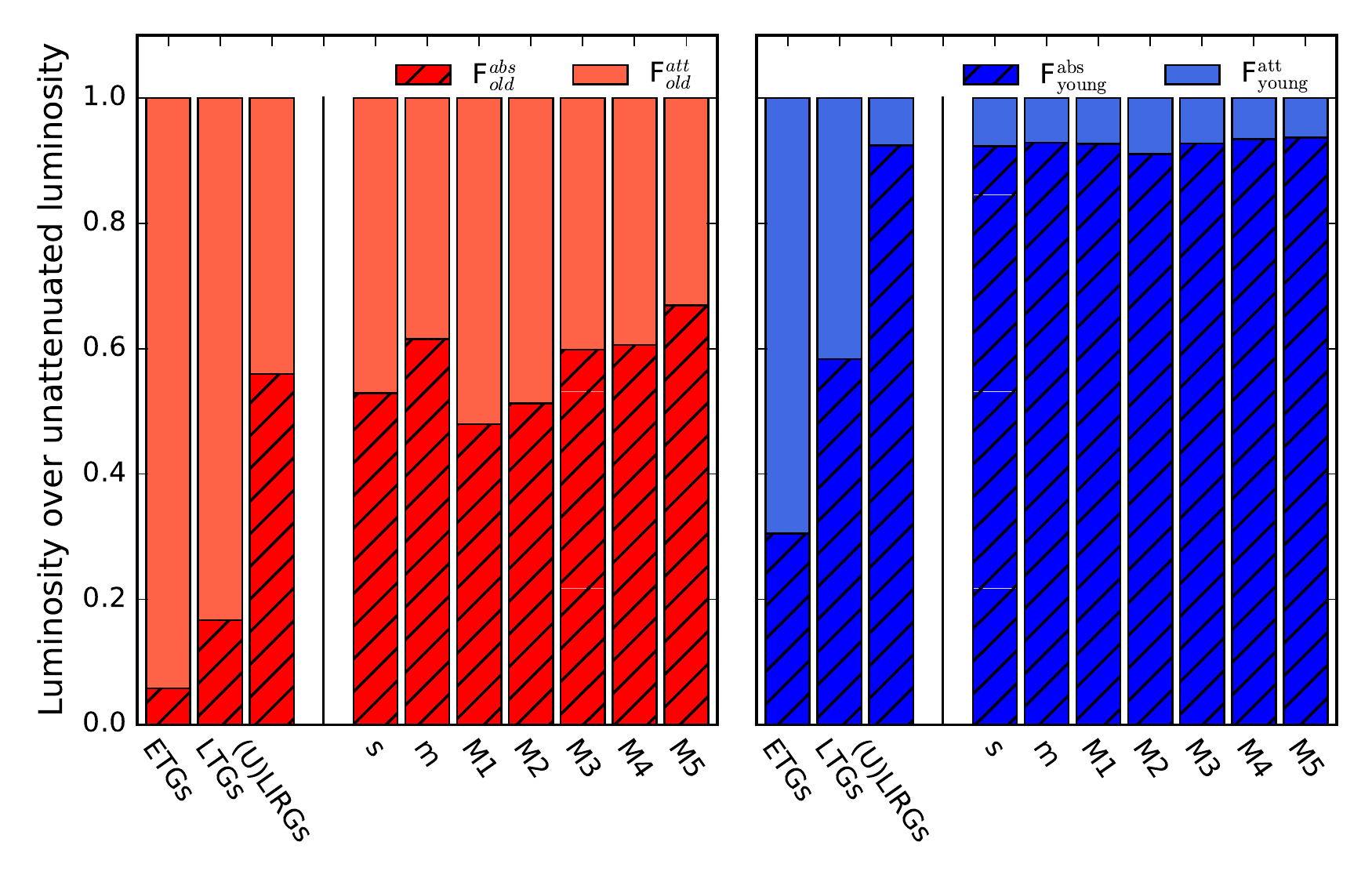}
   \caption{Mean values of the fraction of the luminosity of the old and the young stellar populations (left and right panel, respectively) used for the dust heating. The crossed bars show the mean values of the ratio of the dust-absorbed, luminosity, to the unattenuated luminosity of the corresponding stellar component, while the solid bars are the ratios of the attenuated luminosity of the specific stellar component to its unattenuated luminosity. In each panel the ratios are presented for each of the three galaxy populations (ETGs, LTGs, and (U)LIRGs) in the leftmost sub-panel, and for the four merging classes in the rightmost sub-panel.
   \label{Fabs_Fatt_bars}}
   \end{figure}
   
   As was already discussed above, a large fraction of the energy emitted by the stars, in (U)LIRGs, is absorbed by the dust grains (see the right panel in Fig.~\ref{lum_over_bolo}) resulting in their heating and the production of large amounts of IR radiation in those systems. With our analysis we can not only quantify the total stellar radiation that is absorbed by the dust, but we can also distinguish between the two stellar populations and calculate their efficiency in heating up the dust grains. The quantity that shows the fraction of each stellar population that is absorbed, by the dust, is the ratio of the absorbed luminosity, of each stellar component, to the respective unattenuated stellar component ($F^\text{abs}_\text{old} = L^\text{abs}_\text{old}/L^\text{unatt}_\text{old}$ and $F^\text{abs}_\text{young} = L^\text{abs}_\text{young}/L^\text{unatt}_\text{young}$, for the old and the young stellar component respectively). The remaining luminosity (not absorbed by the dust) is the attenuated luminosity of each stellar component, which, divided by the unattenuated luminosity gives the fraction of the, unaffected, by the dust, luminosity ($F^\text{att}_\text{old} = L^\text{att}_\text{old}/L^\text{unatt}_\text{old}$ and $F^\text{att}_\text{young} =  L^\text{att}_\text{young}/L^\text{unatt}_\text{young}$, for the old and the young stellar component respectively). For each stellar population, the crossed bars in Fig.~\ref{Fabs_Fatt_bars}, show the mean values of the absorbed fraction of the luminosity which contributes to the dust heating ($F^\text{abs}_\text{old, young}$), with the rest, solid bars, being the fraction of the luminosity emitted by stars, without being affected by the dust ($F^\text{att}_\text{old, young}$). The left panel shows the different contributions of the old stars (red colour) and the right panel those of the young stars (blue colour), for each galaxy population (ETGs, LTGs, and (U)LIRGs in the leftmost sub-panels) and for each merging class (rightmost sub-panels). Looking at the different galaxy types, it is evident that it is the young stellar component that offers the larger portion of its total luminosity in the dust heating compared to the old one. In particular, for ETGs, it is 30\% of the young stellar luminosity donated to the dust heating compared to only 6\% which is the case for the old stars, while in LTGs these fractions are 58\% and 17\%, and get extremely high in (U)LIRGs (92\% and 56\% for the young and the old stellar populations, respectively). It is noteworthy that even the old stars in (U)LIRGs have a significant role in the dust heating contributing with more than half of their luminosities. The mean values of the fractions of the dust-absorbed luminosity for each of the two stellar components, have comparable values for the four merging classes with no significant deviations (see the rightmost sub-panels in Fig.~\ref{Fabs_Fatt_bars}). All the related values are summarised in Table~\ref{tab:old_young_dust}.
   
%%%%%%%%%%%%%%%%%%%%%%%%%%%%%%%%%%%%% FIGURE 12
   \begin{figure}[t!]
   \centering
   \includegraphics[width=0.5\textwidth]{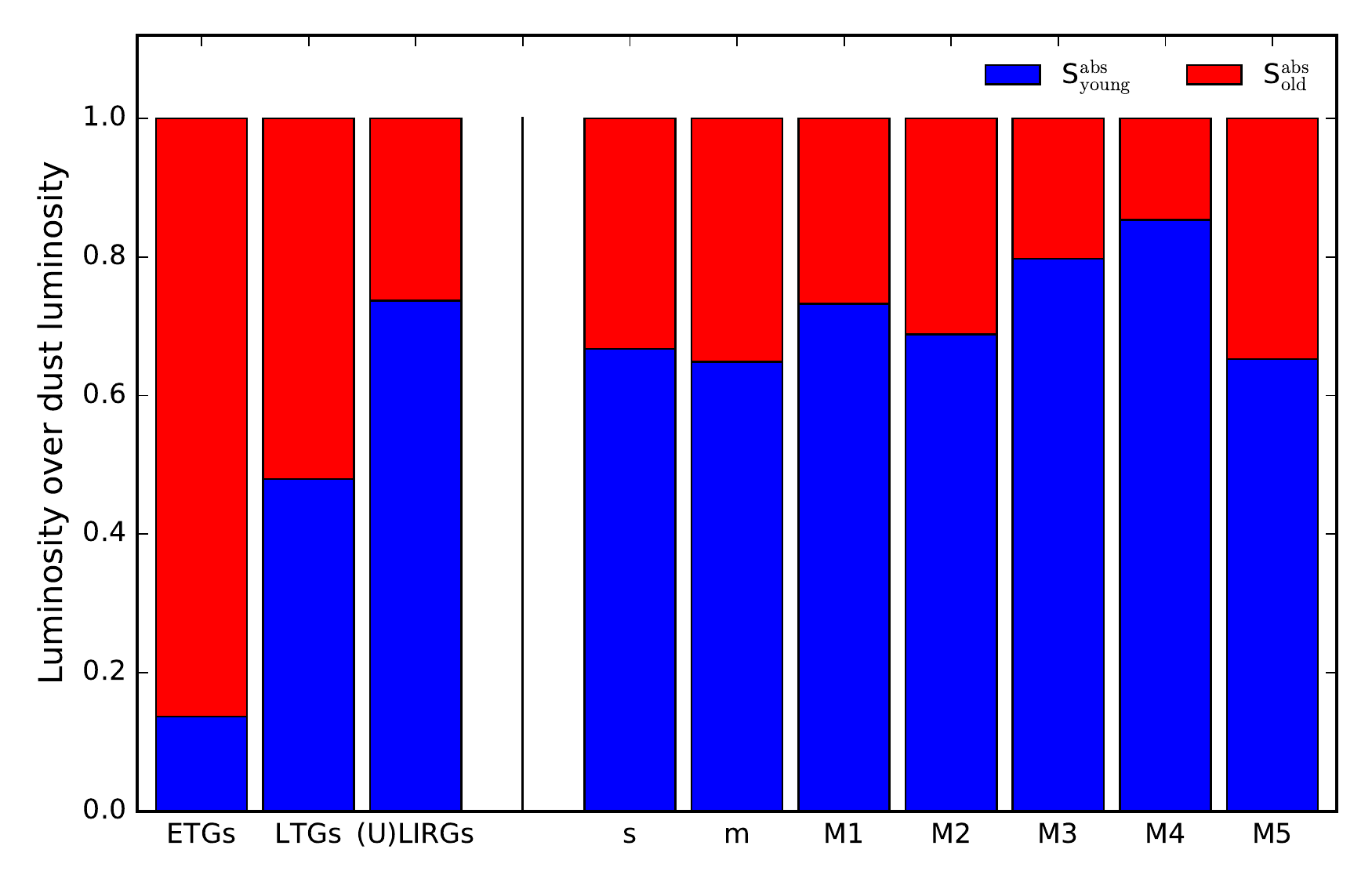}
   \caption{Mean values of the ratios of the dust-absorbed stellar luminosity (originating from old and young stars) to the dust luminosity (red and blue colours, respectively). The ratios are presented for each of the three galaxy populations (ETGs, LTGs, and (U)LIRGs) in the leftmost sub-panel, and for the four merging classes in the rightmost sub-panel. 
   \label{Sabs_bars}}
   \end{figure}
   
   It is also interesting to investigate the relative contribution of the stellar populations to the dust heating. The parameter that indicates this contribution is the ratio of the dust-absorbed luminosity, for each stellar population, to the total dust luminosity ($S^\text{abs}_\text{old} = L^\text{att}_\text{old}/L_\text{dust}$ and $S^\text{abs}_\text{young} = L^\text{att}_\text{young}/L_\text{dust}$ for the old and the young stars respectively). The histograms of the mean values of this parameter are presented in Fig.~\ref{Sabs_bars} with red and blue colours representing the contribution of the old and the young stellar components respectively. In this plot, these contributions in ETGs, LTGs, and (U)LIRGs are plotted in the leftmost part, while the rightmost part shows the relative contribution for the four merging classes. For ETGs, as already described in  \cite{2019A&A...624A..80N}, it is mainly the old stars that contribute more to the dust heating (by 86\%) while, in LTGs, both the old and the young stars contribute almost equally to the dust heating (52\% for the old stellar population). In (U)LIRGs the picture reverses with the young stars taking over the heating of the dust grains with 74\% of the luminosity of this stellar population absorbed by dust.
   
   It is also interesting to investigate the relative contribution of the stellar populations to the dust heating. The parameter that indicates this contribution is the ratio of the dust-absorbed luminosity, for each stellar population, to the total dust luminosity ($S^\text{abs}_\text{old} = L^\text{att}_\text{old}/L_\text{dust}$ and $S^\text{abs}_\text{young} = L^\text{att}_\text{young}/L_\text{dust}$ for the old and the young stars respectively). The histograms of the mean values of this parameter are presented in Fig.~\ref{Sabs_bars} with red and blue colours representing the contribution of the old and the young stellar components respectively. In this plot, these contributions in ETGs, LTGs, and (U)LIRGs are plotted in the leftmost part, while the rightmost part shows the relative contribution for the four merging classes. For ETGs, as already described in  \cite{2019A&A...624A..80N}, it is mainly the old stars that contribute more to the dust heating (by 86\%) while, in LTGs, both the old and the young stars contribute almost equally to the dust heating (52\% for the old stellar population). In (U)LIRGs the picture reverses with the young stars taking over the heating of the dust grains with 74\% of the luminosity of this stellar population absorbed by dust.

%%%%%%%%%%%%%%%%%%%%%%%%%%%%%%%%%%%%% TABLE 4
   \begin{table*}[t]
   \caption{Mean values of the ratios of various combinations of the stellar and dust luminosity components extracted using the \textsc{CIGALE} SED fitting tool. The different ratios \citep[also presented in][]{2019A&A...624A..80N} are defined as
   $f^\text{unatt}_\text{old} = L^\text{unatt}_\text{old}/L_\text{bolo}$, $f^\text{unatt}_\text{young} = L^\text{unatt}_\text{young}/L_\text{bolo}$, $f^\text{att}_\text{old} = L^\text{att}_\text{old}/L_\text{bolo}$, $f^\text{att}_\text{young} = L^\text{att}_\text{young}/L_\text{bolo}$,
   $f_\text{abs} = L_\text{dust}/L_\text{bolo}$,
   $F^\text{att}_\text{old} = L^\text{att}_\text{old}/L^\text{unatt}_\text{old}$,
   $F^\text{abs}_\text{old} = L^\text{abs}_\text{old}/L^\text{unatt}_\text{old}$,
   $F^\text{att}_\text{young} = L^\text{att}_\text{young}/L^\text{unatt}_\text{young}$, $F^\text{abs}_\text{young} = L^\text{abs}_\text{young}/L^\text{unatt}_\text{young}$, $S^\text{abs}_\text{old} = L^\text{att}_\text{old}/L_\text{dust}$, and $S^\text{abs}_\text{young} = L^\text{att}_\text{young}/L_\text{dust}$, 
   where, $L^\text{unatt}_\text{old}$ and $L^\text{unatt}_\text{young}$
   the unattenuated luminosities of the old and the young stars, $L_\text{bolo}$ is the bolometric luminosity of each system ($L_\text{bolo} = L^\text{unatt}_\text{old} + L^\text{unatt}_\text{young}$), $L_\mathrm{dust}$ the dust luminosity, $L_\mathrm{old}^\mathrm{att}$ and $L_\mathrm{young}^\mathrm{att}$ the attenuated luminosity of the old and the young stars, $L_\mathrm{old}^\mathrm{abs}$ and $L_\mathrm{young}^\mathrm{abs}$ the luminosity of the old and the young stars absorbed by dust. These ratios are presented for the three general galaxy populations compared in this study (ETGs, LTGs, and (U)LIRGs), as well as for the four merging classes of (U)LIRGs.}
   \begin{center}
   \scalebox{0.94}{
   \begin{tabular}{c||cc|ccc||cc|cc||cc}
\hline 
\hline 
   Galaxy Type &
   $f_\mathrm{old}^\mathrm{unatt}$ & $f_\mathrm{young}^\mathrm{unatt}$ & $f_\mathrm{old}^\mathrm{att}$ & $f_\mathrm{young}^\mathrm{att}$ & $f_\mathrm{abs}$ & $F_\mathrm{old}^\mathrm{att}$ & $F_\mathrm{old}^\mathrm{abs}$ &  $F_\mathrm{young}^\mathrm{att}$ & $F_\mathrm{young}^\mathrm{abs}$ & $S_\mathrm{old}^\mathrm{abs}$ & $S_\mathrm{young}^\mathrm{abs}$ \\
\hline
   ETGs & 0.96 & 0.04 & 0.91 & 0.02 & 0.07 & 0.94 & 0.06 & 0.70 & 0.30 & 0.86 & 0.14 \\
   LTGs & 0.79 & 0.21 & 0.66 & 0.09 & 0.25 & 0.83 & 0.17 & 0.42 & 0.58 & 0.52 & 0.48 \\
   (U)LIRGs & 0.36 & 0.64 & 0.17 & 0.05 & 0.78 & 0.44 & 0.56 & 0.08 & 0.92 & 0.26 & 0.74 \\
\hline
   s & 0.46 & 0.54 & 0.23 & 0.04 & 0.73 & 0.47 & 0.53 & 0.08 & 0.92 & 0.33 & 0.67 \\
   m & 0.45 & 0.55 & 0.17 & 0.04 & 0.79 & 0.39 & 0.61 & 0.07 & 0.93 & 0.35 & 0.65 \\
   M1 & 0.40 & 0.60 & 0.21 & 0.04 & 0.75 & 0.52 & 0.48 & 0.07 & 0.93 & 0.27 & 0.73 \\
   M2 & 0.44 & 0.56 & 0.22 & 0.05 & 0.73 & 0.49 & 0.51 & 0.09 & 0.91 & 0.31 & 0.69 \\
   M3 & 0.28 & 0.72 & 0.11 & 0.05 & 0.84 & 0.40 & 0.60 & 0.07 & 0.93 & 0.20 & 0.80 \\
   M4 & 0.21 & 0.79 & 0.08 & 0.05 & 0.87 & 0.39 & 0.61 & 0.07 & 0.93 & 0.15 & 0.85 \\
   M5 & 0.43 & 0.57 & 0.14 & 0.04 & 0.82 & 0.33 & 0.67 & 0.06 & 0.94 & 0.35 & 0.65 \\
\hline
\hline
   \end{tabular}}
   \label{tab:old_young_dust}
   \end{center}
   \end{table*}
   
   Concerning the relative contribution of the two stellar populations to the dust heating of the different merging classes we see that it remains close to the mean value of (U)LIRGs (74\% for the young stars) with only small deviations. The largest deviations for the contribution of the young stars to the dust heating are found between classes `s' and `m' (65\% and 67\%) and classes `M3' and `M4' (80\% and 85\%) respectively. All the related values are summarised in Table~\ref{tab:old_young_dust}.
   
\section{Summary}\label{sec:sum_conc}

   In this study we model the SEDs of 67 local (U)LIRGs, using the \textsc{CIGALE} code, to calculate their physical properties. Their stellar mass, dust mass, star-formation rate, dust temperature as well as their luminosity arising from PDR regions are derived and compared to those of 268 ETGs and 542 LTGs \citep[already derived in a similar way in][]{2019A&A...624A..80N}. Furthermore, the (U)LIRGs are categorised in seven classes according to their merging stage (ranging from totally isolated to pre-mergers, mergers, and post-mergers), providing useful information on how their properties depend on the phase of the interaction of the parent galaxies. Finally, the contribution of the two stellar populations (old and young) to the bolometric luminosity of these systems and their role in the dust heating is also explored. Our findings are summarised as follows:
   \begin{itemize}
    
    \item (U)LIRGs occupy the `high-end' on the $M_{\text{dust}}$, $M_{\text{star}}$ and \textit{SFR} plane in the local Universe compared to ETGs and LTG \citep[with the corresponding parameters already calculated in][]{2019A&A...624A..80N}. Their median dust mass is $5.2\times10^7~\text{M}_{\odot}$ compared to $4.6\times10^5~\text{M}_{\odot}$ and $5\times10^6~\text{M}_{\odot}$ for ETGs and LTGs, respectively.
    Their median stellar mass is $6.3\times10^{10}~\text{M}_{\odot}$ compared to $1.5\times10^{10}~\text{M}_{\odot}$, and $4.2\times10^9~\text{M}_{\odot}$ for ETGs and LTGs, respectively. \textit{SFR} in (U)LIRGs gets a much higher median value of $52.0~\text{M}_{\odot}$yr$^{-1}$ compared to 0.01, and $0.4~\text{M}_{\odot}$yr$^{-1}$ for ETGs and LTGs, respectively. The median values of the specific star-formation rate (\textit{sSFR}), on the other hand, ranges from $7\times10^{-4}$ and 0.1 Gyr$^{-1}$, for ETGs and LTGs, to 1.0 Gyr$^{-1}$ for (U)LIRGs, respectively.
    These differences among the three galaxy populations in the local universe can also be traced by carefully examining their median SEDs.
    
    \item The median SEDs show a slight shift (in the wavelength axis) of the dust emission peak   indicating that dust temperature is cooler in LTGs and warmer in (U)LIRGs with median dust temperatures of 28, 22, and 32 K for ETGs, LTGs, and (U)LIRGs respectively. The SEDs also reveal the significance of the dust emission in PDR regions increasing from ETGs to LTGs to (U)LIRGs with the median values of the ratios of the PDR-to-total dust luminosity changing from 1.6\%, to 5.2\%, to 11.7\% respectively. Furthermore, the attenuation effects, caused by the dust, are evident in the median SEDs with the peak of the stellar emission (measured at 1 $\mu$m) being the highest for ETGs, slightly lower (by 0.23 dex) for LTGs and much lower (by 1.95 dex) for (U)LIRGs (compared to ETGs). Comparison of the attenuated and unattenuated curves of the SEDs indicates that the attenuation by the dust becomes significant shortwards of $\sim0.2~\mu\text{m}$, $\sim1~\mu\text{m}$, and $\sim2~\mu\text{m}$ for ETGs, LTGs, and (U)LIRGs, respectively. 

    \item Small differences, in the derived parameters, are seen for the seven merging classes of our sample of (U)LIRGs. Dust mass is very similar among different merging classes (within the scatter of the measurements), while a mild deficit is seen in the stellar mass for class `M3' and `M4' objects. The most evident change is seen in the \textit{SFR} with the median values computed for class `M4' objects being the highest (99 M$_{\odot}$ yr$^{-1}$) followed by class `M3' (93 M$_{\odot}$ yr$^{-1}$), with the lowest \textit{SFR} occurring at class `s' (26 M$_{\odot}$ yr$^{-1}$) sources. 
    A mild change in the dust temperature is found with an increasing trend of the median value from 27.3 K to 35.5 K from the isolated to the more evolved systems respectively.
    The PDR luminosity is slightly enhanced for classes `M3' and `M4' systems, compared to the rest of the classes, consistent with the higher \textit{SFR} observed in those systems.  
    
    \item In contrast to the local `normal' galaxies where the old stars are the dominant source of the stellar emission (with the fraction of their luminosity over the bolometric luminosity being 96\% and 79\% for ETGs and LTGs, respectively) this picture reverses in (U)LIRGs with the young stars being the dominant source of stellar emission with the fraction of their luminosity being 64\% of the bolometric one. Out of the seven merging classes, classes `M4' and `M3' show the highest such contribution (79\% and 72\%, respectively). The effects of dust in (U)LIRGs, parametrized by the dust-absorbed luminosity, to the bolometric luminosity is extremely high (78\%) compared to 7\% and 25\% in ETGs and LTGs, respectively.
    
    \item The fraction of the stellar luminosity used to heat up the dust grains is very high in (U)LIRGs, for both stellar components (92\% and 56\% for young and old, respectively) compared to 30\% and 6\% for ETGs, and 58\% and 17\% for LTGs, respectively. In (U)LIRGs 74\% of the dust heating comes from the young stars, with the old stars being the dominant source of dust heating contributing with 86\% in ETGs and 52\% in LTGs.
    
    \end{itemize}

\begin{acknowledgements}
   We would like to thank the anonymous referee for providing comments and suggestions, which helped to improve the quality of the manuscript. This research is co-financed by Greece and the European Union (European Social Fund-ESF) through the Operational Programme "Human Resources Development, Education and Lifelong Learning 2014-2020" in the context of the project “Anatomy of galaxies: their stellar and dust content through cosmic time” (MIS 5052455). GM acknowledges support by the Agencia Estatal de Investigación, Unidad de Excelencia María de Maeztu, ref. MDM-2017-0765. DustPedia is a collaborative focused research project supported by the European Union under the Seventh Framework Programme (2007-2013) call (proposal no. 606847). The participating institutions are: Cardiff University, UK; National Observatory of Athens, Greece; Ghent University, Belgium; Universit\'{e} Paris Sud, France; National Institute for Astrophysics, Italy and CEA, France. This research has made use of the NASA/IPAC Extragalactic Database (NED), which is operated by the Jet Propulsion Laboratory, California Institute of Technology, under contract with the National Aeronautics and Space Administration. We have also made use of the VizieR catalogue access tool, CDS, Strasbourg, France (DOI : 10.26093/cds/vizier). The original description of the VizieR service was published in A\&AS 143, 23.
\end{acknowledgements}


\bibliographystyle{aa}
\bibliography{References}

%%%%%%%%%%%%%%%%% APPENDIX %%%%%%%%%%%%%%%%%%%%%
\appendix
\section{\textsc{CIGALE} validation (mock analysis)} \label{ap:mock}

In order to examine how well the derived parameters can be constrained from the multi-wavelength SED fitting that CIGALE performs, and to monitor the accuracy and precision expected for each parameter, we made use of the CIGALE module that performs a mock analysis. This module creates a mock SED for each galaxy based on the best fitted parameters, allowing the fluxes to vary within the uncertainties of the observations by adding a value taken from a Gaussian distribution with the same standard deviation as indicated by the observations. By modelling these mock SEDs with CIGALE we can then retrieve the best set of the mock fitted parameters and compare them with those used as an input. This provides us with a direct measure of how accurately one can retrieve specific parameters for a specific sample of galaxies. 

\begin{figure}
\centering
\includegraphics[width=0.5\textwidth]{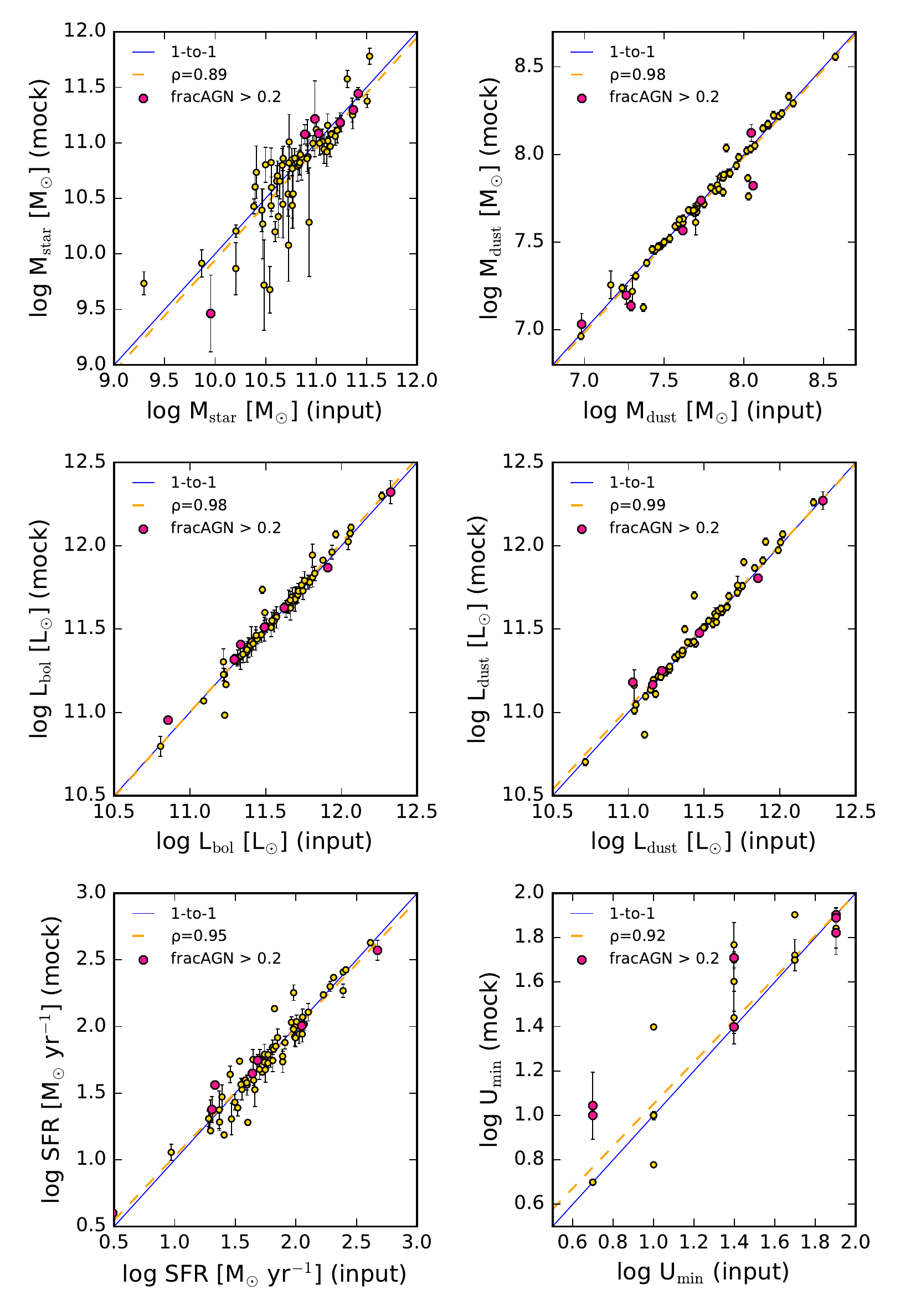}
\caption{Best fitted parameters (input values; x-axis) versus the mock parameters (mock values; y-axis) as derived by the mock analysis performed with the CIGALE. Each circle corresponds to an individual galaxy with the red circles indicating the strongest AGNs in the sample with $frac_{\rm AGN} > 0.2$. The solid blue line corresponds to the one-to-one relation while the orange dashed line the best linear fit to the data. The relevant value of the Spearman's correlation coefficient ($\rho$) is also indicated in each panel.
\label{mock}}
\end{figure}

The results of the mock analysis are presented in Fig.~\ref{mock} with the best fitted values of each parameter (input values; x-axis) compared to the mock values of the each parameter (mock values; y-axis). The red circles indicate the strongest AGNs in the sample with $frac_{\rm AGN} > 0.2$. The solid blue line corresponds to the one-to-one relation while the orange dashed line the best linear fit to the data. The relevant value of the Spearman's correlation coefficient ($\rho$) is also indicated in each panel.

The parameters presented in the mock analysis, are the ones used in this work or have been used for the calculation of other quantities (e.g., $U_{\text{min}}$ for the calculation of $T_\text{dust}$). It is evident that all the mock-derived values have a strong correlation with the input parameters with the Spearman's correlation coefficient being more than $\sim0.9$ in all cases. This indicates that CIGALE can adequately well calculate the true value. 
%The only exception is with the calculation of the stellar mass with relatively large deviations seen in the low-end of the stellar masses (give some explanation?). 
The only exception is with the calculation of $U_{\text{min}}$ showing some deviant points, especially for galaxies with strong AGNs, but, overall, the input and mock data are in a good agreement with $\rho=0.92$.

%%%%%%%%%%%%%%%%%%%%%%%%%%%%%%%%%%%%% FIGURE A.1

\section{Best-fit SED models} \label{appendix_figs}

   The SEDs of the 67 (U)LIRGs, analysed in this study, are presented here. The goodness of the fit for each source is indicated with the reduced $\chi^2$ of the fit listed in Table~\ref{tab:properties2}. For a few exceptional cases, e.g., F12224-0624, F15107+0724 and F15327+2340, CIGALE gives a poor fit close to the FIR peak resulting in relatively large reduced $\chi^2$ values. This may have some effect on derived dust parameters for these systems.

%%%%%%%%%%%%%%%%%%%%%%%%%%%%%%  FIG. B.1
\clearpage
\begin{figure}
\includegraphics[height=0.68\columnwidth]{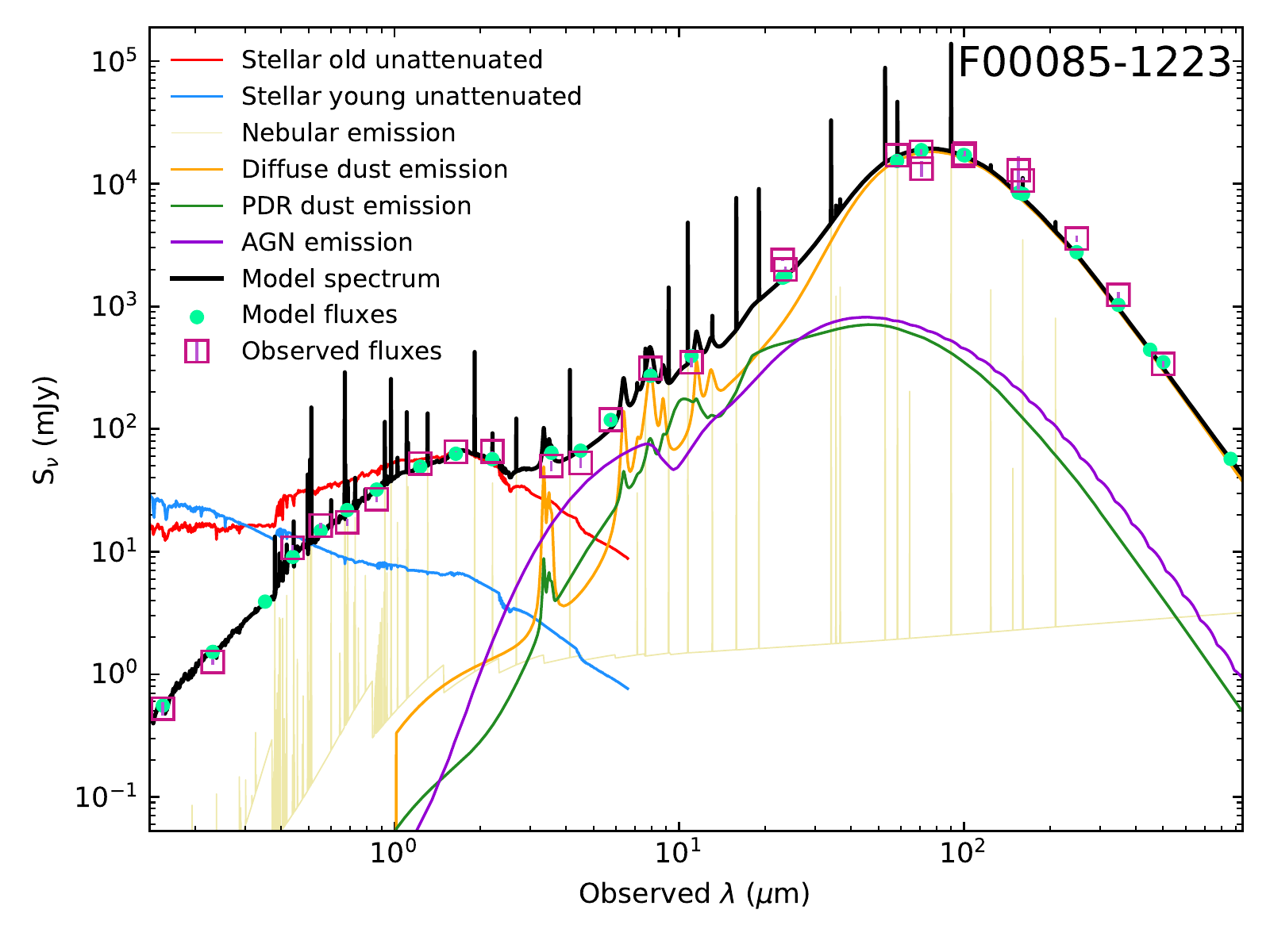}
\includegraphics[height=0.68\columnwidth]{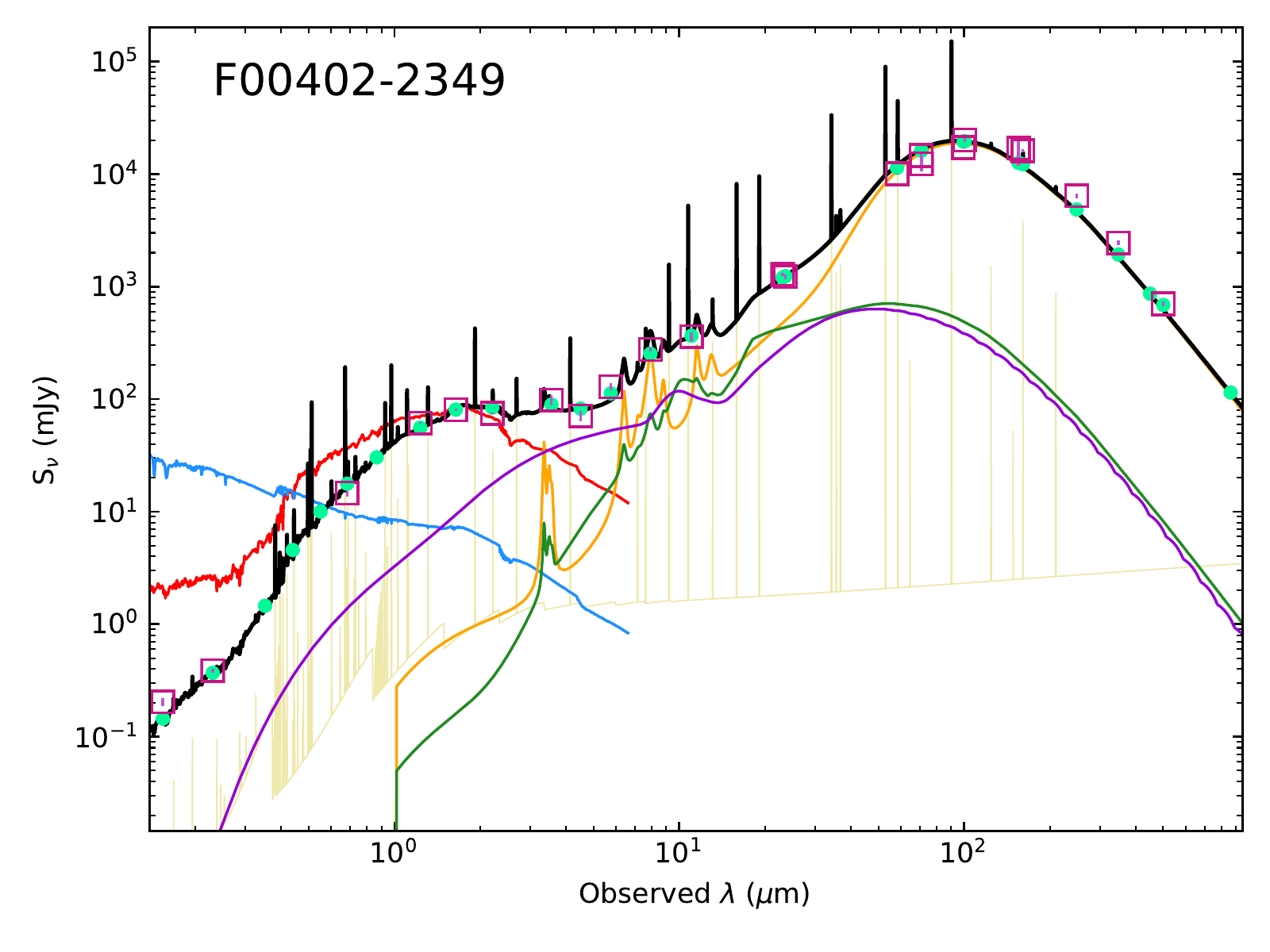}
\includegraphics[height=0.68\columnwidth]{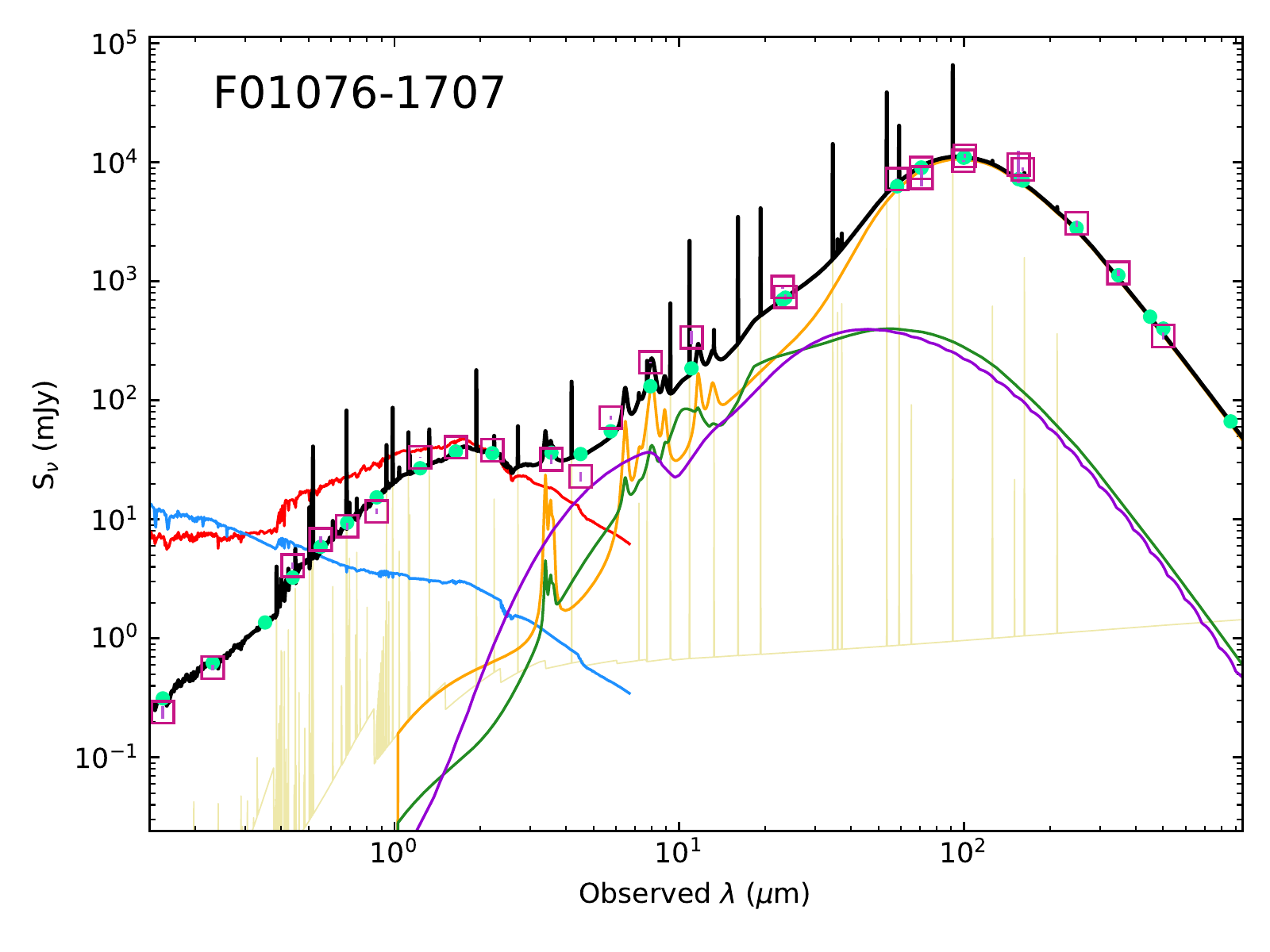}
\includegraphics[height=0.68\columnwidth]{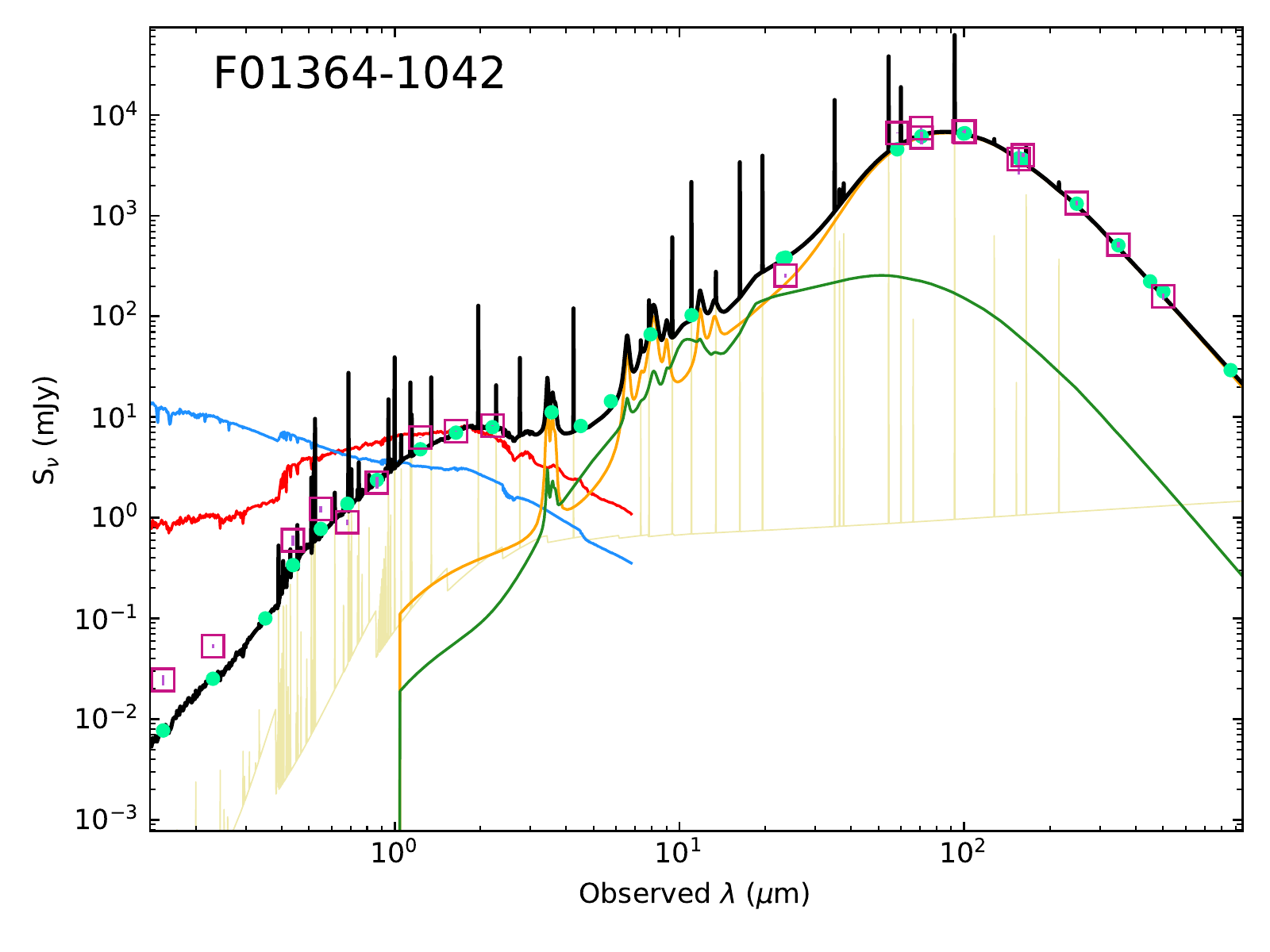}
\label{SEDs}
\end{figure}
\begin{figure}
\includegraphics[height=0.68\columnwidth]{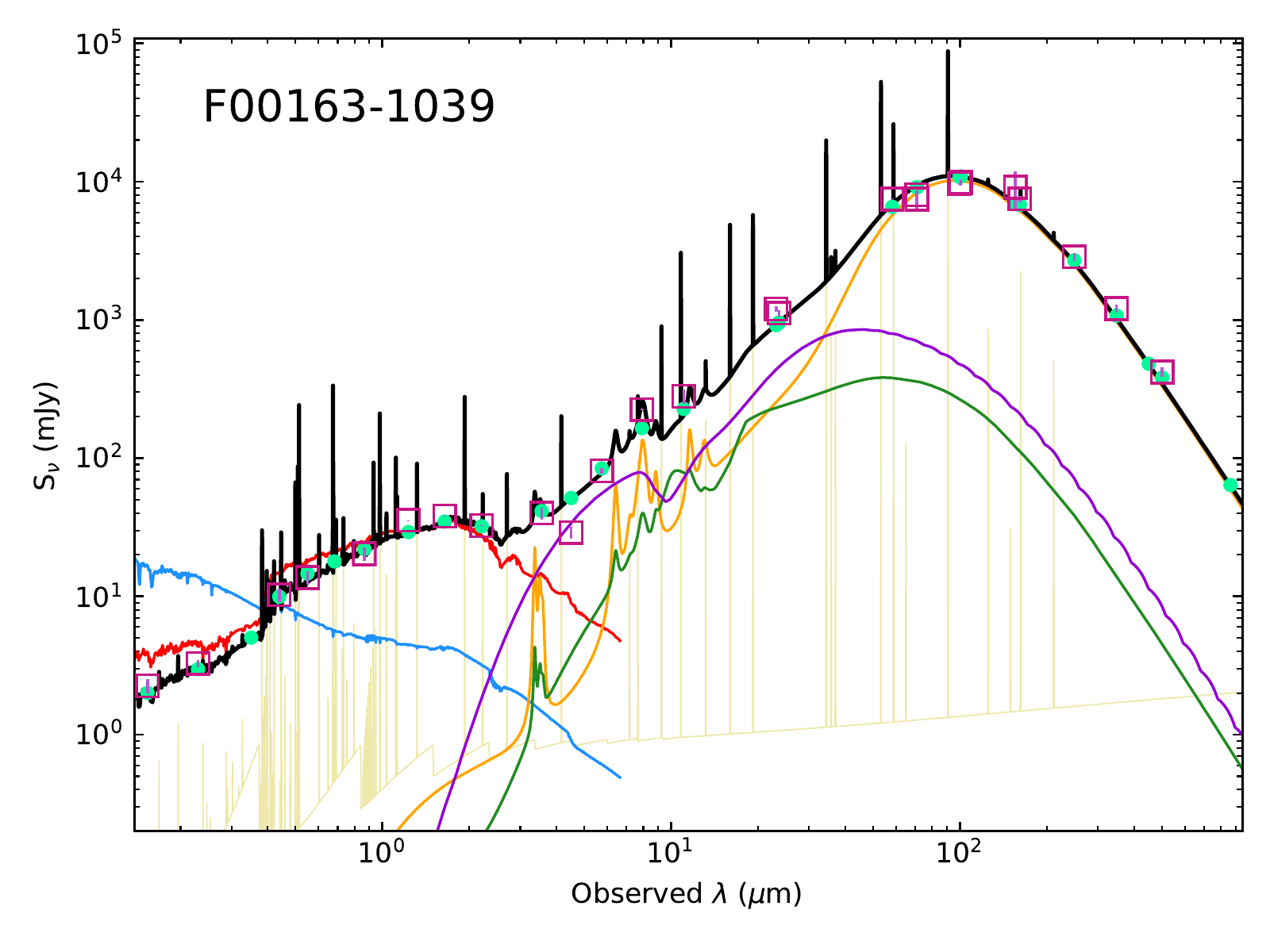}
\includegraphics[height=0.68\columnwidth]{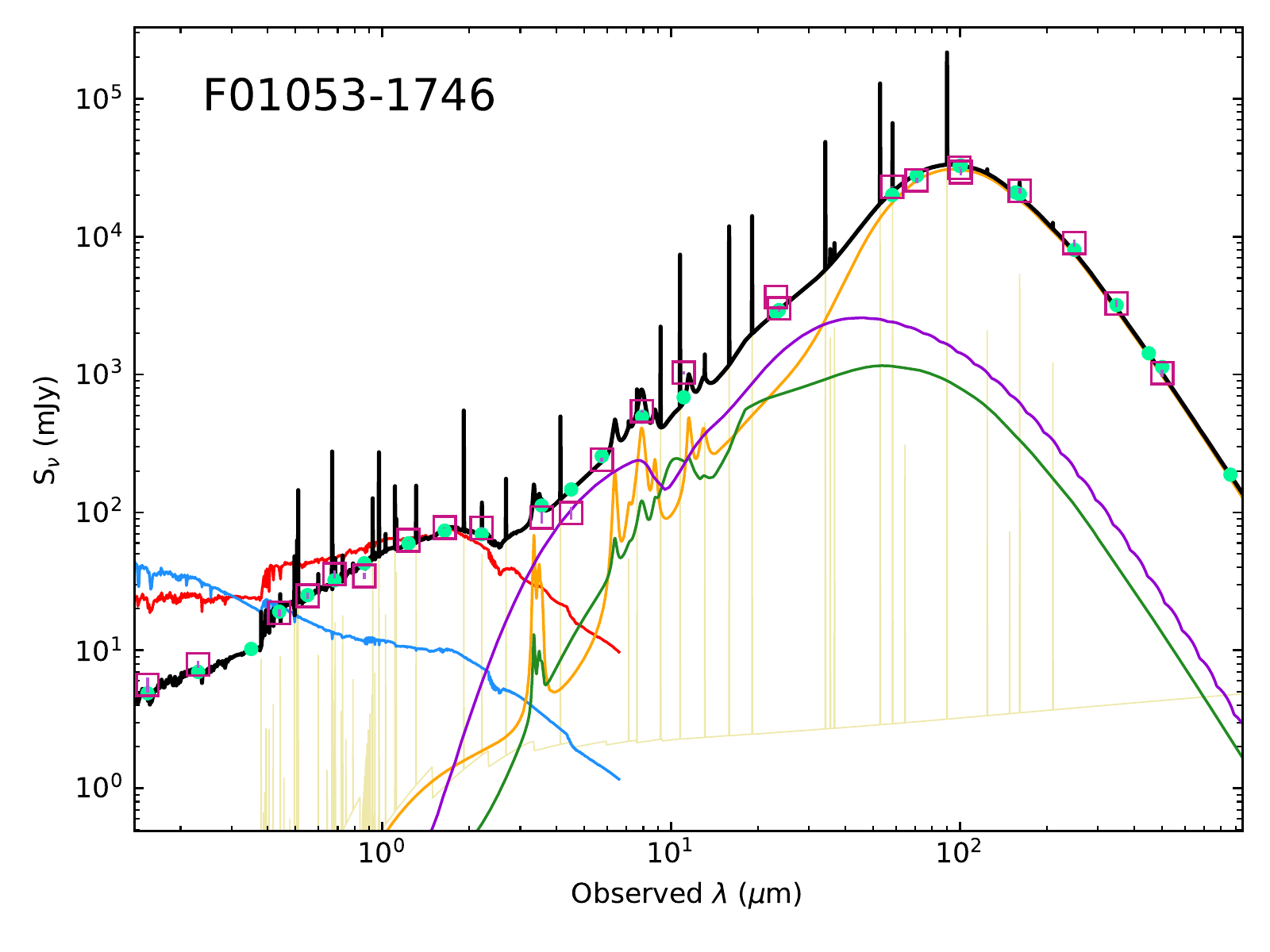}
\includegraphics[height=0.68\columnwidth]{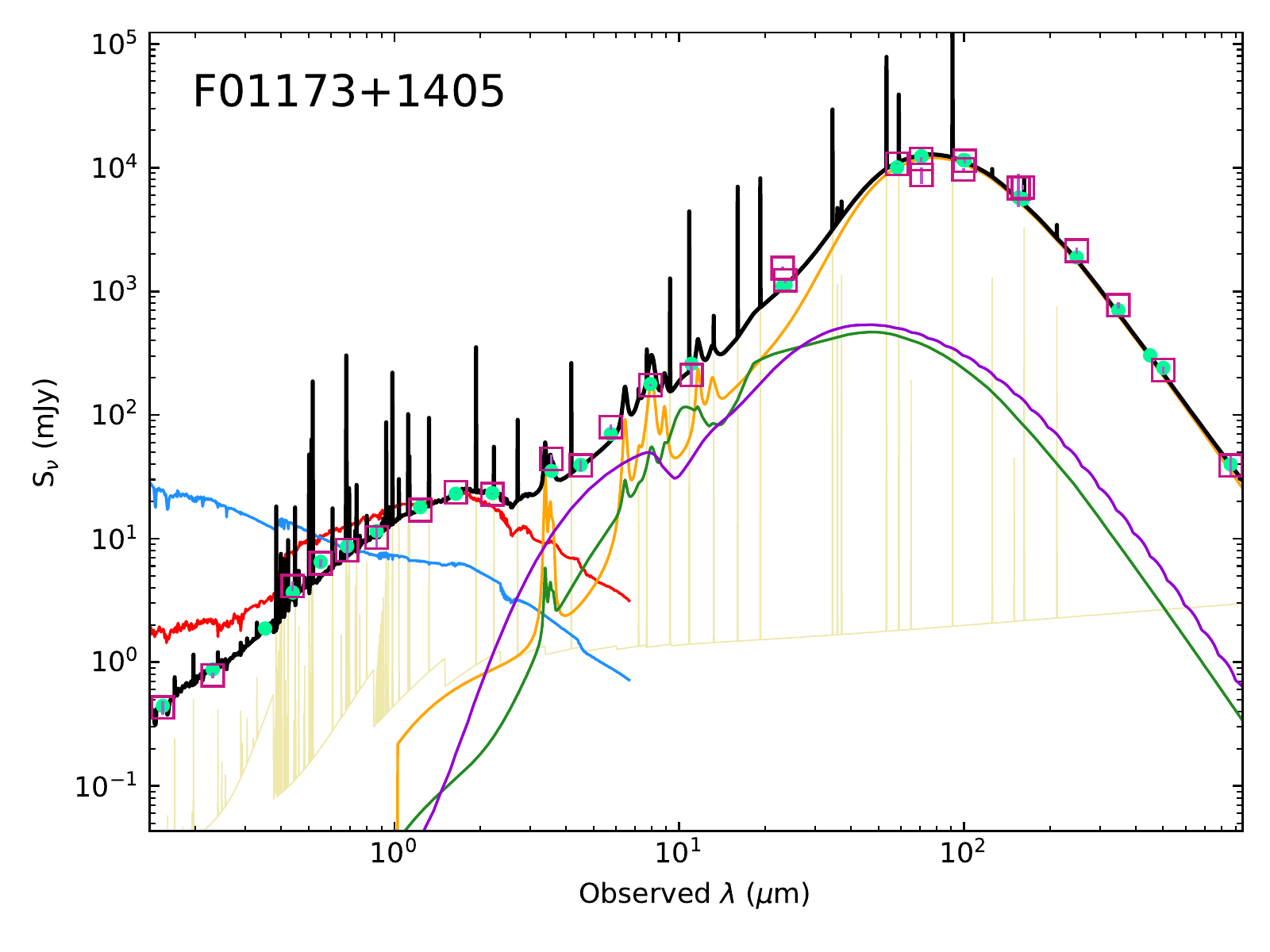}
\includegraphics[height=0.68\columnwidth]{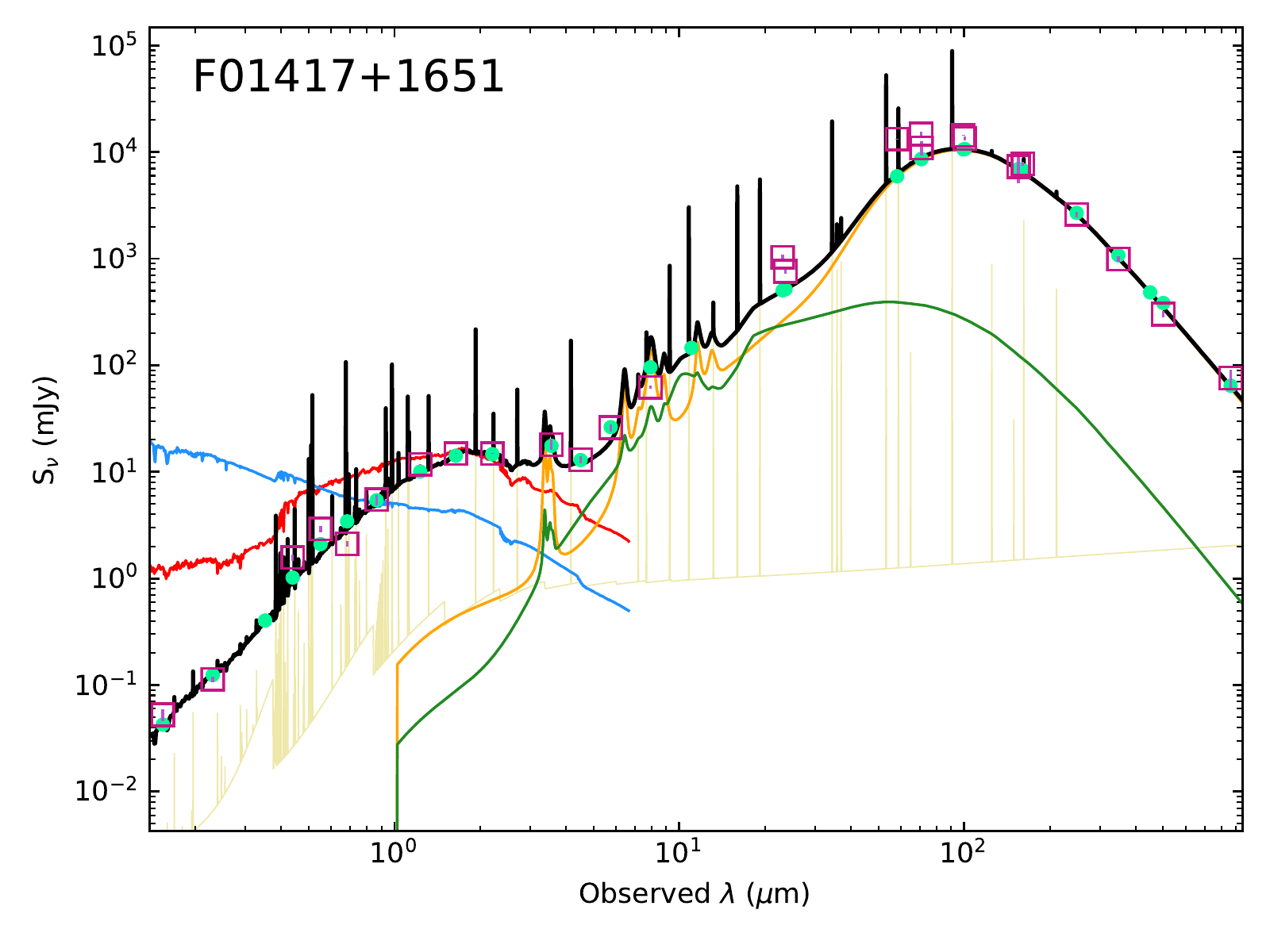}
\label{}
\end{figure}

\begin{figure}
\includegraphics[height=0.68\columnwidth]{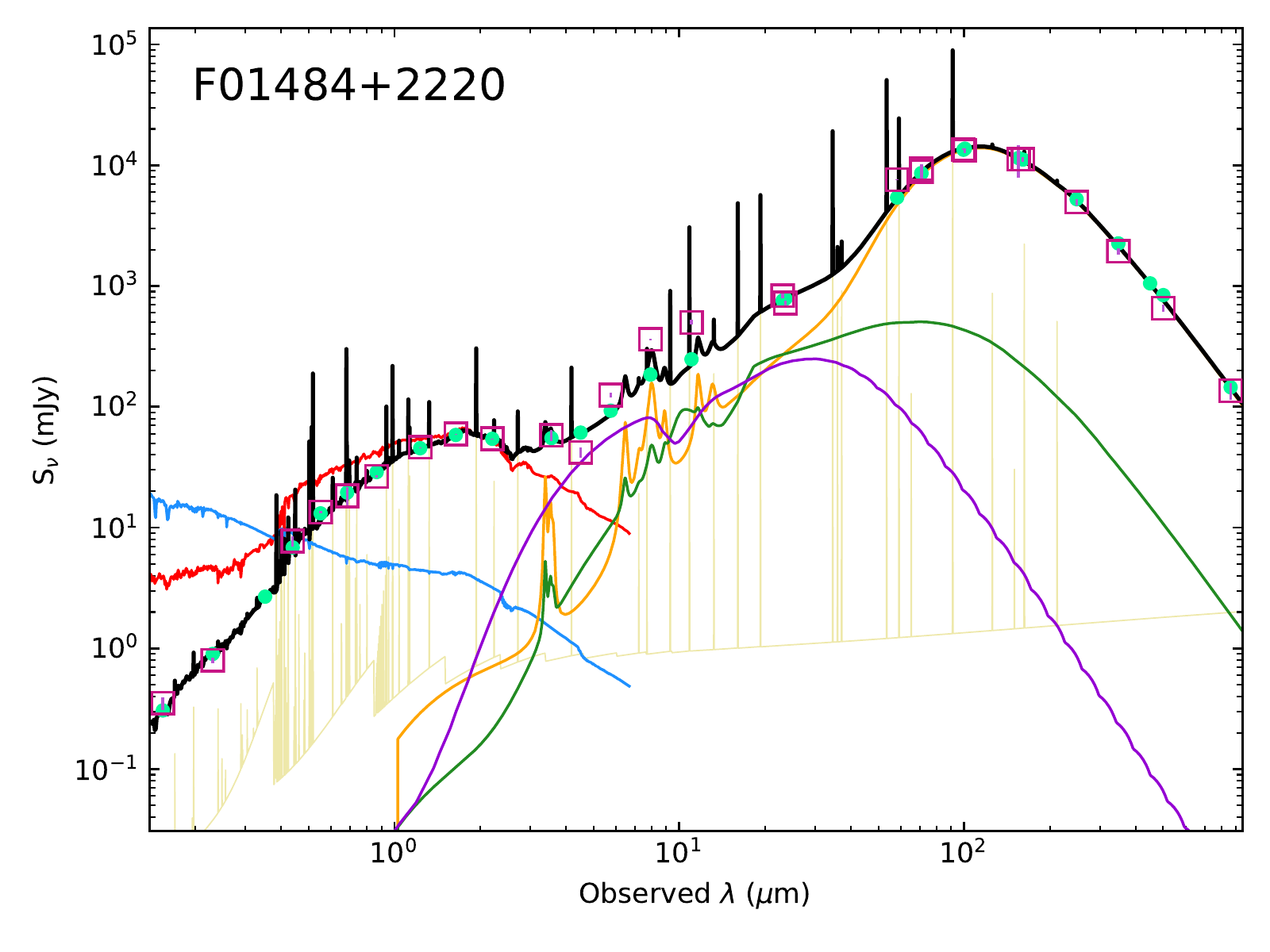}
\includegraphics[height=0.68\columnwidth]{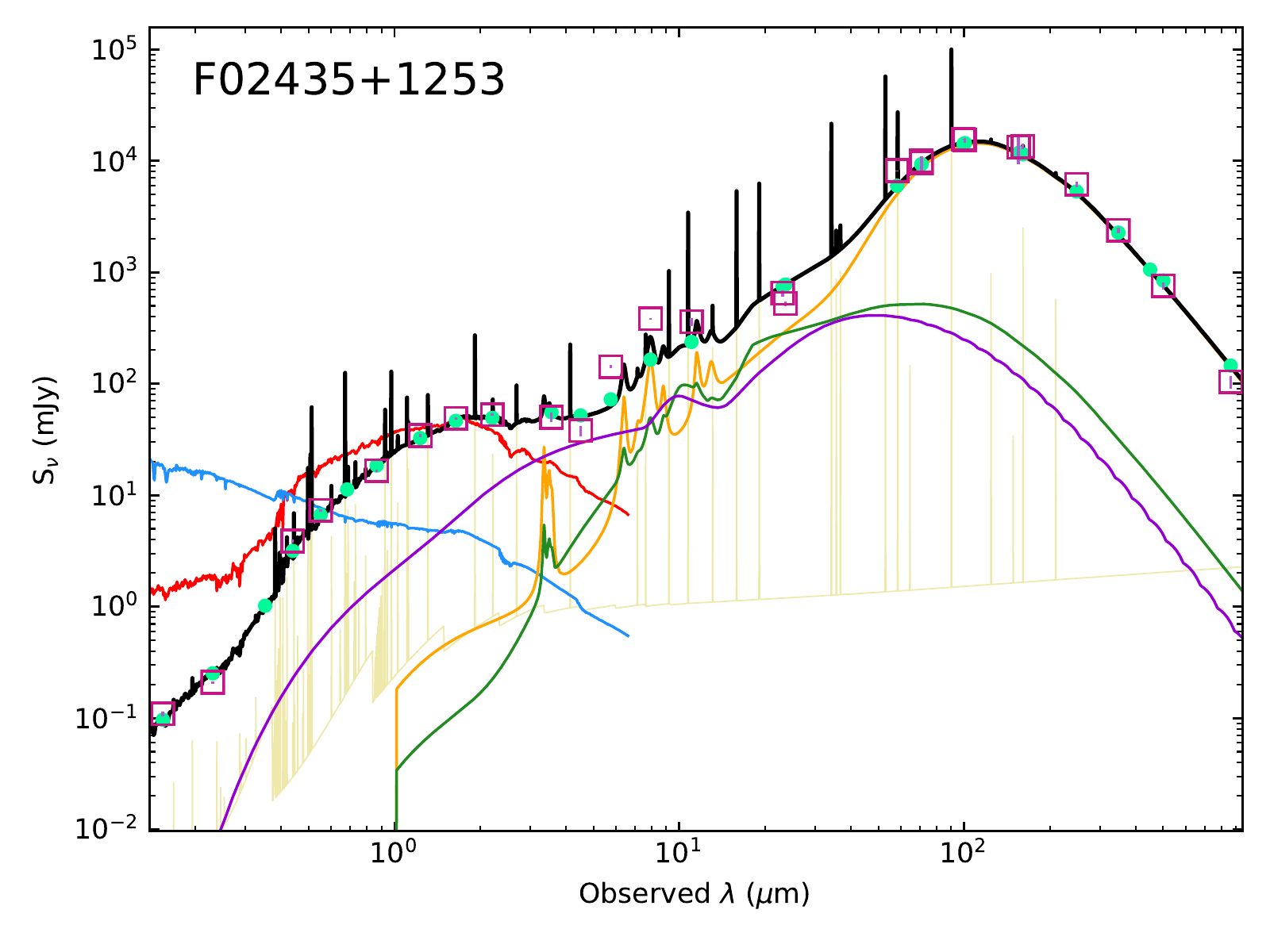}
\includegraphics[height=0.68\columnwidth]{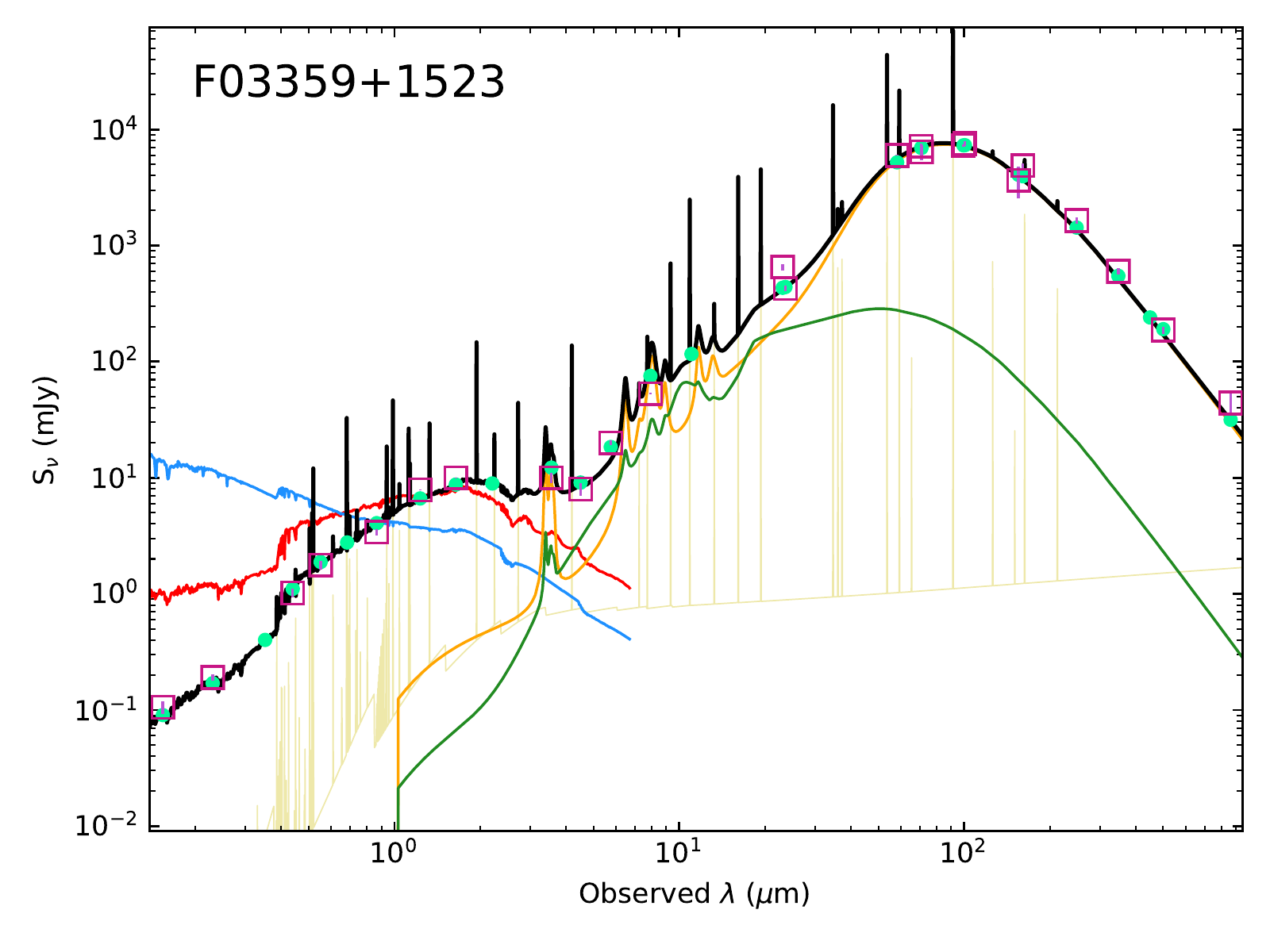}
\includegraphics[height=0.68\columnwidth]{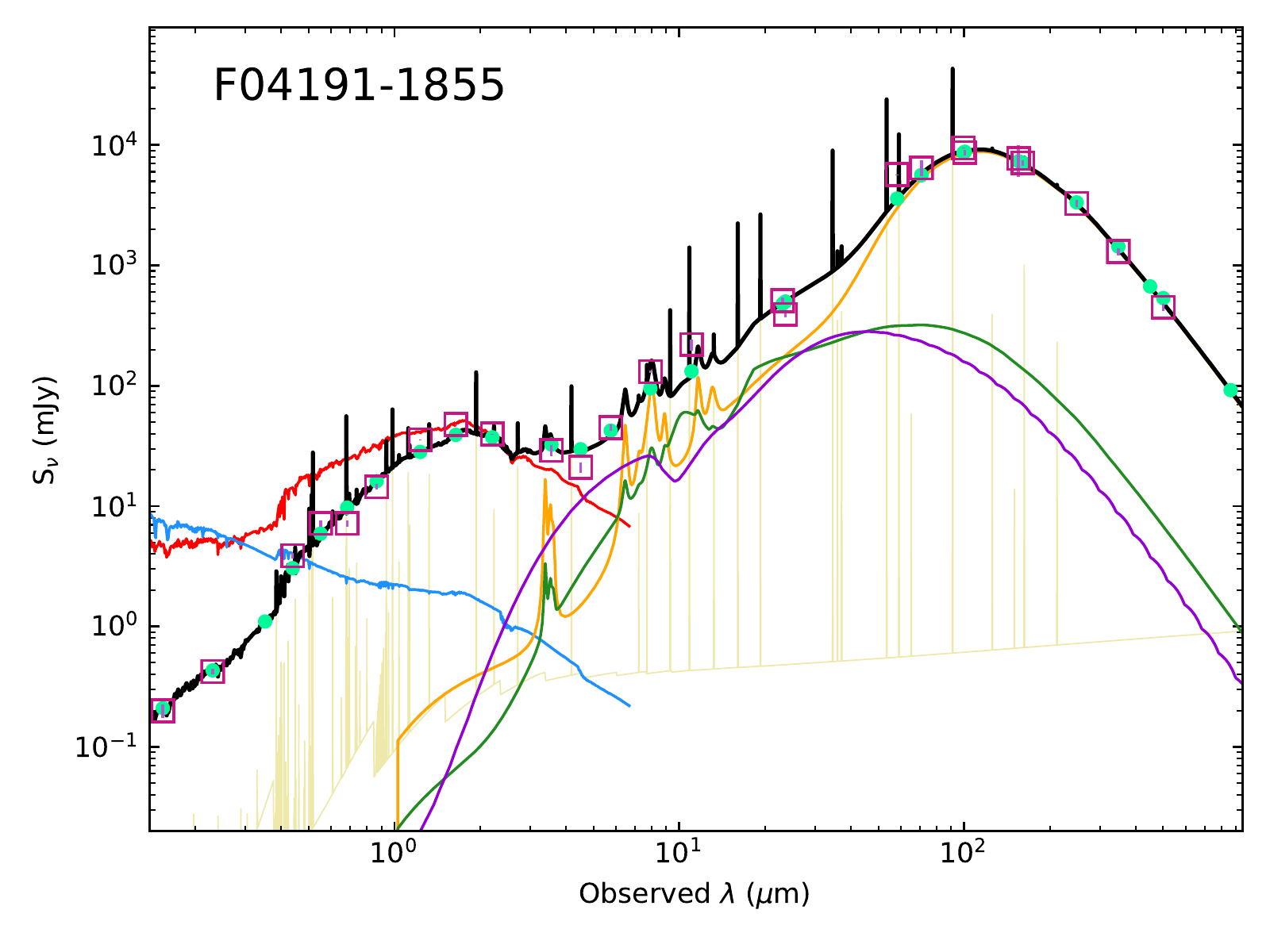}
\label{4}
\end{figure}
\begin{figure}
\includegraphics[height=0.68\columnwidth]{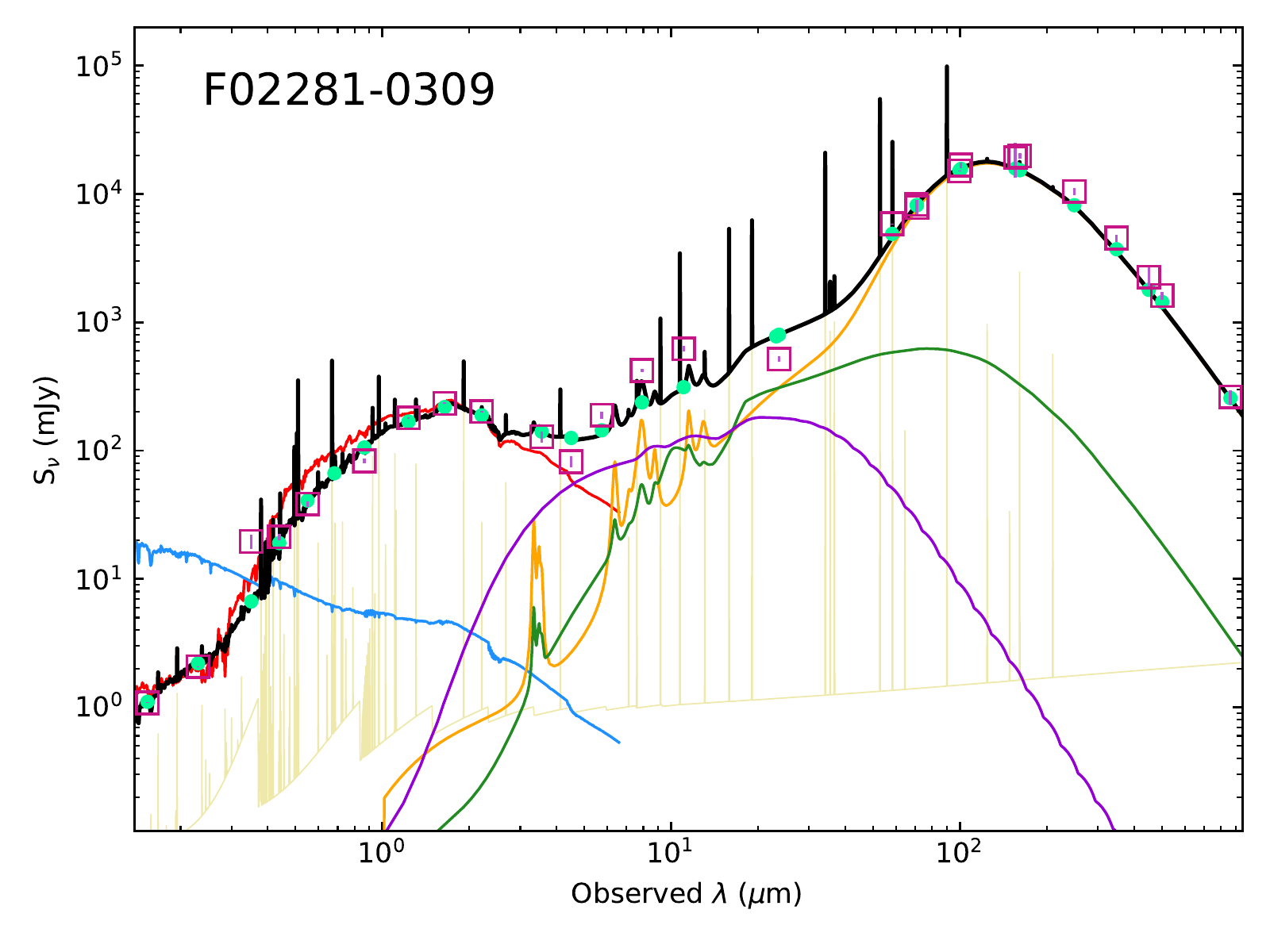}
\includegraphics[height=0.68\columnwidth]{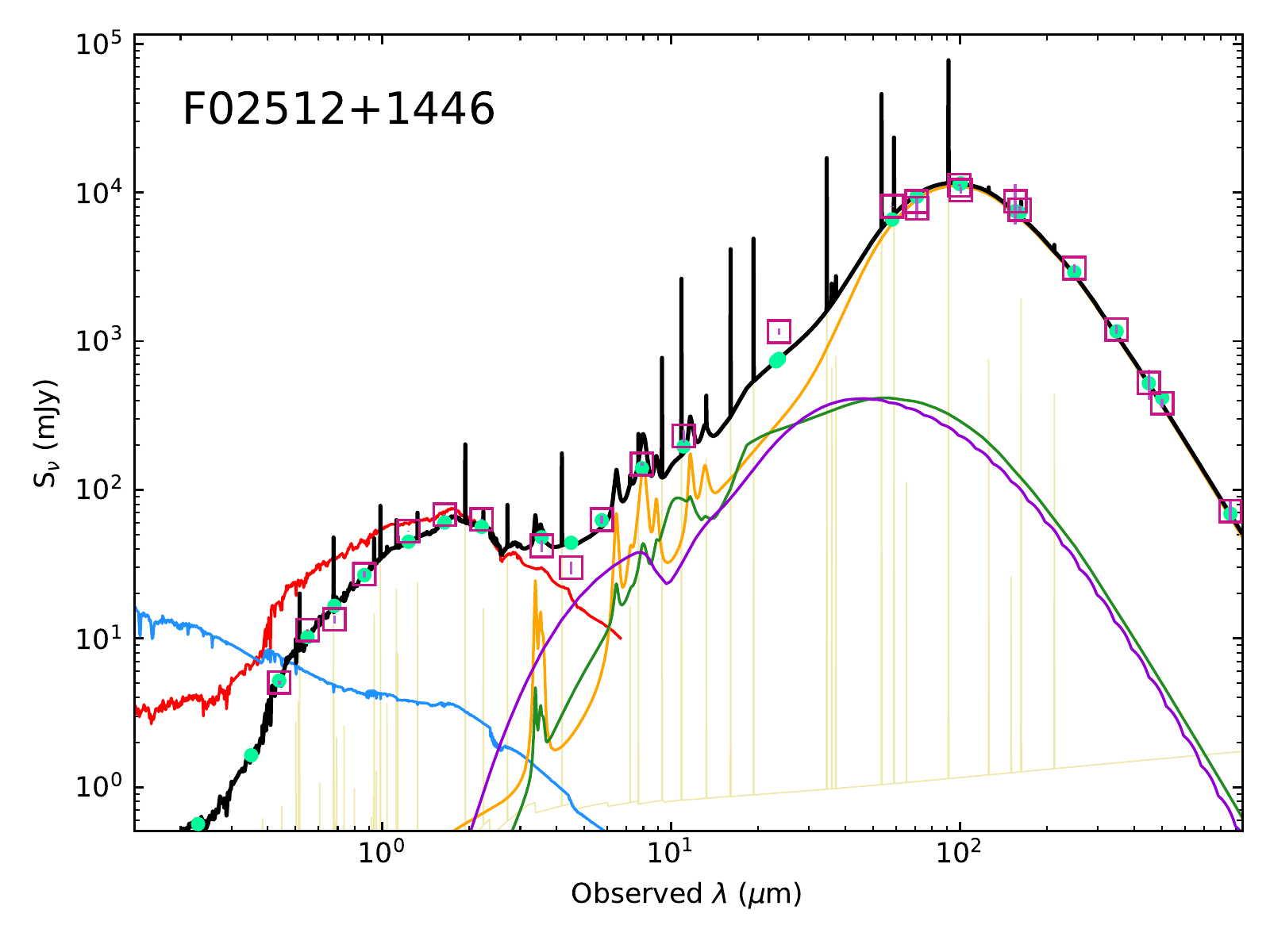}
\includegraphics[height=0.68\columnwidth]{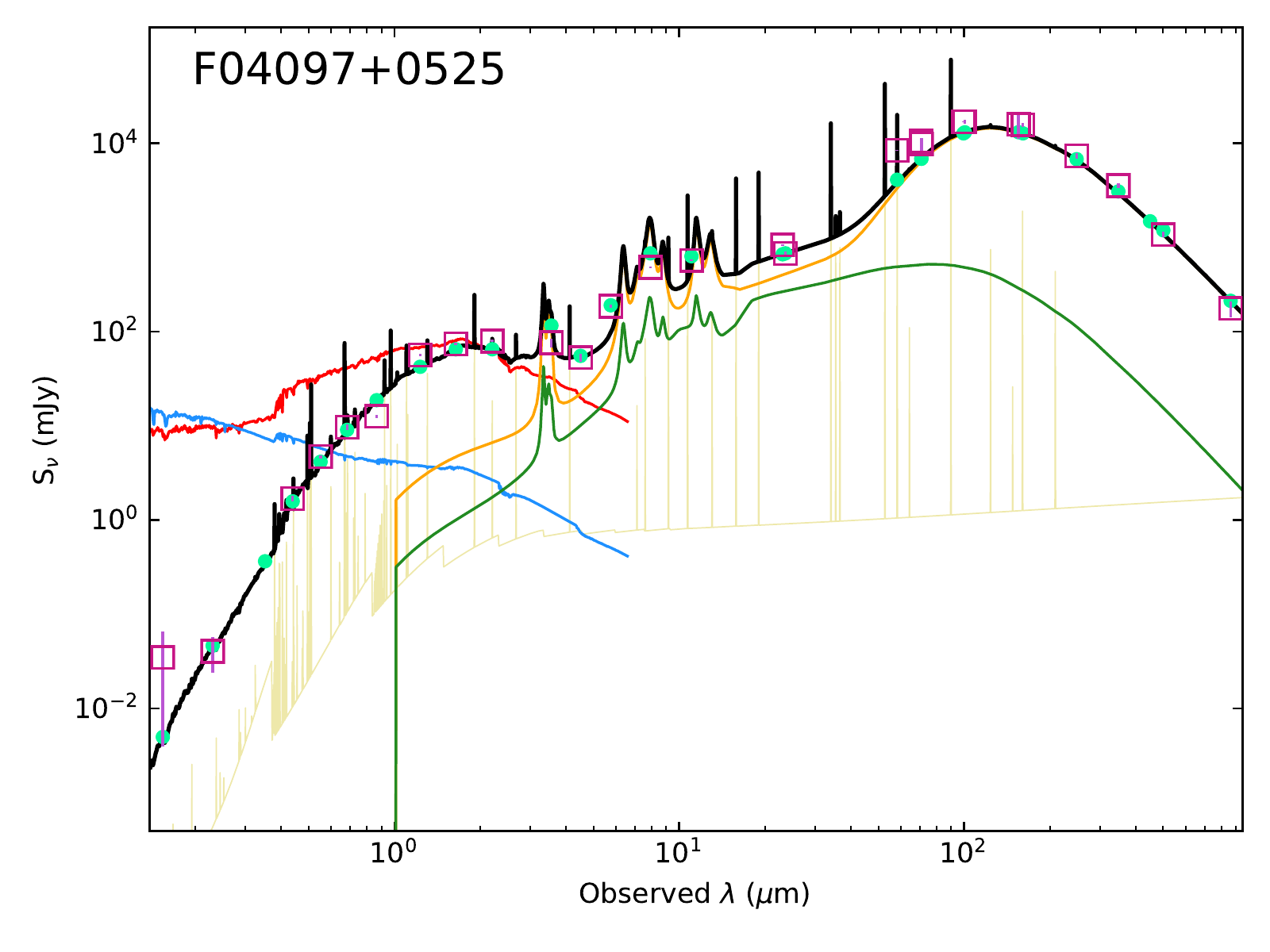}
\includegraphics[height=0.68\columnwidth]{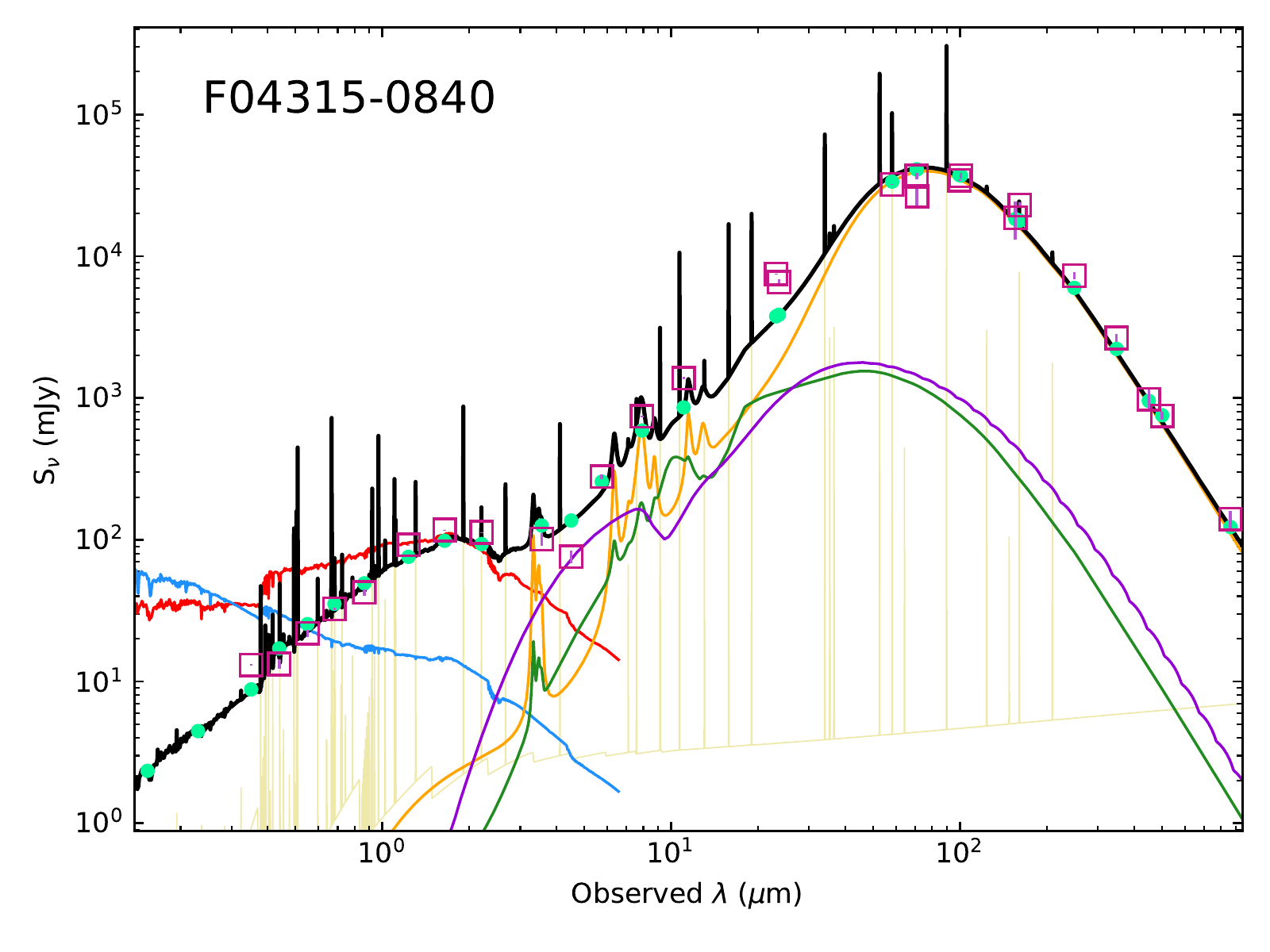}
\label{}
\end{figure}

\begin{figure}
\includegraphics[height=0.68\columnwidth]{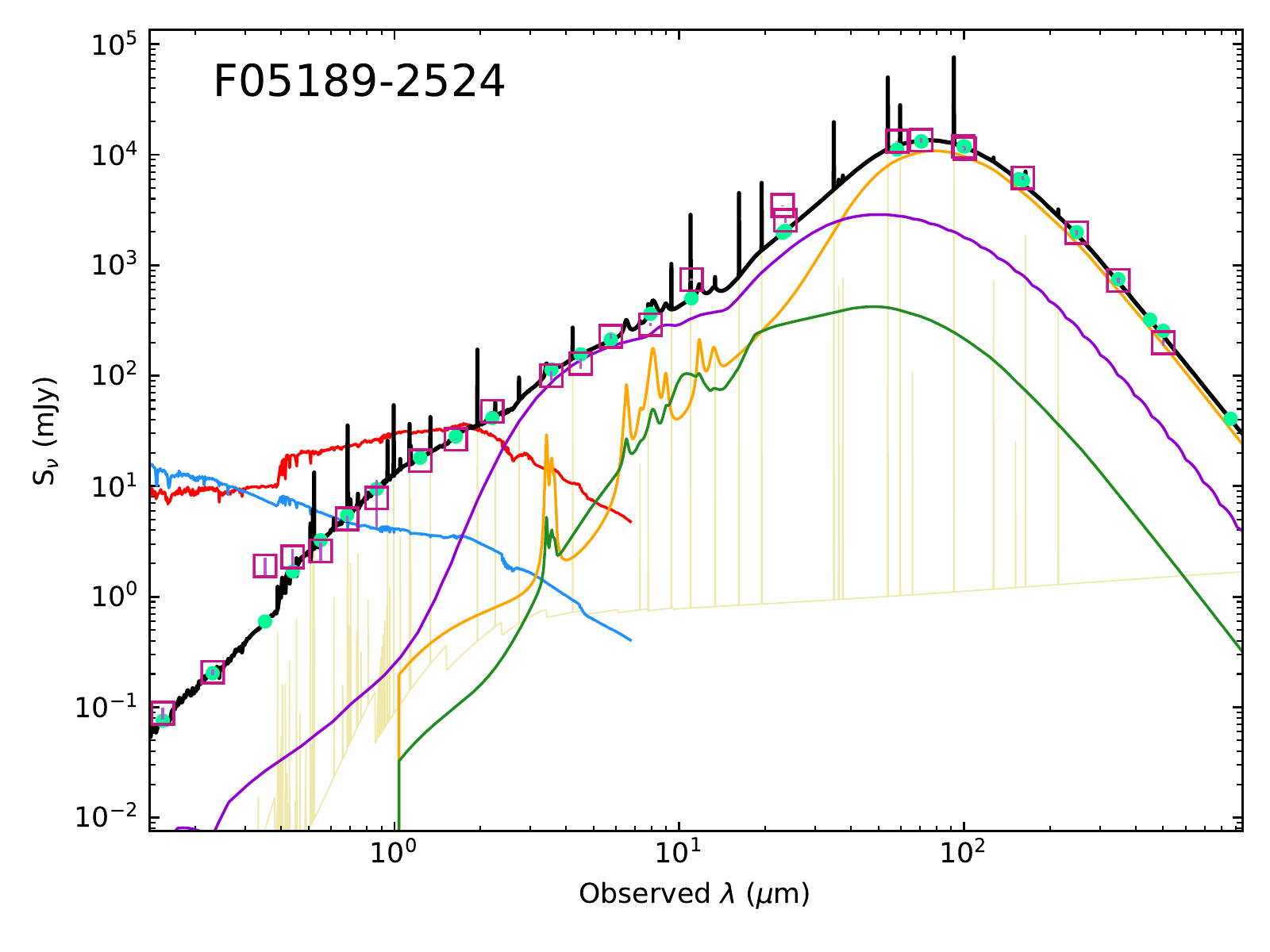}
\includegraphics[height=0.68\columnwidth]{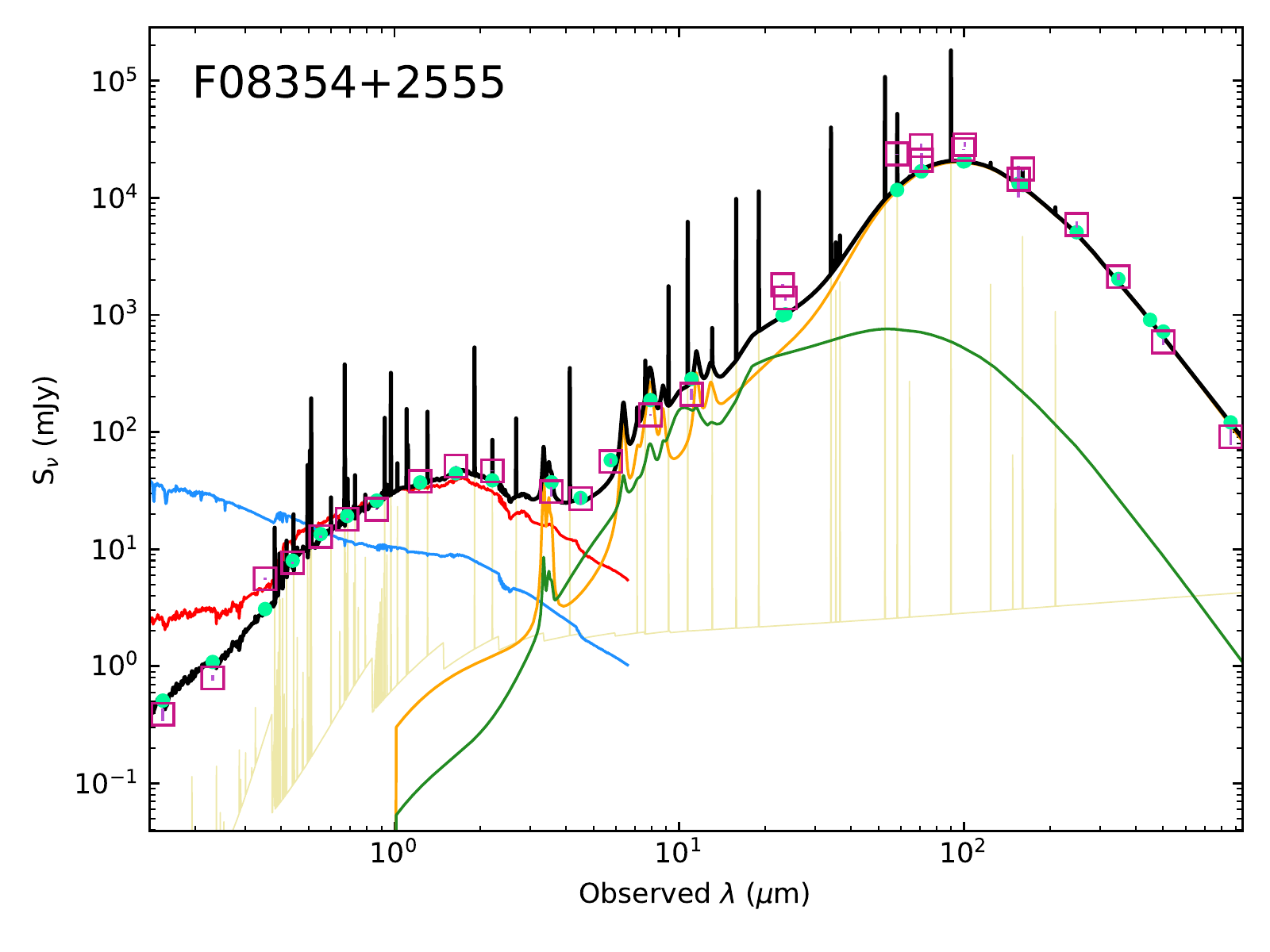}
\includegraphics[height=0.68\columnwidth]{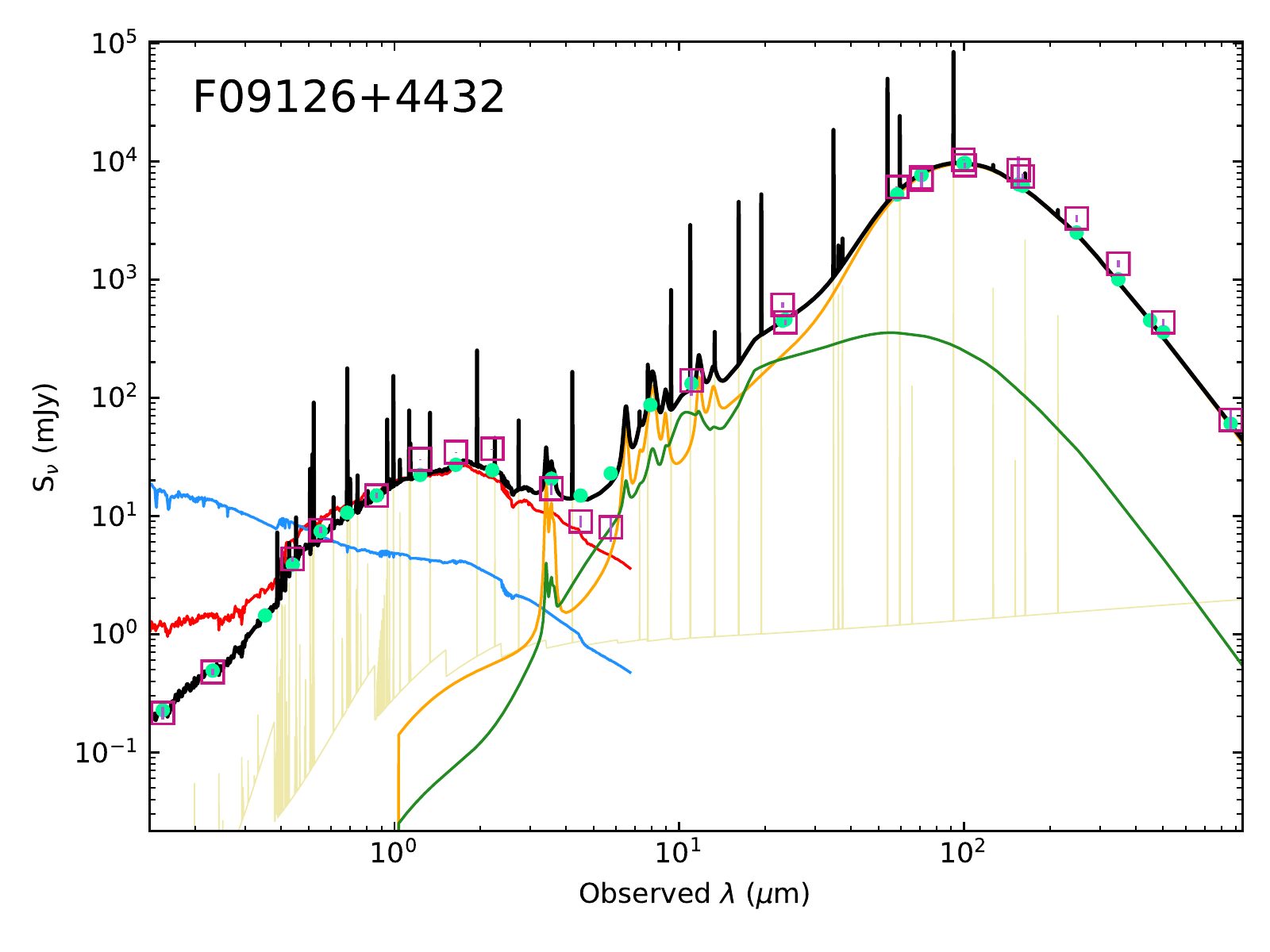}
\includegraphics[height=0.68\columnwidth]{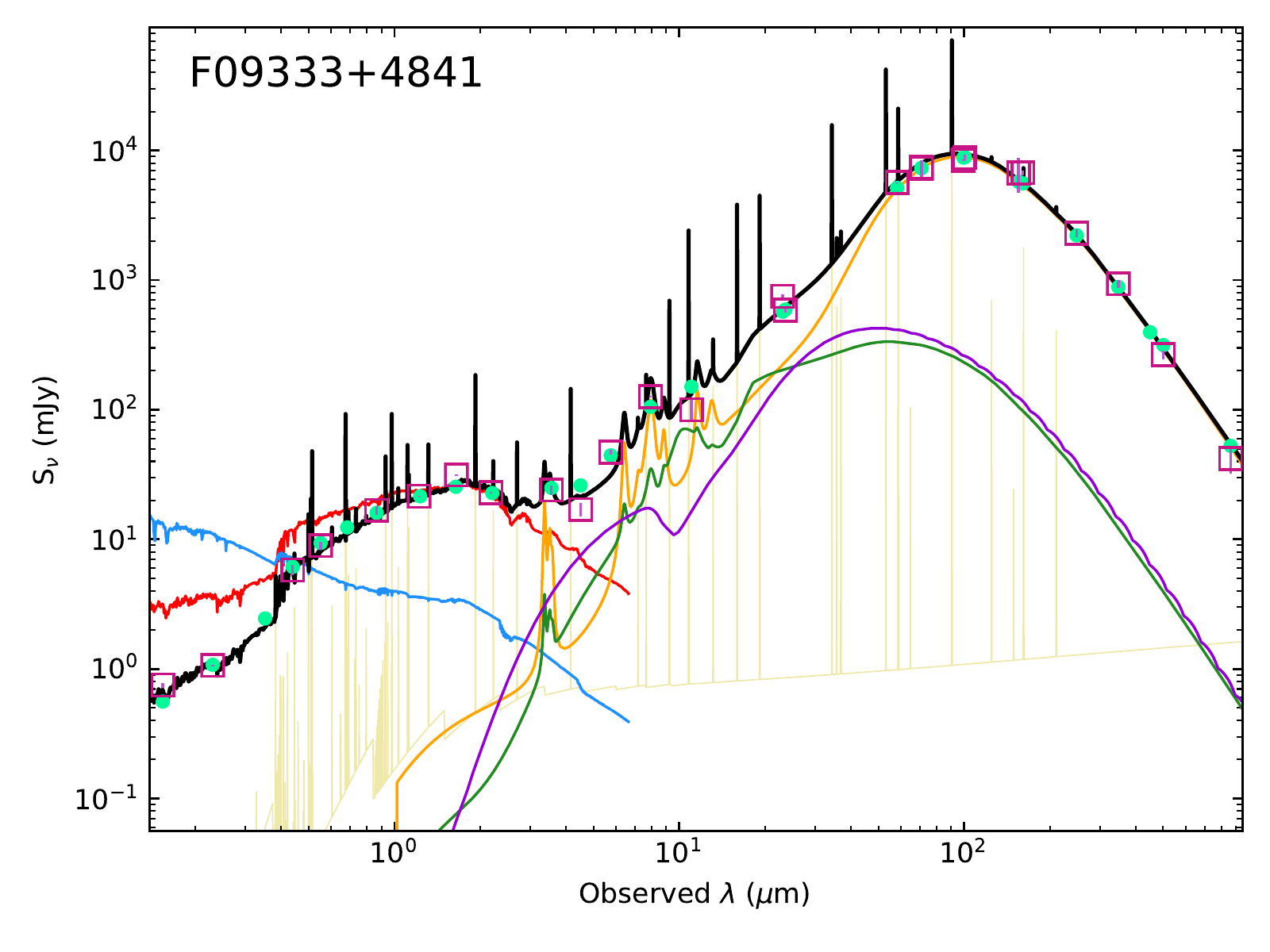}
\label{}
\end{figure}
\begin{figure}
\includegraphics[height=0.68\columnwidth]{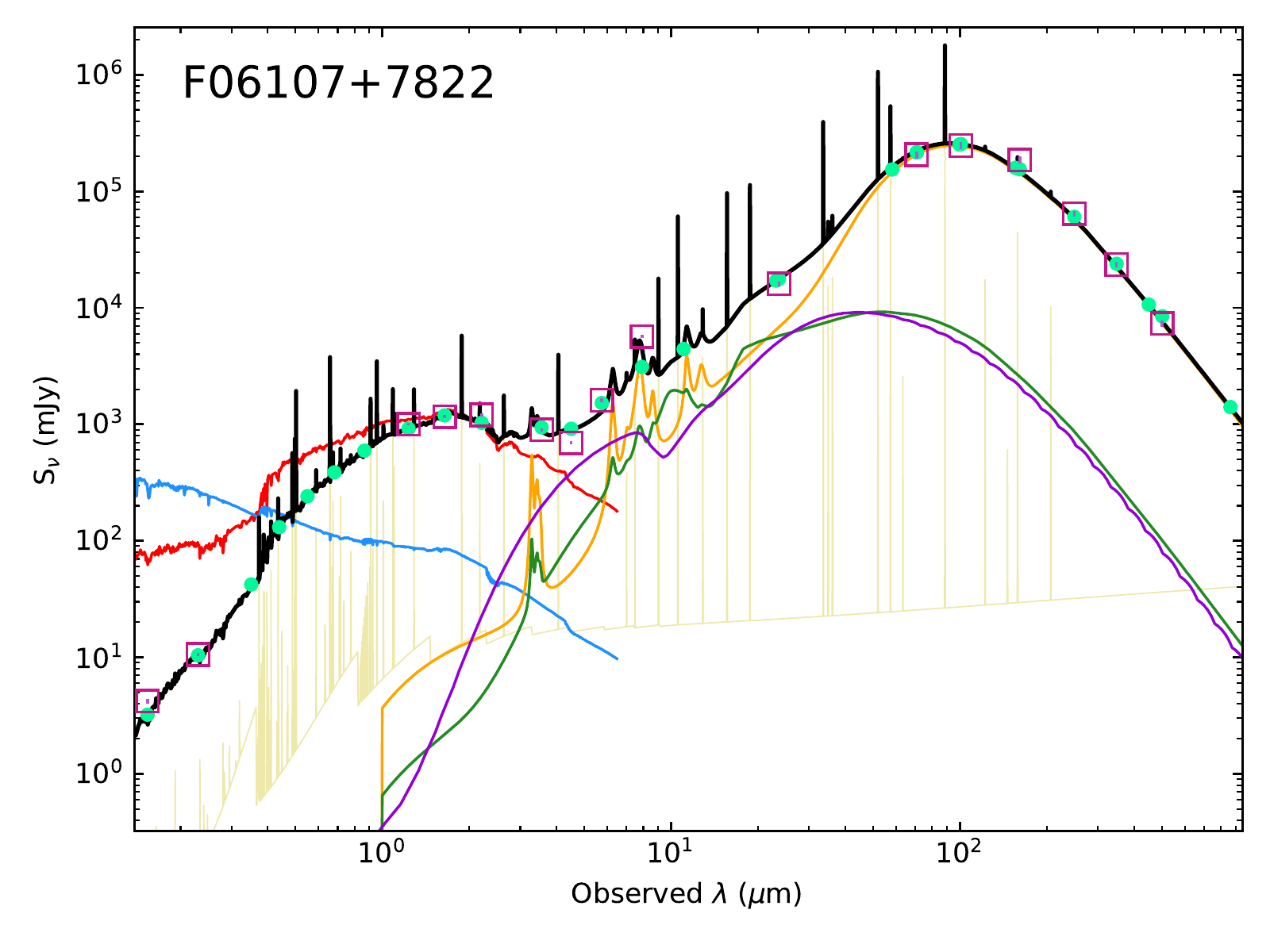}
\includegraphics[height=0.68\columnwidth]{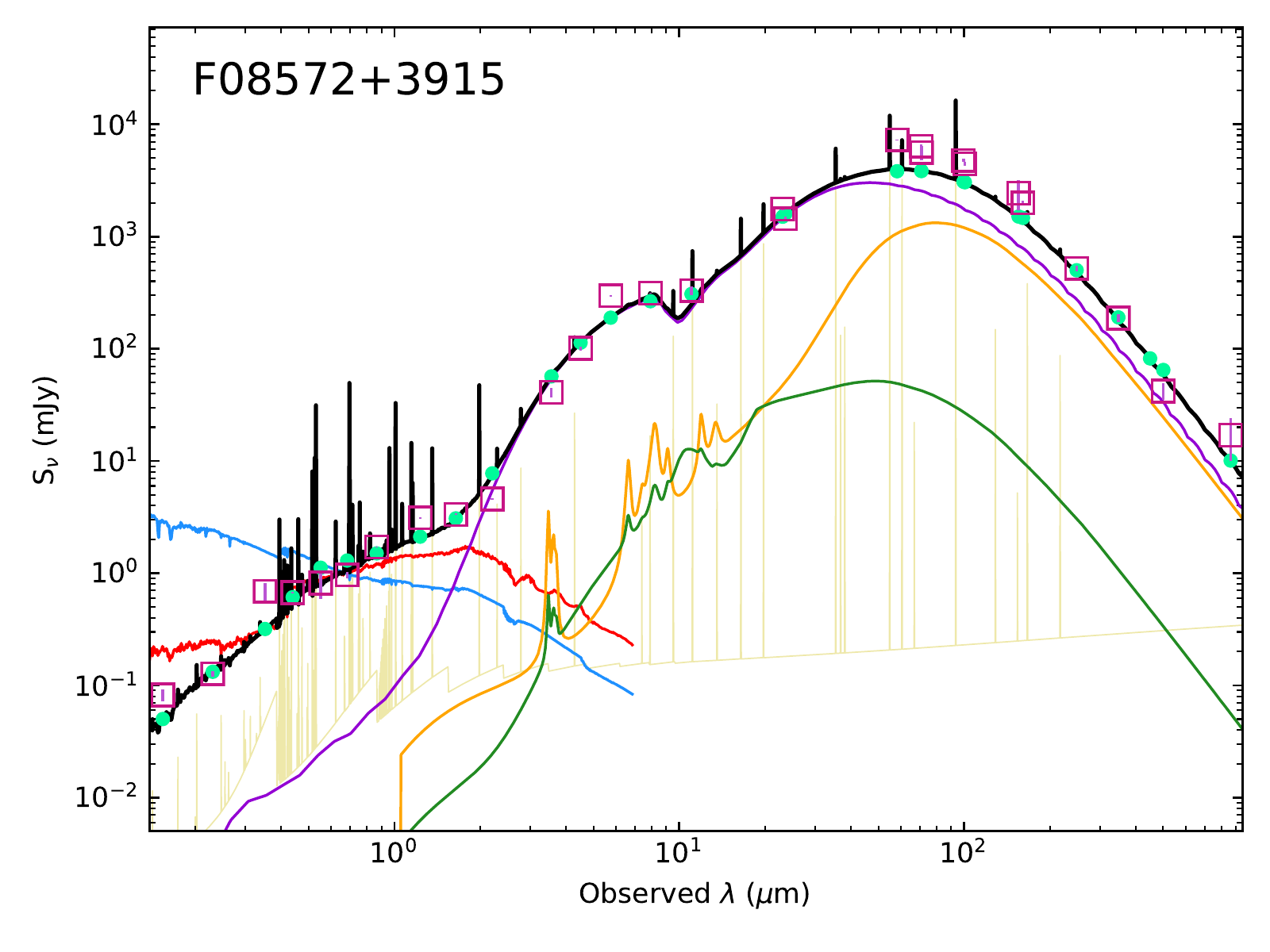}
\includegraphics[height=0.68\columnwidth]{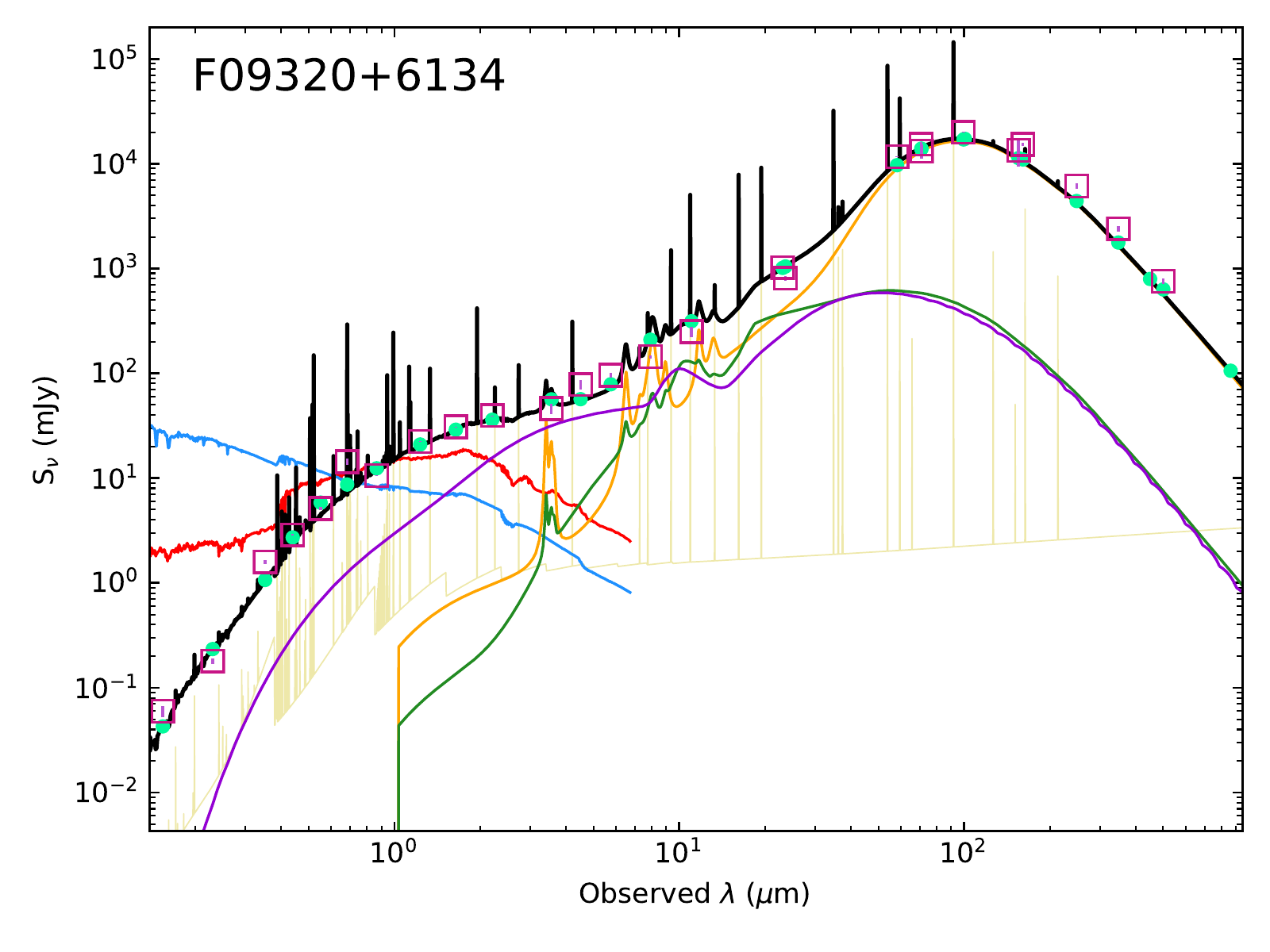}
\includegraphics[height=0.68\columnwidth]{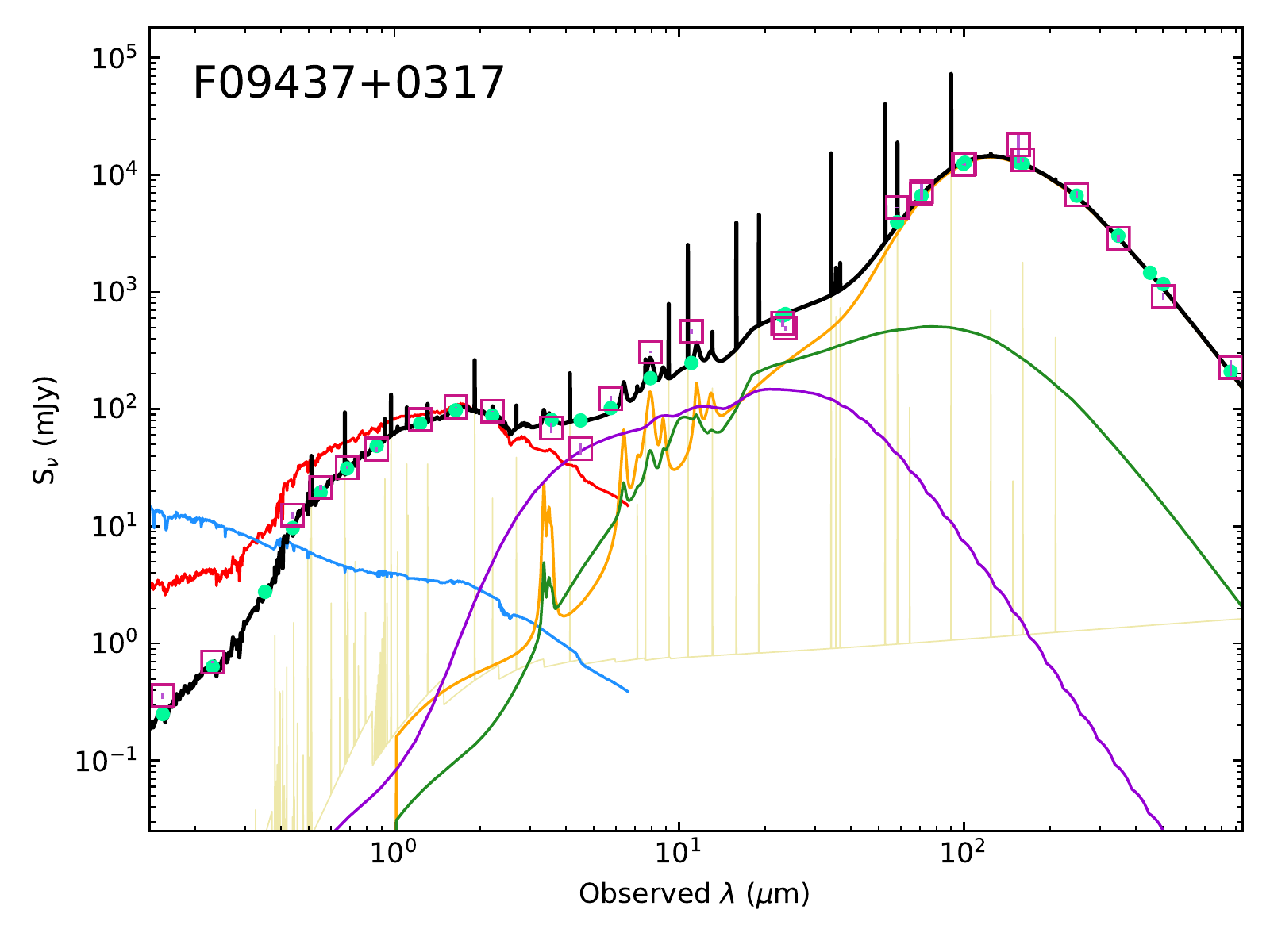}
\label{4}
\end{figure}

\begin{figure}
\includegraphics[height=0.68\columnwidth]{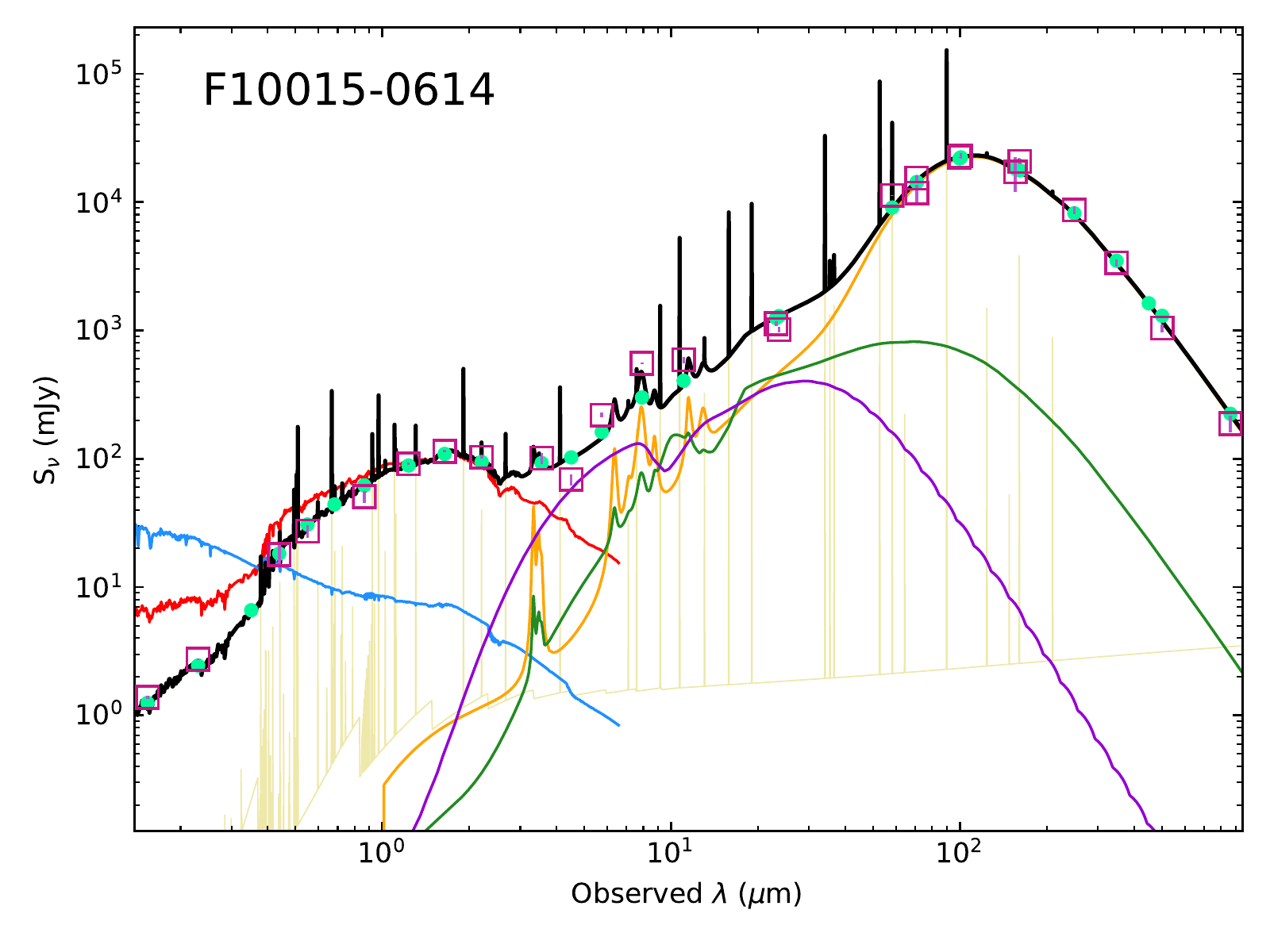}
\includegraphics[height=0.68\columnwidth]{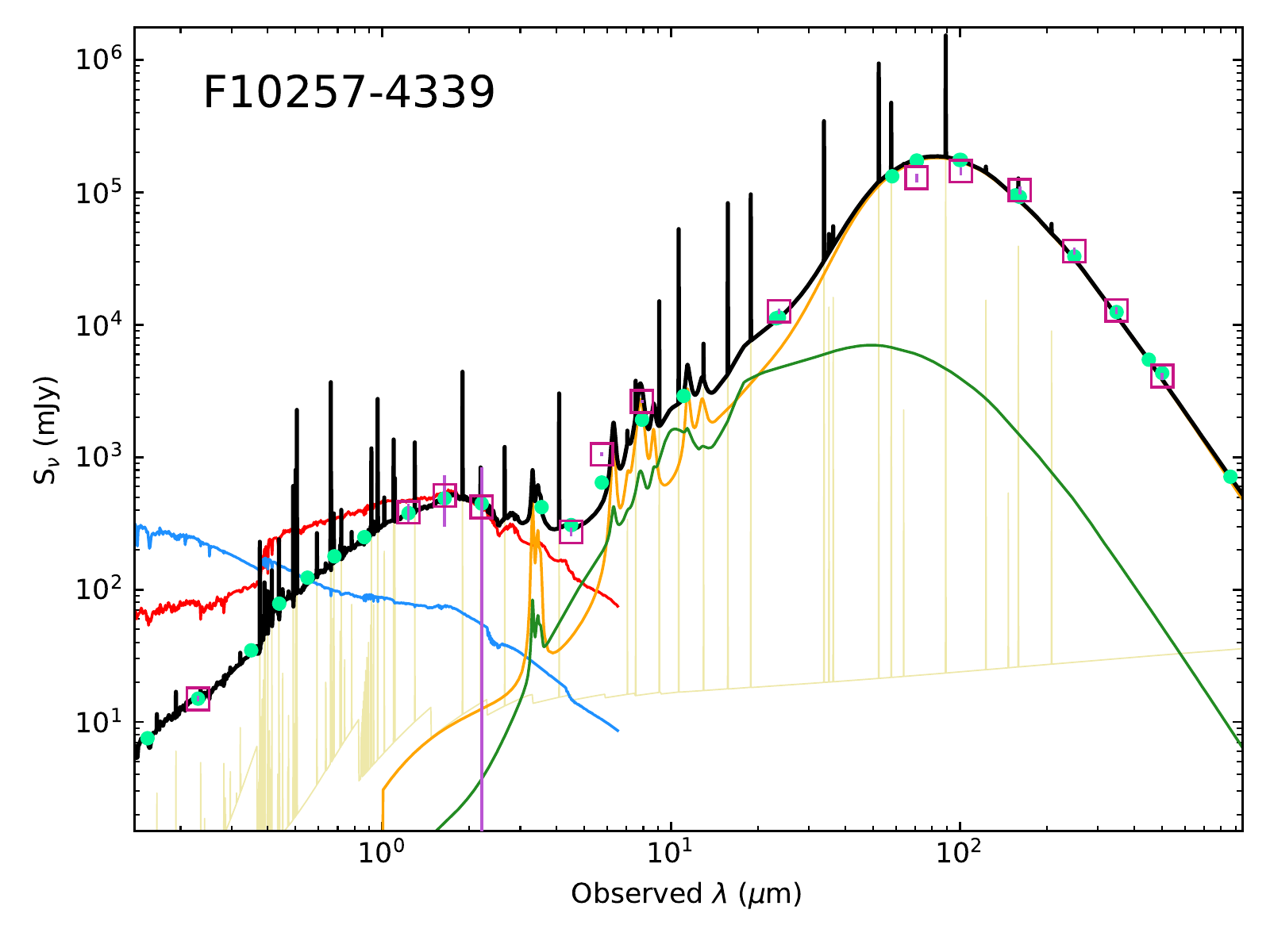}
\includegraphics[height=0.68\columnwidth]{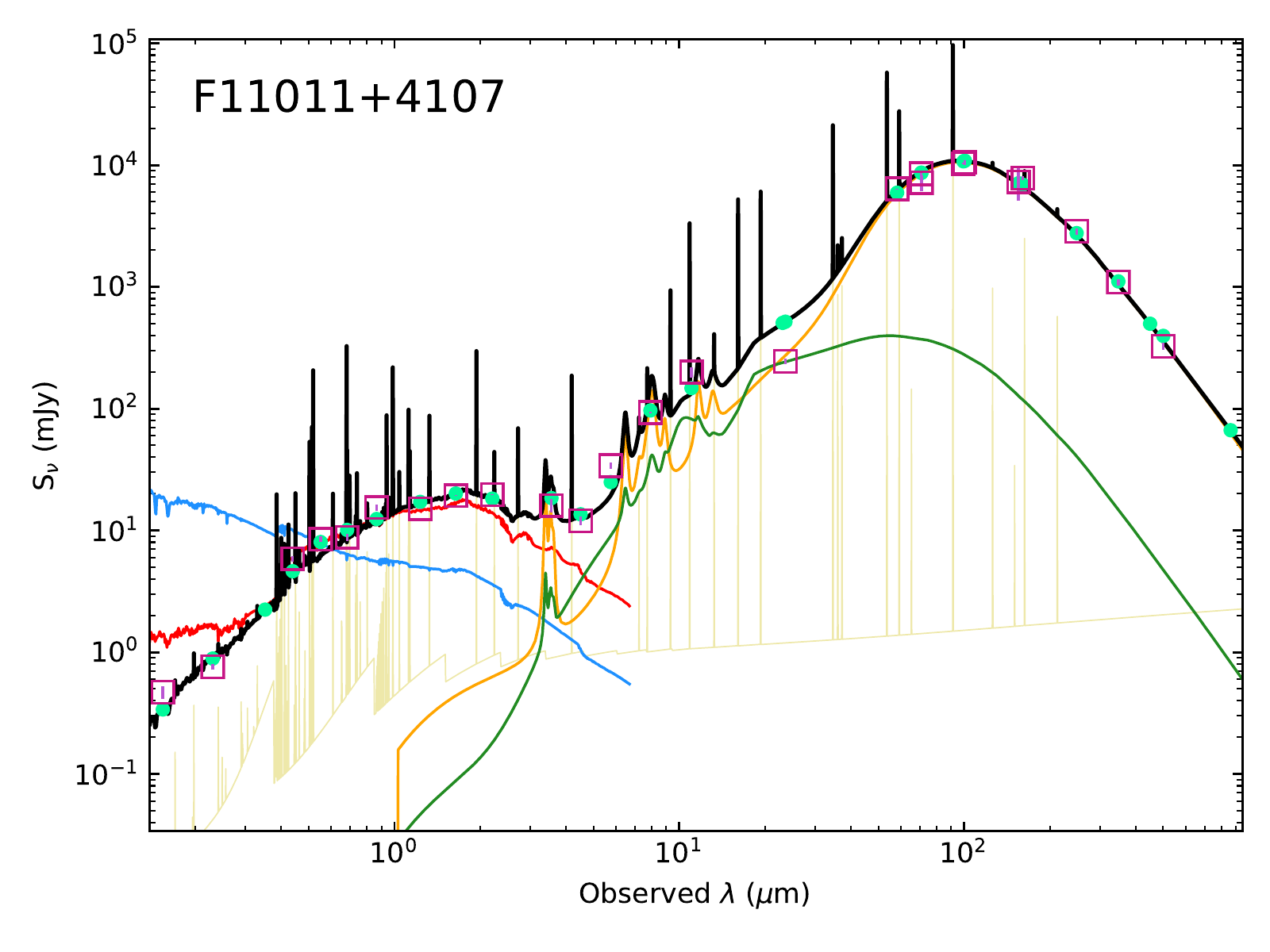}
\includegraphics[height=0.68\columnwidth]{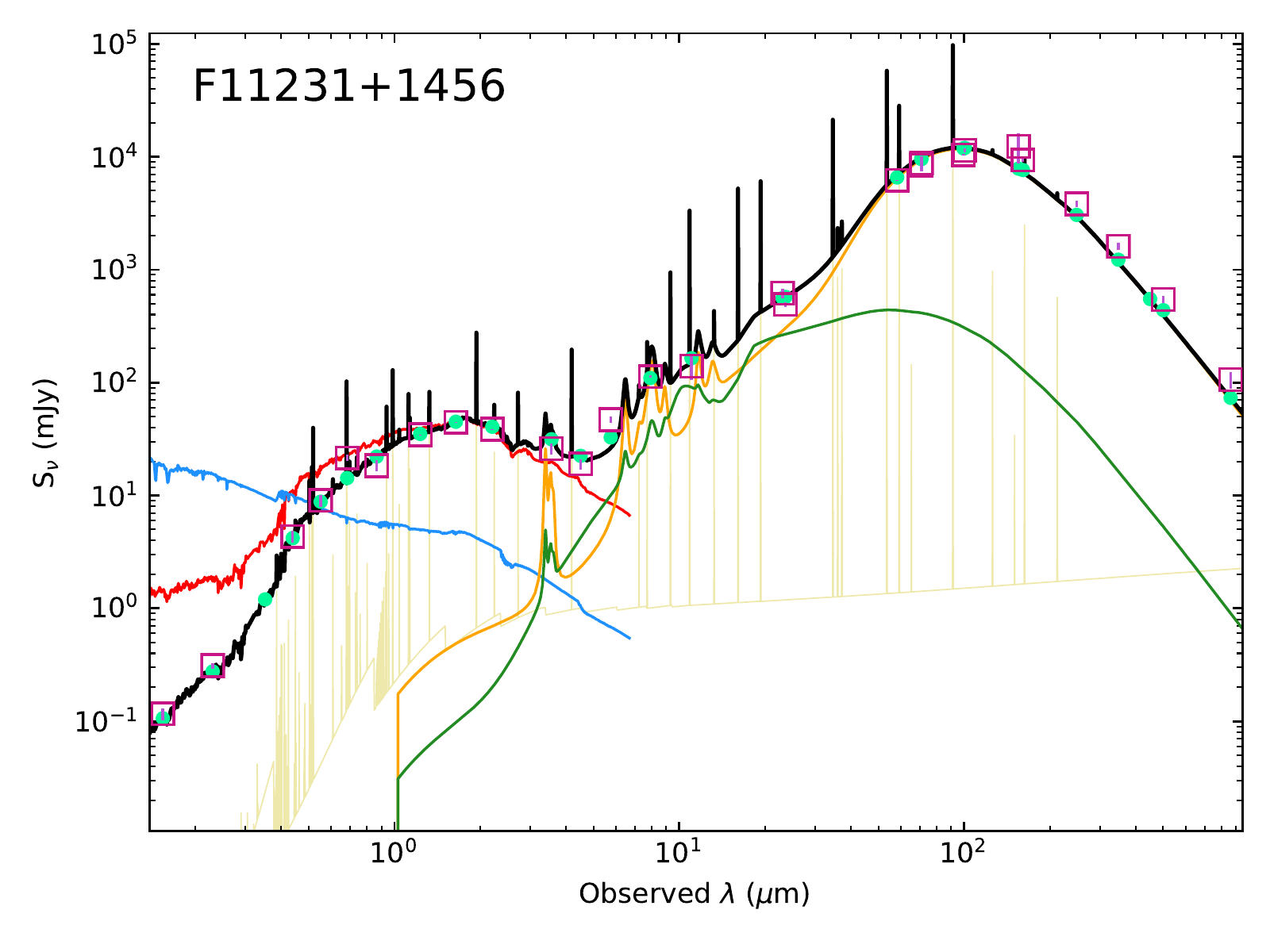}
\label{4}
\end{figure}
\begin{figure}
\includegraphics[height=0.68\columnwidth]{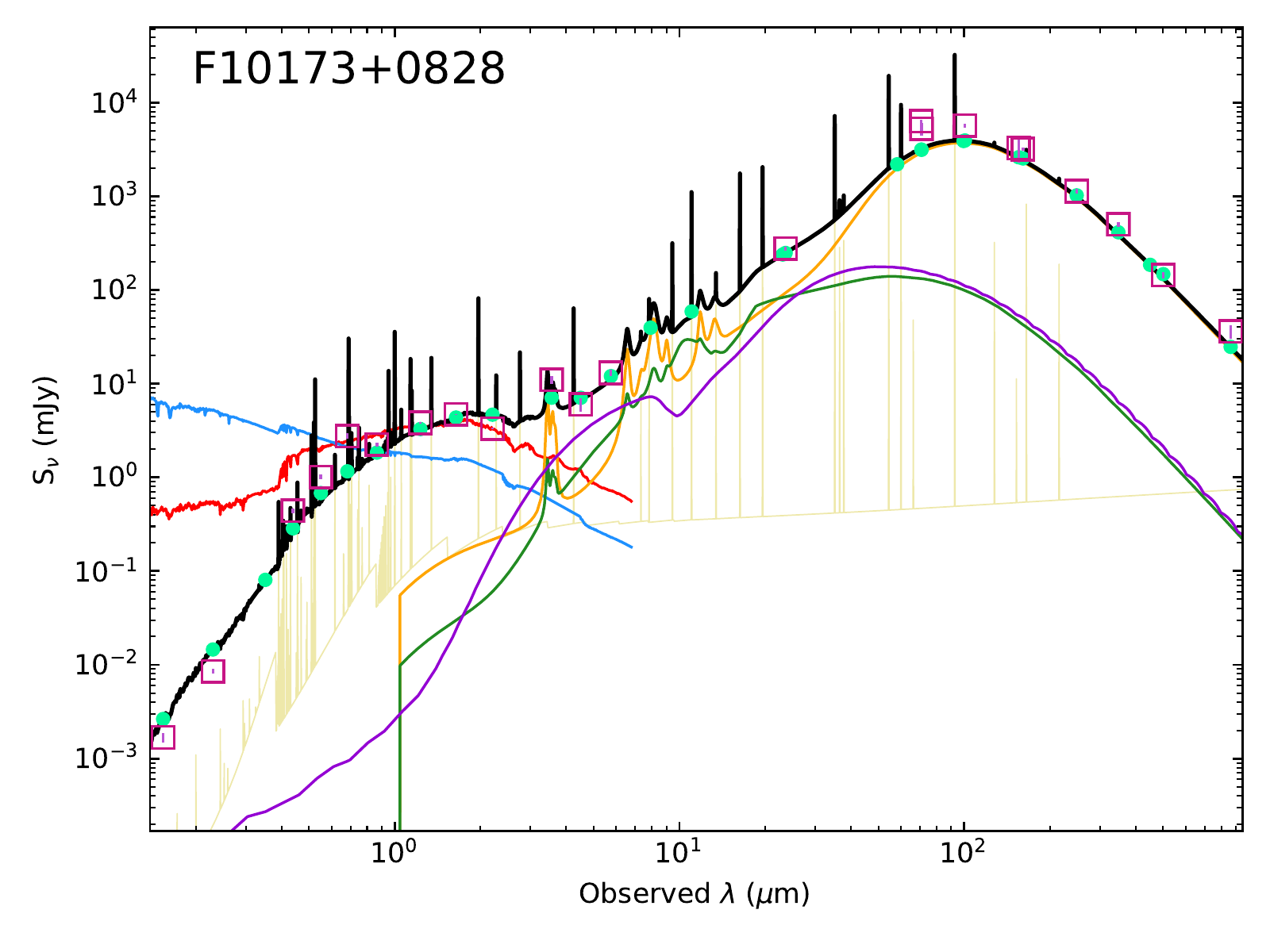}
\includegraphics[height=0.68\columnwidth]{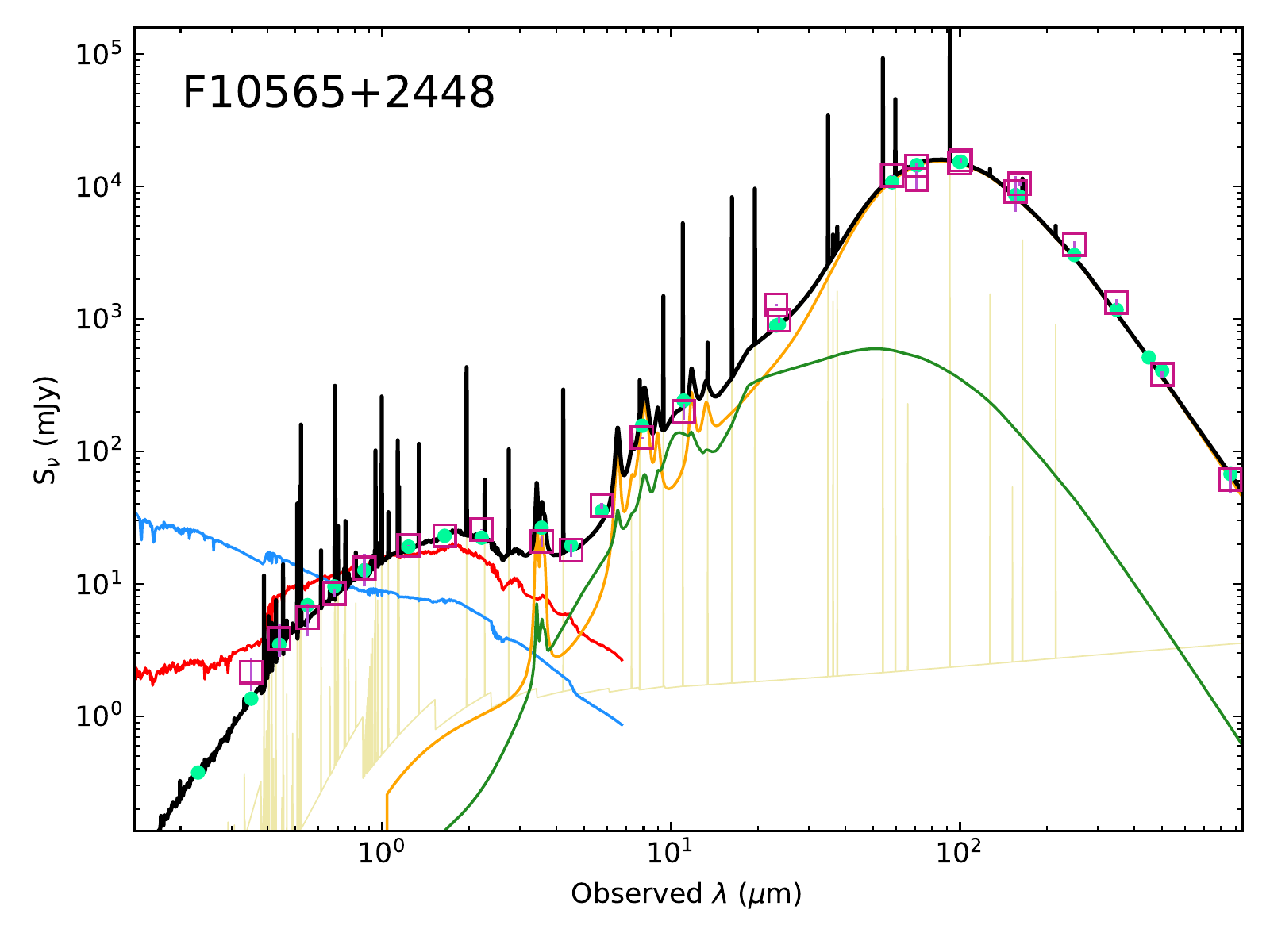}
\includegraphics[height=0.68\columnwidth]{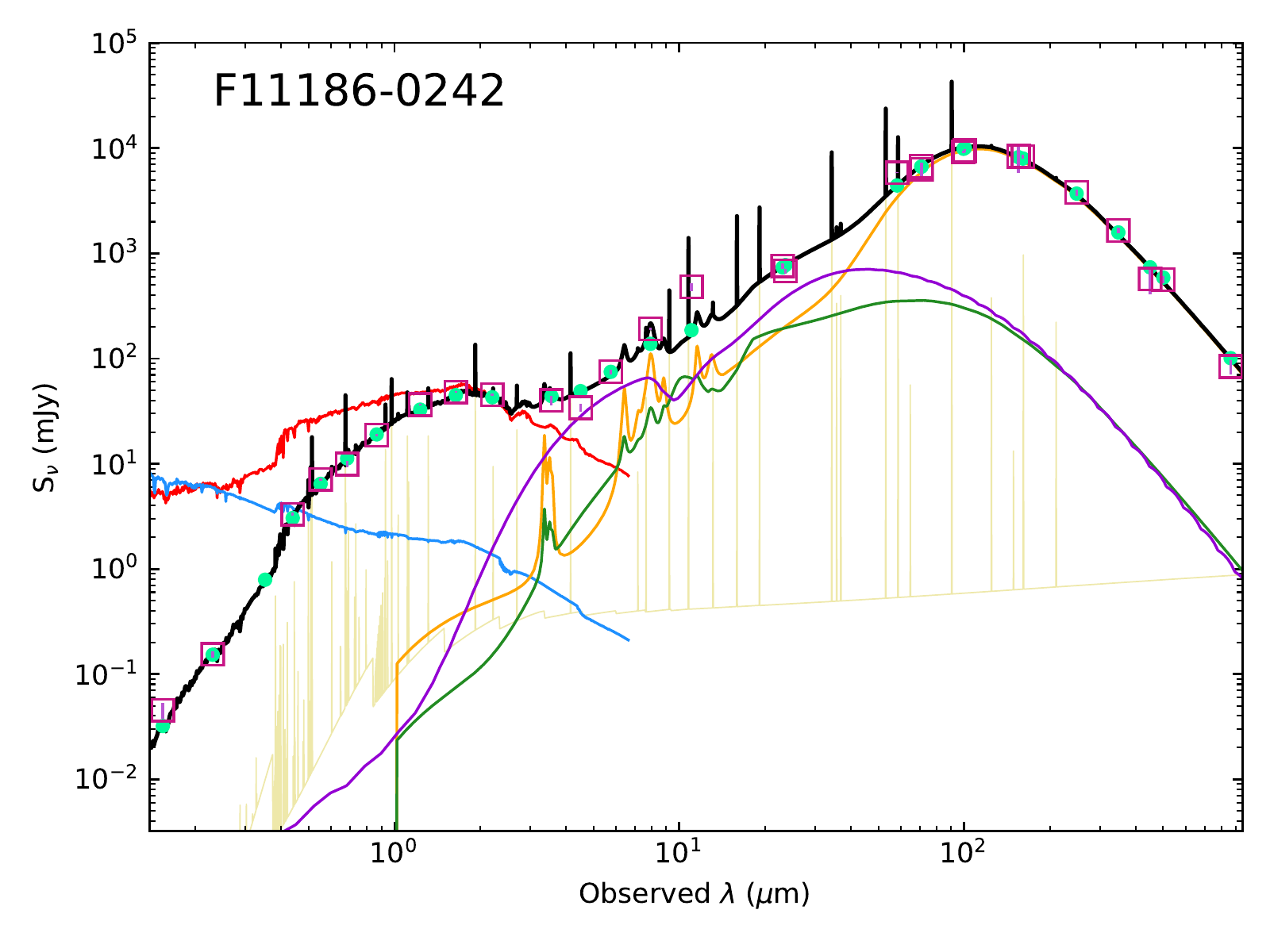}
\includegraphics[height=0.68\columnwidth]{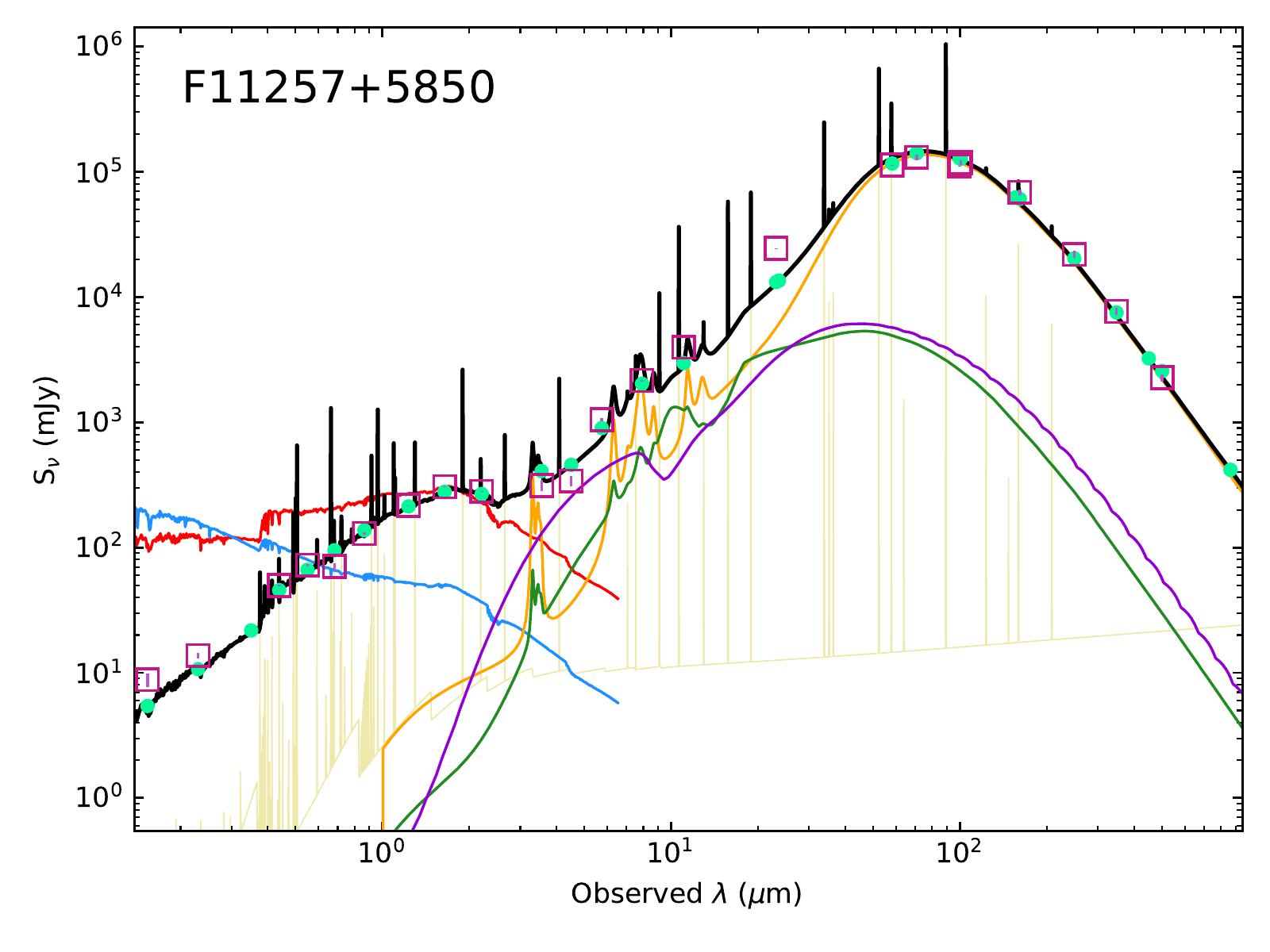}
\label{}
\end{figure}

\begin{figure}
\includegraphics[height=0.68\columnwidth]{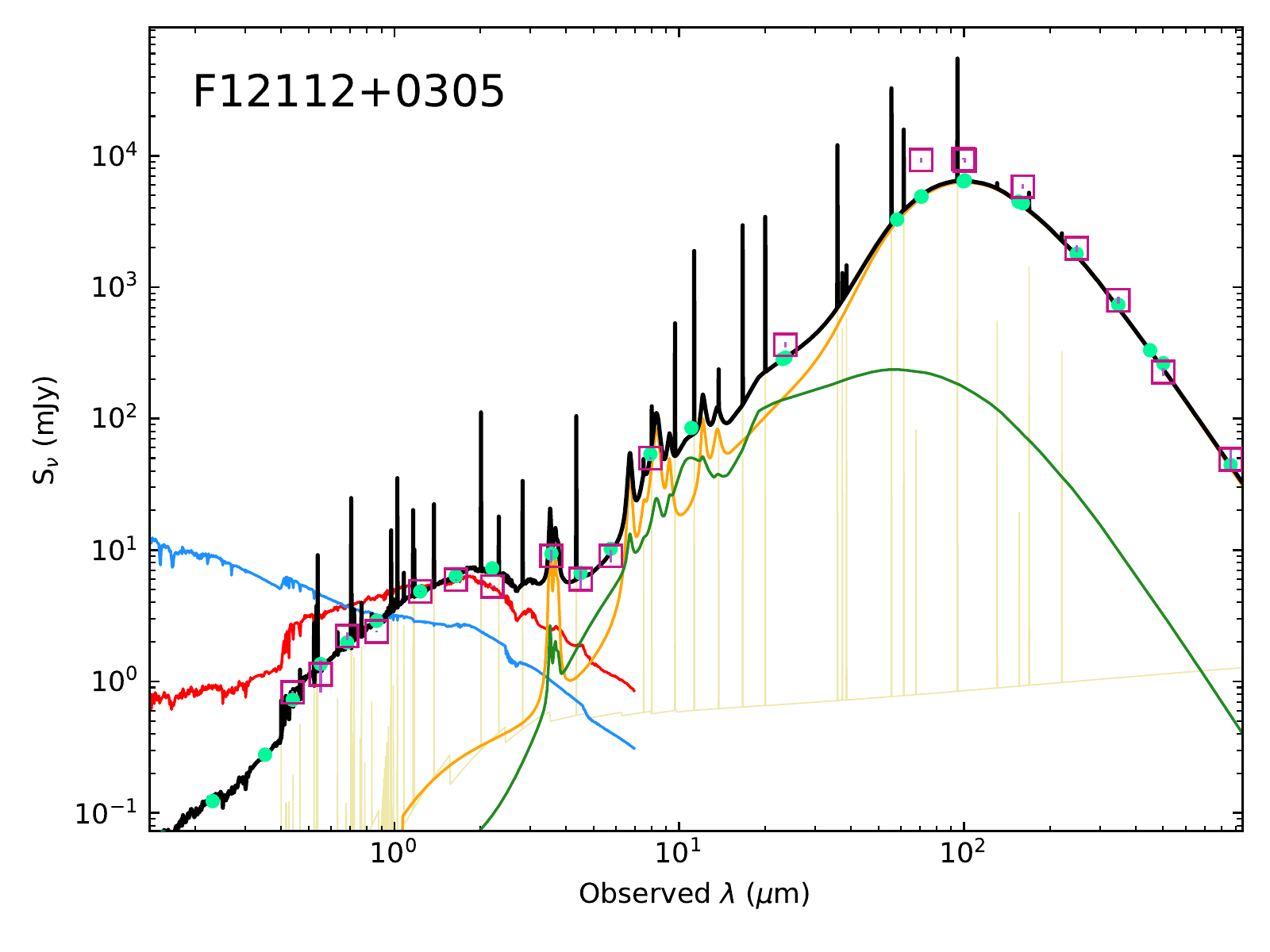}
\includegraphics[height=0.68\columnwidth]{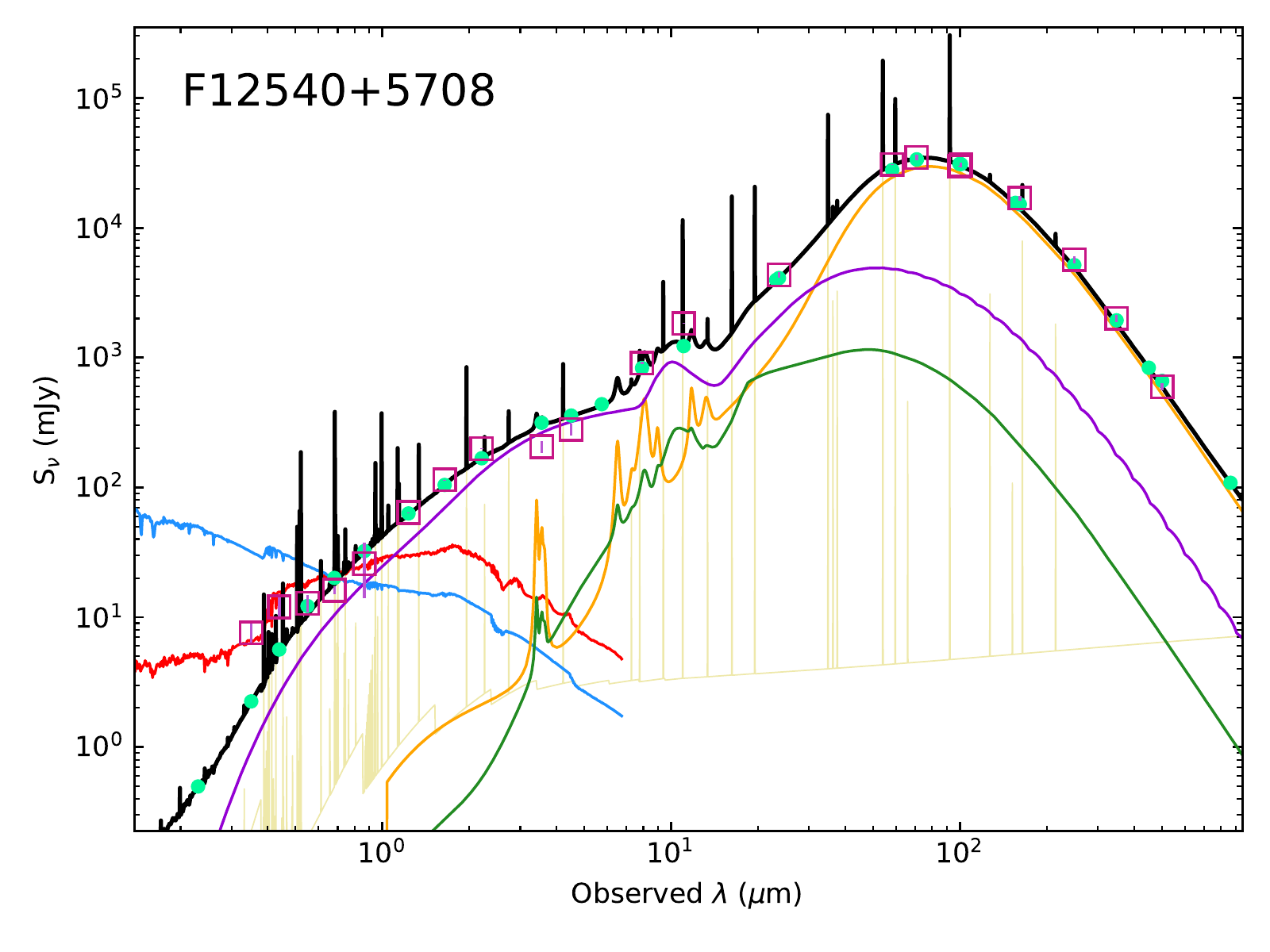}
\includegraphics[height=0.68\columnwidth]{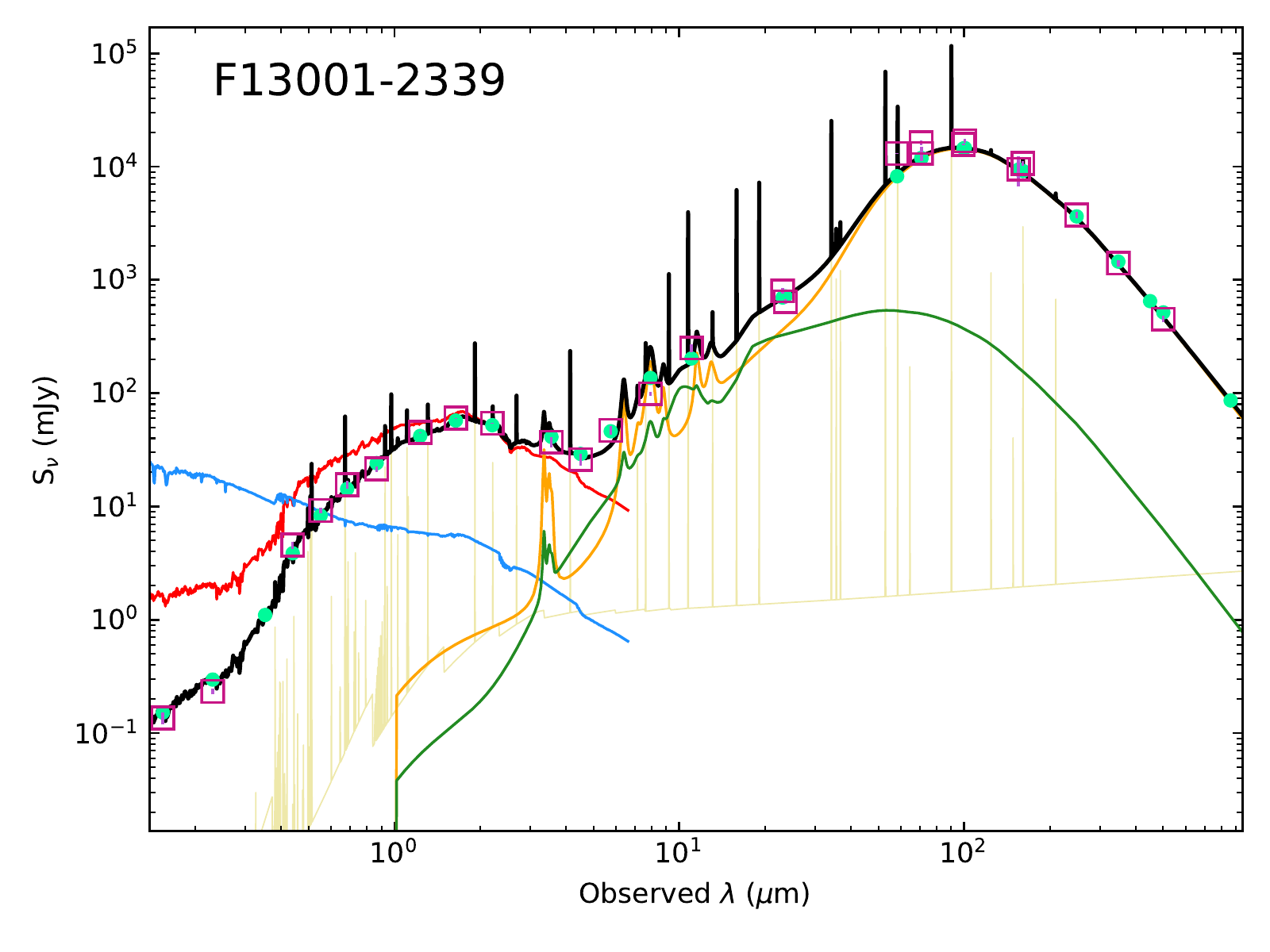}
\includegraphics[height=0.68\columnwidth]{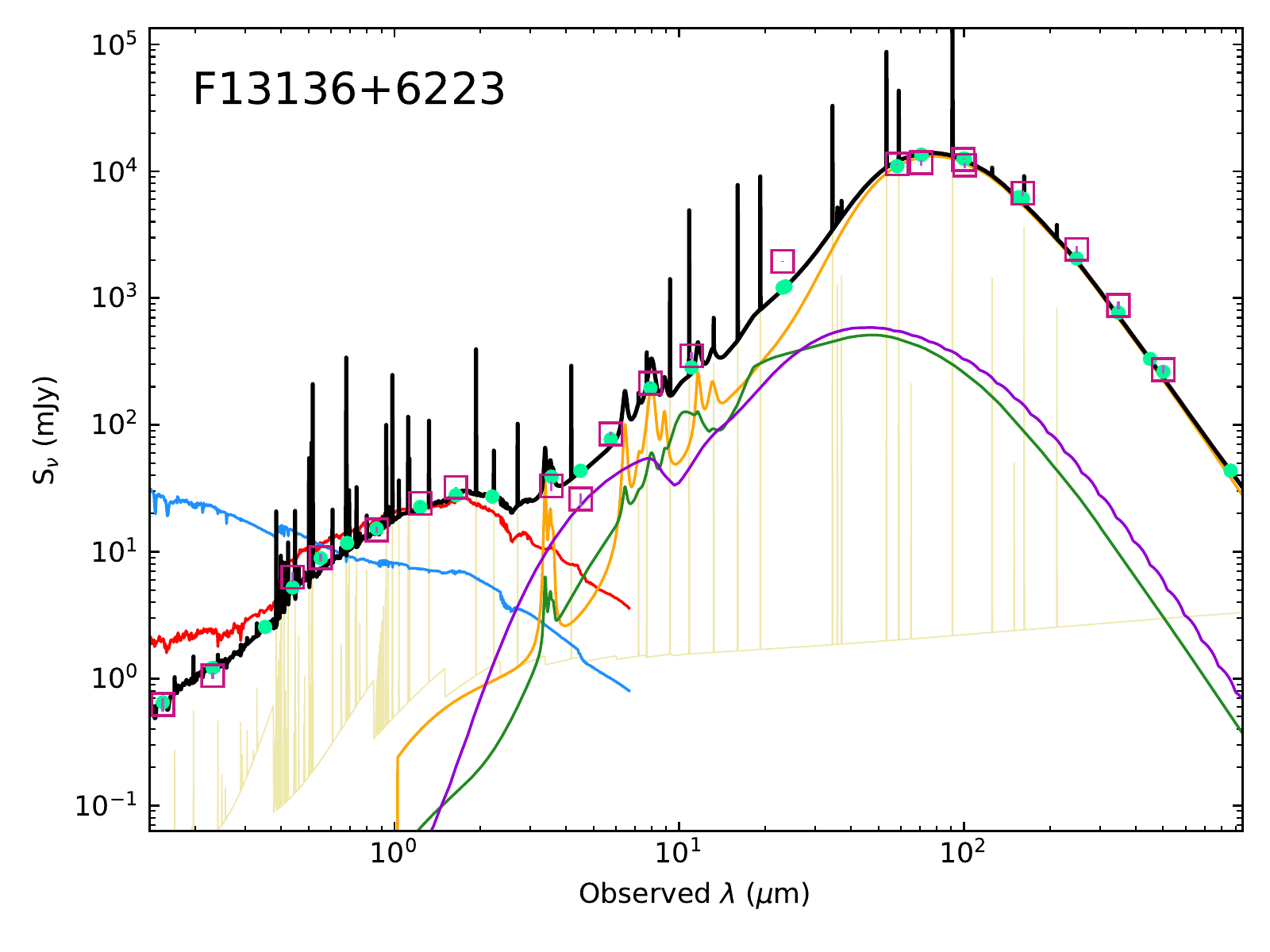}
\label{4}
\end{figure}
\begin{figure}
\includegraphics[height=0.68\columnwidth]{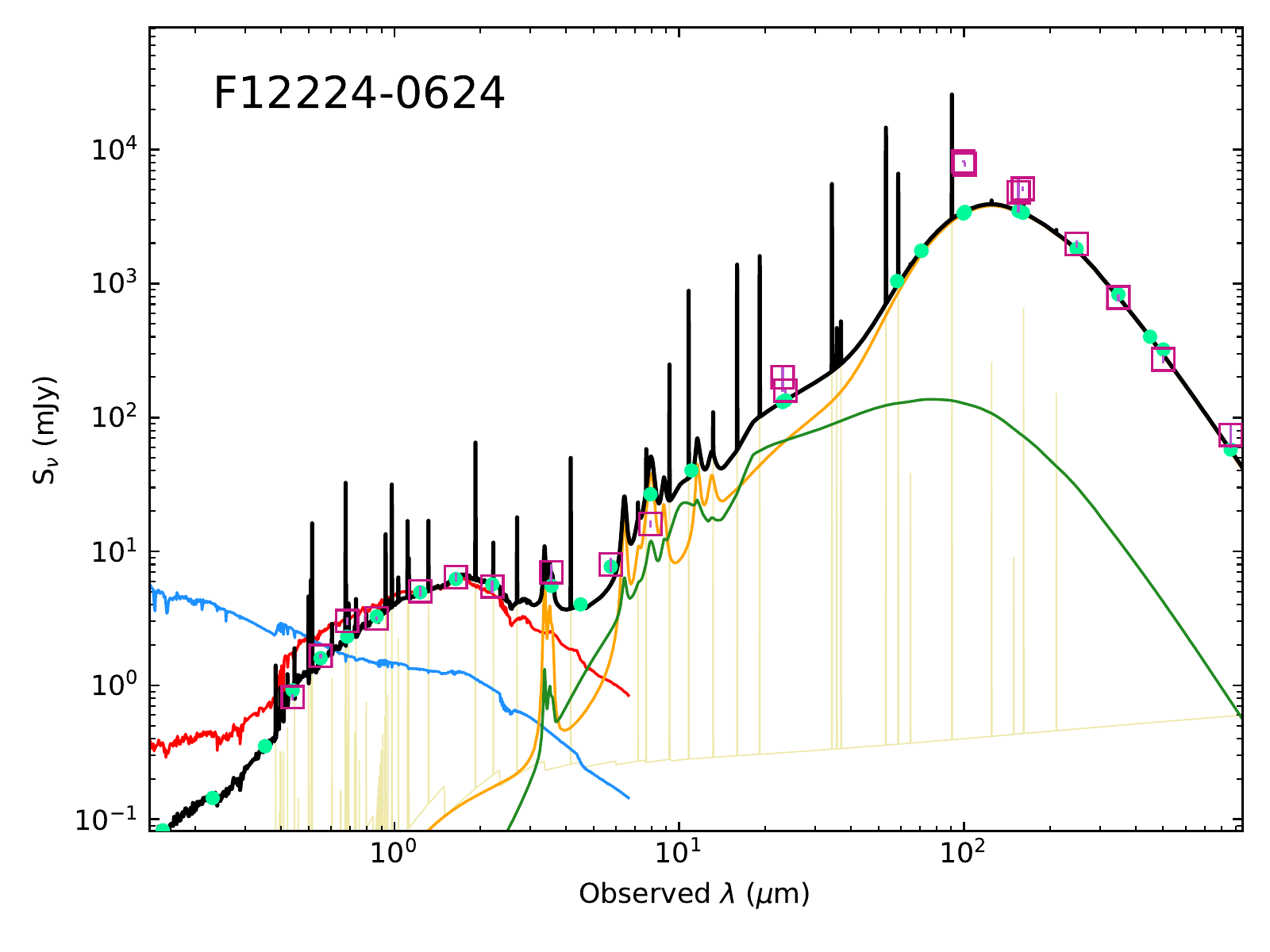}
\includegraphics[height=0.68\columnwidth]{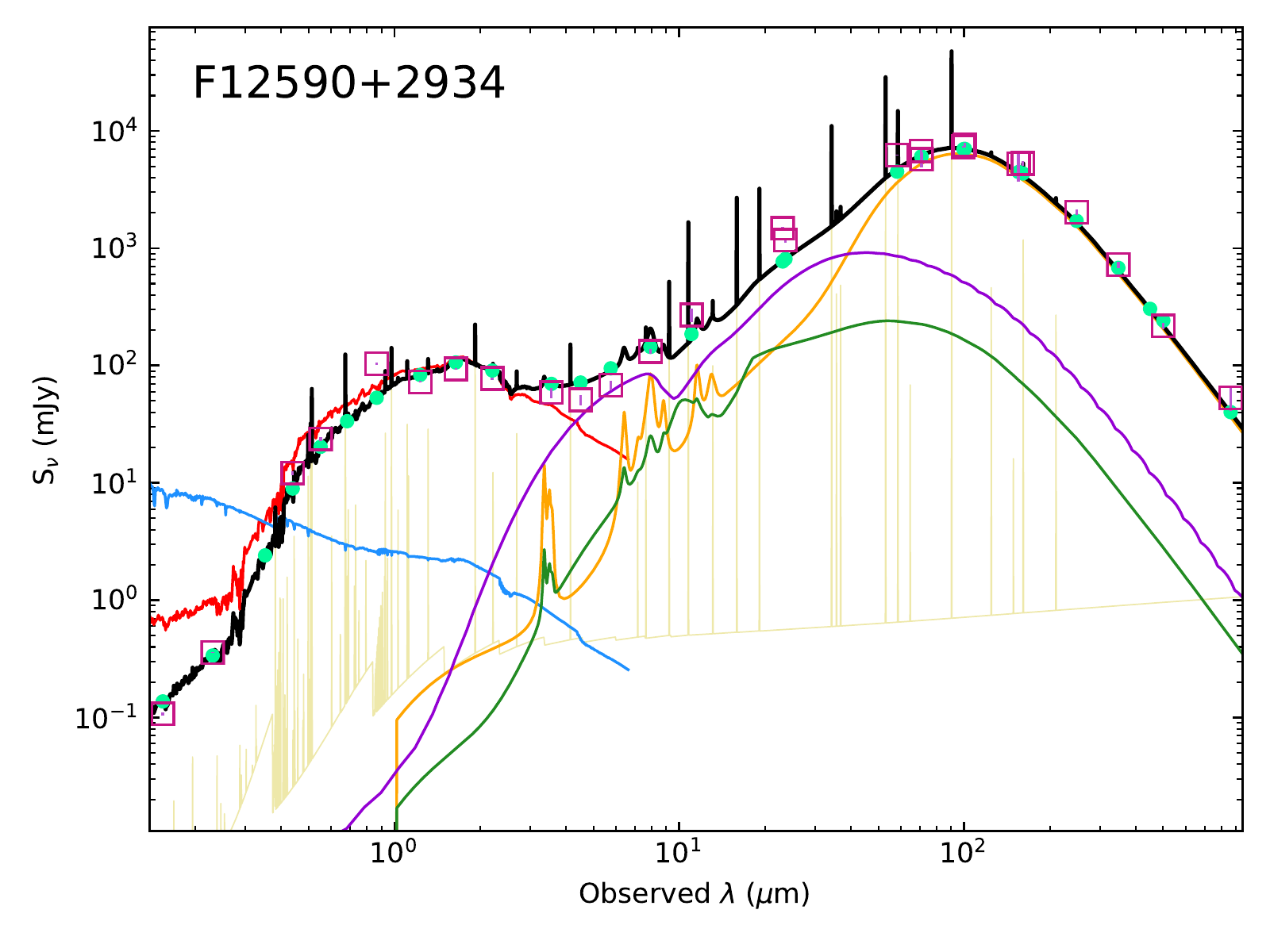}
\includegraphics[height=0.68\columnwidth]{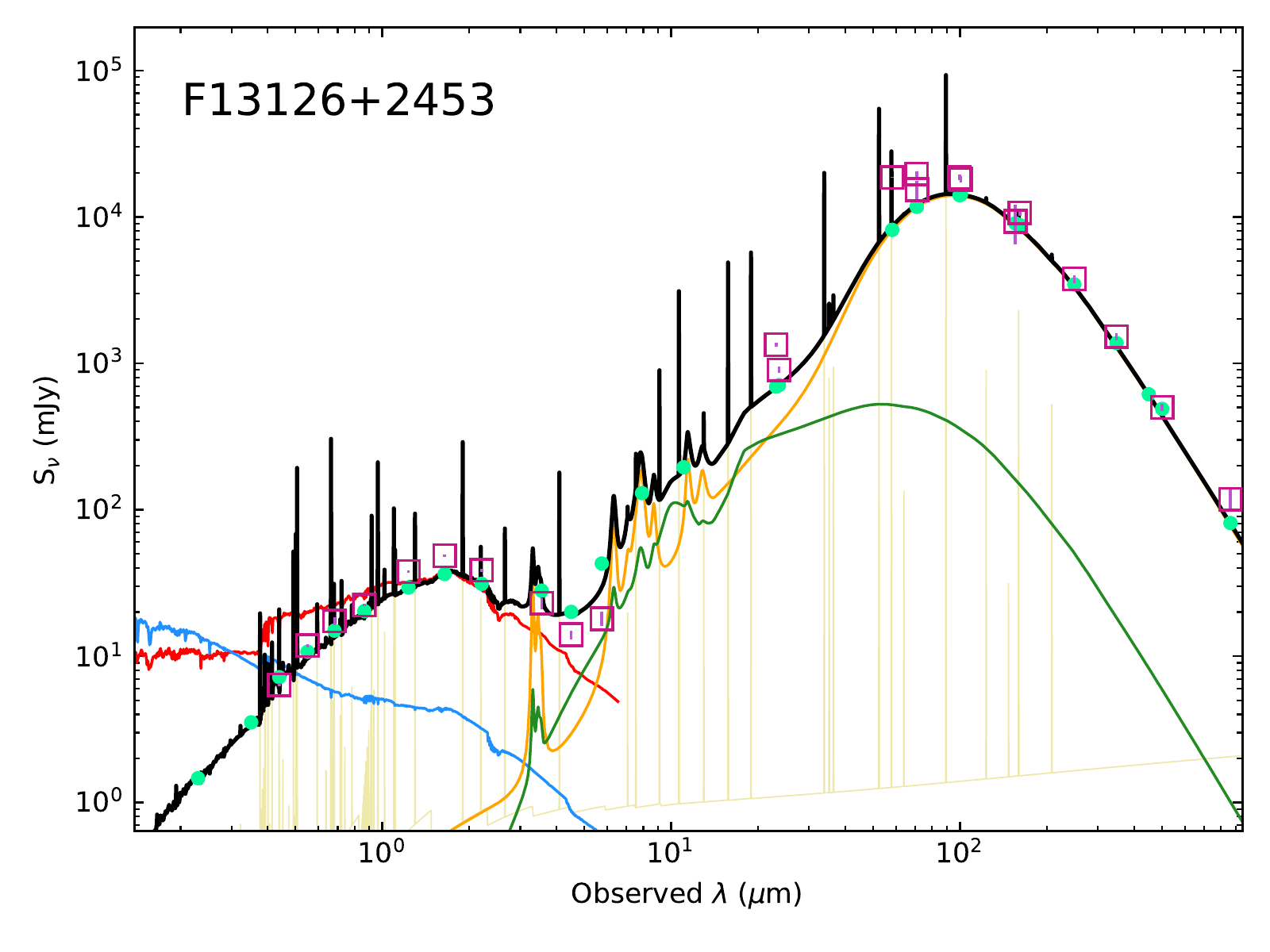}
\includegraphics[height=0.68\columnwidth]{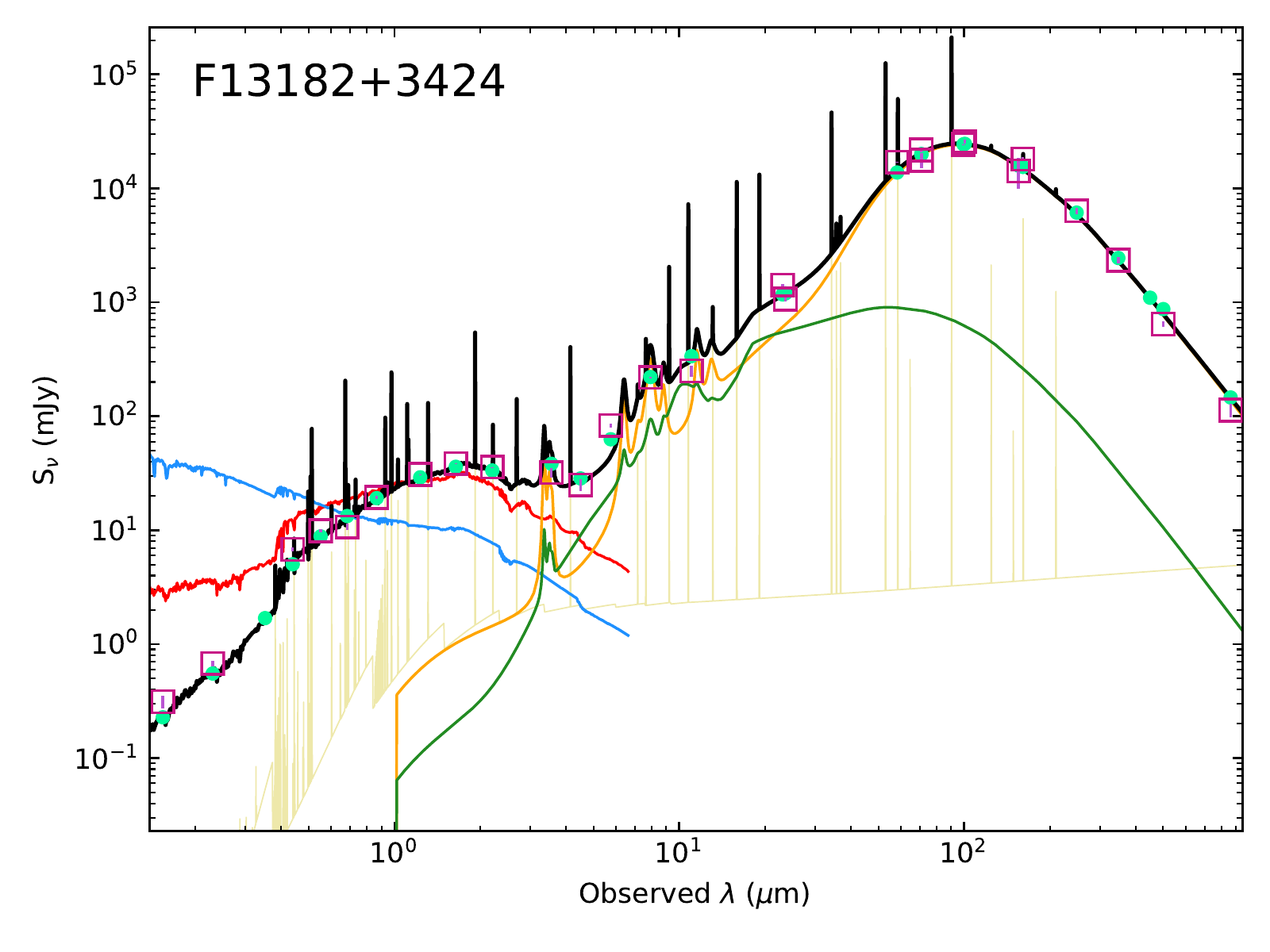}
\label{}
\end{figure}

\begin{figure}
\includegraphics[height=0.68\columnwidth]{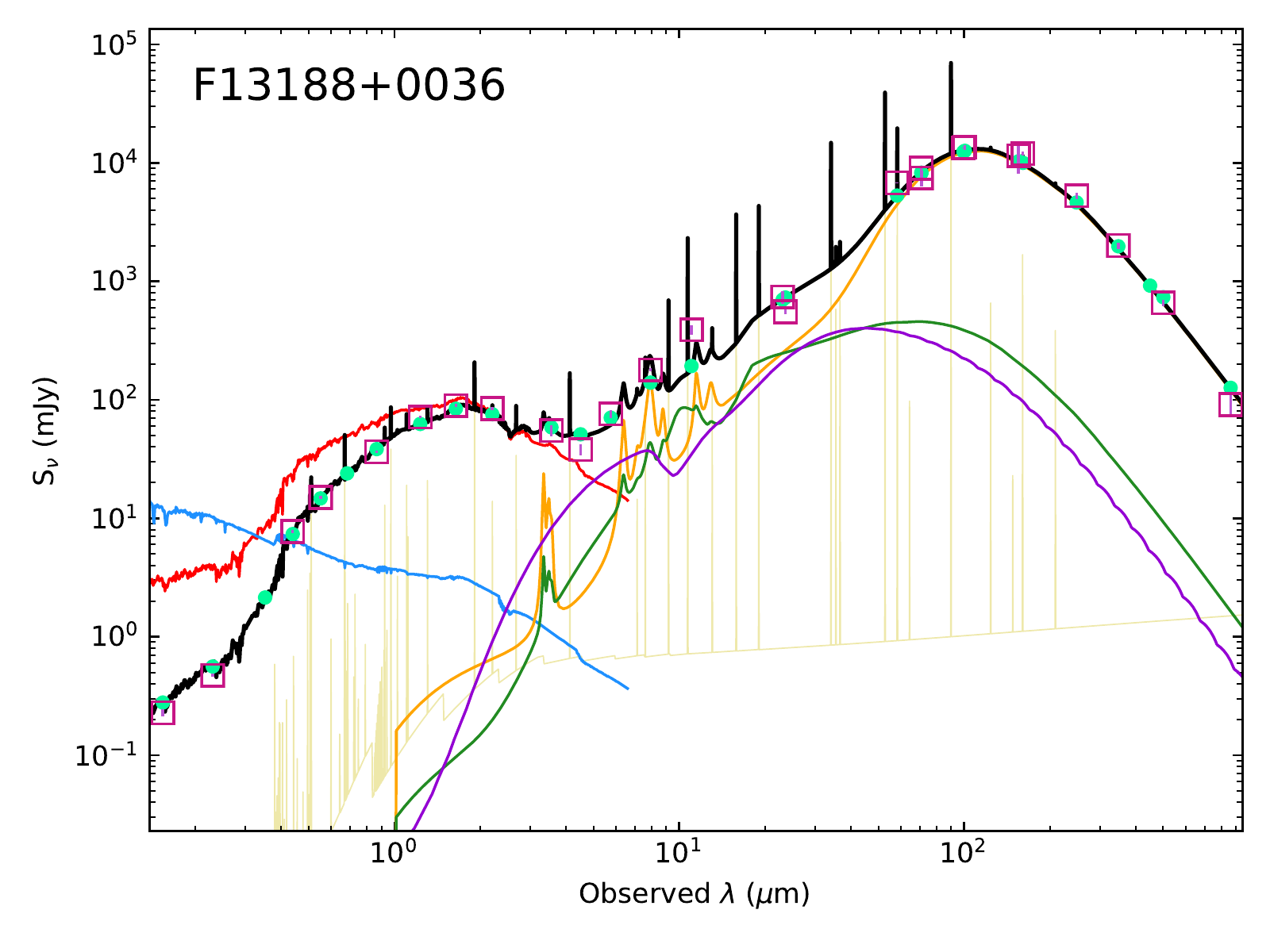}
\includegraphics[height=0.68\columnwidth]{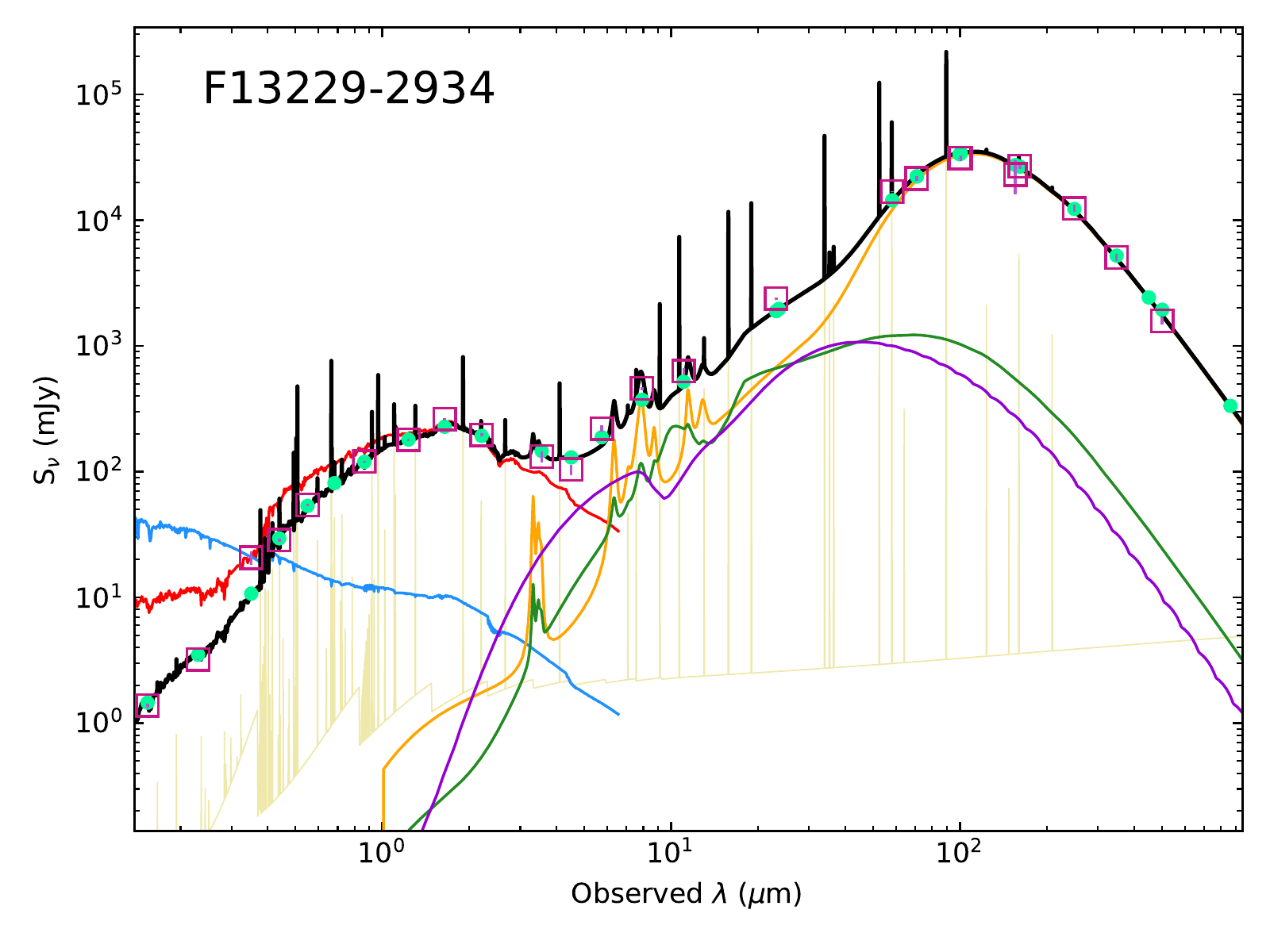}
\includegraphics[height=0.68\columnwidth]{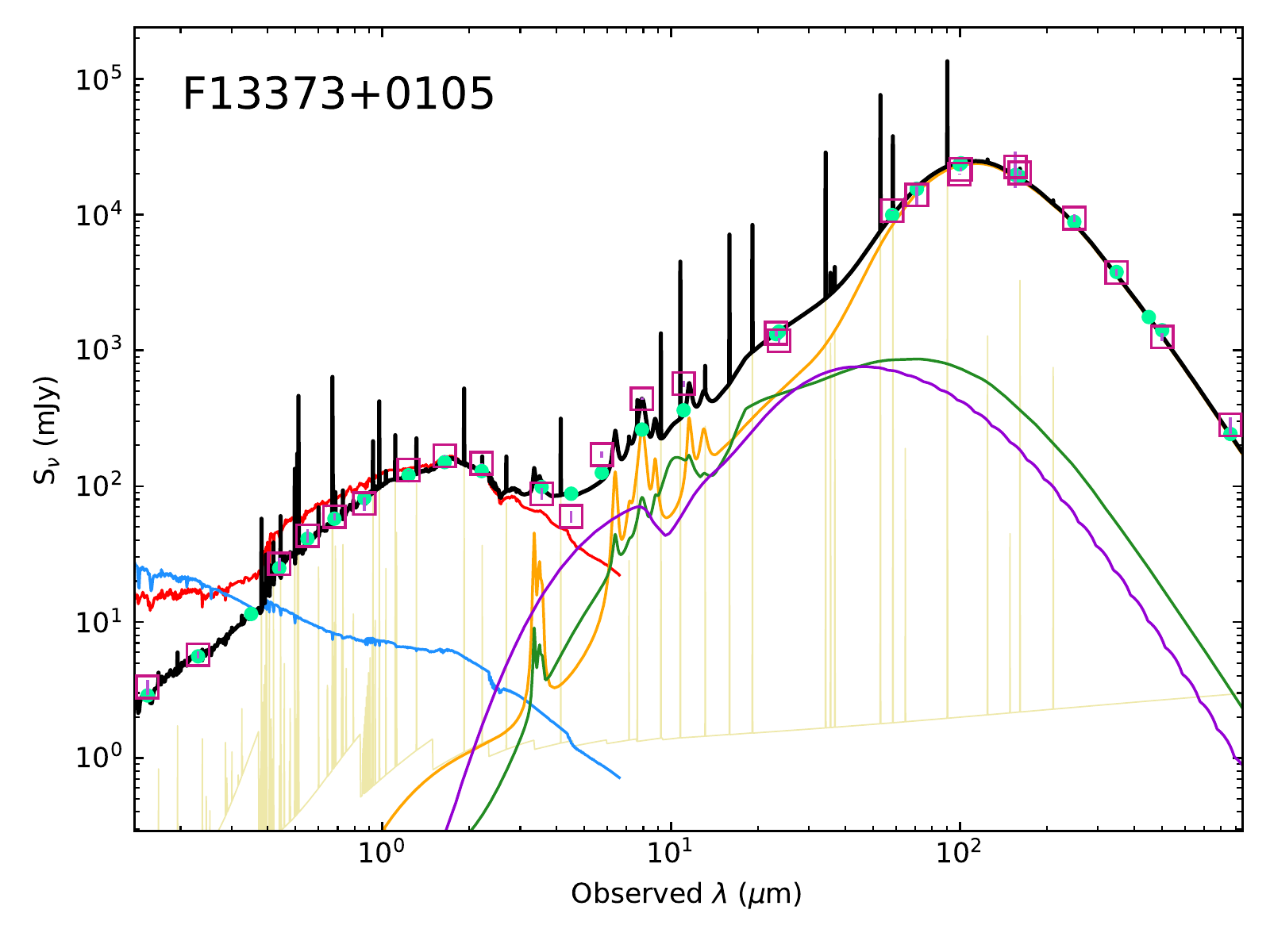}
\includegraphics[height=0.68\columnwidth]{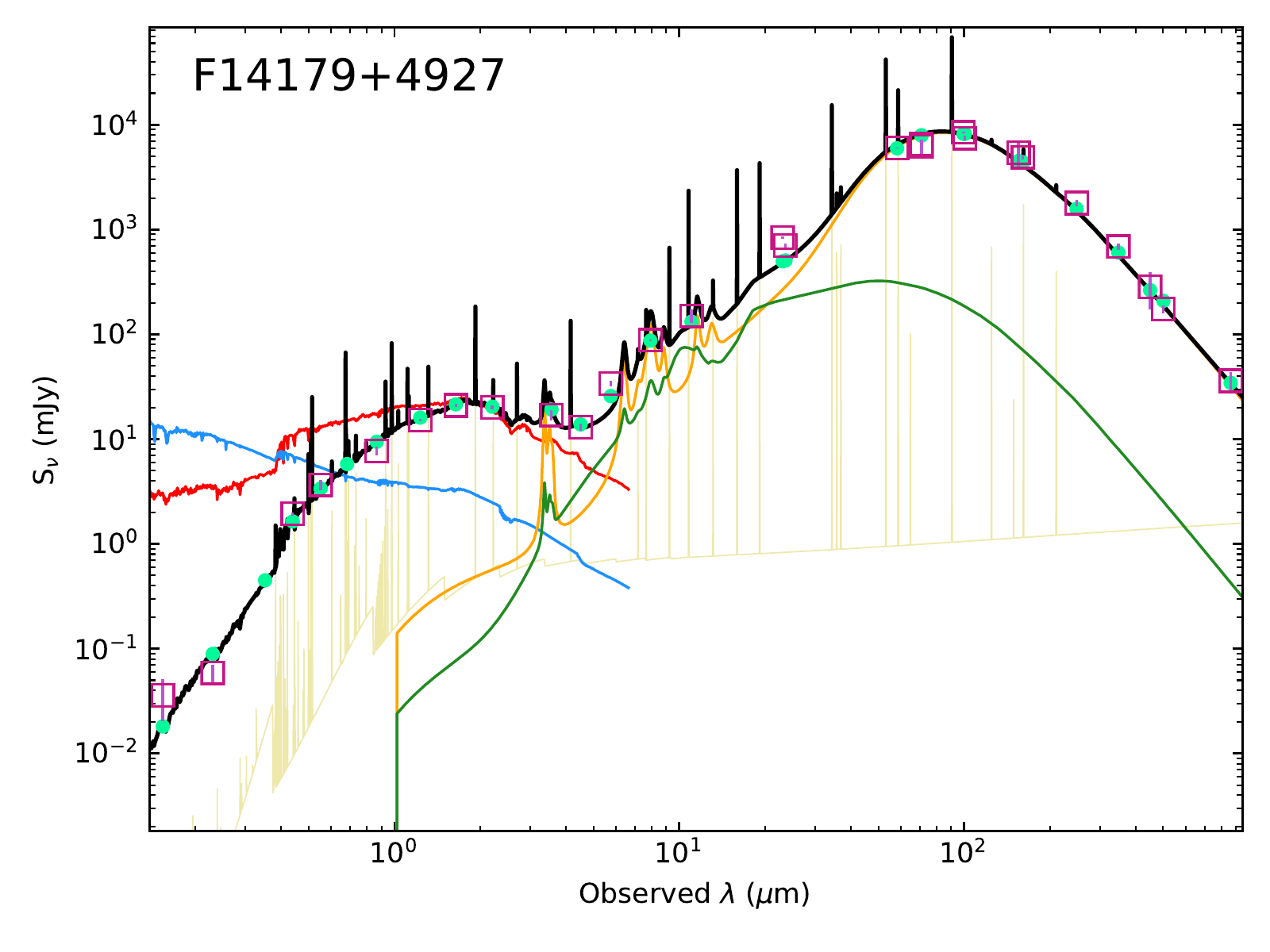}
\label{4}
\end{figure}
\begin{figure}
\includegraphics[height=0.68\columnwidth]{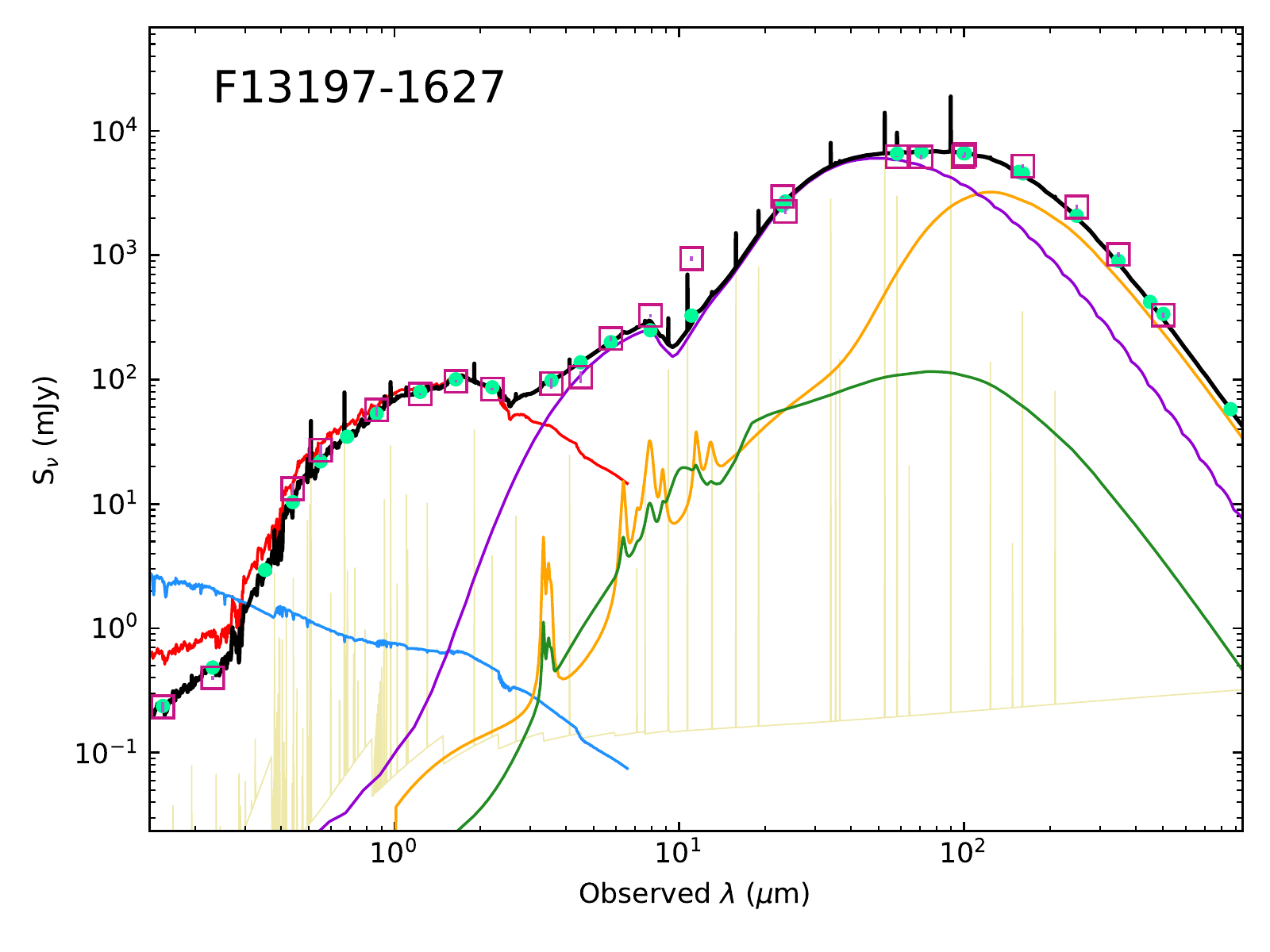}
\includegraphics[height=0.68\columnwidth]{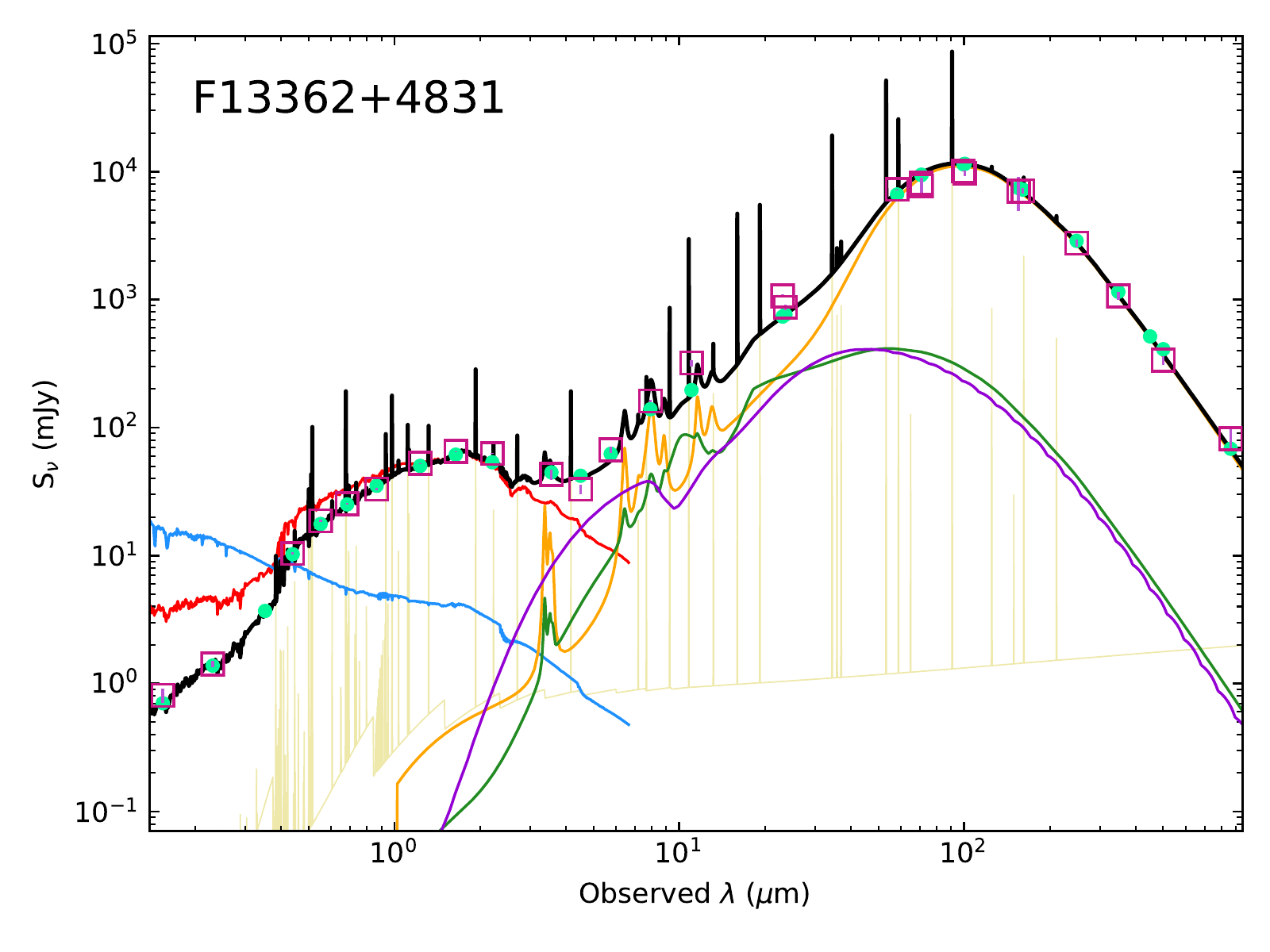}
\includegraphics[height=0.68\columnwidth]{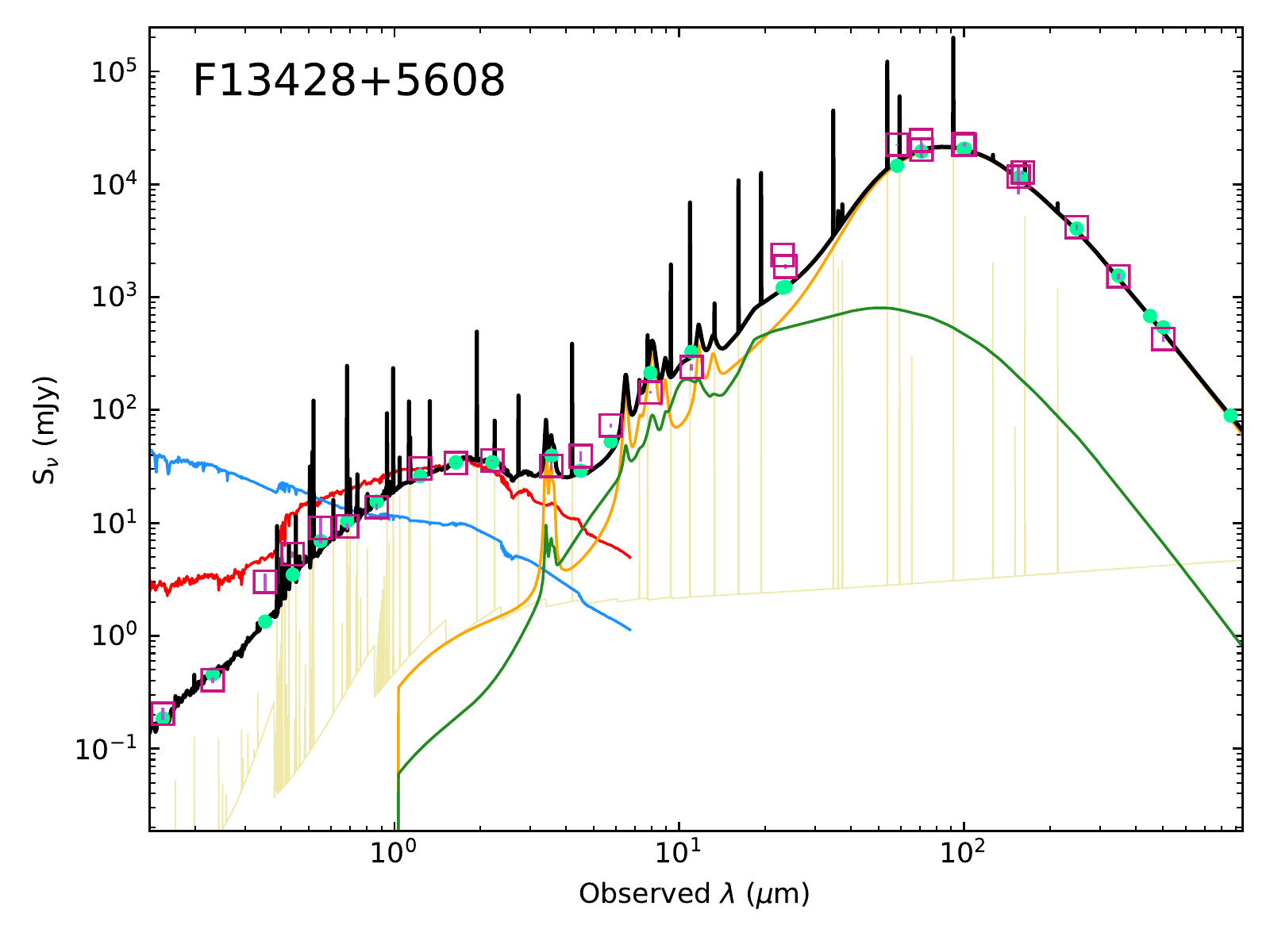}
\includegraphics[height=0.68\columnwidth]{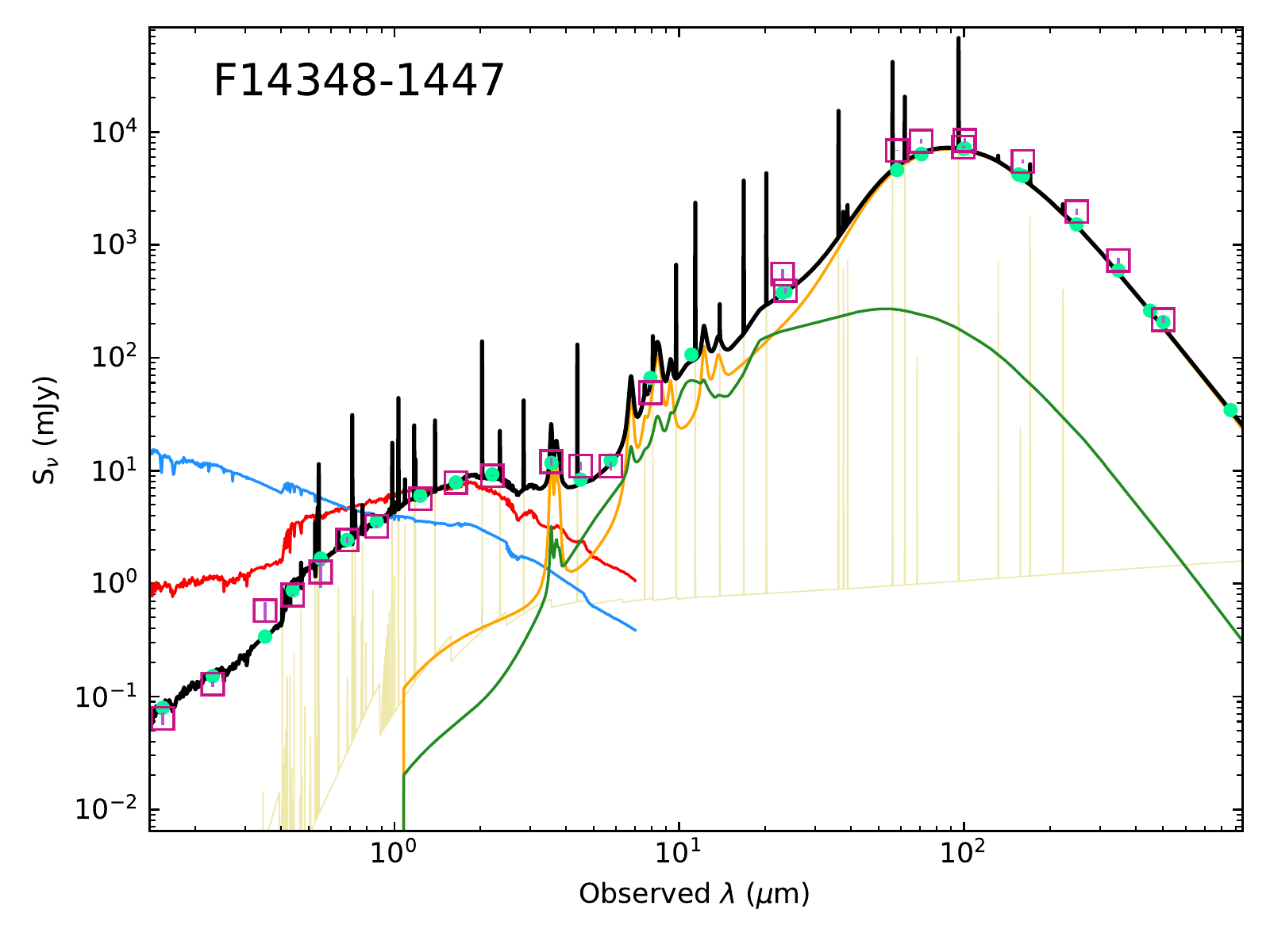}
\label{}
\end{figure}

\begin{figure}
\includegraphics[height=0.68\columnwidth]{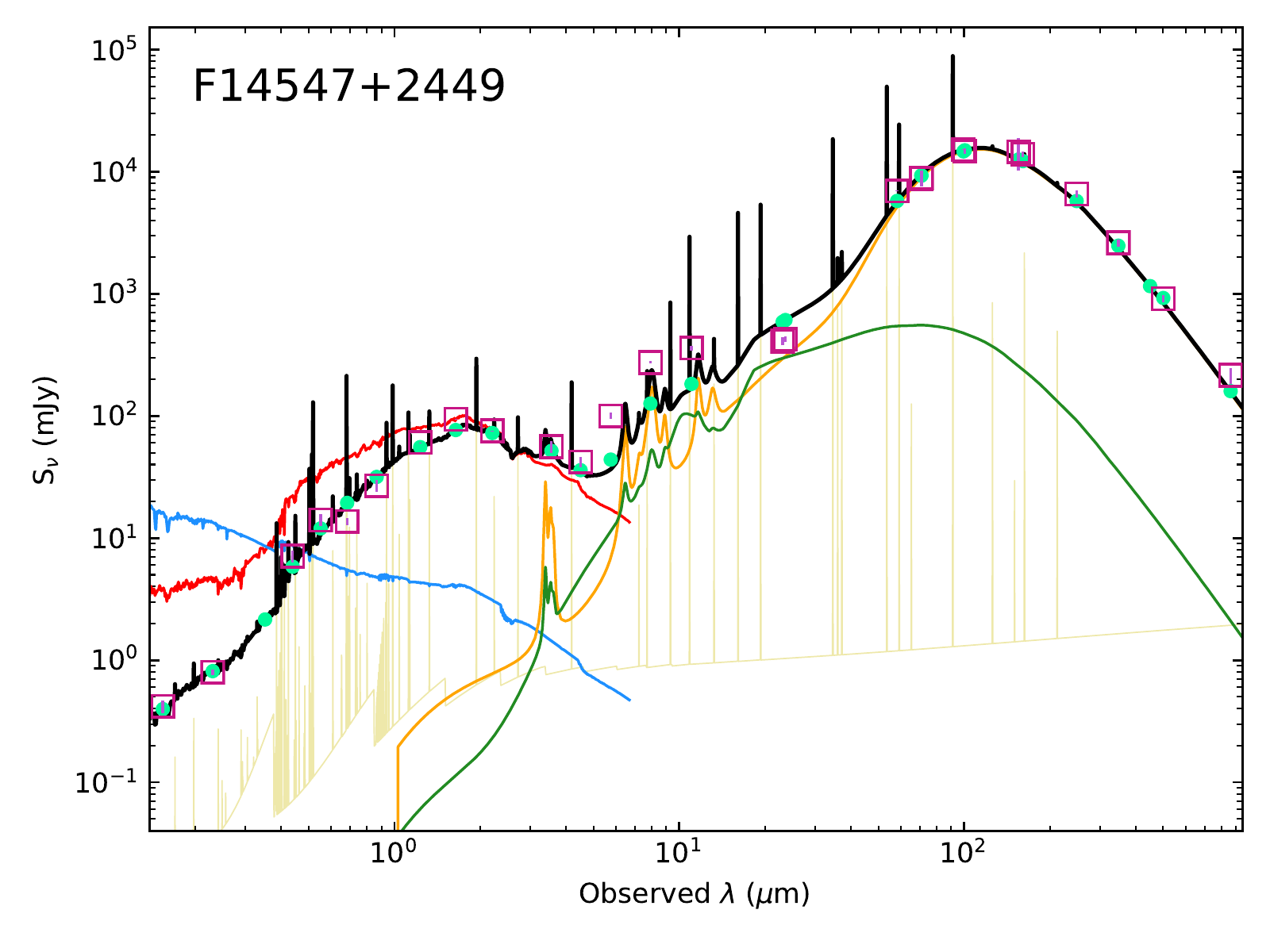}
\includegraphics[height=0.68\columnwidth]{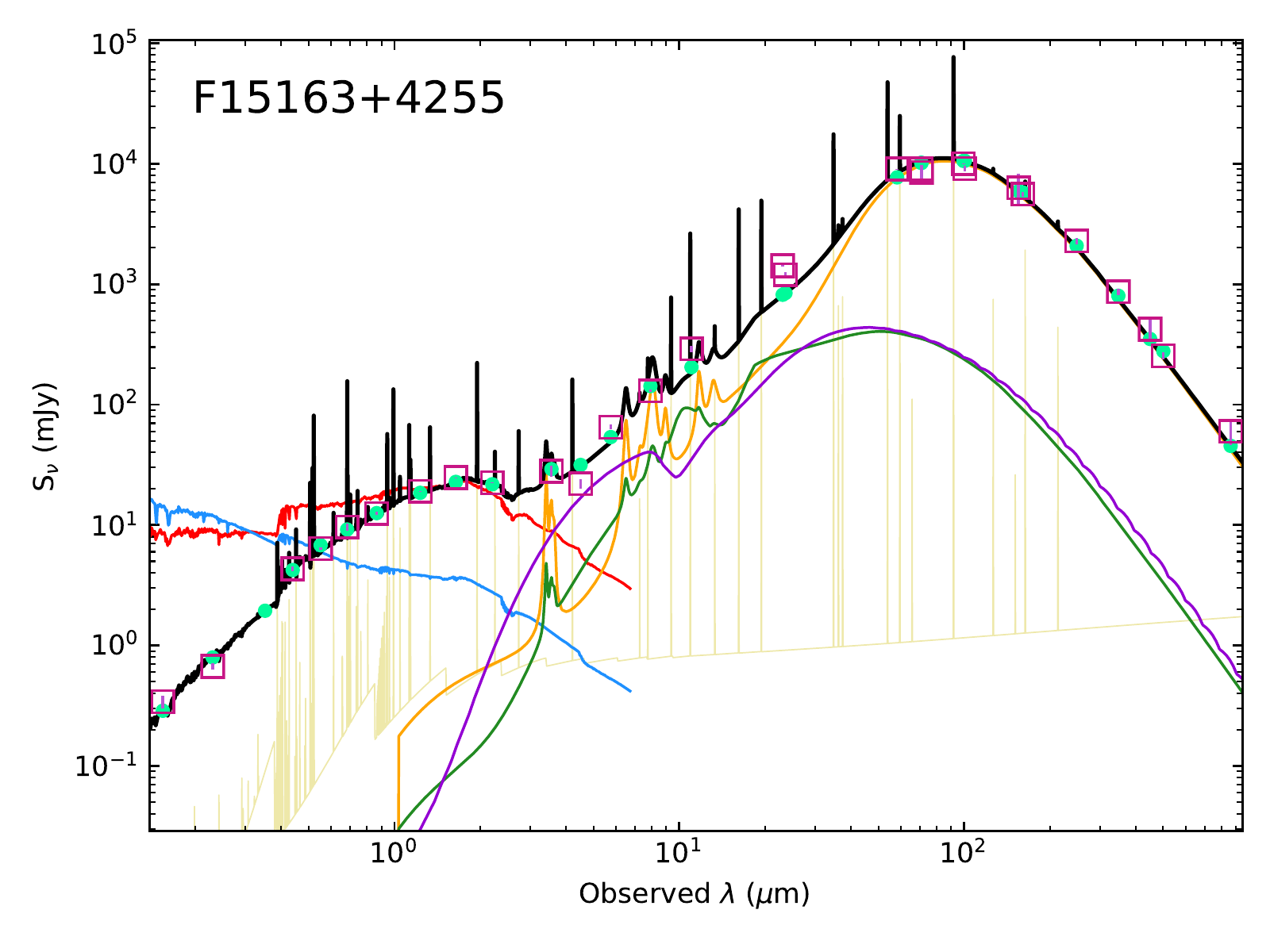}
\includegraphics[height=0.68\columnwidth]{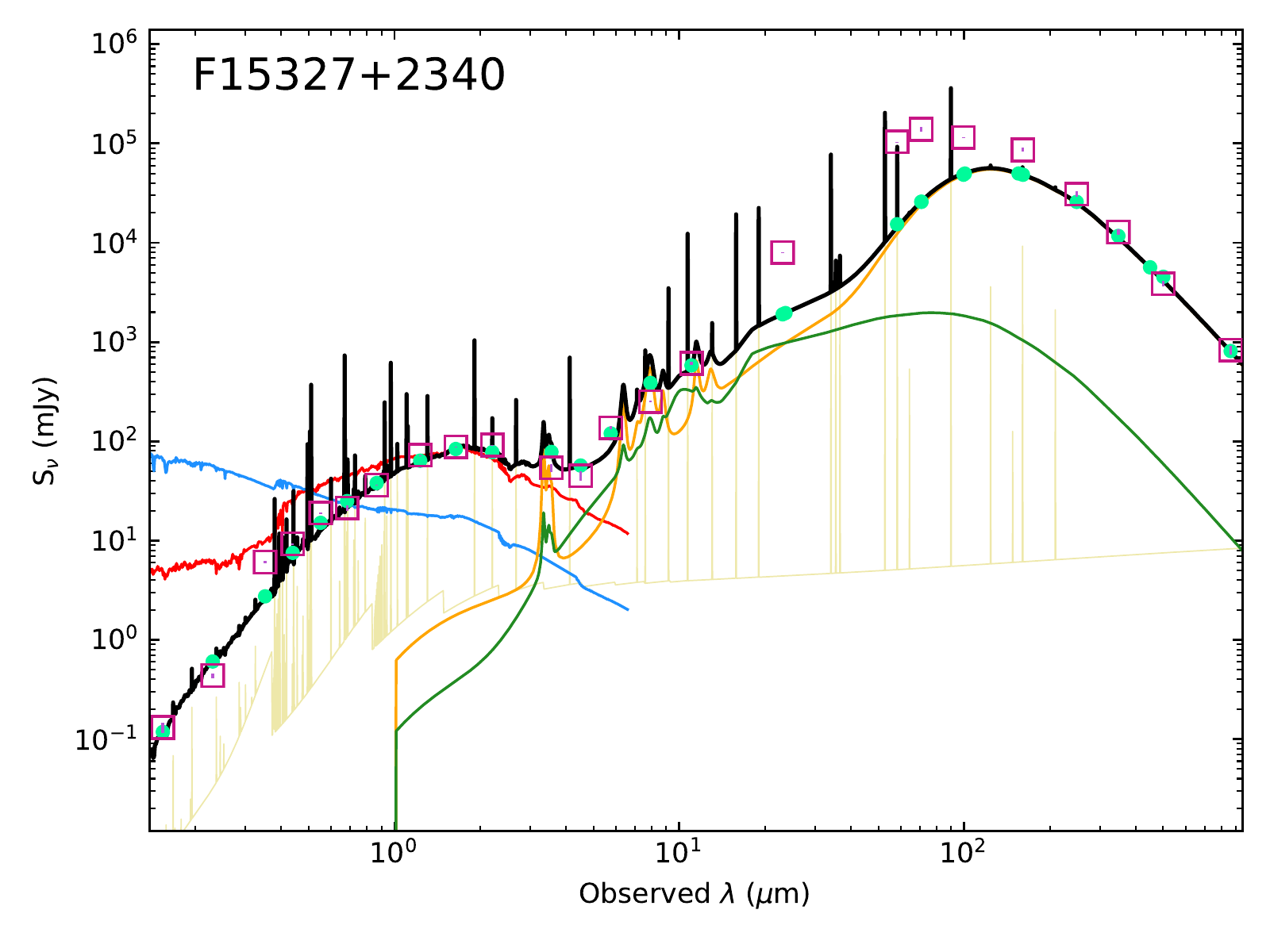}
\includegraphics[height=0.68\columnwidth]{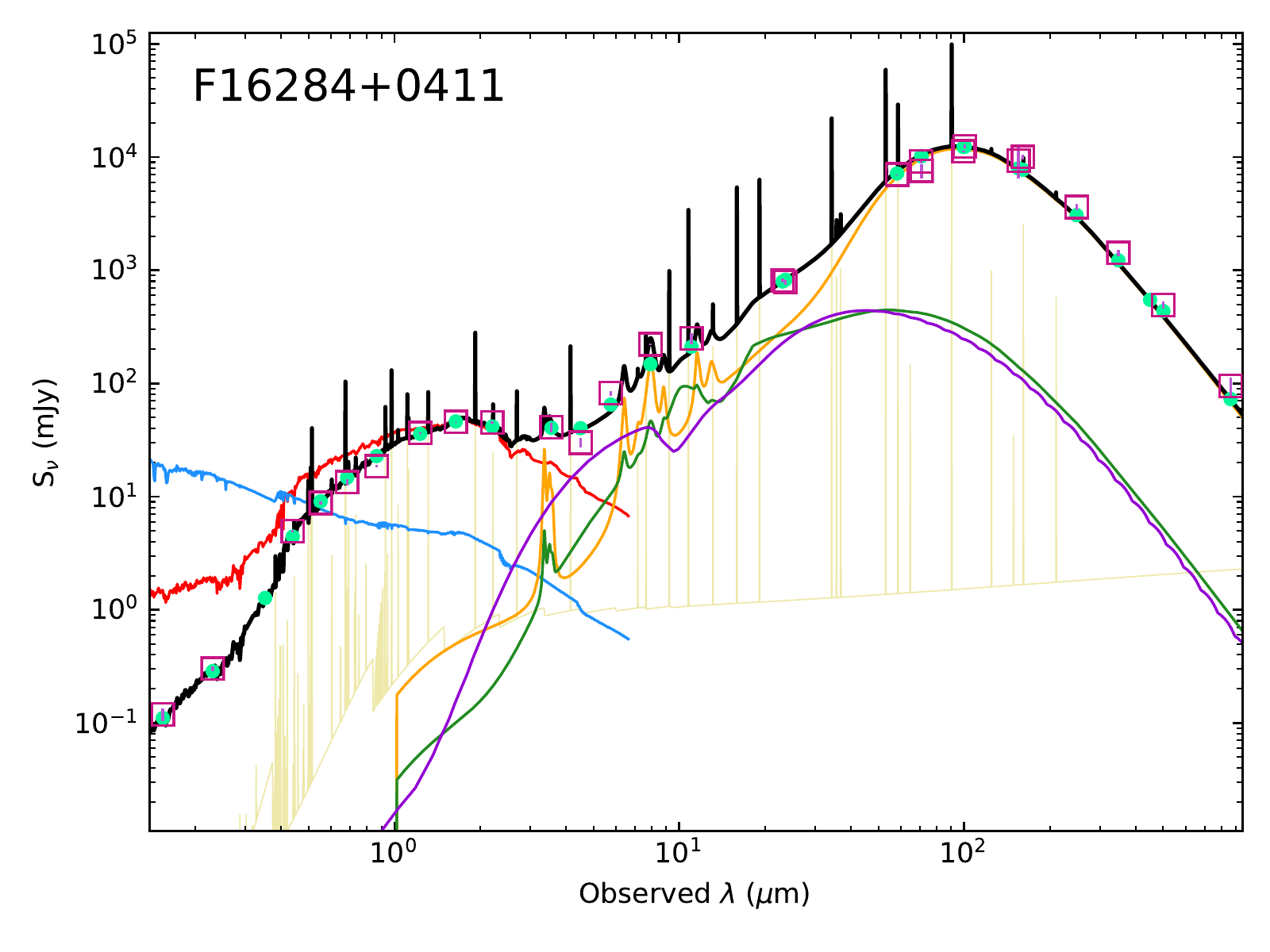}
\label{4}
\end{figure}
\begin{figure}
\includegraphics[height=0.68\columnwidth]{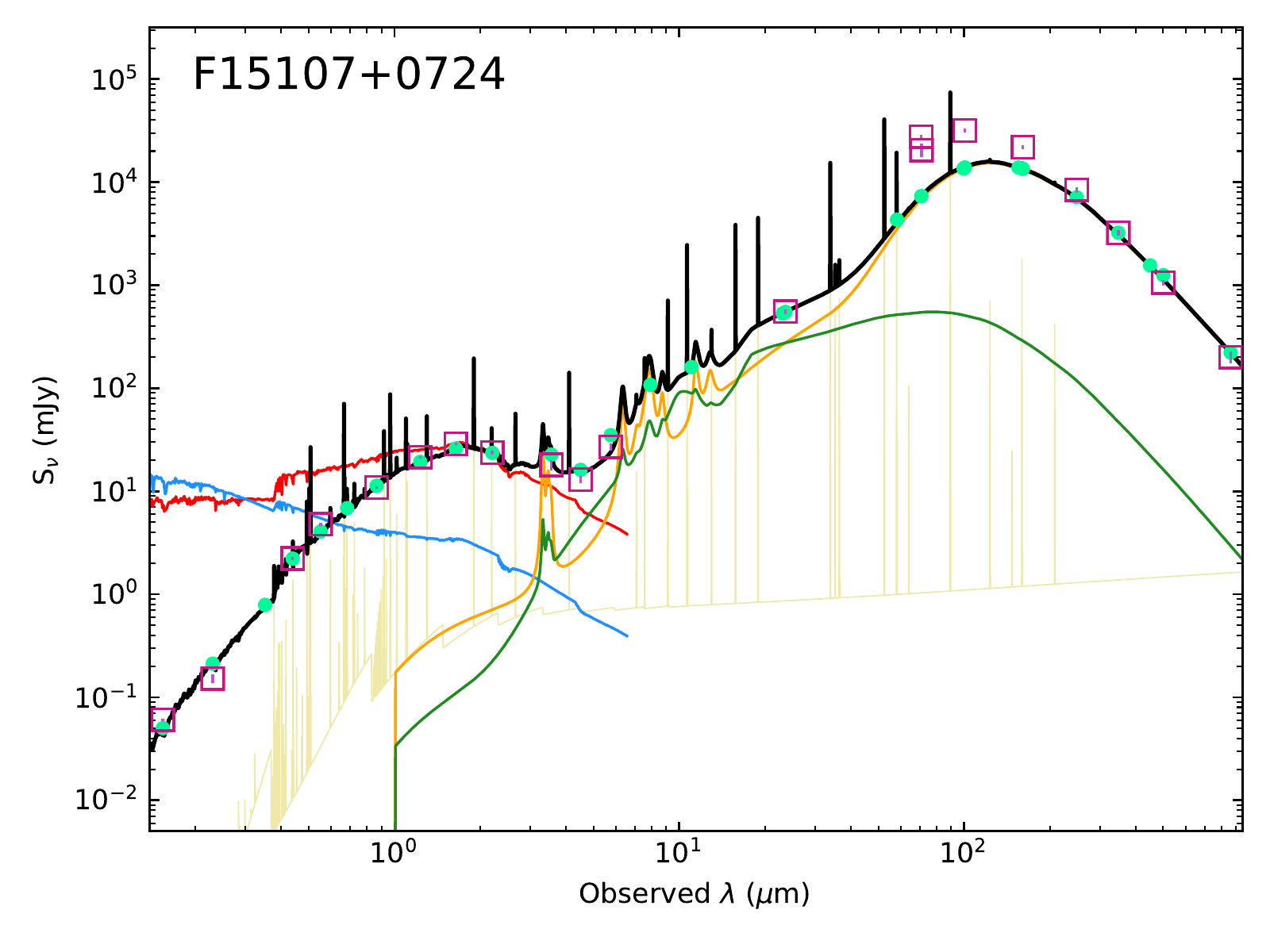}
\includegraphics[height=0.68\columnwidth]{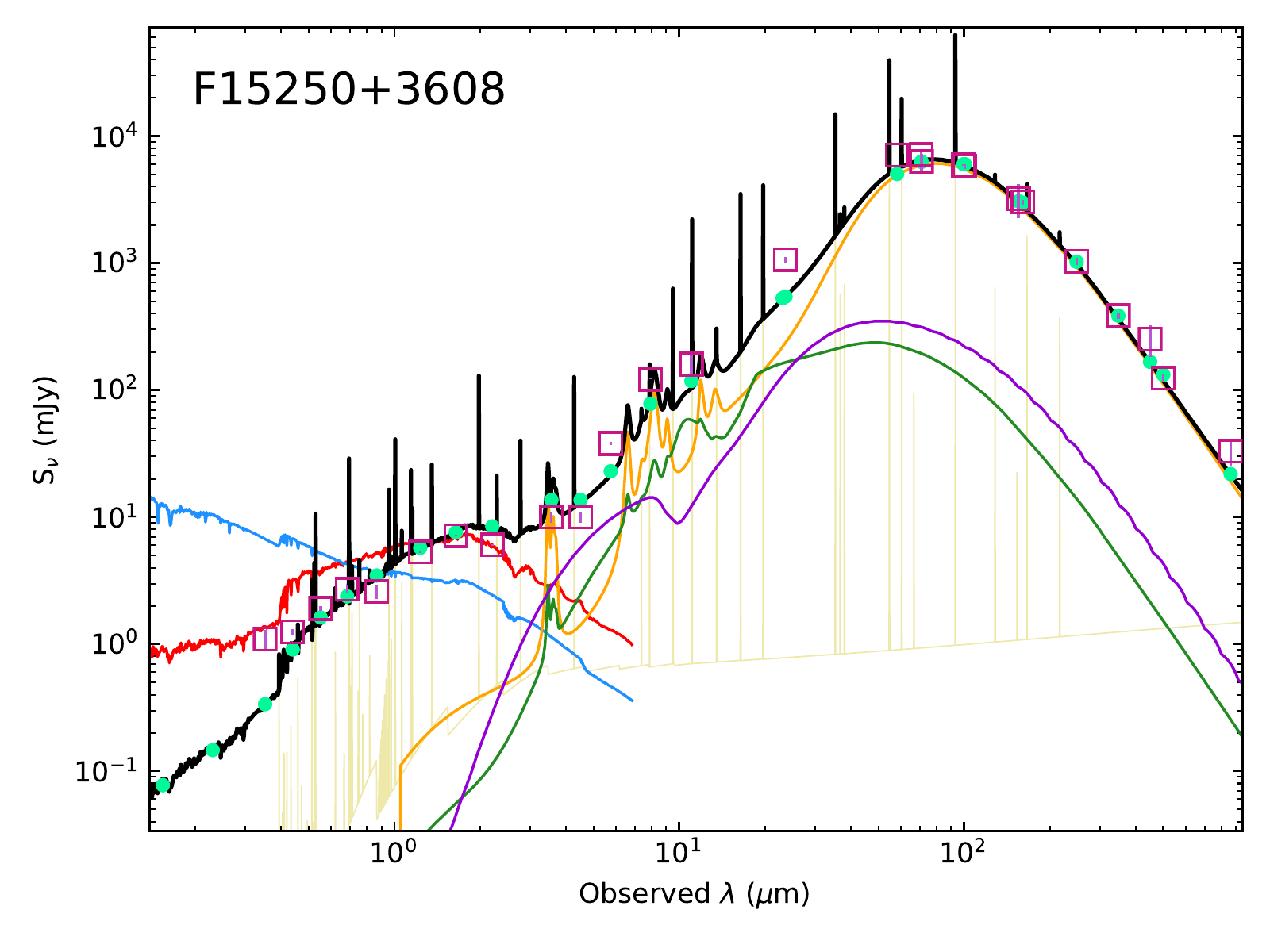}
\includegraphics[height=0.68\columnwidth]{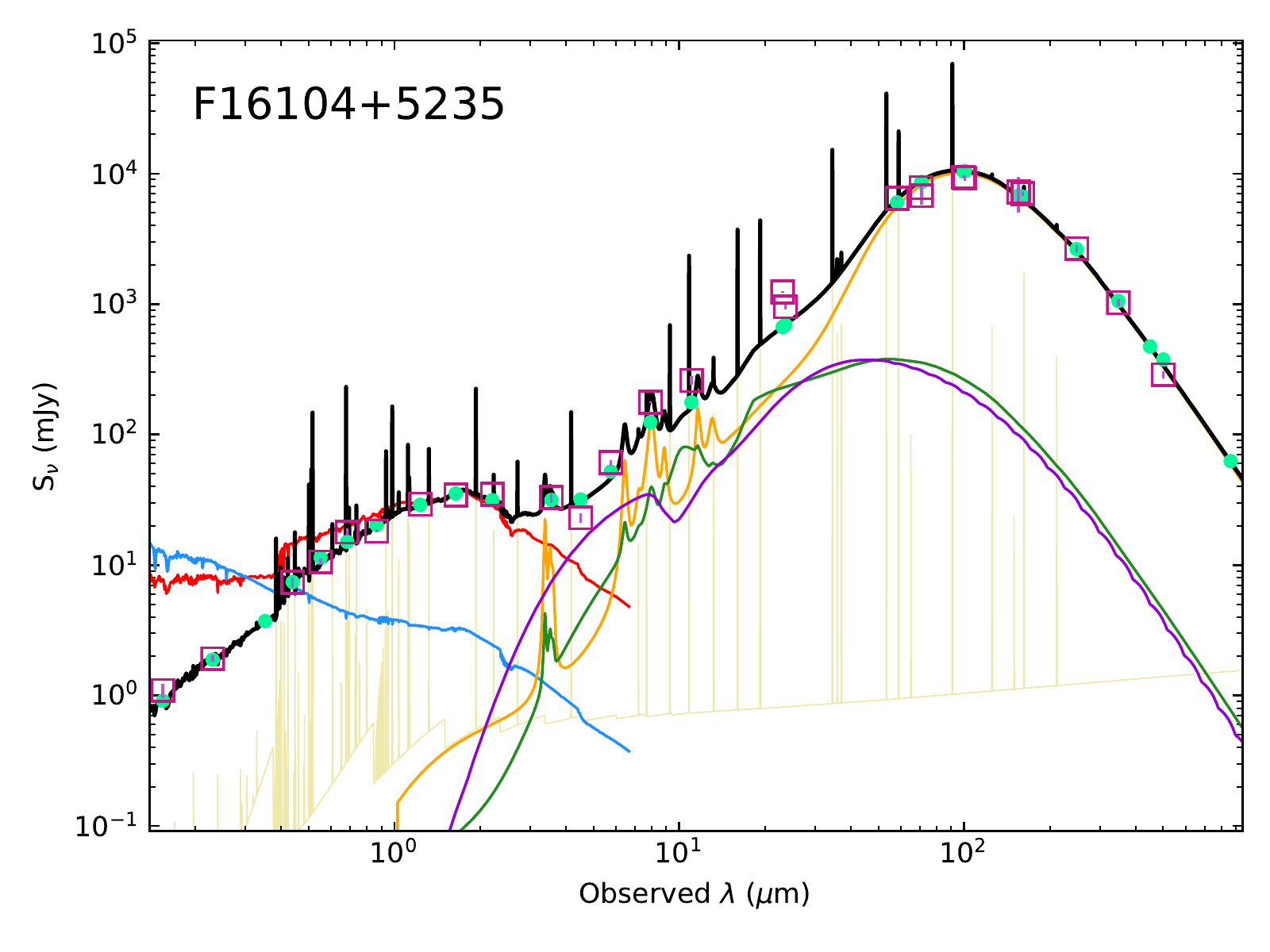}
\includegraphics[height=0.68\columnwidth]{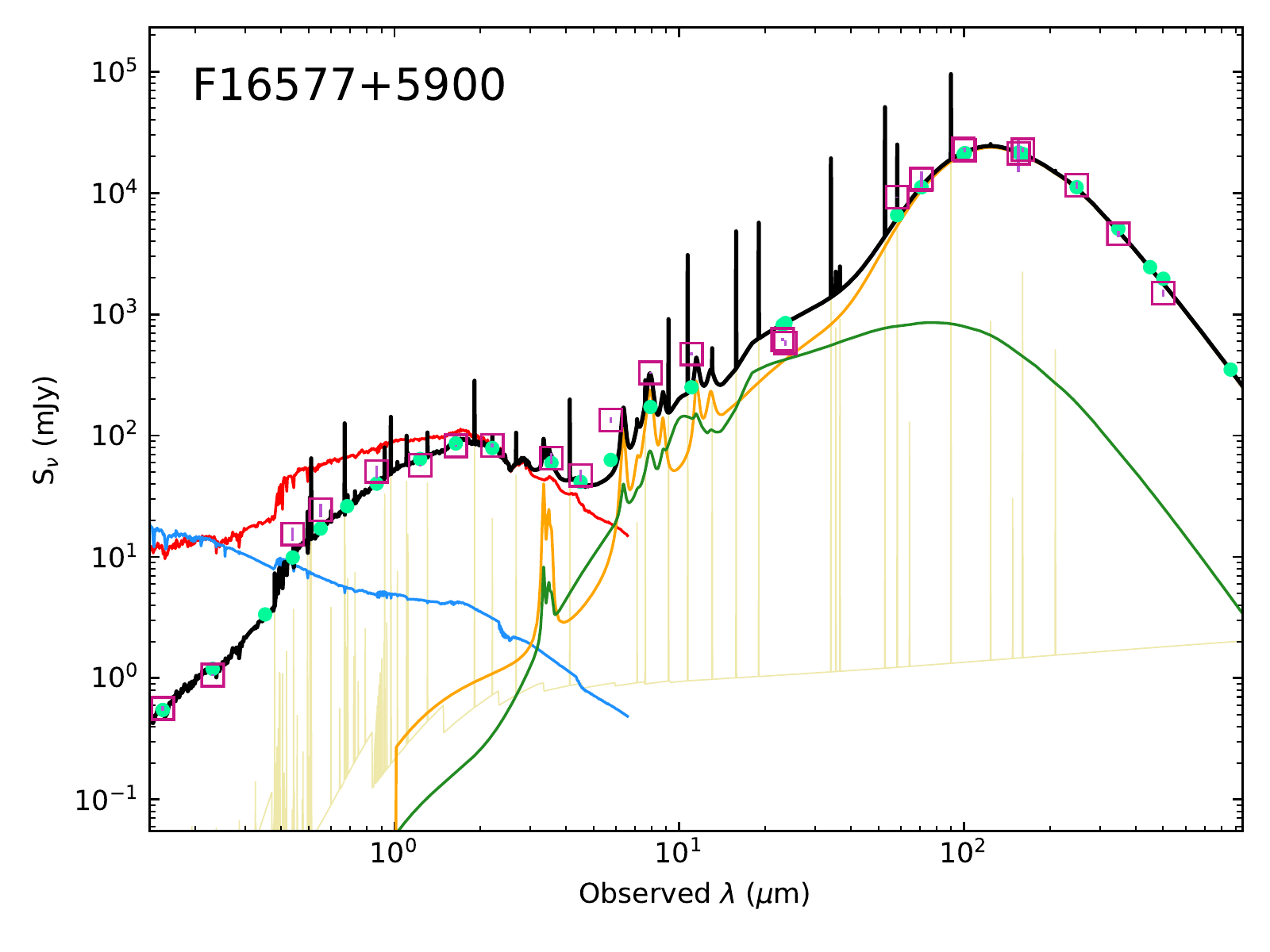}
\label{}
\end{figure}

\begin{figure}
\includegraphics[height=0.68\columnwidth]{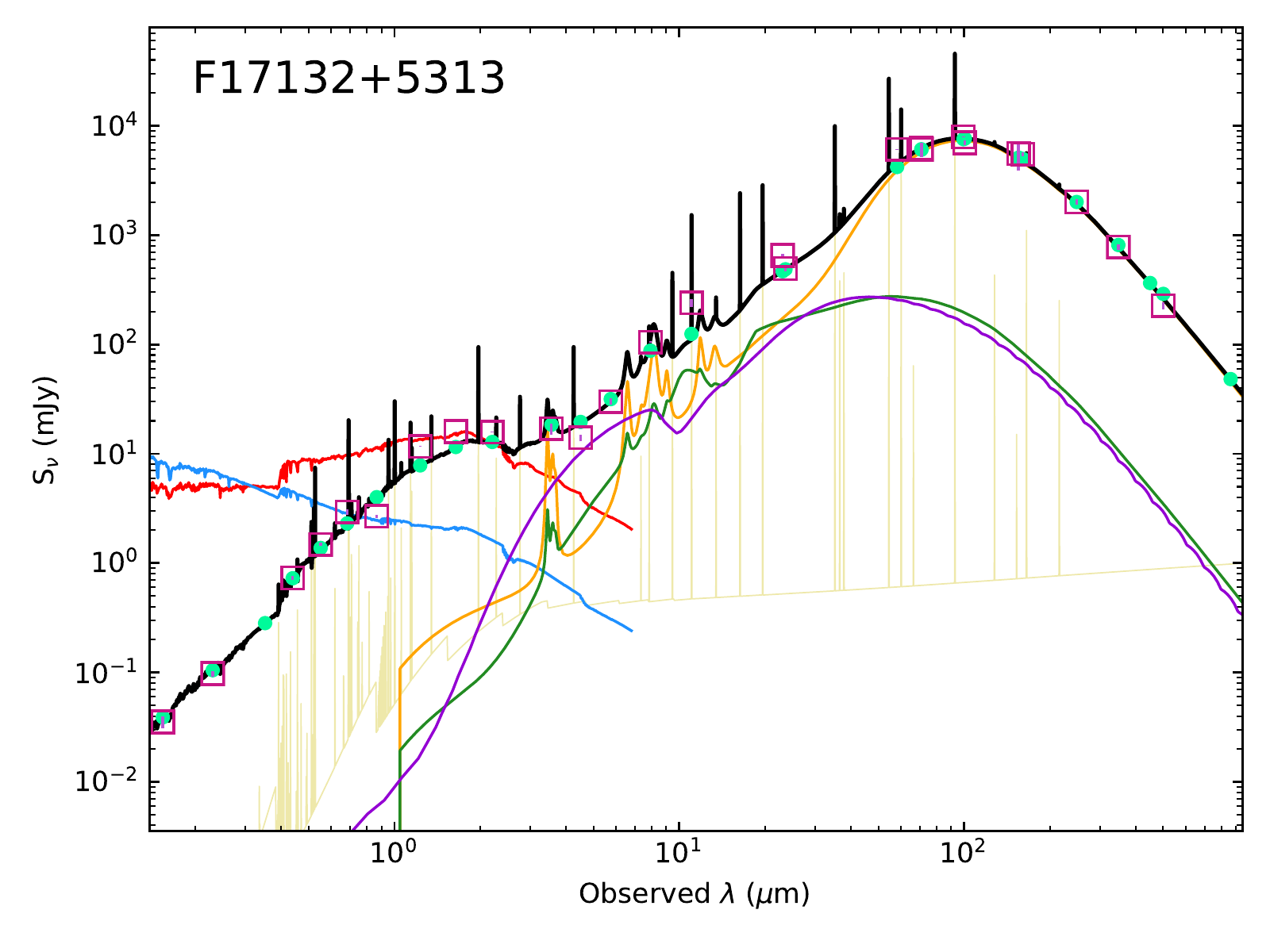}
\includegraphics[height=0.68\columnwidth]{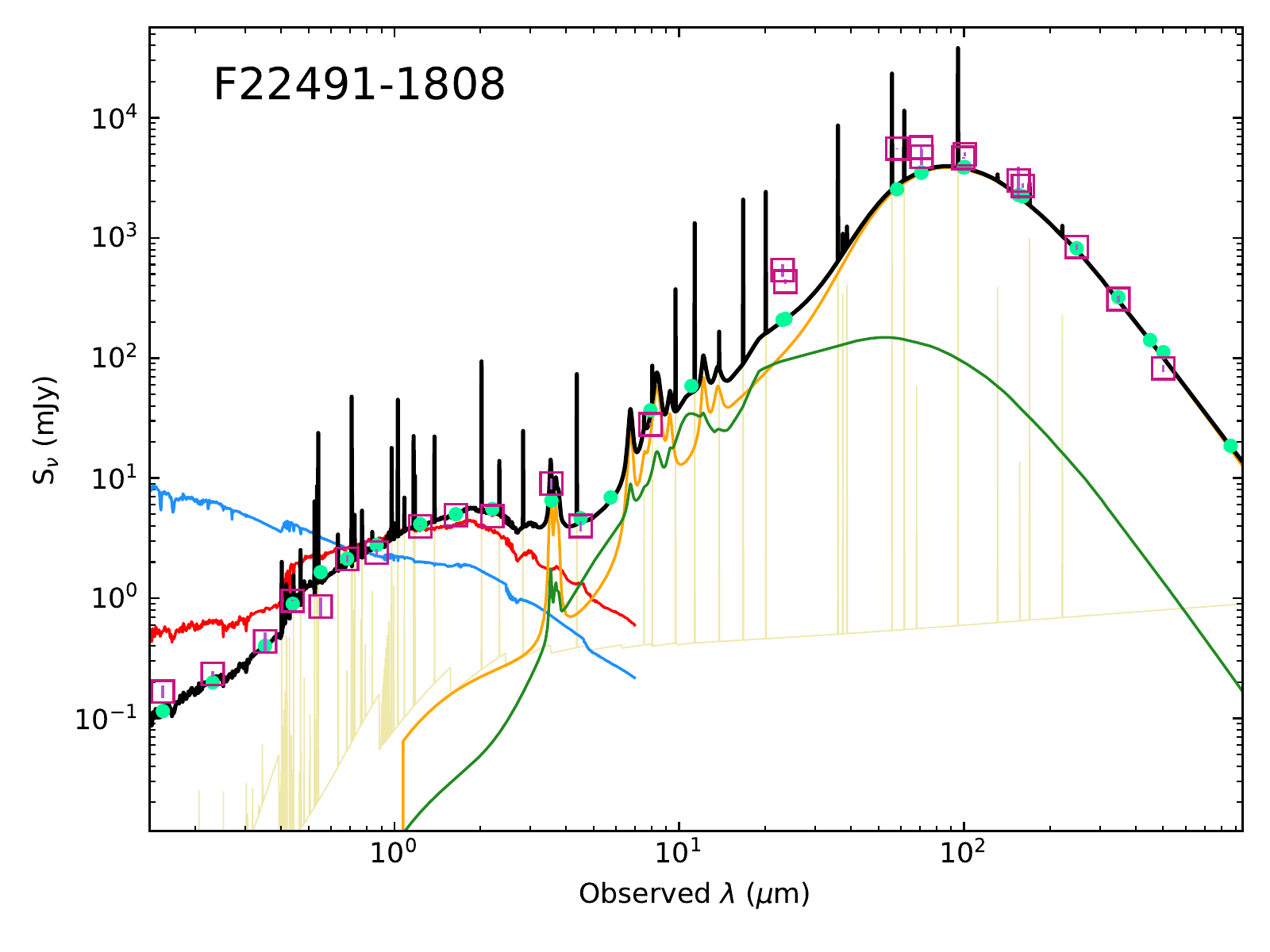}
\includegraphics[height=0.68\columnwidth]{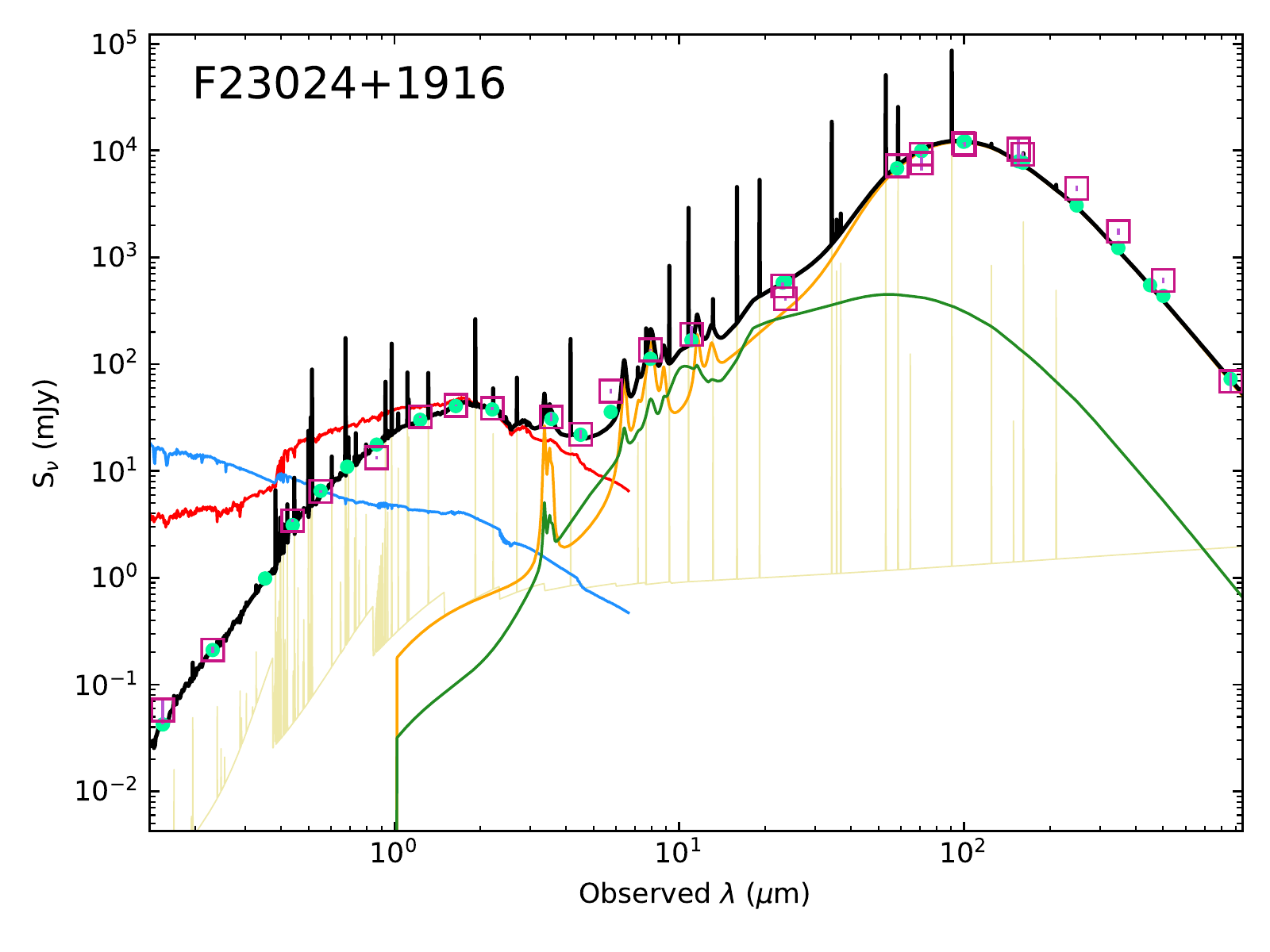}
\includegraphics[height=0.68\columnwidth]{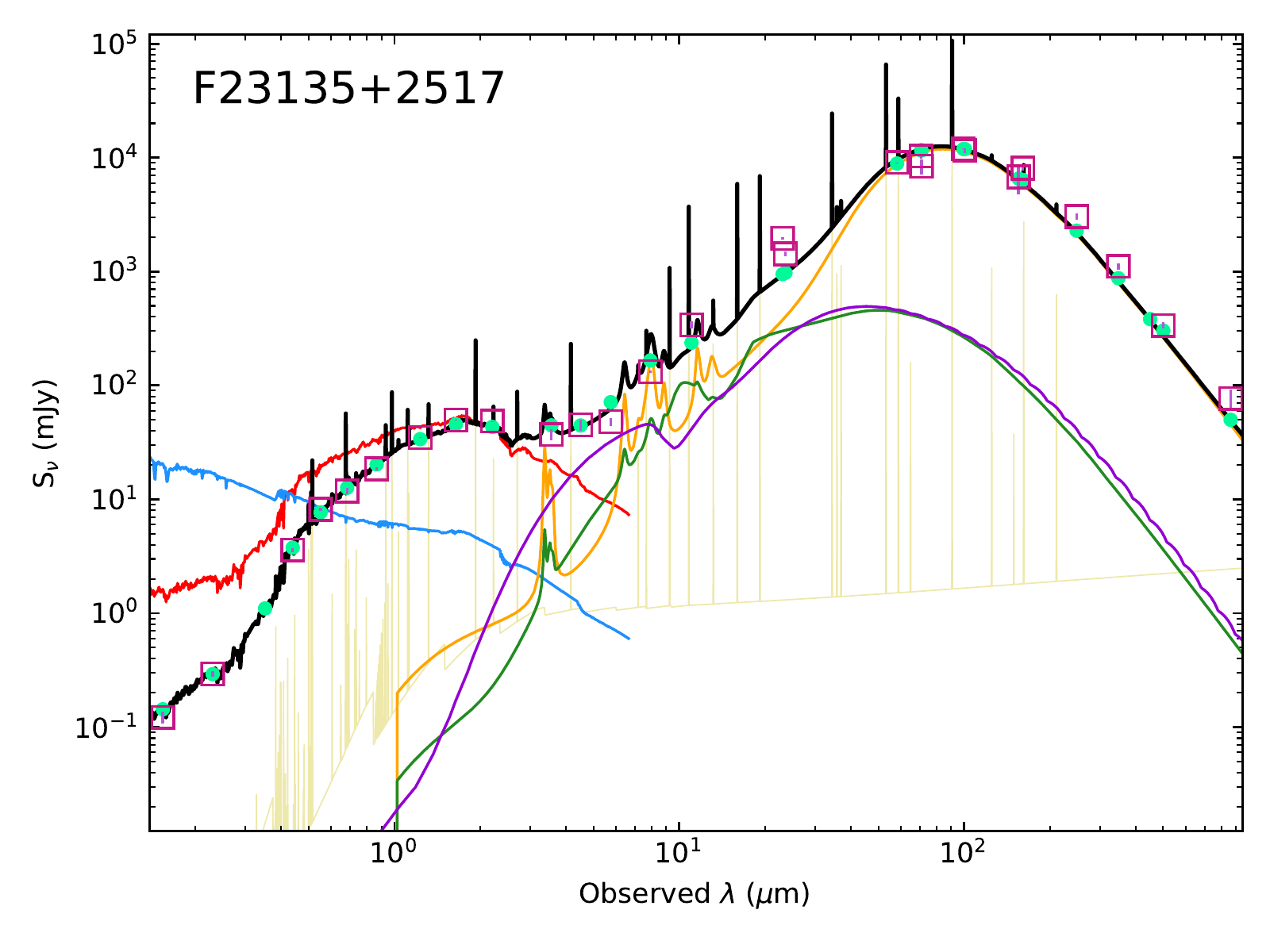}
\label{4}
\end{figure}
\begin{figure}
\includegraphics[height=0.68\columnwidth]{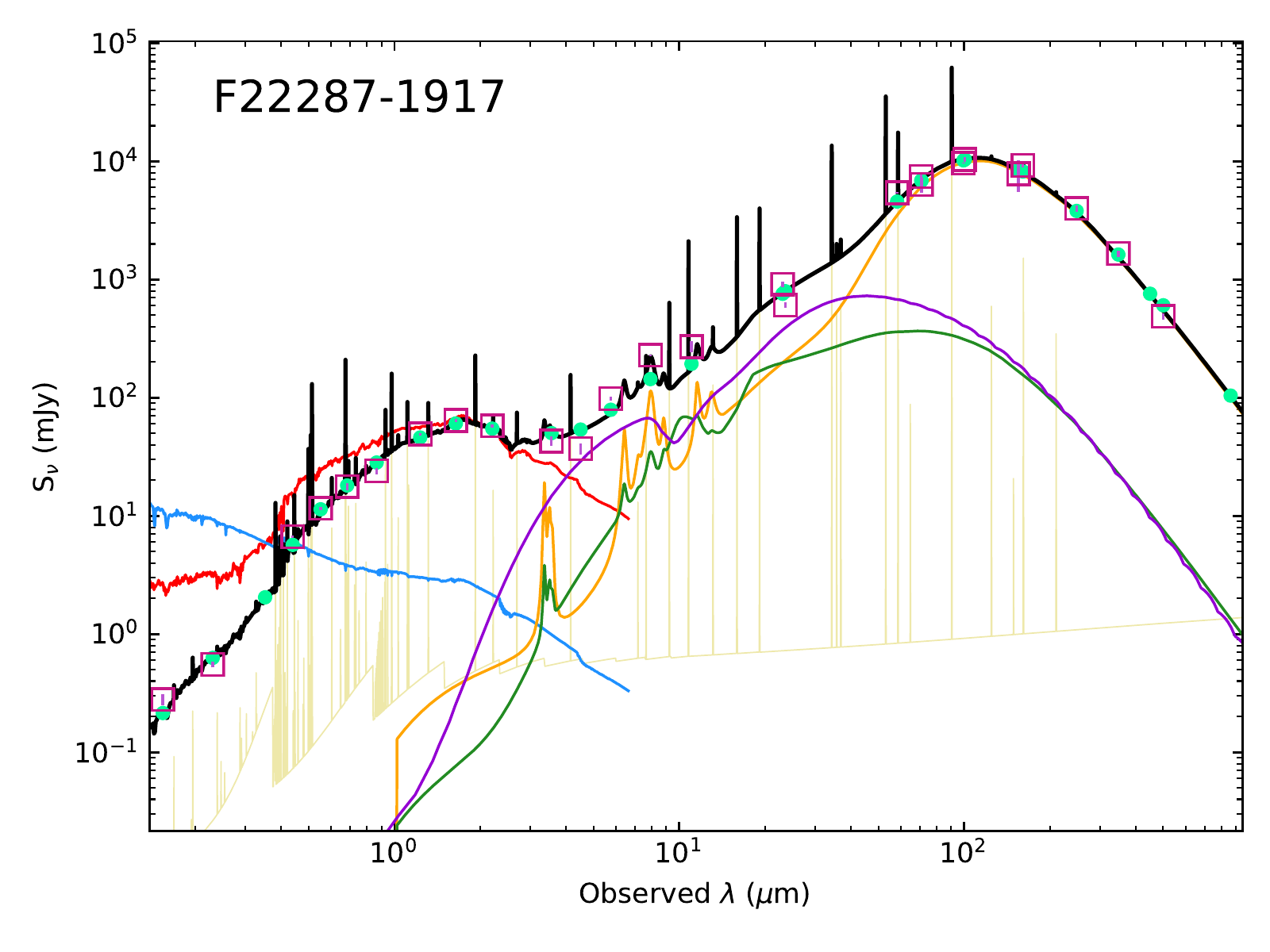}
\includegraphics[height=0.68\columnwidth]{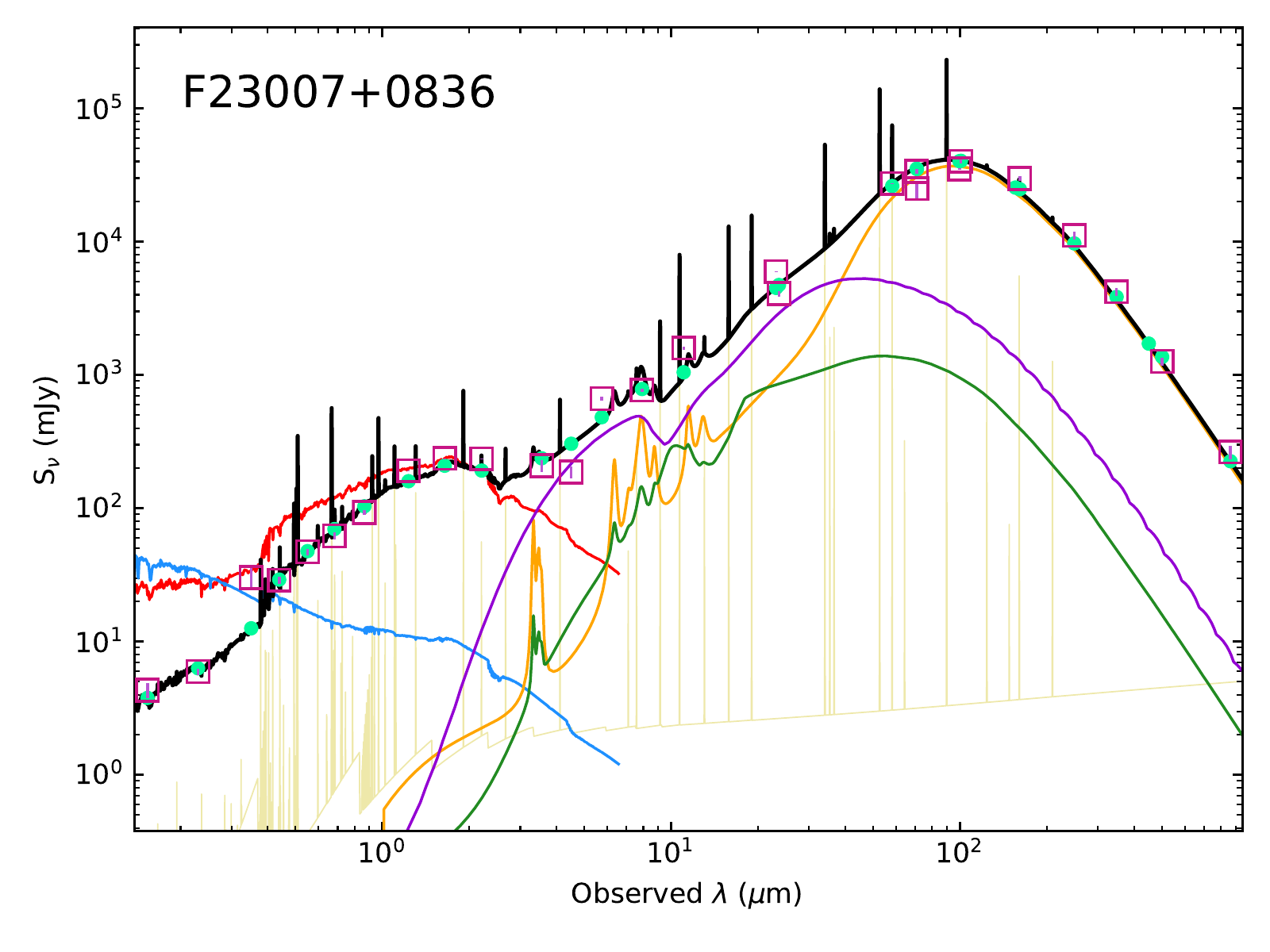}
\includegraphics[height=0.68\columnwidth]{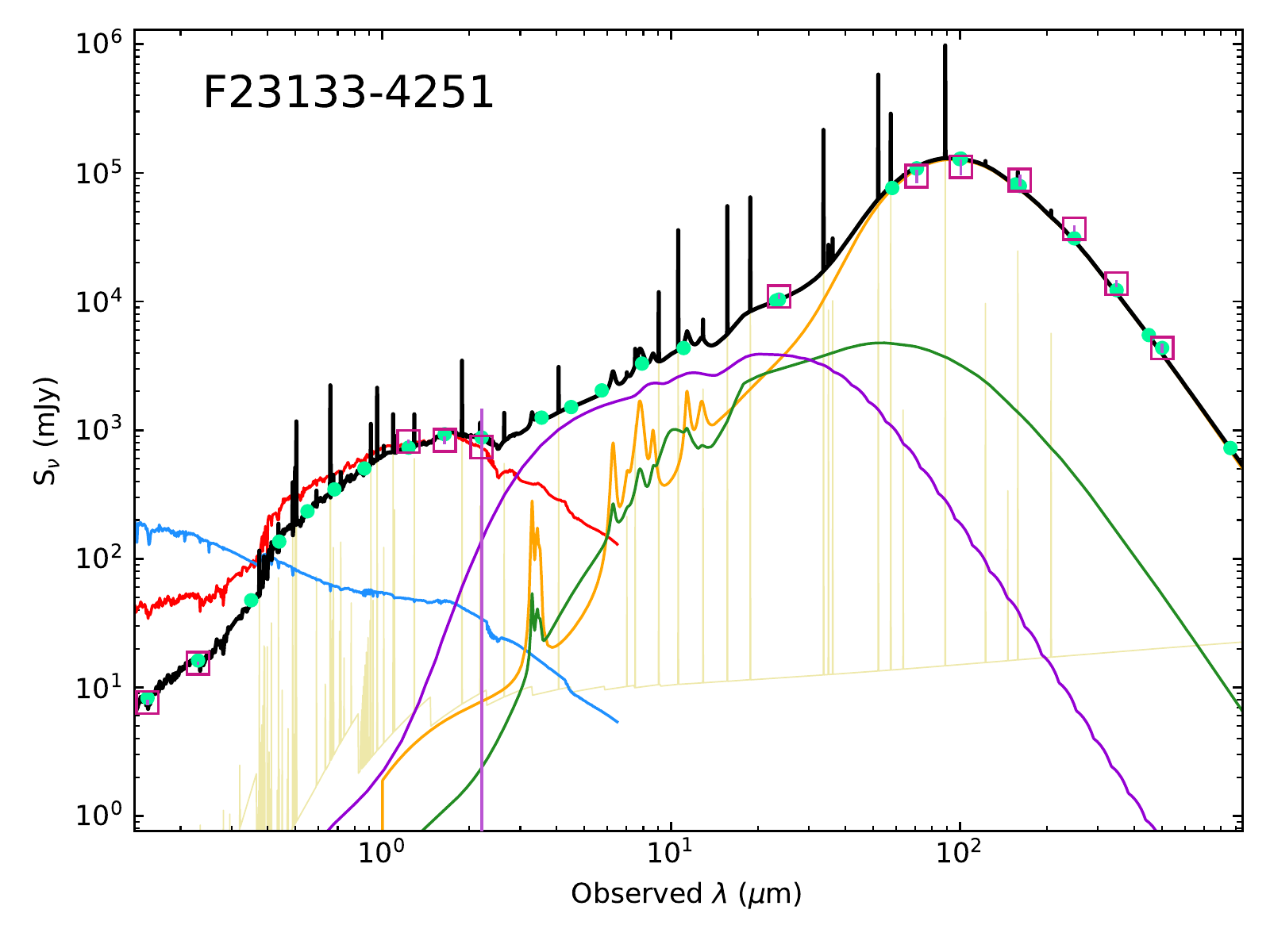}
\includegraphics[height=0.68\columnwidth]{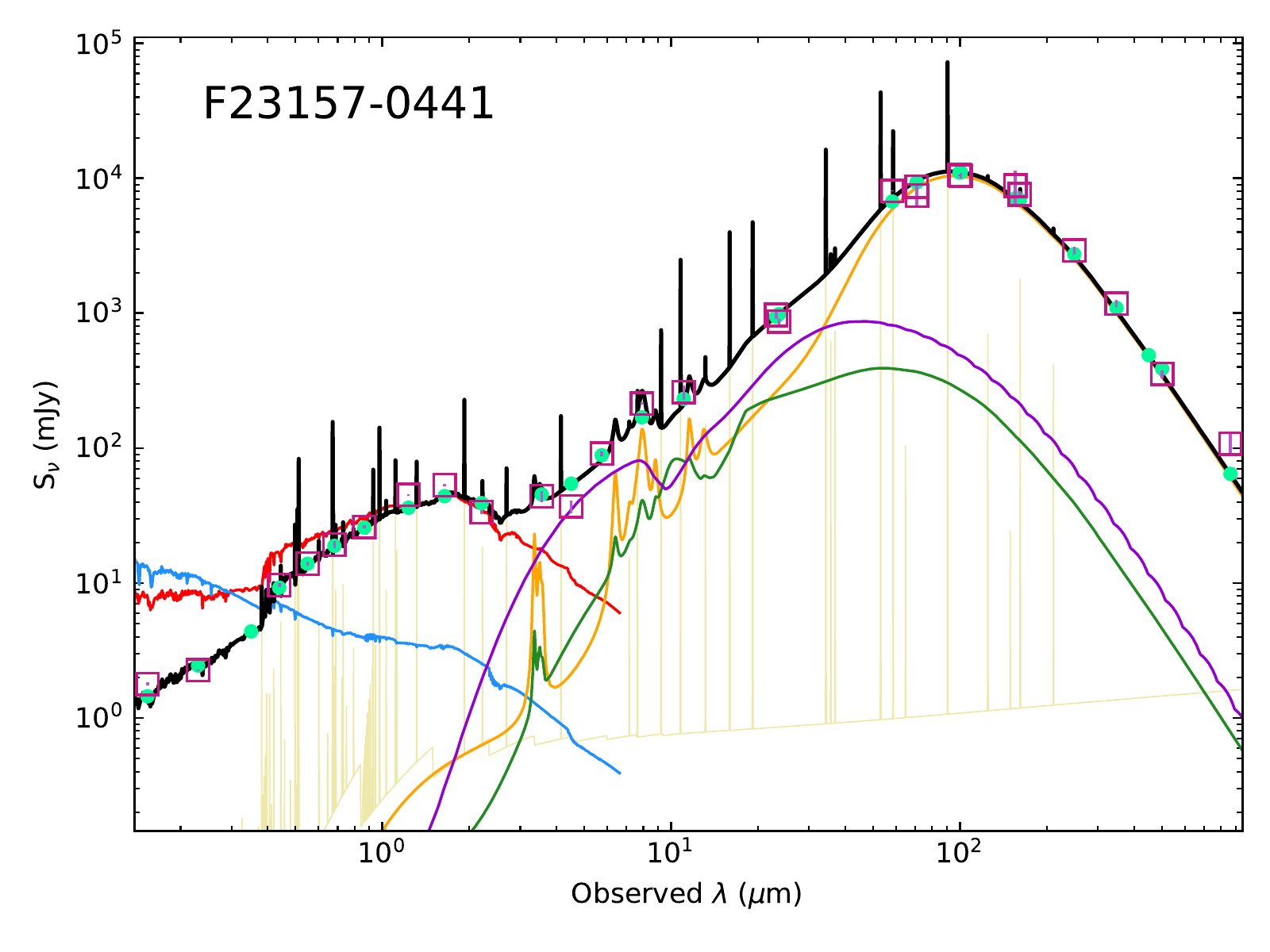}
\label{}
\end{figure}

\begin{figure}
\includegraphics[height=0.68\columnwidth]{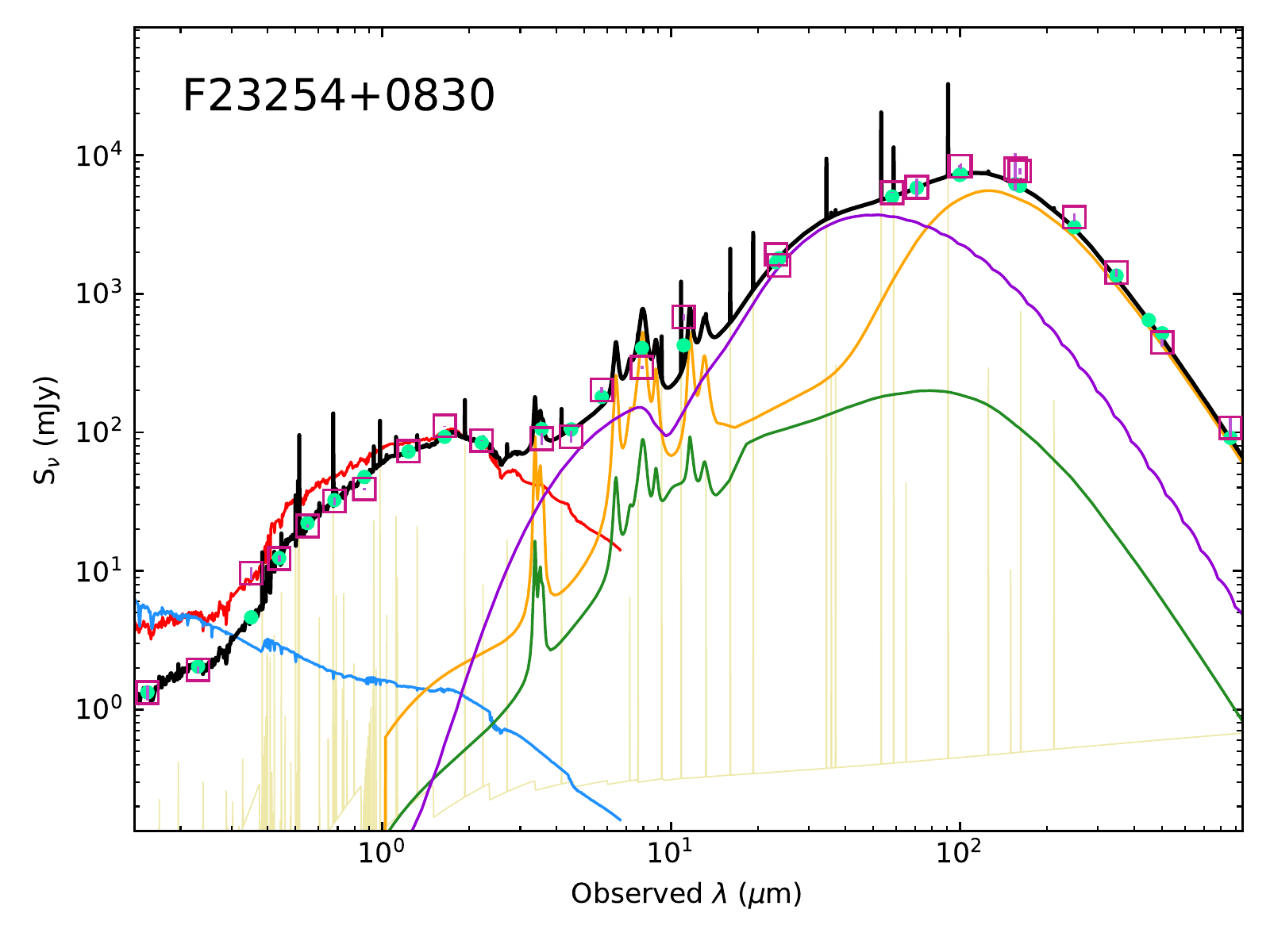}
\includegraphics[height=0.68\columnwidth]{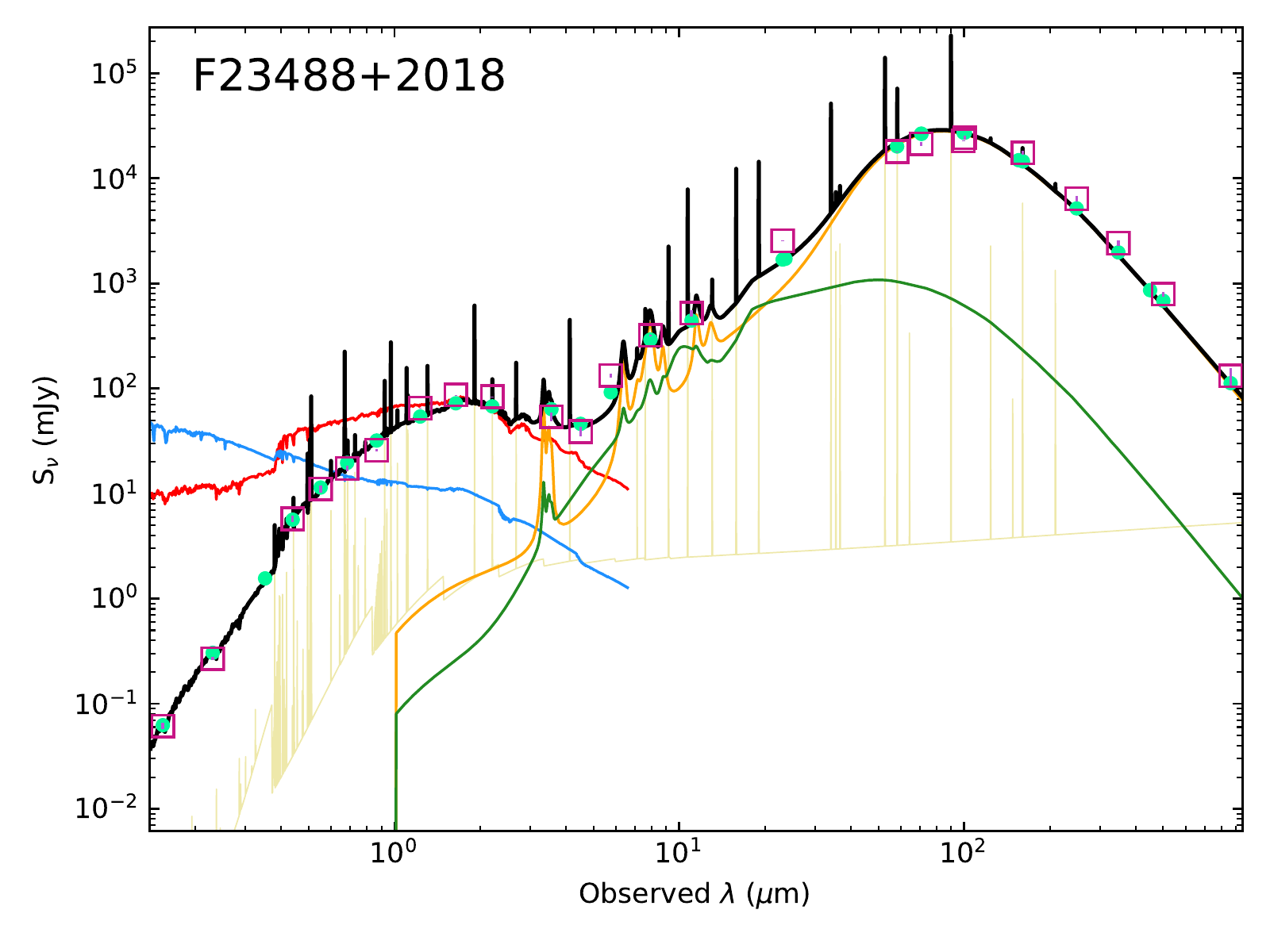}
\label{4}
\end{figure}

\begin{figure}\vspace{-12.7cm}\hspace{9.35cm}
\includegraphics[height=0.68\columnwidth]{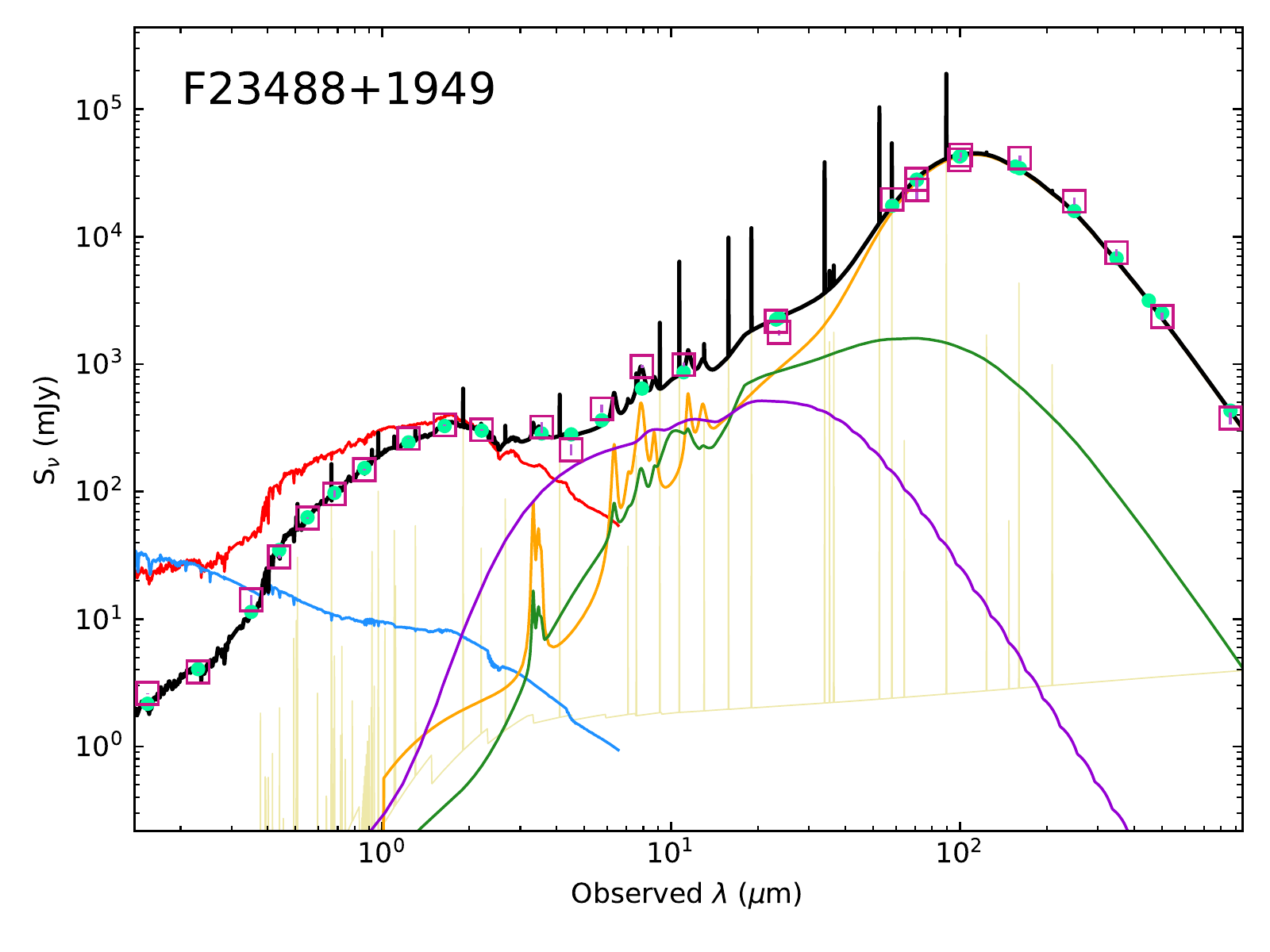}
\\\\\\\\\\\\\\\\\\\\\\\\\\\\\\\\
\caption{Best-fit SED models for the 67 galaxies in the current sample containing the unattenuated emission by the old (red) and the young (blue) stellar populations, the nebular line emission (dark-yellow), the diffuse dust (orange), the emission from the PDR regions (green), as well as the AGN emission (purple) are also presented. The best-fit SED is indicated as a black curve while the observations, for each galaxy, along with their uncertainties, are indicated by violet open squares. Light-green dots stand for the best-model flux densities.
\label{all_SEDs}}
\end{figure}

\clearpage

\section{Comparison with other studies}\label{appC:comparison}

   The (U)LIRGs sample under consideration has already been studied by others and some of the physical parameters discussed in this paper have been computed using either a similar approach or a totally different methodology. In what follows we will compare the parameters provided in the literature with what is computed in the current work and try to explain any differences.

\begin{figure}[t!]
   \centering
   \includegraphics[width=0.50\textwidth]{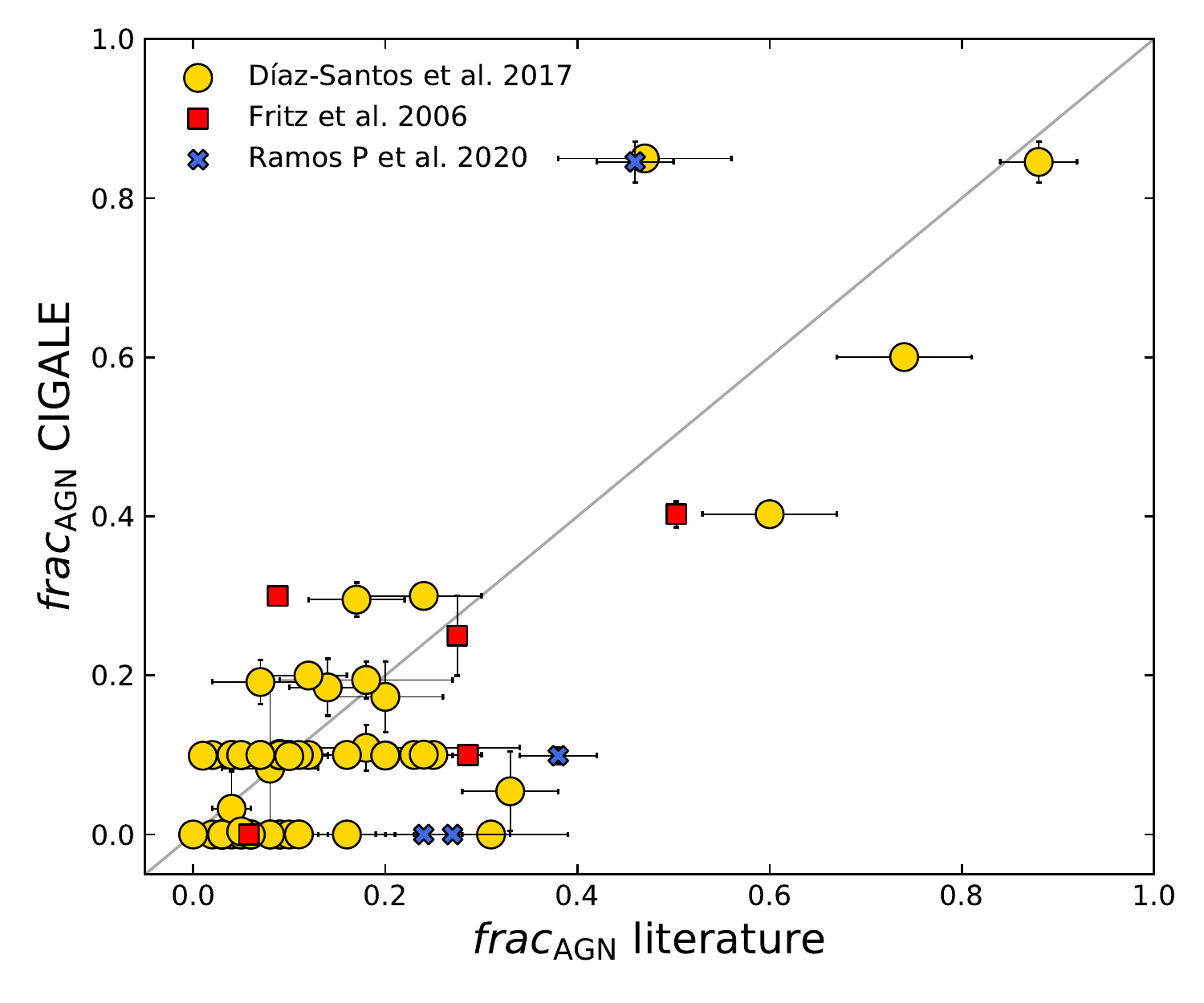}
   \caption{CIGALE-derived AGN fractions of this study, compared to the corresponding fractions calculated in previous works. Yellow circles depict the comparison with AGN fractions by \citet{2017ApJ...846...32D}, red squares to the ones by \citet{2006MNRAS.366..767F}, and filled blue "X"s to the ones by \citet{2020MNRAS.tmp.2637R}. All the data come with their error-bars (black), while the grey solid line stands for the one-to-one relation.
   \label{comp_fagn}}
   \end{figure}

   The AGN is inevitably a very important contributor to the energetics of the galaxies which can shape a large part of their SED (at least for the galaxies that host strong AGNs). The contribution of this component can be parametrized by the fraction of its emitted luminosity to the bolometric luminosity emitted. This fraction ($frac_{\rm{AGN}}$) can be estimated with various methods exploiting the parameter space of observables like X-rays, emission lines at MIR, or MIR colors, but also through SED modelling like the one we are considering in this study. Due to its multi-component nature the SED modelling is not always a robust way to estimate $frac_{\rm{AGN}}$. In \citet{2015A&A...576A..10C} it is found that only strong AGNs (with $frac_{\rm{AGN}}$$>$$0.5$) can be well retrieved. On the other hand, other  methods, like, e.g., MIR emission line ratios, the 6.2 $\mu$m PAH EW, the S$_{30}$/S$_{15}$ dust continuum slope as well as MIR diagnostic diagrams provide a more robust indication of the strength of the AGN. Such a study is presented in \citet{2017ApJ...846...32D} where the average fractional luminosity contribution of the AGN to the bolometric luminosity of the galaxies in the GOALS survey, based on the above methods, is provided, using the Kaplan-Meier (KM) maximum likelihood estimator. In Fig.~\ref{comp_fagn} we compare the values of $frac_{\rm{AGN}}$ derived in the current study with CIGALE with those calculated in \citet{2017ApJ...846...32D} for the galaxies in common (yellow points). We see that, despite the large scatter of the measurements, especially in the low-$frac_{\rm{AGN}}$ end there is an overall agreement with stronger AGNs showing higher $frac_{\rm{AGN}}$ with both methods. Some of the scatter seen in this plot arises from the fact that the values derived with CIGALE come from a parameter grid where discrete values have been pre-selected. In \citet{2017ApJ...846...32D}, on the other hand, a continuous range for $frac_{\rm{AGN}}$ is available. Apart from \citet{2017ApJ...846...32D} we indicate the resulting parameters (for the common galaxies) from two more studies which use an approach similar to what we use in our study. These studies are \citet{2006MNRAS.366..767F} (red boxes), and \citet{2020MNRAS.tmp.2637R} (blue "X"s). In \citet{2006MNRAS.366..767F} $frac_{\rm{AGN}}$ is calculated (in the range 5-1000 ${\mu}$m) by introducing a smoothly distributed, toroidal-like, dusty structure around the galaxy's nucleus, heated by a central source. In \citet{2020MNRAS.tmp.2637R} an approach similar to one presented in our work, is used, with CIGALE SED modelling performed using a different parameter grid, with the most obvious differences being the use of the \citet{2006MNRAS.366..767F} AGN module and the \citet{2014ApJ...784...83D} dust model. We see that, despite the small number of the galaxies in common, the findings broadly agree between the two methods (\citealt{2006MNRAS.366..767F} and \citealt{2020MNRAS.tmp.2637R}).        

\begin{figure*}[t!]
   \centering
   \includegraphics[width=\textwidth]{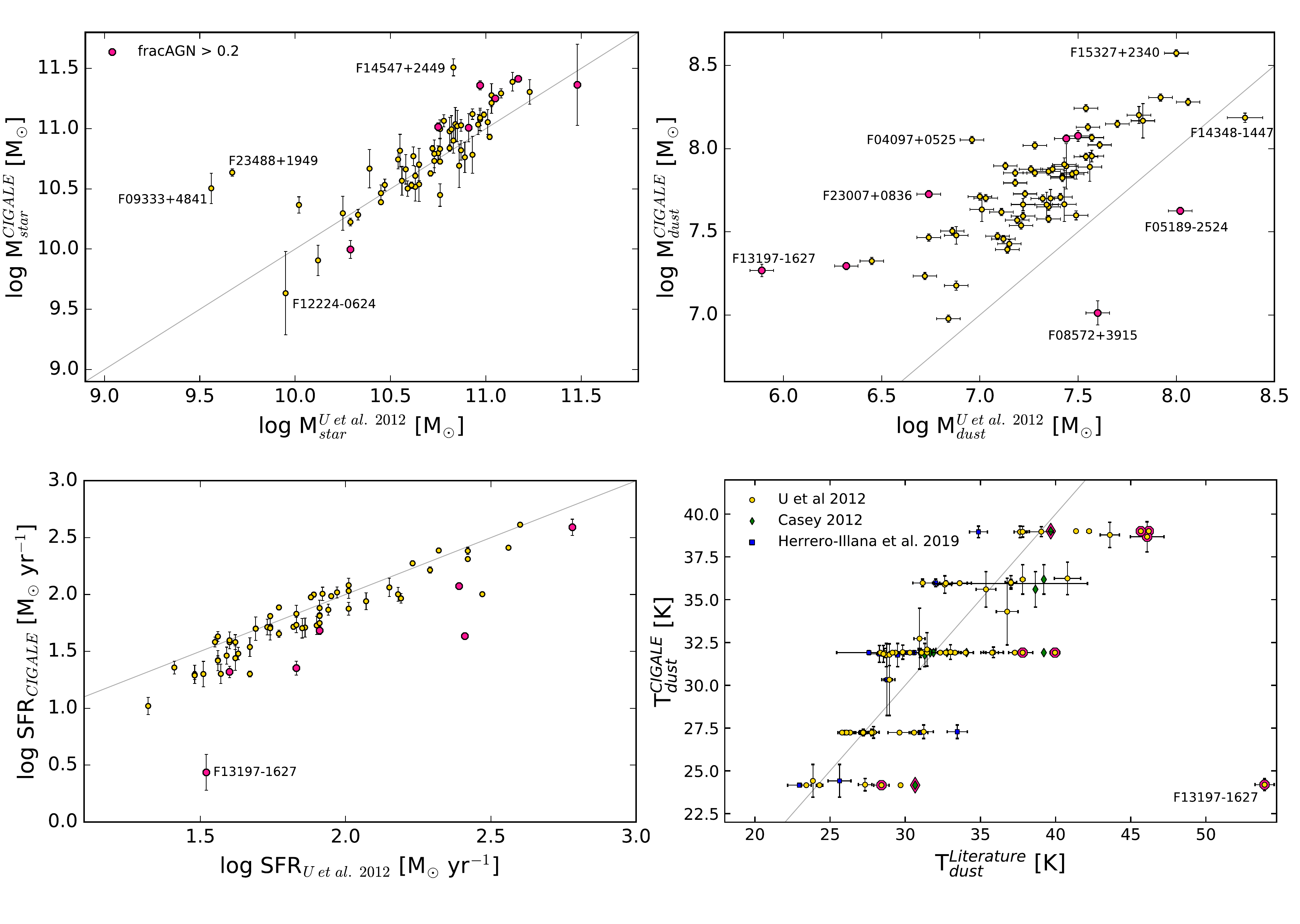}
   \caption{ Comparison between the CIGALE derived properties (this work), $M_\text{star}$ (top-left panel), $M_\text{dust}$ (top-right panel), \textit{SFR} (bottom-left panel), and $T_\text{dust}$ (bottom-right panel) and the corresponding properties presented in \citet{2012ApJS..203....9U}. Pink octagons indicate sources with strong AGN ($frac_\text{AGN} > 0.2$). Wherever available, values come with their uncertainties. The grey solid line corresponds to the one-to-one relation. Extreme outliers are indicated with their IRAS names. In the bottom-right panel, green diamonds correspond to the comparison with the dust temperatures derived in \citet{2012MNRAS.425.3094C}, while blue boxes correspond to the comparison with the ones derived in \citet{2019A&A...628A..71H}. In this panel, sources with strong AGN are indicated by pink edge-colour.
   \label{comp_sfr}}
   \end{figure*}

   In Fig.~\ref{comp_sfr} we present the comparison of four basic parameters for the galaxies (namely, $M_\text{star}$, $M_\text{dust}$, \textit{SFR}, and $T_\text{dust}$) between the values derived in \citet{2012ApJS..203....9U} and the current work (with additional information from \citealt{2012MNRAS.425.3094C} and \citealt{2019A&A...628A..71H} for the case of $T_\text{dust}$).

%%%%%%%%%%%%%%%%%%%%%%%%%%%%%%%%%%%%% FIGURE C.1.

%%%%%%%%%%%%%%%%%%%%%%%%%%%%%%%%%%%%% FIGURE C.2

   In the top-left panel of Fig.~\ref{comp_sfr} we compare the stellar masses between the two studies. In \citet{2012ApJS..203....9U} two different IMFs have been considered \citet{1955ApJ...121..161S}, and \citet{2003PASP..115..763C}. We compare with the Salpeter IMF since this is the one considered in the current study. The stellar masses derived in \citet{2012ApJS..203....9U} are computed with two methods, by performing an optical-NIR SED fitting, adopting the \citet{2003MNRAS.344.1000B} stellar population synthesis model, or by scaling from the H-band luminosity. Since the SED modelling method adopted in \citet{2012ApJS..203....9U} differs from the one used in the current work, we choose to compare our results with the scaling, from the H-band luminosity, method. The H-band, being largely unaffected by dust attenuation but also from contamination by hot dust emission from AGNs \citep{2011ApJ...740...96H} is a reliable tracer of the stellar mass. From Fig.~\ref{comp_sfr} we see an overall good agreement between the two studies (we find an average difference of 0.096 dex), with only a few exceptions deviating substantially from the one-to-one relation (the difference becomes 0.077 dex when the outliers are not considered). For F09333+4841, the JHK band fluxes reported in \citet{2012ApJS..203....9U} appear to be an order of a magnitude lower when compared to the optical and MIR fluxes. This makes the stellar mass calculated in \citet{2012ApJS..203....9U} underestimated. In the relevant SED presented in Fig.~\ref{all_SEDs} we have rescaled the JHK fluxes by an order of a magnitude higher (although not taken into account in the fit). F14547+2449, another outlier on this relation, appears as VV340a in \citet{2012ApJS..203....9U} indicating that only one member of this double system is treated (although not explicitly explained in the text). This could explain the lower stellar mass found in that study. Furthermore, the SED in their Fig.~2 appears twice, which makes it difficult to judge if their fit was successful, or not. Finally, concerning F23488+2018, appearing as MRK0331 in \citet{2012ApJS..203....9U}, we believe that their fitted SED (see their Fig.~2) might have been overestimating the FUV fluxes resulting in lower stellar mass. F12224-0624, on the other hand, although the stellar masses are marginally consistent, within the errors, is also deviating from the one-to-one relation. Since this is a galaxy that accounts for the highest dust-to-stellar mass ratio (0.012) we believe that the differences are due to the significant effects of the dust on the stellar mass computation. In this plot we also indicate the galaxies with the largest AGN fractions ($frac_{\rm{AGN}}$$>$$0.2$; pink octagons). It is evident that the majority of the stronger AGNs in our sample are hosted in galaxies on the high-end of the stellar masses.   

   In the top-right panel of Fig.~\ref{comp_sfr} we present the comparison of the dust masses derived in \citet{2012ApJS..203....9U} and in our work. We find that, apart from some outliers, which are going to be discussed later, the derived masses between the two methods agree with a systematic offset of 0.46 dex (0.45 when the outliers are not considered). This can be explained with  the use of the different dust absorption coefficients ($\kappa{_{850}}$$=$$0.15~\rm{m}^2~\rm{kg}^{-1}$ and $\kappa{_{250}}$$=$$17.3~\rm{cm}^2~\rm{g}^{-1}$; assuming an emissivity index of $\beta$$=$$2$) in the case of \citet{2012ApJS..203....9U} compared to $\kappa{_{250}}$$=$$6.4~\rm{cm}^2~\rm{g}^{-1}$ dictated by the THEMIS model (used in the current study) which translates to a 0.43 dex difference. Concerning the obvious outliers in this plot we see that the majority of them host a luminous AGN (with $frac_{\rm{AGN}}$$>$$0.2$; pink octagons) making the estimation of the dust mass quite uncertain if no AGN component is considered in the modelling. In the case of F04097+0525, we believe that in \citet{2012ApJS..203....9U} the dust is underestimated due to the fact that the Reyleigh-Jeans part of the SED is not well fitted (only constrained by the $850~\mu$m observation) while in the current work, the Herschel data fill this gap and a more accurately determination of the dust mass is achieved. A similar case is F14348-1444 with the dust emission being constrained only by the IRAS 60 and $100~\mu$m observation in the case of \citet{2012ApJS..203....9U} compared to our analysis where Herschel observations are also available.   

   In the bottom-left panel of Fig.~\ref{comp_sfr} we present the comparison of the \textit{SFR} derived in \citet{2012ApJS..203....9U} and in our work. In \citet{2012ApJS..203....9U} the \textit{SFR} is derived by combining the monochromatic UV luminosity at 2800$\AA$ and IR luminosity, using the \citet{2011ApJ...738..106W} recipe. Since the IMFs used in the two methods are different, the values derived with the Chabrier IMF were divided by a constant scaling factor of 0.63 \citep{2014ARA&A..52..415M}. We see that there are some obvious outliers (with F13197-1627 being the most extreme case) though all are amongst the strongest AGNs in our sample with $frac_{\rm{AGN}} > 0.2$ (pink octagons). Since the IR luminosity may be largely affected by the presence of the AGN (see, e.g., the SED of F13197-1627) it is expected that the \textit{SFR} is overestimated when the IR luminosity is used as an \textit{SFR} tracer. The difference between the two methods is 0.12 dex (with 0.08 dex when the outliers are not considered). 

   In the bottom-right panel of Fig.~\ref{comp_sfr} the comparison among the values of $T_{\text{dust}}$ derived in this work and the studies of \citet{2012ApJS..203....9U}, \citet{2012MNRAS.425.3094C}, and \citet{2019A&A...628A..71H} (yellow circles, green diamonds, and blue boxes respectively) is presented. The strongest AGNs in our sample (those with $frac_{\rm{AGN}} > 0.2$) are indicated in pink while the line is the one-to-one relation. In all three literature studies mentioned above, the dust temperature was estimated by fitting a single temperature modified black-body to observations landwards of $\sim 70 \mu$m. One thing to notice is that the resulting values of our study come in discrete ranges in dust temperature. This is due to the discrete nature of the parameter space used by CIGALE and in particular for the parameters that define $U_\text{min}$ (see Eq.~\ref{eq:aniano}). Even if the scatter is large it is evident that the values derived in this work follow the general trend observed in other studies. Indicative of the scatter is the average difference and the standard deviation of the differences between our values and the literature values which is $-0.96\pm4.3$ K. As can be seen, though, from Fig.~\ref{comp_sfr}, there are several outliers, the majority of them being AGNs (see the data points covered with pink colour). If these outliers are omitted from the statistics the average difference and the standard deviation of the differences in dust temperature then drops to $0.09\pm2.5$ K. The most extreme example is F13197-1627 which shows the largest deviation. This is the strongest AGN in our sample ($frac_{\rm{AGN}}$=0.85) with the AGN component occupying a large part of the FIR emission (see Fig.~\ref{all_SEDs}) resulting in misleading results if a pure dust emission model is only fitted.

\section{Cumulative distributions}\label{cumulative}

   Examining the cumulative distributions of the various physical parameters is a powerful tool that may indicate if two population of objects can originate from the same parent population or not. The Kolmogorov-Smirnov (\cite{smirnov1948}) test is a well known non-parametric statistical method that compares distributions by measuring a ``difference'' between the distributions and report a p-value which shows a statistical significance of the result. Non-zero p-value of less than 0.15 means that the null hypothesis that the distributions come from the same parent distribution can be rejected with 85\% probability (see, e.g., \citet{2011AJ....141..100H}).

\begin{figure}[t!]
   \centering
   \includegraphics[width=0.50\textwidth]{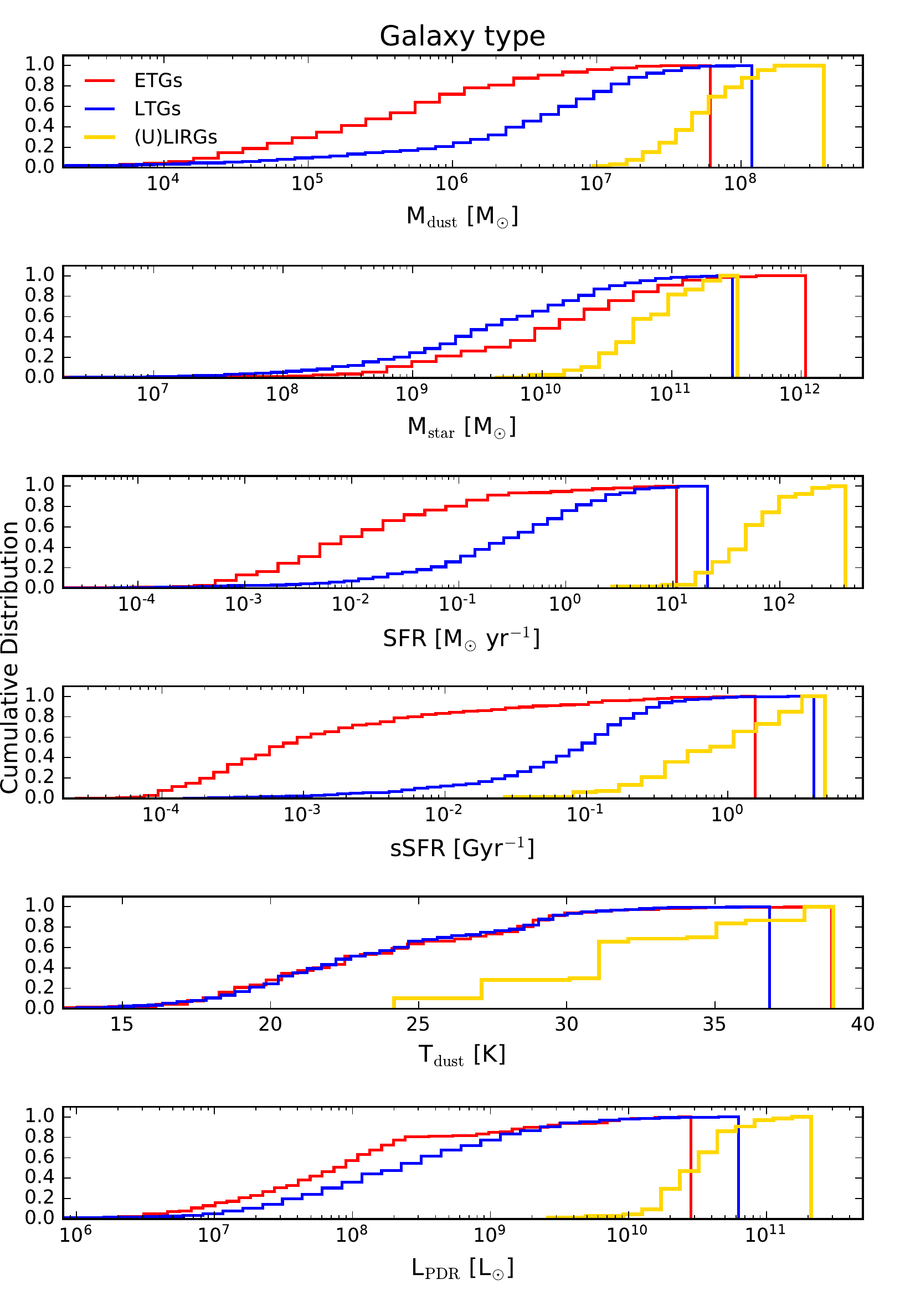}
   \caption{Cumulative distributions of the physical parameters discussed in this study, for each galaxy type. The colouring is identical to Fig.~\ref{fig:type-props}.
   \label{fig:cum_types}}
   \end{figure}

   In Fig.~\ref{fig:cum_types} we present the cumulative distributions of the physical parameters presented in Sect.~\ref{place} (see also Fig.~\ref{fig:type-props}) for the three different populations of ETGs, LTGs, and (U)LIRGs (red, blue, and yellow color respectively). As can be seen from the plots the relevant distributions are very different among the three galaxy populations with p-values less than 0.15. The only exception being the comparison of the temperature distributions between ETGs and LTGs which give a p-value of 0.82. A relevant discussion on the results of the KS tests is presented in Sect.~\ref{place}.

\begin{figure}[t!]
   \centering
   \includegraphics[width=0.50\textwidth]{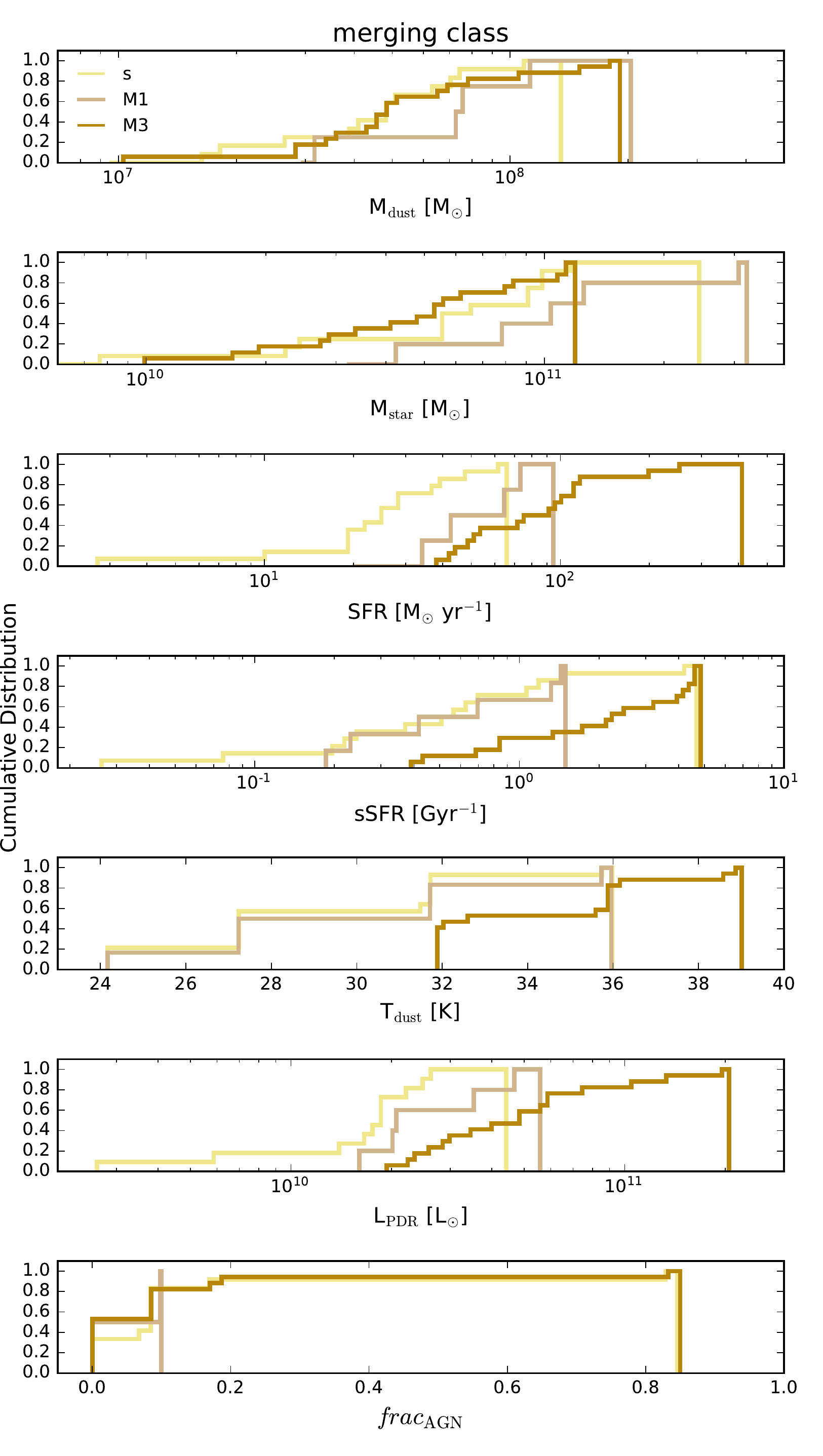}
   \caption{Cumulative distributions of the physical parameters discussed in this study, for galaxies in merger stages, `s', `M1' and `M3'. The colouring is identical to Fig.~\ref{sfr_per_class}.
   \label{fig:cum_merger}}
   \end{figure}

   In Fig.~\ref{fig:cum_merger} the cumulative distributions of the physical parameters presented in Sect.~\ref{sec:merg-evol} (see also Figs.~\ref{sfr_per_class}, \ref{fagn_per_class}, \ref{fig:clas-props}) for the different merging stages. The p-values drawn from the cumulative distributions of all the combinations of merging stages for each physical parameter are given in Table~\ref{tab:KStest}. A relevant discussion on the results of the KS tests is presented in Sect.~\ref{sec:merg-evol}.

%%%%%%%%%%%%%%%%%%%%%%%%%%%%%%%%%%%%% FIGURE D.1
   
%%%%%%%%%%%%%%%%%%%%%%%%%%%%%%%%%%%%% FIGURE D.2
   
%%%%%%%%%%%%%%%%%%%%%%%%%%%%%%%%%%%%% TABLE D.1
   \begin{table}
   \caption{Kolmogorov-Smirnov (KS) test p-values of the physical properties ($M_\text{dust}$, $M_\text{star}$, \textit{SFR}, \textit{sSFR}, $T_\text{dust}$, $L_\text{PDR}$ and $frac_\text{AGN}$) for all the merging stage combinations. All p-values $\leq$ 0.15 are underlined.}
   \label{tab:KStest}
   \tiny
   \begin{center}
   \begin{tabular}{l||c|c|c|c|c|c|c}
\hline
   \multicolumn{1}{c||}{} &
   \multicolumn{1}{c|}{$M_\text{dust}$} &
   \multicolumn{1}{c|}{$M_\text{star}$} &
   \multicolumn{1}{c|}{\textit{SFR}} &
   \multicolumn{1}{c|}{\textit{sSFR}} &
   \multicolumn{1}{c|}{$T_\text{dust}$} &
   \multicolumn{1}{c|}{$L_\text{PDR}$} &
   \multicolumn{1}{c}{$frac_\text{AGN}$} \\
\hline
\hline
   s-m & 0.97 & 0.92 & \underline{0.07} & 0.75 & 0.54 & \underline{0.05} & 0.92\\
   s-M1 & 0.27 & 0.56 & \underline{0.11} & 0.99 & 0.89 & \underline{0.11} & 0.35\\
   s-M2 & 0.64 & 0.16 & \underline{0.02} & 0.92 & 0.16 & \underline{0.01} & 0.64\\
   s-M3 & 0.73 & 0.39 & \underline{0.00} & \underline{0.01} & \underline{0.00} & \underline{0.00} & 0.26\\
   s-M4 & 0.36 & 0.85 & \underline{0.00} & \underline{0.01} & \underline{0.01} & \underline{0.00} & 0.75\\
   s-M5 & 0.35 & 0.70 & 0.70 & 0.97 & \underline{0.05} & 0.70 & 0.70\\
   m-M1 & 0.70 & 1.00 & 0.70 & 0.92 & 0.55 & 0.70 & 0.55\\
   m-M2 & 0.92 & 0.54 & 0.41 & 0.54 & 0.92 & 0.75 & 0.92\\
   m-M3 & 0.80 & 0.41 & \underline{0.14} & \underline{0.11} & 0.30 & 0.23 & 0.23\\
   m-M4 & 0.69 & 0.76 & \underline{0.08} & \underline{0.13} & 0.92 & 0.19 & 0.41\\
   m\_M5 & 0.93 & 1.00 & 0.93 & 1.00 & 0.93 & 0.93 & 0.93\\
   M1-M2 & 0.92 & 0.66 & 0.81 & 0.81 & 0.71 & 0.56 & 0.35\\
   M1-M3 & 0.29 & 0.45 & 0.22 & \underline{0.03} & \underline{0.02} & 0.25 & 0.36\\
   M1-M4 & 0.96 & 0.31 & \underline{0.09} & \underline{0.03} & 0.51 & 0.25 & 0.51\\
   M1-M5 & 0.43 & 0.86 & 0.86 & 1.00 & 0.43 & 0.86 & 0.86\\
   M2-M3 & 0.39 & \underline{0.03} & 0.21 & \underline{0.02} & \underline{0.03} & \underline{0.07} & 0.26\\
   M2-M4 & 0.98 & \underline{0.11} & \underline{0.11} & \underline{0.01} & 0.64 & \underline{0.03} & 0.47\\
   M2-M5 & 0.25 & 0.47 & 0.87 & 0.70 & 0.17 & 0.87 & 0.87\\
   M3-M4 & 0.49 & 0.93 & 0.84 & 0.87 & 0.51 & 0.84 & 0.95\\
   M3-M5 & 0.25 & 0.33 & 0.63 & 0.25 & 0.79 & 0.63 & 0.53\\
   M4-M5 & 0.30 & 0.46 & 0.76 & 0.18 & 0.76 & 0.76 & 0.76\\
\hline
   \end{tabular}
   \end{center}
   \end{table}

\end{document}